\documentclass[10pt]{article}
\usepackage{caption}
\usepackage[T1]{fontenc}
\usepackage[margin=1.8cm]{geometry}
\usepackage{tikz-cd}
\usepackage{tikz}
\usetikzlibrary{calc,shapes.geometric,shadows,backgrounds, positioning}
\usepackage{booktabs}
\usepackage{mathtools}
\usepackage{color}
\definecolor{figInk}{HTML}{243448}
\definecolor{figMuted}{HTML}{627183}
\definecolor{figLine}{HTML}{D8DFE6}
\definecolor{dataTeal}{HTML}{258C92}
\definecolor{clockGold}{HTML}{B98B2E}
\definecolor{auxViolet}{HTML}{7662AD}
\definecolor{loadRose}{HTML}{BE617F}
\tikzset{
  mode/.style={circle, minimum size=8.5mm, inner sep=0pt,
    line width=.7pt, font=\normalsize, text=figInk},
  data mode/.style={mode, draw=dataTeal!85, fill=dataTeal!11},
  clock mode/.style={mode, draw=clockGold!90, fill=clockGold!14},
  auxiliary mode/.style={mode, draw=auxViolet!90, fill=auxViolet!13,
    circular glow={fill=auxViolet!17, shadow scale=1.24}},
  loading mode/.style={mode, draw=loadRose!90, fill=loadRose!13,
    circular glow={fill=loadRose!17, shadow scale=1.24}},
  panel/.style={draw=figLine, fill=white, rounded corners=5pt,
    line width=.6pt},
  register/.style={draw=figMuted!60, densely dotted,
    line width=.65pt, fill=none},
  loading link/.style={draw=loadRose!85, line width=.85pt,
    dash pattern=on 2.5pt off 2.2pt},
  computation link/.style={draw=auxViolet!90, line width=1.15pt},
  group title/.style={font=\sffamily\bfseries\small, text=figInk,
    align=center, inner sep=0pt},
  group count/.style={font=\sffamily\footnotesize, text=figMuted,
    align=center, inner sep=0pt},
  small label/.style={font=\sffamily\scriptsize, text=figMuted,
    inner sep=0pt},
}

\usepackage{booktabs}
\usepackage{makecell}

\usepackage{caption}
\usepackage{bbm}
\usepackage{amssymb}
\usepackage{microtype}
\usepackage{xcolor}
\usepackage{enumitem}
\usepackage{amsthm}
\usepackage{physics}
\usepackage{proof}
\usepackage{thmtools,thm-restate}
\usepackage{graphicx}
\usepackage{algorithmicx}
\usepackage{algpseudocode}
\usepackage{qcircuit}
\usepackage{braket}
\usepackage{float}
\usepackage{subcaption}
\usepackage{multirow}
\usepackage{nicefrac}

\usepackage{lmodern}

\usepackage{mathrsfs}

\usepackage{soul}
\usepackage{comment}

\usepackage{amsmath}
\usepackage{amsfonts}
\usepackage{url}
\DeclareMathAlphabet{\mathpzc}{OT1}{pzc}{m}{it}
\definecolor{quantumviolet}{HTML}{53257F}
\usepackage[
    colorlinks=true,
    linkcolor=quantumviolet,
    citecolor=quantumviolet,
    urlcolor=quantumviolet,
    linktoc=page
]{hyperref}
\usepackage[capitalize, noabbrev]{cleveref}

\usepackage{amsmath,amssymb}
\usepackage{lmodern}
\usetikzlibrary{arrows.meta,calc,patterns.meta}

\definecolor{axisgray}{HTML}{263238}
\definecolor{easyblue}{HTML}{247BA0}
\definecolor{easyfill}{HTML}{DDF3F7}
\definecolor{fixedgold}{HTML}{D79B16}
\definecolor{fixedfill}{HTML}{FFF0BF}
\definecolor{hardred}{HTML}{C83E4D}
\definecolor{hardpurple}{HTML}{7251B5}
\definecolor{boundary}{HTML}{566573}

\numberwithin{equation}{section}
\newtheorem{theorem}{Theorem}[section]
\newtheorem{lemma}[theorem]{Lemma}
\newtheorem{definition}[theorem]{Definition}
\newtheorem{proposition}[theorem]{Proposition}
\newtheorem{corollary}[theorem]{Corollary}
\newtheorem{example}[theorem]{Example}

\newtheorem{remark}[theorem]{Remark}

\newcounter{prob}
\newtheorem{problem}[prob]{Problem}

\newcommand{\eq}[1]{\hyperref[eq:#1]{(\ref*{eq:#1})}}
\renewcommand{\sec}[1]{\hyperref[sec:#1]{Section~\ref*{sec:#1}}}
\newcommand{\thm}[1]{\hyperref[thm:#1]{Theorem~\ref*{thm:#1}}}
\newcommand{\lem}[1]{\hyperref[lem:#1]{\cref*{lem:#1}}}
\newcommand{\cor}[1]{\hyperref[cor:#1]{\cref*{cor:#1}}}
\newcommand{\itm}[1]{\hyperref[itm:#1]{\ref*{itm:#1}}}
\newcommand{\app}[1]{\hyperref[app:#1]{Appendix~\ref*{app:#1}}}
\newcommand{\dfn}[1]{\hyperref[dfn:#1]{Definition~\ref*{dfn:#1}}}
\newcommand{\fig}[1]{\hyperref[fig:#1]{Figure~\ref*{fig:#1}}}
\newcommand{\clm}[1]{\hyperref[clm:#1]{Claim~\ref*{clm:#1}}}
\newcommand{\alg}[1]{\hyperref[alg:#1]{Algorithm~\ref*{alg:#1}}}
\newcommand{\stp}[1]{\hyperref[stp:#1]{Step~\ref*{stp:#1}}}
\newcommand{\asm}[1]{\hyperref[asm:#1]{Assumption~\ref*{asm:#1}}}
\newcommand{\prot}[1]{\hyperref[prot:#1]{Protocol~\ref*{prot:#1}}}
\newcommand{\prob}[1]{\hyperref[prob:#1]{Problem~\ref*{prob:#1}}}
\newcommand{\rmk}[1]{\hyperref[rmk:#1]{Remark~\ref*{rmk:#1}}}
\newcommand{\cons}[1]{\hyperref[cons:#1]{Construction~\ref*{cons:#1}}}
\newcommand{\conj}[1]{\hyperref[conj:#1]{Conjecture~\ref*{conj:#1}}}
\newcommand{\tbl}[1]{\hyperref[tbl:#1]{Table~\ref*{tbl:#1}}}
\usepackage{tabularx}

\newcommand{\proj}[1]{\vert #1\rangle\!\langle #1\vert}

\let\originalleft\left
\let\originalright\right
\renewcommand{\left}{\mathopen{}\mathclose\bgroup\originalleft}
\renewcommand{\right}{\aftergroup\egroup\originalright}

\DeclareMathOperator{\poly}{poly}

\newcommand{\class}[1]{\mathsf{#1}}
\renewcommand{\P}[0]{\class{P}}

\newcommand{\BPP}[0]{\class{BPP}}

\newcommand{\BQP}[0]{\class{BQP}}
\newcommand{\QMA}[0]{\class{QMA}}

\newcommand{\NP}[0]{\class{NP}}

\newcommand{\PH}[0]{\class{PH}}

\newcommand{\PostBQP}{\mathsf{PostBQP}}
\newcommand{\PostBPP}{\mathsf{PostBPP}}
\newcommand{\PP}{\mathsf{PP}}
\newcommand{\PostLO}{\mathsf{PostLO}}

\renewcommand{\Re}{\mathop{\mathrm{Re}}}

 \newcommand{\nai}[1]{} 
\usepackage{authblk}

\usepackage{qcircuit}

\providecolor{FPTableSamplingRed}{RGB}{190,45,45}
\providecolor{FPTableDecisionBlue}{RGB}{35,95,180}
\providecolor{FPTableREOrange}{RGB}{195,105,20}

\definecolor{AIAuditBlue}{HTML}{075C80}
\newif\ifshowAIcomments
\showAIcommentstrue

\newcounter{AIunresolved}
\providecommand{\AIunresolvedLastRun}{?}

\makeatletter
\AtEndDocument{%
  \immediate\write\@auxout{%
    \string\gdef\string\AIunresolvedLastRun
    {\number\value{AIunresolved}}%
  }%
}
\makeatother

\begin{document}

\title{The computational complexity of bosonic Lindbladian fixed points}

\author[1]{En-Jui Kuo}
\affil[1]{\footnotesize Department of Electrophysics, National Yang Ming Chiao Tung University, Hsinchu, Taiwan, R.O.C.}

\author[2]{Arsalan Motamedi}
\affil[2]{\footnotesize University of Toronto, Ontario, Canada}

 \author[3]{Mohammad Hafezi}
 \affil[3]{\footnotesize Joint Quantum Institute, University of Maryland, College Park, MD 20742, USA.}

 \author[4]{Saeed Mehraban}
 \affil[4]{\footnotesize Tufts University, Medford, MA, USA}

\date{}
\maketitle

\begin{abstract}

We study the computational complexity of estimating
observables and sampling from stationary states, or
fixed points, of bosonic Lindbladian dynamics with  bounded-degree and  local interactions.
Previous work has investigated the computational
complexity of transient dissipative bosonic dynamics
(e.g.,~\cite{yuan2025universal,gonzalez2025dynamical}),
while other works have established that universal
quantum computation can be encoded into the fixed
points of Lindbladians acting on qubits
\cite{verstraete2009quantum}.
Extending such complexity results to bosonic fixed
points requires addressing subtleties associated
with the long-time behavior of infinite-dimensional
systems, particularly the existence, uniqueness,
and convergence properties of stationary states.

In this work, we establish a broad range of complexity results
for bosonic Lindbladian fixed points, spanning
universal quantum computation, efficient classical simulation, sampling hardness,
 and undecidability.

\begin{itemize}

\item \textbf{Universal computation in bosonic fixed points.}
Our first result establishes that universal quantum
computation can be embedded into the problem of
estimating observables at the unique fixed point
of a bosonic Lindbladian with a Hamiltonian of
degree at most twelve and only linear photon loss.
Furthermore, we show that the fixed point can be
prepared from the global vacuum in polynomial time.
Our construction embeds a variant of the construction
in \cite{verstraete2009quantum} into dissipative
bosonic dynamics and uses an adiabatic elimination
technique to replace higher-degree jump operators
with Hamiltonian interactions and linear loss.
A central technical challenge is establishing that
the resulting infinite-dimensional Lindbladian
dynamics is well-defined and converges to the
desired unique fixed point.

\item \textbf{Classical simulation.}
We next investigate the complexity of fixed points
under more restricted Hamiltonian and dissipation
models. We identify Gaussian regimes with incoherent
pumping that admit efficient classical photon-number
sampling and show that classical approximability
of their fixed points persists under sufficiently
weak non-Gaussian Kerr perturbations.
A key technical contribution is
establishing stability of these stationary states despite
the challenges posed by the long-time behavior of
infinite-dimensional dynamics.

\item \textbf{Sampling hardness and non-Gaussian regimes.}
Conversely, in the absence of incoherent pumping,
we establish hardness of exact sampling from
Gaussian Lindbladian fixed points with uniform
linear loss at a constant rate, as well as
hardness of approximate sampling at
inverse-polynomial loss rates under standard
boson-sampling conjectures.
We further investigate non-Gaussian constructions
involving higher-degree Hamiltonians and
dissipative operators, relating the complexity
of sampling from their fixed points to
different variants of boson sampling.

\item \textbf{Undecidability of fixed-point existence.}
Finally, we show that deciding whether a bosonic
Lindbladian admits a stationary density operator
is $\mathsf{RE}$-hard, even for purely dissipative
dynamics with jump operators of degree at most
five, each acting on at most four modes.

\end{itemize}
\end{abstract}

\tableofcontents

\section{Introduction}

Quantum information processing devices based on bosonic degrees of freedom, such as quantum states of light, have gained much interest recently. These devices offer potentially scalable platforms operating at room temperature \cite{aghaee2025scaling}.

The standard model of quantum computation is based on the circuit model corresponding to the complexity class $\BQP$. One can in principle embed universal quantum computation in a bosonic computer \cite{knill2001scheme}, with proposed schemes for robust fault-tolerant quantum computation~\cite{gottesman2000encoding, matsuura2026continuous}. However, embedding quantum computation into bosonic
systems introduces additional challenges associated
with their infinite-dimensional Hilbert spaces, such as energy and complexity divergences~\cite{chabaud2026bosonic, chabaud2025energy, brenner2025factoring}. As a result quantum computation over bosonic modes require special care for energy management and other subtleties appearing in infinite-dimensional Hilbert spaces.

In this work, we study the computational complexity
of stationary states, or fixed points, of bosonic
systems undergoing continuous-time dissipative dynamics
governed by Lindblad master equations.
Our central question is how the computational complexity
of these fixed points depends on the structure of the
underlying Lindbladian, particularly when they can
encode universal quantum computation and when they
admit efficient classical simulation.

\paragraph{Quantum computing using dissipative systems:} A natural framework for studying continuous-time
dissipative quantum dynamics is the theory of
open quantum systems, which models interactions
between a quantum system and its environment.
Under a Markovian approximation, such dynamics
are described by the Lindblad master equation
(see~\cite{rivas2012open} for a review). In particular, for a quantum system with Hamiltonian $H$ the Lindblad master equation is according to 
\begin{equation}
\frac{d\rho}{dt} = - i [H,\rho] + \sum_j (L_j \rho L_j^\dagger - \frac{1}{2}\{L_j^\dagger L_j, \rho\})
\label{eq:basic-Lindbladian}    
\end{equation}
where $\rho$ is the density matrix of the system, and $L_j$, known as jump operators, model interaction with the surrounding environment. For instance, if the loss mechanism corresponds to a bosonic particle dissipating to the environment with rate $\gamma$, one can take $L_j = \sqrt{\gamma} \hat a_j$, where $\hat a_j$ is the bosonic annihilation operator acting on bosonic mode $j$, mapping $\hat a_j \ket {n}_j = \sqrt{n}\ket{n-1}_j$, where $\ket{n}_j$ is the state of having $n$ bosonic particles in mode $j$. One can also consider incoherent pumping of bosonic particles to the system with rate $\eta$ modeled by jump operators $L_j = \sqrt{\eta} \hat a^\dagger_j$, where $\hat a^\dagger_j$ is the creation operator mapping $\hat a^\dagger_j \ket{n} = \sqrt{n+1} \ket{n+1}$. 

We often simplify the master equation using the notation $\frac{d\rho}{dt} = \mathcal{L} [\rho]$ which implies
$$
\rho (t) = e^{t\mathcal{L}} [\rho(0)].
$$

\paragraph{Transient v.s. stationary states:} Intuitively, one can consider two regimes for the evolution of a quantum system undergoing dissipative dynamics: (1) transient states where the state of the system evolves, and (2) stationary regime where the quantum state does not evolve under Lindbladian dynamics. Mathematically, stationary distribution is defined as the fixed point of $e^{t \mathcal{L}}$, i.e., $\mathcal{L}[\rho] = 0$. We hence use the terms ``stationary state'' and ``fixed point'' interchangeably. For infinite-dimensional systems, even for well-defined $\mathcal L$, $e^{t \mathcal{L}}$ may not be well-defined on the entire Hilbert space and one need to specify domains within the Hilbert space where this evolution is well-defined. That is one of the main complications in studying bosonic Lindbladian systems. 

A finite-dimensional  Lindbladian system may have one or many fixed points. For bosonic systems it may not have a fixed point (e.g. when pumping rate exceeds that of loss). The fixed point may be unique or correspond to a higher-dimensional subspace corresponding to the kernel of $\mathcal{L}$. Starting from an initial quantum state, the state of the system may converge to a specific fixed point within the kernel subspace. The time required for such convergence is called the mixing time of the system. 

If the Lindbladian satisfies a condition called detailed balance with respect to a Hamiltonian, the corresponding fixed point corresponds to equilibrium Gibbs states corresponding to that Hamiltonian. While the computational complexity of equilibrium
states, particularly quantum Gibbs states, has
received considerable attention
(e.g.,~\cite{chen2023efficient,bergamaschi2024quantum, rajakumar2026gibbs}), in this work
we mainly investigate  bosonic fixed points without imposing structures such as
detailed balance. The fixed point for such Lindbladians are sometimes called out-of-equilibrium fixed points. However,
our constructions assume a bounded degree structure: Hamiltonians
and jump operators are constant degree polynomials in canonical bosonic operators, with individual
interactions often acting on only a constant number of modes. We place particular
emphasis on keeping the dissipative terms as simple
and low-degree as possible, notably considering
linear uniform photon loss. Notably Lindbladians satisfying detailed balance do not necessarily have local interactions.

It is natural to ask how much computational power one obtains if we define a model of quantum computation based on Lindbladian open quantum system dynamics. Importantly, for finite-dimensional systems, one can show that the power of such model is equivalent to that of universal quantum computation. In particular, \cite{verstraete2009quantum} showed that we can encode universal quantum computation in the fixed point of a qubit Lindbladian system using a Lindbladian analog of circuit to Hamiltonian constructions. \cite{kliesch2011dissipative} furthermore proved that local Lindbladian systems can be simulated on quantum computers. We emphasize for bosonic systems extra care needs to be taken into account due to infinite-dimensional Hilbert spaces, because as we indicated sometimes a bosonic system may not have a fixed point, or even when the fixed point is unique, fast convergence to the fixed point may not be guaranteed even for single mode systems. In this work, we study complexity and mixing times alongside each other as separate attributes of a fixed point. 

\paragraph{The complexity of bosonic Lindbladian fixed points:}
Previous work has studied the complexity of transient dynamics for the Lindbladian systems. For instance, \cite{yuan2025universal} demonstrated universal control of quantum operations using transient dissipative Lindbladian dynamics via Hamiltonians with time dependent coefficients. \cite{gonzalez2025dynamical} specified regime of classical simulation for sampling from temporal dynamics for dissipative Lindbladian systems with on-site Kerr nonlinearity when incoherent pumping rate exceeds a certain amount. 

Many results have also studied the behavior of fixed points solutions (see, e.g., \cite{diehl2008quantum,
prosen2010quantization,
azouit2016well,barthel2022solving}). For instance, exact solution for single-mode driven dissipative dynamics undergoing Gaussian and Kerr interaction are known~\cite{drummond1980quantum} corresponding to single photon loss jump terms $\sqrt{\gamma}\hat a_j$. Symmetries and structure of fixed points for higher degree loss terms (such as double photon loss corresponding to jump terms $\sqrt{\gamma}\hat a_j^2$) have been furthermore studied in depth~\cite{albert2014symmetries}. Importantly, many of these previous works motivate the study of fixed points in terms of state preparation applications.

In this work, we mainly focus the computational complexity of fixed points of bosonic Lindbladian dynamics.
There are several reasons to perform this study. First, when the Lindbladian mixes rapidly to a unique fixed point, the fixed point is arguably a more experimentally natural object to access than a prescribed transient state, since its preparation does not require precise control of the evolution time beyond waiting for equilibration. Furthermore, the details of the initial state does not matter and one does need to worry about precise state preparation. This motivates asking whether computationally nontrivial states—and ultimately the output of a universal quantum computation—can arise as rapidly accessible fixed points of simple dissipative bosonic dynamics. Second, from a theoretical perspective, this setting appears to be relatively less explored, with much of the existing literature focusing instead on state preparation such as steady-state engineering~\cite{ikeda2013deterministic}, or analytically tractable models admitting exact solutions~\cite{prosen2008third, prosen2010quantization,
barthel2022solving, karevski2013exact}. Third, from a computer science perspective one can view the complexity of the fixed point for Lindbladians (i.e., sampling from a quantum state satisfying $\mathcal{L}[\rho] \approx 0$) as a dissipative analog of the local Hamiltonian problem which is known to be equivalent to the complexity class $\mathsf{QMA}$~\cite{kempe2006complexity}. As also pointed out in~\cite{firanko2024area} it is desirable to develop complexity theoretic foundations for a local Lindbladian problem studying the complexity of fixed points. We emphasize recent work of \cite{borras2026complexity} studies the complexity of detecting if the fixed point of a finite-dimensional Lindbladian is pure and prove equivalence with $\QMA$. The setting we study in this paper is about measuring the fixed point. Finally, this study will address foundational questions about the complexity of open quantum systems. For instance a natural question in this context is whether there exist nontrivial Lindbladian systems whose transient dynamics is hard to simulate classically, while their fixed points are classically tractable; in other words, understanding when the computational complexity of transient Lindbladian dynamics carries over to the fixed point, and when it does not.

\subsection{Overview of results}

We develop the following results about the computational complexity of bosonic Lindbladian fixed points (see \Cref{sec:problem-definition} for details).

\paragraph{Embedding universal quantum computation in the fixed point problem:}
It is desirable to embed universal quantum computation in the fixed point of a Hamiltonian system undergoing loss ideally using a Hamiltonian system with natural non-Gaussian interactions such as degree $4$ cross-Kerr (corresponding to terms such as $\hat n_j \hat n_k$ where $\hat n_j$ is the bosonic number operator on mode $j$) undergoing a basic loss model such as linear loss corresponding to jump terms $\sqrt{\gamma} \hat a_j$ (see problem 1 in \Cref{sec:problem-definition}). We make progress towards achieving this goal by embedding a universal quantum computation (corresponding to complexity class $\BQP$) in the unique fixed point of a family of Lindbladians with Hamiltonian of degree $12$ and nonuniform linear loss (meaning loss rates on different sites are different and the construction includes some lossless modes). The construction is by first slightly revising the construction of \cite{verstraete2009quantum} which embed the history state of a $\BQP$ computation in the fixed point of a qubit Lindbladian system to have better control over mixing time. We then use standard qubit to boson encoding to obtain a bosonic Lindbladian with jump degree $6$ and no Hamiltonian terms. We then apply a tool which in physics is known as  adiabatic elimination technique (which replaces high rate interactions with effective Lindbladian with lower rate interactions; also analogous to perturbation gadgets used in Hamiltonian complexity \cite{kempe2006complexity}) to replace high degree jump terms with linear loss and Hamiltonian terms. The resulting Lindbladian system has Hamiltonian degree $7$ and loss degree $1$. It however does not have a unique fixed point. We then add additional Hamiltonian terms to obtain a Lindbladian system with Hamiltonian degree $12$ and loss degree $1$ with a unique fixed point. We furthermore show that starting from vacuum state the Lindbladian trajectory flows to the fixed point encoding the history state in polynomial time. This result is established in \Cref{sec:loss-only-bqp}.

We emphasize that while $\BQP$ is a lower bound for the fixed point problem described above, in general, without specifying restrictions such as mixing time, it might not be possible to establish a decidable upper bound on the Lindbladian fixed point problem. To justify this we show that deciding if a Lindbladian system consisting of jump terms of degree at least $5$ has a fixed point or not is undecidable via reduction from solving diophantine equations (See \Cref{subsec:uncomputability}).

\paragraph{Regime of classical easiness for sampling:}
We then turn to the opposite end of the complexity landscape, identifying regimes that admit efficient classical simulation. Previous work of \cite{nicacio2016determining} has established that Gaussian dynamics subject to incoherent pumping undergoing single-photon loss yield a classical fixed point corresponding to a positive Glauber--Suddarshan $P$ function representation (implying a separable state) if the strength of pumping exceeds a threshold value. We observe that the same condition implies positive $P$ representation throughout the trajectory.
Intuitively, we can understand this condition as the strength of pumping dominating the squeezing contribution.  Sometimes nonzero loss and pumping are considered as model for system interacting with environment at non-zero temperature.  So one can view this result as disappearance of quantumness at high enough temperature.

Our main contribution is to show that this classical simulation remains reliable even at the fixed point if we allow non-Gaussian Kerr interaction with inverse polynomial strength. The proof is based on considering non-Gaussian interactions as perturbation. A key technical challenge is that due to infinite-dimensional Hilbert spaces slight non-Gaussian perturbation can in principle alter mixing time significantly and thereby the fixed point would become unstable to perturbations. We overcome this challenge by developing a framework to apply perturbation directly to the fixed point in the specific setting we consider (related frameworks studied in \cite{gondolf2024energy}). We emphasize that stability of perturbation in our analysis heavily relies on a dissipative contraction condition, which ensures uniform control of the photon-number moments and rapid mixing of the unperturbed Gaussian dynamics. This result is established in \Cref{sec:classical-algorithms}.

 We note that one can show that there are examples of dynamics which are not separable initially but the fixed point is classically tractable. This specifies a natural example of instance when complexity does not survive in the stationary regime suggesting the existence of regimes where classical simulation technique fails for transient regime while it works for the stationary distribution. 

We emphasize that while we showed $\BQP$ can  in principle be embedded into Lindbladian fixed points undergoing linear loss, the constructions we used applies loss to a constant fraction of modes and relies on high rate of dissipation on some modes and zero loss on other modes. Based on numerical insights (see, e.g., \Cref{sec:sampling-hardness-vs-observables}) we leave open the possibility that if loss is applied to all modes uniformly for basic Hamiltonian systems such those with cross-Kerr non-Gaussian interaction the fixed point becomes classically simulable (by satisfying criteria such as Wigner non-negativity). It might however be possible that lossless modes can be replaced with fault-tolerant lossy modes and high loss rates can be spread uniformly by adding additional modes. Such a construction can be an initial starting point for more robust universal computation using continuous time and space. We leave this study to future direction.

\paragraph{Hardness of sampling from Gaussian fixed points:} We next complement the classical simulation results by specifying the regime where sampling from the fixed point becomes hard for Gaussian Lindbladians. In particular, we show that assuming Gaussian Hamiltonians undergoing pure loss at a constant rate exact sampling from the fixed point is hard for classical computers assuming non-collapse of the polynomial hierarchy. The main insight is that upon removing pumping from the system one should expect hardness of sampling as a generic phenomenon. As a matter of fact, zero pumping seems essential because we have instances where hardness construction fails even if an exponentially small amount of incoherent pumping is added.
The proof is based on engineering careful thermal squeezed states relating the sampling problem to complexity of hafnians. We furthermore show that at inverse polynomial rate of loss sampling within total variation distance is hard for classical computer assuming the same set of conjectures as~\cite{aaronson2011computational}. The proof is via encoding approximate sampling from the bipartite Gaussian boson sampling of \cite{grier2022complexity} in the approximate sampling from the fixed point. This result is established in \Cref{sec:Gaussian-hardness-sampling}.

\paragraph{Hardness of sampling for specific low-degree instances:}
Next we extend hardness of sampling results from \Cref{sec:Gaussian-hardness-sampling} to basic non-Gaussian Hamiltonian dynamics such as those with degree $4$ interactions. We show that allowing non-Gaussian interactions of the form $\hat a_j \hat a_k \hat a^\dagger_l \hat a^\dagger_s$, sampling from the fixed point is as hard as sampling from noisy boson sampling with constant fraction of photons lost. This instance is not known to be easy or hard (see~\cite{go2026polynomial} for recent progress). Therefore a classical algorithm for the instance considered in this work would yield an algorithm for this regime which was not known before.  
We also construct two-mode quartic gadgets with lossless
data modes and lossy auxiliary modes, which prepare the
required single-photon inputs on specified invariant
subspaces. Together with additional vacuum data modes,
these gadgets yield a reduction from approximate standard
boson sampling, with hardness under the conjectures
of \cite{aaronson2011computational}.
We also construct two-mode quartic gadgets with lossless
data modes and lossy auxiliary modes, which prepare the
required single-photon inputs on the specified invariant subspaces.

Finally, we study models with higher-degree dissipation.
For quadratic (two-photon) loss, we begin with jump operators
of the form $\sqrt{\kappa}(\hat a_j^2-\alpha^2)$,
equivalently combining two-photon loss with a coherent
two-photon drive. Building on previous work on dissipative
cat-state preparation~\cite{albert2014symmetries,
mirrahimi2014dynamically,lingenfelter2021unconditional},
we show that evolution from vacuum prepares states
from which standard boson sampling can be recovered
through postselection. This establishes hardness of exact
photon-number sampling from vacuum-accessible fixed points
of Lindbladians with quadratic Hamiltonians and quadratic
jump operators, unless the polynomial hierarchy collapses.
 We also give a finite-cutoff single-photon preparation
construction based on a degree-five Hamiltonian and quadratic loss. Importantly without finite cut-off we don't expect initial states such as vacuum to flow in finite time to the fixed point. For cubic incoherent terms we show that one can design jump terms that isolate single photon sources without Hamiltonian terms, implying hardness for standard boson sampling. 
These results are established in \Cref{sec:Hardness-specific-instances}.

\subsection{Model}
\label{sec:model}

We study open quantum systems governed by Lindbladian dynamics according to \Cref{eq:basic-Lindbladian}. 
We consider the infinite-dimensional separable Hilbert space of $m$ bosonic particles spanned by the Fock (aka number) basis $\{ \ket{s_1, \ldots, s_m} : s_1, \ldots , s_m \geq 0\}$, where $\ket {s_1, \ldots, s_m}$, which sometimes we denotes with $\ket{\mathbf{s}}$, corresponds to having $s_j$ particles in mode $j$. 

We consider Hamiltonians that are bounded-degree polynomials
in canonical bosonic creation and annihilation operators
$\hat a_1, \ldots, \hat a_m, \hat a^\dagger_1, \ldots,
\hat a^\dagger_m$, with the degree bound specified separately
for each result. We sometimes represent operators in terms
of the number operators $\hat n_j$ or the canonical position
and momentum operators $\hat q_1, \ldots, \hat q_m,
\hat p_1, \ldots, \hat p_m$.
See \Cref{sec:preliminaries} for complete definitions.

\paragraph{The (coherent) Hamiltonian terms.} In this paper, we consider Hamiltonians that are constant-degree polynomial in canonical bosonic operators. Broadly speaking for a polynomial Hamiltonian $H$ we can separate quadratic (Gaussian) and superquadratic terms (non-Gaussian) and write $H = H_G + H_{NG}$, where $H_G$ is a Gaussian part and $H_{NG}$ is the non-Gaussian part. 
 
Let $\hat{\mathbf r}=(\hat q_1,\hat p_1,\ldots,\hat q_m,\hat p_m)^T$.
We use the convention
\[
H_G=\tfrac12\hat{\mathbf r}^{T}\mathbf M\hat{\mathbf r}
       +\boldsymbol (\boldsymbol{\Omega}^{-1}d)^{T} \hat{\mathbf r},
\qquad \mathbf M=\mathbf M^T\in\mathbb R^{2m\times2m}.
\]
$\mathbf d$ is called the phase-space drive, and
\begin{align}
\boldsymbol{\Omega} = \bigoplus_{j=1}^m \begin{pmatrix}0 & 1 \\ -1 & 0 \end{pmatrix}
\end{align}
is called the symplectic form. 
Examples include
drive terms $F_j\hat a_j+F_j^*\hat a_j^\dagger$, squeezing terms
$\nu_j\hat a_j^2+\nu_j^*\hat a_j^{\dagger2}$, and Hermitian sums of hopping
terms $J_{jk}\hat a_j\hat a_k^\dagger$.

An example of non-Gaussian Hamiltonian with degree $4$ has the following form
$$
H_{\mathrm{NG}} = \sum_{i,j,k,l} U_{i,j,k,l} \hat a^\dagger_i \hat a^\dagger_j \hat a_k \hat a_l.
$$
Important special cases include the on-site Kerr interaction of the form $\sum_j U_j \hat n_j^2$ and cross-Kerr interactions $\sum_{j,k} U_{j,k} \hat n_j \hat n_k$, where $\hat n_j = \hat a^\dagger_j \hat a_j$ is known as the number operator.

\paragraph{The incoherent jump terms.} For jump terms we consider polynomial operators of a given degree in creation and annihilation operators. An important instance of jump operators modeling simple photon loss is linear loss corresponding to jump terms $\sqrt{\gamma} \hat a_j$. We call such loss models uniform if the same rate $\gamma$ is applied for all sites. Another important instance is incoherent pumping corresponding to jump term $\sqrt{\eta} \hat a_j^\dagger$ with rate $\eta$. In this work we consider other higher degree loss terms as well.

\subsection{The Lindbladian fixed-point problem}
\label{sec:problem-definition}

 We consider Hamiltonians and jump terms as polynomials in creation and annihilation operators of a given degree as discussed in the previous section.   Before stating the Lindbladian fixed-point computational problem we need some terminology in order to suitably specify inputs to the problem.

\begin{definition}[Specification of a stationary state]
Let $\mathcal L$ be a Lindbladian admitting a closed
realization on the trace-class operators that generates
a strongly continuous, completely positive,
trace-preserving semigroup $(\mathcal T_t)_{t\geq 0}$ (see \Cref{sec:preliminaries} for more details). In short, we call such Lindbladian ``well defined''.

A stationary state (fixed point) $\rho_*$ satisfies
$\mathcal T_t(\rho_*)=\rho_*$ for every $t\geq 0$.
We consider the following conventions for specifying
a stationary state:
\begin{enumerate}
    \item \textbf{Unique fixed point:}
    $\rho_*$ is the unique stationary density operator
    of $\mathcal L$.

    \item \textbf{Unique fixed point in an invariant sector:}
    An efficiently specified closed subspace
    $\mathcal H_{\mathrm{inv}}$ of the bosonic Hilbert
    space is given, such that states supported on
    $\mathcal H_{\mathrm{inv}}$ remain supported there
    under $\mathcal T_t$. The state $\rho_*$ is the
    unique stationary density operator supported on
    $\mathcal H_{\mathrm{inv}}$.

    \item \textbf{Initial-state-selected fixed point:}
    An efficiently preparable initial state $\rho_0$
    is specified, and
    \[
       \lim_{t\to\infty}
       \|\mathcal T_t(\rho_0)-\rho_*\|_1=0.
    \]
\end{enumerate}
If the fixed point is specified using either of the ways above, we denote it as ``specified''.
In presenting results we sometimes explicitly specify the relevant (uniqueness, invariance,
or convergence) convention of efficient specification of the fixed point. 
\label{def:fixed-conventions}
\end{definition}

The main problem studied in this paper is the following
\begin{problem}[Complexity of Lindbladian fixed points]

Let $\mathcal L$ be a well-defined Lindbladian (see the definition above)
with Hamiltonian degree at most $k$ and jump-operator
degree at most $\ell$, whose coefficients are specified
with polynomially many bits of precision.

We consider the following computational problems.

\begin{enumerate}
    \item \textbf{Observable estimation
    ($\textsc{FP}^{k,\ell}_{\varepsilon}$):}
    Given an observable $O$ of constant degree, with  polynomially bounded sum of the absolute values of the monomial coefficients,
    and a stationary state $\rho_*$ specified according
    to one of the conventions above, estimate
    \[
        \Tr(\rho_* O)
    \]
    to within additive error $\varepsilon$.

    We promise that
    \[
        \Tr\big[
        (\mathbb I+\hat N)^{\deg(O)}\rho_*
        \big]
        \leq \poly(n),
    \]
    where $n$ denotes the input size and $\hat N$
    is the total photon-number operator.

    \item \textbf{Photon-number sampling
    ($\textsc{SampFP}^{k,\ell}_{\varepsilon}$):}
    Given a stationary state $\rho_*$ specified
    according to one of the conventions above,
    sample from its photon-number distribution
    to within total variation distance $\varepsilon$.

    \item \textbf{Stationary-state existence
    ($\exists\textsc{FP}^{k,\ell}$):}
    Given $\mathcal L$, determine whether there
    exists a stationary density operator $\rho_*$
    satisfying
    \[
        \mathcal T_t(\rho_*)=\rho_*
        \qquad \text{for all }t\geq 0.
    \]
\end{enumerate}

\paragraph{Reachability:} As a separate attribute to the problem definition we specify if fixed point is accessible from a given initial state. This is particularly relevant to item 3 of the conventions in \Cref{def:fixed-conventions}. Given an efficiently preparable initial
state $\rho_0$ satisfying
\[
    \Tr\big[(\mathbb I+\hat N)^{d/2}\rho_0\big]
    \leq \poly(n),
    \qquad
    d=\max\{k,\ell,\deg(O)\},
\]
we say that $\rho_*$ is efficiently accessible
from $\rho_0$ if there exists
a time $t\leq\poly(n)$ such that
\[
    \|\mathcal T_t(\rho_0)-\rho_*\|_1
    \leq o(1).
\]

\label{prob:FP}
\end{problem}

The condition $
        \Tr\big[
        (\mathbb I+\hat N)^{\deg(O)}\rho_*
        \big]
        \leq \poly(n)$ is specified to make sure the expectation and variance of measurements from $O$ are polynomially bounded. 
        Note also in the definition, we demand polynomially bounded energy for $\rho_0$ because in the infinite-dimensional setting mixing time can be arbitrarily slow if we consider a high enough energy state. 

 We note that, in this work, most instances involve fixed points that are efficiently accessible from a simple initial state. There are some instances in \Cref{sec:Hardness-specific-instances} where the complexity of a specified fixed point can be characterized while the corresponding mixing time remains unknown or is not expected to be polynomial (e.g., the Lindbladian used to show undecidablity of existence of fixed point in \Cref{subsec:uncomputability}). We therefore treat the mixing time as a separate attribute of the computational problem.

In the problem definition, we only specified the degree for Hamiltonian and jump terms. However, one could argue that some form of Hamiltonian/jump terms are more natural than others. For instance for Hamiltonian terms we would like the degree of active terms to be as low as possible (ideally linear). For loss jump terms, the most interesting form is the simple uniform loss which for degree $k$ uses jump terms of the form $\sqrt{\gamma} \hat a^{k}$. We however sometimes (for quadratic above) consider other jump terms as polynomials in creation and annihilation operators or use non-uniform coefficients.

 We call a loss model to be uniform if every mode experiences the same loss model (e.g. $\sqrt{\gamma} \hat a_j$ for all $j$). In general we call the incoherent model to be uniform if jump terms are the same for each mode. In our constructions we sometimes make crucial use of loss-less modes which lead to non-uniform loss models.

\subsection{Technical overview and statement of the main results}
\label{sec:summary-of-results}

The central question studied in this paper is understanding the complexity of $\textsc{FP}^{k,\ell}_{\varepsilon}$, $\textsc{SampFP}^{k,\ell}_{\varepsilon}$ and $\exists \textsc{FP}^{k, \ell}$ defined in \Cref{prob:FP}. In this section we give an overview to the technical details of the main results.

\subsubsection{Encoding universal computation in Lindbladian fixed points with linear loss.} Our first result studies embedding universal $\BQP$ computation into the fixed point of Lindbladian systems undergoing nonuniform linear loss corresponding to jump terms of the form $\sqrt{\gamma}\hat a_j$. We show the following result.

\begin{theorem}
[Embedding $\BQP$ into Lindbladian fixed points
with simple loss;
informal summary of
\Cref{thm:uniq-linear-loss,thm:bqp-mixing-time}]
\label{thm:loss-only-bqp-hardness-inf}

Given an $n$-qubit circuit with $T$ one- and
two-qubit gates and an accuracy parameter
$0<\varepsilon\leq 1/4$, one can efficiently
construct a bosonic Lindbladian
\begin{align}
\mathcal K[\bullet]
=
-i[\hat H,\bullet]
+\gamma\sum_{j=1}^{3T+2n+1}
    \mathcal D(\hat b_j)[\bullet]
+\Gamma\sum_{r=1}^{n+1}
    \mathcal D(\hat s_r)[\bullet],
\label{eq:loss-only-final-generator}
\end{align}
with a unique stationary state $\rho_*$. Here $\hat b_j$ and $\hat s_r$ are annihilation operators on disjoint set of modes. 

The reduced stationary state on the data and
clock modes is within trace distance
$\varepsilon$ of the encoded history state
of the circuit. Furthermore, $\rho_*$ can be
prepared from the global vacuum to within
trace distance $\varepsilon$ in time
$\poly(n,T,\varepsilon^{-1})$.

The Hamiltonian $\hat H$ has degree at most
$12$, and all jump operators are linear
single-photon loss operators. The construction
uses $5n+4T+3$ bosonic modes, of which
$3n+3T+2$ undergo photon loss. All parameters
can be chosen polynomially bounded in
$n,T,\varepsilon^{-1}$.

Consequently, estimating suitable
constant-degree observables at the unique
stationary state to inverse-polynomial
additive accuracy is $\BQP$-hard, i.e.,
\[
\textsc{FP}^{12,1}_{O(1)}
\geq \BQP.
\]
\end{theorem}

The proof is by combining three ingredients into
a bosonic fixed-point reduction. Note due to infinite-dimensional Hilbert spaces one needs to be very careful about operator domain issues when combining different constructions together.

\begin{proof}[Proof outline of \Cref{thm:loss-only-bqp-hardness-inf}]

We first consider the construction of
\cite{verstraete2009quantum}, which embeds the history state
of a $\BQP$ circuit into the stationary state of a qubit
Lindbladian. We revisit this construction to obtain
quantitative control over the relaxation time of the
Lindbladian, which will be needed in the remainder of
the proof. This construction is purely dissipative:
the Lindbladian consists only of jump terms.

We then use dual-rail encoding \cite{knill2001scheme}
to embed the data qubits into bosonic modes, and
single-photon unary encoding for the clock register.
In particular, the logical states $\ket{0}$ and
$\ket{1}$ of each data qubit are represented by the
two-mode states $\ket{10}$ and $\ket{01}$, respectively.
These encodings identify a finite-dimensional invariant
sector of the bosonic dynamics, in which the original
qubit Lindbladian is faithfully reproduced.
The resulting bosonic Lindbladian has no Hamiltonian terms
and only consists of jump operators of degree at most six.

Our next step is to replace these high-degree jump
operators with Hamiltonian interactions and simple
linear photon loss. To achieve this, we develop an
adiabatic elimination technique
(\Cref{thm:loss-only-stationary-elimination}),
building on previous work such as
\cite{reiter2012effective,kessler2012generalized,
azouit2016adiabatic}.

In physics, \emph{adiabatic elimination} is a technique
for eliminating rapidly evolving degrees of freedom
and describing a physical system in terms of effective
interactions among its slow degrees of freedom.
As an intuitive example, consider two people speaking
over the phone. The transmission of speech signals
through the telecommunication network occurs on a
much shorter timescale than the conversation itself.
We can therefore eliminate the intermediate
transmission process and effectively describe the
interaction as two people talking directly to each
other. In the context of computational complexity,
adiabatic elimination can be viewed as a dissipative
analogue of the perturbation gadgets of
\cite{kempe2006complexity}. We use
this technique to simulate Lindbladian jump operators
of degree $k$ using Hamiltonian interactions of degree
$k+1$ and ``rapid'' simple linear photon loss acting on
auxiliary modes.

The underlying idea is that the ancillary modes
undergo dissipation at a much faster rate than the
effective dynamics of the system. Eliminating these
rapidly relaxing modes generates the desired
dissipative evolution on the slower system degrees
of freedom. This construction is perturbative,
with the effective dynamics approaching the
target Lindbladian in the large-loss limit.

To elaborate, consider a finite-dimensional Lindbladian
of the form
\begin{align}
\mathcal L
=
\sum_{j=1}^{m}\mathcal D(\hat F_j).
\label{eq:pure-loss-Lindbladian}
\end{align}
We introduce one ancillary bosonic mode $\hat b_j$
for every jump operator $\hat F_j$, and consider
the Lindbladian
\begin{align}
\mathcal L_\gamma[\bullet]
&=
-i\left[
\frac12\sum_{j=1}^{m}
\left(
\hat b_j^\dagger\hat F_j
+
\hat b_j\hat F_j^\dagger
\right),
\bullet
\right]
+
\gamma\sum_{j=1}^{m}
\mathcal D(\hat b_j)[\bullet].
\label{eq:adiabatic-ell}
\end{align}

Here the ancillary modes undergo single-photon loss
at rate $\gamma$, while the original system modes
remain lossless. For sufficiently large $\gamma$,
the ancillary modes relax rapidly, inducing an
effective dissipative evolution on the system modes.

Under suitable assumptions on the relaxation of
the effective Lindbladian and the photon-number
moments of the stationary state, the reduced
stationary state of $\mathcal L_\gamma$ approximates
the stationary state of $\mathcal L$, with an error
decreasing as $O(1/\gamma)$.
More precisely, we establish the following result.

\begin{theorem}
[Informal restatement of
\Cref{thm:loss-only-stationary-elimination}]
\label{thm:temp-loss-only}

Suppose $\mathcal L$ is a finite-dimensional
Lindbladian of the form
\Cref{eq:pure-loss-Lindbladian} with a unique
stationary state $\rho_f$ and finite trace-norm
relaxation time $\tau(\mathcal L)$
(see \Cref{eq:loss-only-integrated-mixing}).

Let
\[
f_1:=\sum_{j=1}^{m}\norm{F_j},
\]
and assume $\gamma\geq f_1$.
Let $\sigma_\gamma$ be a stationary state
of $\mathcal L_\gamma$ satisfying
\begin{align}
\Tr\bigl[
\sigma_\gamma(\hat N_B+I)^3
\bigr]
<\infty,
\qquad
\overline\rho_\gamma
:=
\Tr_B\sigma_\gamma,
\end{align}
where $\hat N_B$ is the total photon-number
operator of the ancillary modes.
Then
\begin{align}
\norm{\overline\rho_\gamma-\rho_f}_1
\leq
\frac{
200\,\tau(\mathcal L)f_1^3
}{\gamma}.
\label{eq:loss-only-stationary-error-simple}
\end{align}
\end{theorem}

The finite third photon-number moment is needed
to control the unbounded creation and annihilation
operators appearing in the perturbative analysis.
For the Lindbladian constructed in our reduction,
we establish a constant bound on this quantity.

We now apply this adiabatic elimination technique
to the boson-encoded Lindbladian described above.
Since the encoded jump operators have degree
at most six, the corresponding Hamiltonian
interactions have degree at most seven.
The resulting fully bosonic Lindbladian preserves
the encoded subspace, and its stationary states
within this sector approximate the encoded
history state of the original computation,
as established in \Cref{thm:loss-only-bqp-hardness}.

In this intermediate construction, photon loss
acts only on the elimination ancillas, which
constitute a constant fraction of the total
number of modes. The data and clock modes
remain lossless. Similar constructions involving
couplings between lossy and lossless modes
will be used later in
\Cref{prop:robust-hardness-gbs}.
We also note that the constructions of
\cite{ikeda2013deterministic}, which we use
later for Gaussian dynamics, can be viewed
as another application of adiabatic elimination techniques
for dissipative state generation.

The resulting degree-seven Lindbladian does not
yet establish uniqueness of the stationary state
on the full bosonic Fock space.
Indeed, the desired stationary state has so far
been characterized only within the invariant
encoding sector. The goal is to penalize space orthogonal to this subspace.

To establish global uniqueness, we introduce
additional ancillary modes, with annihilation
operators $\hat s_r$, together with Hamiltonian
couplings that select the desired encoding sector.
These additional interactions have the form
\[
\hat s_r^\dagger\hat a_{r,0}^\dagger
(\mathbb{I}-\hat N_r)+\mathrm{h.c.},
\]
where $\hat a_{r,0}$ is one of the two modes encoding
the $r$-th data qubit, and
$\hat N_r=\hat a_{r,0}^\dagger\hat a_{r,0}
+\hat a_{r,1}^\dagger\hat a_{r,1}$
counts the total photons in its dual-rail register.
An analogous interaction is introduced for the clock
register. The full construction is given in \Cref{sec:unique-fixed-point-full-domain}.

These couplings, together with loss on the
new ancillary modes, ensure that the stationary
state contains exactly one photon in each
dual-rail data register and one photon in the
clock register. We also add a diagonal
Hamiltonian term of the form
\[
\eta W
=
\eta\left(
\mathbb{I}+\sum_i(\hat a_i^\dagger\hat a_i)^6
\right),
\]
where the sum runs over the data and clock modes.
This term allows us to establish that the
Lindbladian generates a well-defined,
conservative quantum Markov semigroup on
the full Fock space. It also raises the
Hamiltonian degree to twelve.

The final Lindbladian therefore contains two
groups of lossy ancillary modes: the elimination
ancillas $\hat b_j$, with loss rate $\gamma$,
and the additional loading ancillas $\hat s_r$,
with loss rate $\Gamma$.
We fix $\Gamma>0$ and choose $\gamma$
sufficiently large to control the approximation
error. The resulting dynamics has a unique
stationary state on the full Fock space,
whose reduced data-clock state approximates
the encoded circuit history state
(\Cref{thm:uniq-linear-loss}).
We refer to \Cref{fig:BQP-hard} for an
illustration of the complete construction.

Finally, we prove that this unique stationary
state can be prepared from the global vacuum
in polynomial time
(\Cref{thm:bqp-mixing-time}).
This requires an additional argument because
the vacuum lies outside the encoded subspace,
so invariance of that subspace alone does not
establish efficient preparation.

The proof proceeds in two stages.
First, the additional Hamiltonian couplings
and lossy loading ancillas populate the data
and clock registers, bringing the system close
to the initialized state within the encoded
subspace. We bound the time required for
this initialization process.

Second, starting from the initialized encoded
state, we use the mixing properties of the
effective circuit Lindbladian together with
a finite-time adiabatic elimination bound
to show that the actual bosonic dynamics
approaches the encoded history state.
Combining these two stages with the
stationary-state uniqueness and approximation
established in \Cref{thm:uniq-linear-loss},
we conclude that the unique stationary
state can be prepared from the global
vacuum in polynomial time.

\end{proof}

\subsubsection{Classical algorithms for the fixed point problem}

We next study the regime of parameters for which classical simulation is possible for the fixed point problem.

\paragraph{Gaussian dissipative dynamics and positive-\(P\) classicality.}

We first focus on the dynamics under Gaussian Hamiltonians with loss and pump. Starting from a Gaussian state, the quantum state remains a Gaussian state throughout its trajectory. Standard results using the symplectic formalism~\cite{bartlett2002efficient} imply efficient observable estimation. However, sampling in the photon number basis can still be difficult for a classical computer. We first specify the conditions under which efficient classical sampling is possible and then prove stability under non-Gaussian perturbations.

 We use the centered covariance
$\Sigma_{jk}=\operatorname{Tr}(\rho\{\hat r_j-\mu_j,\hat r_k-\mu_k\})$,
where $\mu_j=\operatorname{Tr}(\rho\hat r_j)$, so the vacuum has covariance
$\mathbb I$. 
The Glauber--Sudarshan $P$ function of a state $\rho$ is defined through the coherent-state decomposition (see \Cref{sec:preliminaries} for more details):
\begin{align}
\label{eq:P-function}
    \rho
    =
    \int_{\mathbb{C}^m}
    P_\rho(\boldsymbol{\alpha})
    \ket{\boldsymbol{\alpha}}\!\bra{\boldsymbol{\alpha}}\,
    d^{2m}\boldsymbol{\alpha}.
\end{align}
A Gaussian state has a nonnegative $P$ representation exactly when
$\boldsymbol\Sigma\geq\mathbb I$.
In this case $P$ is a Gaussian probability measure, possibly singular. Such a state
is separable and can be sampled in the Fock basis by first sampling $P$ and
then independent Poisson variables of means $|\alpha_j|^2$.

Consider a Gaussian Hamiltonian with parameters defined in \Cref{sec:model}. \cite{nicacio2016determining} showed that if the strength of pumping exceeds a certain amount the fixed point of a Gaussian dynamics will satisfy postive $P$ function, and is hence classically sampleable. By slightly extending this result, we observe that under the same criteria the Gaussian dynamics preserves positive $P$ function states throughout the whole time trajectory.

\begin{theorem}[Preservation of Gaussian classicality; reformulation and slight generalization of \cite{nicacio2016determining}]
The Gaussian evolution preserves positivity of the $P$ function for every
Gaussian input if and only if
\[
[\boldsymbol\Omega,\mathbf M]+2\eta\mathbb I\geq0
\]
If, in addition,
\[
\gamma-\eta>2\max_{z\in\sigma(\boldsymbol\Omega\mathbf M)}\operatorname{Re}z,
\]
every input converges to a Gaussian state with positive $P$ function.
\label{thm:gaussian-classical-trajectory-inf}
\end{theorem}

Here $\sigma(A)$ denotes the spectrum of matrix $A$.
Importantly in \Cref{sec:non-classical-trajectory-vs-end-points} we find examples of instances where the full trajectory does not demonstrate classicality (i.e. the $P$ funcation can take negative values) but the fixed point nevertheless satisfies classicality. This suggests that there exist families of dissipative processes
whose fixed points can be sampled efficiently even though
the nonnegative-$P$ sampling criterion fails during
their transient evolution making classical simulation by directly evolving the Lindblad trajectory potentially infeasible.

Note that for loss and pumping alone, with $\gamma>\eta$, the unique fixed point of each mode is a thermal state. We however know that boson sampling on thermal states is classically easy~\cite{rahimi2016sufficient, bressanini2024gaussian}. We can therefore view \Cref{thm:gaussian-classical-trajectory-inf} as an extension of this result to Gaussian Linbdladian systems.

Our main contribution in this section is to show that classical simulations based on this result is stable under non-Gaussian perturbations. In particular consider the Kerr perturbation
\[
    H_{\rm Kerr}
    =
    \sum_{(j,k)\in E}U_{jk}\hat n_j\hat n_k .
\]
Let
$$
U := \sum_{j,k} |U_{j,k}|.
$$
Applying perturbation to finite time trajectory can be achieved with a basic application of the Duhamel identity:
\begin{theorem}[Finite-time stability under weak Kerr nonlinearities; Informal restatement of \Cref{thm:finite-time-stability}]
\label{thm:kerr-stability}
Let $\rho_0(t)$ be a stable Gaussian trajectory and let $\rho_U(t)$ be the
trajectory obtained by adding Kerr interactions $ H_{\rm Kerr}
    =
    \sum_{(j,k)\in E}U_{jk}\hat n_j\hat n_k.$
Then
\begin{align}
    \|\rho_U(t)-\rho_0(t)\|_1
    =O(UE_G^2t),
\end{align}  
Here $E_G$ only depends on the Gaussian parameters and not $U$.
\end{theorem}

We now note that if
the nonlinear dynamics admits a unique fixed point and mixes within time
$t_{\rm mix}^U(\epsilon)$, then the Gaussian trajectory provides an
$\epsilon$-accurate approximation of the nonlinear steady-state sampler
whenever
\[
  U
    \le
   O( \frac{\epsilon}
    {E_G^2t_{\rm mix}^U(\epsilon/2)}).
\]
However this bound explicitly depends on the mixing time and uniqueness of fixed point for the perturbed model. This mixing time can in general be very sensitive to $U$. It is desirable to phrase the stability criteria solely based on the parameters of the unpurturbed system. The main result of this section is achieving this by applying perturatbion directly to the fixed point.

Let $\rho_0$ be the fixed point of the Gaussian dynamics corresponding to interaction matrix $\mathbf{M}$, phase space drive $\mathbf{d}$ (see \Cref{sec:model}) and $\rho_U$ be the fixed point of the perturbed dynamics. As before, let $\gamma$ be the strength of uniform loss and $\eta$ be the strength of uniform pumping.

\begin{theorem}[Restatement of \Cref{thm:kerr-perturbative}; perturbation at the fixed point]
\label{thm:kerr-perturbation-informal}
Let $D=2m(\eta+\frac12\norm{[\boldsymbol{\Omega}, \mathbf{M}]})+16\|\mathbf d\|_2^2/\delta$, where $\delta = \gamma-\eta-\norm{[\boldsymbol\Omega, \mathbf M]}>0$ is assumed to be positive. Let $s^\ast = \min\{\frac12, \frac{\delta}{8D}\}$. Then, for every $0<\varepsilon<1$, the condition
\begin{align}
U=
O\left(
\frac{\varepsilon \delta (s^{*})^2}
{m\cdot \log(\frac{m(1+D/\delta)}{\varepsilon})}
\right)
\end{align}
is sufficient to achieve
\begin{align}
\norm{\rho_U - \rho_0}_1 \le \varepsilon.
\end{align}
Furthermore, for any such $U$ with $\varepsilon\le 1$, the fixed point ($\rho_U$) is unique.
\end{theorem}
As indicated by the bounds above, the stability of the fixed point under Kerr perturbations relies on the dissipative contraction condition \(\delta>0\). This condition ensures control over the photon-number moments of stationary states and rapid mixing of the unperturbed Gaussian dynamics. Importantly, $\delta > 0$ by itself does not imply that the Gaussian fixed point admits a nonnegative \(P\) representation. 
Combining with \Cref{thm:gaussian-classical-trajectory}, efficient classical sampling for the perturbed model is implied when $\norm{[\mathbf{M}, \boldsymbol{\Omega}]} \leq \min \{2 \eta, \gamma - \eta \}$.

\begin{proof} [Proof sketch of \Cref{thm:kerr-perturbation-informal}]
Let $\mathcal{L}_0$ be the unperturbed Lindbladian and $\mathcal{L}_U$ be the perturbed one with fixed point $\rho_U$, then for any $T \geq 0$, we can write 
\begin{equation}
\norm{\rho_U - \rho_0}_1
\le
\underbrace{\norm{\rho_U - e^{T \mathcal{L}_0}[\rho_U]}_1}_{\varepsilon_1}
+
\underbrace{\norm{e^{T \mathcal{L}_0} [\rho_U] - \rho_0}_1}_{\varepsilon_2}.
\label{eq:rho-U-0-split}
\end{equation}
We bound each term in this equation for a suitable choice of $T$. To bound the first term in \Cref{corol:l0-on-rhoU} we show that
\begin{equation}
\|\mathcal{L}_0 [\rho_U]\|_1 = O( \frac{U}{(s^*)^2}),
\label{eq:L0-rho-U-1}
\end{equation}
This bound implies that 
$
\varepsilon_1 \leq O(\frac{UT}{(s^*)^2})
$.
A major tool in establishing this bound is placing suitable $U$ independent bound on the moments of $\rho_U$ (established in \Cref{thm:bound-rhou}):
$$
\Tr (e^{s \hat N} \rho_U) \leq \exp (O(Ds/\delta)), \quad \delta > 0
$$
where $\hat N$ is the total photon number operator, and using an exponential norm (also studied in \cite{gondolf2024energy}) for operators. See \Cref{def:exp-norm}.

To bound the second term we note that since $\mathcal{L}_0$ has a suitable gap, applying $e^{\mathcal{L}_0 T} [\rho_U]$ for a short time $T = \frac{m}{\delta} \log (\Omega(\frac{m D}{\delta \varepsilon}))$ is enough to reach $\rho_0$ which implies that 
$
\varepsilon_2 \leq O(\varepsilon)
$
(see \Cref{lem:fast-mixing}).  
\end{proof}
\noindent Note the construction above directly reasons about the fixed points and not the whole trajectory.

We also emphasize the recent work of \cite{gonzalez2025dynamical} which gave classical simulation for non-Gaussian dynbamics with on-site Kerr interaction undergoing loss and pump, when the strength of pumping exceeds a value depending on the Gaussian parameters of the Hamiltonian. Our result applies to both on-site and cross-Kerr interactions.

\subsubsection{Establishing hardness of sampling for simpler Lindbladians}

To complement the classical algorithms established in the previous section we next show that removing incoherent pumping yields models that are hard to sample even approximately. We note that the construction in \Cref{thm:loss-only-bqp-hardness-inf} already establishes survival of sampling complexity in the fixed point. However there are two major draw-backs: (1) the construction involves relatively complicated Hamiltonians of degree $12$, and (2) loss is imposed on a fraction of modes. Our next series of results establish hardness of sampling for simpler Hamiltonians based on reduction from different variations of boson sampling \cite{aaronson2011computational}. In several cases, including Gaussian Lindbladians, we are able to establish hardness when loss is uniformly applied to every mode. 

\paragraph{Hardness through Gaussian Lindbladians.}

We first start with Gaussian Lindbladians (see \Cref{sec:Gaussian-hardness-sampling}). Our first result is hardness of exact sampling from the fixed points of Gaussian dynamics undergoing photon loss with a uniform constant rate of noise:
\begin{theorem}[Hardness of exact sampling at constant rate of loss; Restatement of~\Cref{thm:hardness-of-gaussian-lindbladian}]
\label{thm:hardness-of-gaussian-lindbladian-inf}
There is no $\mathsf{BPP}$ algorithm to exactly sample from the fixed point of a Gaussian dynamics undergoing photon loss with a uniform constant rate assuming the polynomial hierarchy ($\mathsf{PH}$) does not collapse.
\end{theorem}
Note that the exact sampling hardness result above holds even if loss rate is exponentially large.

\begin{proof}[Proof outline of \Cref{thm:hardness-of-gaussian-lindbladian-inf}]
    Our starting point is the hardness of Gaussian boson sampling, which relates photon-number measurement probabilities of squeezed Gaussian states to the computation of hafnians. However, this hardness result cannot be directly applied to our setting: we first show that a Gaussian Lindbladian with a quadratic Hamiltonian and uniform linear loss and pumping cannot have a nontrivially squeezed pure state as its fixed point. We therefore develop a different reduction that establishes sampling hardness directly for the mixed Gaussian fixed points of Lindbladians with quadratic Hamiltonians and pure photon loss.

    The main ingredient is a construction that, given an arbitrary symmetric matrix $\mathbf B$, produces a quadratic Hamiltonian whose corresponding pure-loss Lindbladian has a unique Gaussian fixed point with a specially engineered Bargmann matrix and a tunable parameter. By choosing this parameter sufficiently small, while keeping its description length polynomial in the input size, we can suppress unwanted contributions to a particular photon-counting probability and isolate the squared hafnian $|\operatorname{Haf}(\mathbf B)|^2$, whose multiplicative approximation is $\#\mathsf P$-hard. The remainder of the proof follows standard complexity-theoretic arguments based on Stockmeyer's approximate counting theorem, as in \cite{aaronson2011computational}.
\end{proof}

One ideally hopes to prove a robust hardness of sampling from Lindbladian fixed points, i.e., hardness survives even for approximate sampling from the fixed point. Our next result establishes such robust hardness. 
In particular, we show
\begin{theorem}[Robust hardness of sampling at inverse polynomial rate of loss; restatement of
\Cref{thm:robust-hardness-all-modes}]
\label{thm:robust-hardness-all-modes-inf}
There is a family of Gaussian Lindbladians on $m$ modes, with loss rates that are inverse-polynomially large in $m$, such that sampling from the photon-number distribution of the fixed point to within inverse-polynomial total variation distance is classically hard, assuming the Permanent-of-Gaussians Conjecture, the Permanent Anti-Concentration Conjecture (stated in \cite{aaronson2011computational}), and that $\mathsf{PH}$ does not collapse.
\end{theorem}

\begin{proof}[Proof outline of \Cref{thm:robust-hardness-all-modes-inf}]
Our construction first uses a two-mode gadget in which one couples a lossless mode to a lossy mode. This construction is similar to the use of lossless data modes based on \cite{ikeda2013deterministic} in \Cref{thm:loss-only-bqp-hardness-inf} (see \Cref{sec:loss-only-bqp} for details). \cite{ikeda2013deterministic} used this gadget for applications such as state generation via Lindbladians. 

The key observation used in this gadget is that a simple hopping Hamiltonian between these two modes, together with loss on only one of them, has the vacuum as its unique attracting fixed point. 

Since the data mode is lossless, we can conjugate the Lindbladian by an arbitrary Gaussian unitary acting on the data modes without changing the form of the loss operators. This allows us to engineer fixed points of the form $\ket{\psi}\otimes\ket{0^m}$, where $\ket{\psi}$ is an arbitrary pure Gaussian state. In particular, we can choose $\ket{\psi}$ from a family of Gaussian boson sampling instances that are hard to sample even approximately under standard complexity-theoretic conjectures.

We can then use perturbative arguments to introduce a small nonzero loss rate on the previously lossless data modes. The central technical step is to show that this perturbation changes the covariance matrix of the fixed point by a controlled amount. The dynamics being Gaussian plays a crucial role in the success of this perturbation idea. Using bounds relating covariance-matrix distance to trace distance for Gaussian states, we conclude that the perturbed fixed point remains inverse-polynomially close to the desired hard Gaussian state. Consequently, hardness of approximate photon-number sampling is preserved, even though every mode now experiences nonzero loss. 

We now have some modes being more lossy than others. We then use additional gadgets to distribute loss from the more lossy modes in a way that makes all modes lossy with a uniform rate. More precisely, we couple each data mode to a polynomial number of ancillary modes through a passive quadratic Hamiltonian that spreads the data mode over many normal modes. This suppresses the effect of loss on the reduced state of the data modes by a factor proportional to the inverse number of ancillary modes. By choosing polynomially many ancillas, we ensure that the reduced fixed point remains inverse-polynomially close to the target Gaussian state, while every physical mode experiences the same loss rate. 
 
\end{proof}

We emphasize that we were not able to apply the same perturbation and loss-spreading idea to the case of non-Gaussian Hamiltonians such as those used in \Cref{thm:loss-only-bqp-hardness-inf}. While extra careful analysis might achieve the same result on that front, we leave open the possibility that uniform weak loss on every mode would severely limit the fixed point of non-Gaussian Lindbladians such as those with cross Kerr. Understanding the answer to this question is an interesting future direction.

As emphasized before, the two-mode gadget construction used in the proof above is very similar to the construction based on coupling lossy and lossless modes used in the proof of \Cref{thm:loss-only-bqp-hardness-inf}. As a matter of fact we can reproduce the former gadget with the latter: To encode a Gaussian state $\ket{\psi}$ as a fixed point of a Lindbladian with linear jump terms in $\hat a_i,\hat a_i^\dag$, we can readily use its nullifiers. In other words, consider the Lindbladian in \Cref{eq:pure-loss-Lindbladian} with 
$F_j = \sum_{k} (\Lambda_{j,k} a_j + \Gamma_{j,k} a_j^\dag) + \zeta_j\, \mathbb I$ be nullifiers of $\ket\psi$ (i.e., $F_j\ket\psi=0$ for all $j=1,2,\cdots,m$). Then, utilizing \Cref{eq:adiabatic-ell} recovers the construction given in \Cref{prop:robust-hardness-gbs}. Nevertheless, in \Cref{prop:robust-hardness-gbs} we are able to prove that the fixed points match exactly (not just that they are close).

\paragraph{Hardness of sampling with models with non-Gaussian interactions.}
Next, we study the hardness regime with non-Gaussian Lindbladians with simpler structures than those studied in \Cref{thm:loss-only-bqp-hardness-inf}.

We start by Hamiltonian systems with non-Gaussian interactions of the form $\hat a^\dagger_i \hat a^\dagger_j \hat a_k\hat a_\ell$ undergoing single-photon loss. However, we limit the active terms to linear coupling with a magnetic field, i.e., squeezing terms are removed. To address the complexity of fixed point in this setting we first consider a special single-mode Lindbladian:
\begin{align}
H
=
-\Delta\omega \hat n
+
\frac{U}{2}\hat a^{\dagger 2}\hat a^2
+
F\hat a^\dagger+F^*\hat a,
\qquad
\dot\rho
=
-i[H,\rho]+\gamma\mathcal{D}(\hat a)\rho .
\label{eq:basic-Kerr-single-mode}
\end{align}
Previous work of \cite[Sec.~III.A]{le2014bose} indicates that, for a fixed positive integer $q$, when $\Delta\omega=(q-1)U/2$, $|F|/U\to0$, and $(\gamma/U)/(|F|/U)^q\to0$, the fixed point of the Lindbladian converges to the binomial density matrix
$$
\sigma_q := \frac{1}{2^q} \sum_{j = 0}^q {q \choose j} \ketbra{j}{j}.
$$
In \Cref{app:nk} we first make this bound explicit using asymptotic bounds
\begin{theorem} [Convergence to binomial distribution; Consequence of \Cref{thm:explicit-binomial-bound}]
Assume $q\ge1$ is a fixed integer, and pick $U=1$, $\Delta\omega=(q-1)/2$, and $F=\gamma^{1/(q+1)}$. As $\gamma\to0^+$, the fixed point $\rho$ of the Lindbladian in \Cref{eq:basic-Kerr-single-mode} satisfies:
\begin{align}
    \mathcal{F}(\rho,  \sigma_q) \geq 1-O_q(\gamma^{2/(q+1)})
\end{align}
Here $\mathcal{F} (\rho, \sigma) := \|\sqrt{\sigma}\sqrt\rho\|_1$ is the fidelity between density matrices $\rho$ and $\sigma$.
\end{theorem}
We then consider the Lindladian $\mathcal{L} = \sum_{j=1}^m\mathcal{L}_j$, where each $\mathcal{L}_j$ is a Lindbladian of the form \Cref{eq:basic-Kerr-single-mode} applied to mode $j$ (assuming the same uniform loss rate on each site). We then make the simple observation that by rotating the Lindbladian with a linear optical operation to obtain
$$
\mathcal{L}_C [\bullet] := C \mathcal{L}[C^\dagger \bullet C] C^\dagger
$$
the loss term remains invariant, the Hamiltonian term takes the quartic shape we described, and $\rho$ is a fixed point of $\mathcal{L}$ if and only if $C \rho C^\dagger$ is a fixed point of $\mathcal{L}_C$. Choosing $q = 1$ for all $j$, we obtain an instance of approximate boson sampling within total variation distance $O(m\sqrt{\gamma})$ where each photon at the very beginning layer has $1/2$ probability of being lost to the environment.

For a sufficiently small inverse-polynomial choice of
$\gamma$, the approximate sampling complexity of the
resulting noisy boson-sampling instances is not fully
understood. Hardness under standard conjectures is
established when only $O(1)$ photons are lost
\cite{aaronson2016bosonsampling}. On the other hand,
recent work of \cite{go2026polynomial} establishes
polynomial-time classical approximate sampling for
standard boson sampling with any constant photon-survival
fraction, with high probability over Haar-random
passive linear-optical unitary and under suitable assumptions on the
number of modes. This result does not establish
efficient simulation in the worst case.

Consequently, hardness of approximate sampling for the
corresponding binomial boson-sampling family would imply
hardness of $\textsc{SampFP}^{4,1}_{1/\poly}$, provided
the fixed-point approximation error is sufficiently
small. Conversely, an efficient approximate classical
sampler for the same binomial family would yield an efficient sampler for
the particular family of conjugated Lindbladian
fixed points constructed here.

More fundamentally, the effective loss rate of $1/2$
is not merely a consequence of our particular choice
of parameters. As we prove in Appendix~\ref{app:nk},
there is an intrinsic limitation to the single-mode
gadget construction: the effective photon-loss
fraction cannot be reduced below $1/2$.
Consequently, this approach cannot realize
boson-sampling instances with arbitrarily small
constant photon loss, even if the stationary-state
approximation is made arbitrarily accurate.
Overcoming this obstruction requires going beyond
the single-mode gadget framework, for instance
by introducing interactions between multiple modes which we will establish later.

Finally, exact sampling from the binomial
boson-sampling distributions with inputs
$\sigma_q^{\otimes m}$, for any fixed positive integer
$q$, is hard unless the polynomial hierarchy
collapses, as shown in \Cref{thm:PostLOk}.
However, this exact-sampling hardness does not
imply hardness of approximate sampling from the
corresponding Lindbladian fixed points, which are
only guaranteed to approximate the ideal binomial
distributions in total-variation distance.

We still need to show fast mixing time for this model. For a fixed finite-dimensional relaxing Lindbladian, exponential convergence gives a mixing time $O(\log(1/\varepsilon))$ with constants depending on the generator. A uniform efficiency claim additionally requires control of its decay rate and prefactor; a nonzero gap alone does not supply these bounds.
 The Lindbladian in \Cref{eq:basic-Kerr-single-mode} has a unique fixed point~\cite{casteels2016powerlaws, drummond1980quantum}. However since the model is infinite dimensional, the Lindbladian may still have infinitely many eigenvalues getting arbitrarily close to $0$. To give evidence that this does not happen we show that starting from the vacuum state the whole quantum trajectory remains effectively finite-dimensional (photon-number sectors far
above the natural scale $\mu=4|F|^2/\gamma^2$ have exponentially small weight). Therefore finite Fock-space truncations are plausibly useful when studying the fixed
point and its low-degree observables. We leave a rigorous derivation of mixing time to future work.

\begin{proposition}[Uniform trajectory tail bound from vacuum; Informal restatement of \Cref{prop:kerr-vacuum-trajectory-tail}]
Let $\rho(t)$ be the quantum state evolved according to \Cref{eq:basic-Kerr-single-mode} starting from $|0\rangle\langle0|$.
Define
$$
    \mu
    :=
    \frac{4|F|^2}{\gamma^2},
$$ 
then for all \(t\ge0\) and all \(M\ge\mu\),
\begin{align}
    \Pr_{\rho(t)}(\hat n\ge M)
    \le
    \exp\!\left(
        -M\log\frac{M}{\mu}
        +M-\mu
    \right).
\end{align}
\end{proposition}

\paragraph{Additional notes on the solution to \Cref{eq:basic-Kerr-single-mode}.} One can heuristically consider a classical limit solution to \Cref{eq:basic-Kerr-single-mode} by replacing $\hat a$ with a complex number $\alpha$, $\hat a^\dagger$ with $\alpha^*$ and $\hat n$ with $|\alpha|^2$. We can write the fixed point condition based on this classical limit and obtain
$$
\dot{\alpha} = (i \Delta \omega - \gamma/2) \alpha - i U |\alpha|^2\alpha - i F = 0 \implies |F|^2
=
n\left[
(\Delta-Un)^2+\frac{\gamma^2}{4}
\right].
$$
which is a polynomial equation of degree $3$ in $n$, revealing three solutions usually drawn as S shape curve (see for example \cite[Figure 1]{drummond1980quantum}). Two out of the three solutions are stable and another is unstable with switching between them without additional fluctuations. So one would immediately suspect that including quantum fluctuations, i.e., solving the actual Lindbladian fixed point problem we would get three different fixed points. However, interestingly we get only one unique fixed point~\cite{drummond1980quantum}. \cite{drummond1980quantum} used a $P$ function representation to reduce the fixed point problem to Fokker--Planck differential equation with a unique solution. One way to make sense of this is that the fixed point solution $\rho_{\rm ss}$ can be approximately written as $p_1 \rho_1 + p_2 \rho_2$, where $\rho_1$ and $\rho_2$ intuitively correspond to the two stable classical points.

Interestingly, once the incoherent pumping term is added, the equations describing the fixed point become more strongly coupled. In the zero-pump case, the equations have a special structure that allows them to be solved exactly, but the pumping term generally breaks this structure. As a result, it is no longer obvious how to obtain a closed-form steady-state solution by the same method.

At $q$-photon resonance, in the regime $|F|/U\to0$ and $(\gamma/U)/(|F|/U)^q\to0$, \cite{le2014bose} showed that the fixed-point can be approximated by the binomial distribution before bistability induced by larger values of $F$ appears. Bistability specifically can dramatically slow down convergence to the fixed point. For instance \cite{casteels2017critical} studied the spectral gap of the Lindbladian in the regime $U \rightarrow 0$ and $F \rightarrow \infty$ (where bistability is expected), while keeping $UF^2$ a constant, and argued that the spectral gap closes as $e^{- O(1/U)}$. However, faster mixing in the weak $F, \gamma$ regime used to approximate noisy boson sampling in this work is still plausible.

\paragraph{Embedding standard boson sampling using $2$-mode non-Gaussian Gadgets with linear non-uniform loss.} 
As indicated in \Cref{thm:rho-11-Kerr} reliable approximation of the $\ket{1}$ with fidelity $> 1/2$ is not possible using single mode Lindbladian in \Cref{eq:basic-Kerr-single-mode}. A natural question is if we can engineer $\ket{1}$ using two mode non-Gaussian gadgets with linear loss. This is what we achieve in our next result. 

In \Cref{ex:2-mode-1-engineering} we repeat the two-mode gadget construction used in \Cref{thm:loss-only-bqp-hardness-inf} and  \Cref{thm:robust-hardness-all-modes-inf}, but this time with quartic Hamiltonians. In particular, on the invariant subspace reached from vacuum,
the unique fixed point of the restricted Lindbladian
\begin{equation}
    \mathcal{L}[\bullet] = -i[\hat a^\dag_2 \hat a^\dag_1(1-\hat n_1)+(1-\hat n_1)\hat a_1 \hat a_2, \bullet] + \mathcal{D}(\hat a_2)[\bullet]
\label{eq:quartic-loss-2-mode-gadget}
\end{equation}
is $\ket{10}$. 
 We then use the standard method we used before: define the Lindbladian $\mathcal{L} = \sum_{j=1}^m \mathcal{L}_j$ over $2m$ modes (the first $m$ being noisy and the second $m$ being noiseless) and let $C$ be a linear optical network applied to the first $m$ modes. The unique fixed point of $\mathcal{L}_C$ is $(C \ket{1^m}) \otimes \ket{0^m}$ in which we can embed standard boson sampling. We leave bounding the mixing time for this model to future work. We note that this construction fundamentally uses coupling with loss-less modes. We were unable to apply the perturbative toolsets which allowed us to distribute loss on every mode in \Cref{thm:robust-hardness-all-modes-inf} to the non-Gaussian case.

\paragraph{The complexity for higher loss mechanisms.}

We next consider dissipation mechanisms with higher degrees starting from quadratic dissipation.

For quadratic loss, we establish sampling hardness
by using postselection to recover boson sampling
with odd-cat inputs.
Our starting point is the two-photon
driven-dissipative dynamics under the Lindbladian
\[
\mathcal G_\alpha
=
\kappa\mathcal D[\hat a^2-\alpha^2],
\]
which combines a quadratic-loss jump with a two-photon
drive. An important feature of this dynamics is that
its stationary-state subspace contains the even and odd
cat states. The stationary state generally
depends on the initial condition and starting from vacuum
selects the pure even cat state
$\ket{C_\alpha^+}\bra{C_\alpha^+}$
\cite{mirrahimi2014dynamically,azouit2016well}.

Our construction exploits this dissipative preparation
mechanism to embed boson sampling into the fixed point
of a Lindbladian with a quadratic Hamiltonian and
quadratic jump operators. The key idea is to convert
the even cats into odd cats using photon subtraction.
For each active mode, we introduce an auxiliary
vacuum mode and apply a passive beam splitter.
Conditioning on detecting one photon in the auxiliary
output prepares an odd cat:
\[
{}_b\langle 1|B_{ab}(\theta)
\ket{C_\alpha^+}_a\ket{0}_b
\propto
\ket{C_{\alpha\cos\theta}^-}_a.
\]
Importantly, this conditioning can be performed
entirely through photon-number measurements, without
introducing qubits or adaptive operations.

We incorporate these beam splitters and the passive
linear-optical transformation used in the boson-sampling
reduction into a single passive unitary $V$, and
conjugate the original Lindbladian by $V$.
Since passive unitaries preserve the global vacuum,
the resulting dynamics still starts from vacuum and
converges to the corresponding conjugated stationary
state. Moreover, unitary conjugation preserves the
polynomial degree of the Hamiltonian and jump operators.
Thus, the final construction has a Gaussian Hamiltonian
and quadratic loss, with no linear loss required.

We establish that exact photon-number sampling from
this vacuum-selected stationary state is classically
hard unless the polynomial hierarchy collapses
(\Cref{thm:postselected-hardness}). The reduction
first postselects one photon in each auxiliary output,
thereby preparing odd-cat inputs, and then conditions
on the minimum possible total photon number in the
remaining modes. This isolates the single-photon
component of each odd cat and recovers ordinary
boson sampling under the chosen passive unitary.

We also establish a quantitative preparation guarantee.
For fixed model parameters, the full bosonic state
converges from vacuum to the selected stationary state
in trace norm within time
\[
O\bigl(\log n+\log(1/\varepsilon)\bigr)
\]
(\Cref{thm:fast-mixing-nmodes}).
This guarantee concerns the unconditioned state and
does not require uniqueness of the stationary state.
The hardness reduction, however, relies on an
exponentially unlikely postselection event.
Consequently, the exact-sampling hardness result
does not directly extend to approximate sampling
with inverse-polynomial total-variation error.

We emphasize that the conjugated quadratic jump
operators generally act collectively on the physical
modes. Thus, this construction establishes hardness
for quadratic dissipation, but not for independent
local two-photon losses in the physical output basis.

We next consider a different single-photon preparation
construction using a degree-five Hamiltonian and quadratic
loss. Our result concerns the dynamics at a finite photon
cutoff. Define
\[
H=g\left((\hat n-I)\hat a^3
+(\hat a^\dagger)^3(\hat n-I)\right).
\]
For each integer $R\ge3$, let
\[
\mathcal H_R=\operatorname{span}\{\ket0,\ldots,\ket R\},
\qquad
P_R=\sum_{n=0}^R\ket n\bra n,
\]
and consider the compressed Lindbladian on $\mathcal H_R$,
\[
\mathcal L_R[\rho]
=-i[P_RHP_R,\rho]
+\gamma\mathcal D(P_R\hat a^2P_R)[\rho],
\qquad \gamma>0.
\]

\begin{proposition}[Finite-cutoff single-photon preparation;
restatement of \Cref{prop:finite-cutoff-one-photon}]
For every integer $R\ge3$, real $g\neq0$, and $\gamma>0$,
the Lindbladian $\mathcal L_R$ has the unique fixed point
\[
\rho_\ast=\ket1\bra1.
\]
Moreover, every initial state supported on $\mathcal H_R$
converges to $\rho_\ast$ in trace norm.
\end{proposition}
This establishes convergence for each fixed cutoff.
We emphasize that the conclusion above heavily relies on finite truncation. 
It does not provide a mixing-time bound uniform in $R$
or a preparation guarantee for the untruncated dynamics.

Next, we can consider cubic jump operators of the form
$\hat a^\dagger(\hat n-I)$ on each mode with no Hamiltonian which preserve the subspace
$\operatorname{span}\{\ket0,\ket1\}$.
The resulting Lindbladian has the unique fixed point
$\ket1 ^{\otimes n}$ and we can show the vacuum state $\ket{0^n}$ flows to an inverse polynomial neighborhood of this fixed point in polynomial time.
This gives a single-photon preparation construction using
a structured degree-three jump rather than simple
three-photon loss $\hat a^3$.
 Note this Lindbladian is directly related to the Lindbladian in \Cref{eq:quartic-loss-2-mode-gadget} via adiabatic elimination tool of \Cref{eq:adiabatic-ell}. We however study the mixing time directly in
\Cref{sec:cubic-loss}.

\paragraph{Undecidability of existence of fixed point:}Finally, we show that the fixed-point problem can become
undecidable when higher-degree dissipation is allowed.
In particular, we prove that deciding whether a bosonic
Lindbladian admits a stationary density operator is
$\mathsf{RE}$-hard, even with no Hamiltonian and jump
operators of degree at most five, each acting on at most
four modes (\Cref{thm:finite-energy-fixed-point-re-hard}).
The proof reduces the existence of a common nonnegative
integer root of a system of quadratic Diophantine equations
to the existence of a stationary state. The construction
uses occupation-number-dependent photon gain and linear
loss: if the polynomial constraints admit a common root,
the dynamics has a finite-energy stationary state;
otherwise, no normalizable stationary state exists.
Thus, undecidability persists even in well-defined,
trace-preserving Markovian dynamics with low-degree
dissipative interactions.

\begin{table*}[t]
\centering
\footnotesize
\caption{Summary of the main results. $\textsc{SampFP}^{k,l}$ is the problem of sampling from a particular fixed point of a Lindbladian with coherent degree $k$ and incoherent degree $l$. $\textsc{FP}^{k,l}$ is the same problem for observable estimation. See \cref{prob:FP} for more details. Where present, $\varepsilon=1/\poly$; subscript $0$
denotes exact sampling.}
\label{tab:main-results}
\renewcommand{\arraystretch}{1.15}
\begin{tabular}{p{2cm} p{6cm} p{5.1cm}}
\toprule
\textbf{Problem} &  \textbf{Setting/Notes} & \textbf{Complexity} \\
\midrule
\midrule

 \textsc{FP}$^{12,1}_{O(1)}$ 
&  
Unique fixed point accessible from vacuum, nonuniform loss & $\geq$ BQP 
 \\

 \midrule
\textsc{SampFP}$^{2,1}_{\varepsilon}$ 
& Allowing squeezing, uniform loss  
&  $\geq$ Gaussian Boson Sampling \\

\midrule
\textsc{SampFP}$^{4,1}_0$ 
& Allowing cross Kerr, no squeezing, uniform loss
 &$\geq$ Noisy Boson Sampling (approximate)
 \\
\midrule

\textsc{SampFP}$^{4,1}_{\varepsilon}$ 
& Nonuniform loss, accessible from vacuum
 &$\geq$  Boson Sampling
 \\
\midrule

{\textsc{SampFP}$^{2,2}_{0}$}
& {Nonuniform quadratic loss, accessible from vacuum}
& {$\geq$ Cat Boson Sampling (exact)} 
 \\
\midrule

\textsc{SampFP}$^{0,3}_\varepsilon$ 
& Structured jump terms, accessible from vacuum state 
& $\geq$ Boson Sampling 
 \\
\midrule
 $\exists$\textsc{FP}$^{0,5}$ 
& Deciding the existence of fixed point 
& $\geq$ RE
 \\
 \midrule
\end{tabular}
\par\smallskip
\begin{minipage}{\linewidth}
\end{minipage}
\end{table*}

\subsection{Discussions, related work and open questions}
\label{sec:discussions}

In this work, we study the computational complexity of fixed points of bosonic Lindbladians in two computational settings: observable estimation and sampling in the photon-number basis. A central takeaway is that these problems exhibit a rich complexity landscape, ranging from efficient classical simulability to universal quantum computation and, ultimately, undecidability. This landscape depends on the degree of the Hamiltonian and jump operators, as well as the assumptions imposed on the mixing time of the dissipative dynamics.

Many open questions and directions remain from this work.
\begin{enumerate}

\item \textbf{Encoding robust universal quantum computation in the fixed point of the driven-dissipative Bose--Hubbard model.}
In this work, we embed $\mathsf{BQP}$ computation into the unique fixed point (reachable rapidly from the vacuum state) of a bosonic Lindbladian with a Hamiltonian of degree $12$ and linear loss. A natural question is whether a similar construction can be achieved using lower-degree Hamiltonians, ideally within standard physical models such as the driven-dissipative Bose--Hubbard model. This model is described by the Lindblad master equation
\begin{equation}
    \frac{d\rho}{dt}
    = -i[H_{\mathrm{BH}},\rho]
    + \sum_j \gamma_j \mathcal{D}[\hat a_j](\rho),
\end{equation}
where
\begin{equation}
    H_{\mathrm{BH}}
    = -\sum_{\langle i,j\rangle} J_{ij}
    (\hat a_i^\dagger \hat a_j + \hat a_j^\dagger \hat a_i)
    + \sum_j \left(
        \Delta_j \hat n_j
        + \frac{U_j}{2}\hat n_j(\hat n_j-1)
        + F_j \hat a_j^\dagger
        + F_j^* \hat a_j
    \right),
\end{equation}
In particular, this model involves only quartic Hamiltonian
interactions and linear loss. Can universal quantum computation
be encoded into its fixed point while retaining polynomial
mixing time? We believe that the adiabatic degree-elimination
techniques developed in this work provide a natural route
toward achieving this goal. 

\item \textbf{Robust hardness results at constant parameter strengths:} Can we establish hardness of the fixed point problems when all parameters, including loss rates or Gaussian and non-Gaussian couplings are constant? We want such construction ideally for a uniform linear loss model. In our constructions we often choose a parameter such as loss rate as inverse polynomial value (for instance sampling hardness results) or polynomially large in our BQP embedding in \Cref{sec:loss-only-bqp}. 
    \item \textbf{Incoherent pumping as the main source of classicality.} Our result isolates incoherent pumping as the main source of classicality. We note a closely related line of work is the recent study of dynamical complexity in dissipative bosonic systems \cite{gonzalez2025dynamical}. In particular, \cite{gonzalez2025dynamical} establishes that for a class of bosonic systems with linear loss, linear incoherent pumping, Gaussian interaction and on-site cross Kerr interactions, the time-evolved state $\rho(t)$ admits an efficient classical simulation. More precisely, their Theorem~2 shows that there exists a randomized classical algorithm that can sample from $\rho(t)$ in the Fock basis up to total variation error $\epsilon$ in time $\poly(n, t, 1/\epsilon)$. Crucially, they consider on-site self-Kerr or cross-Kerr non-Gaussian terms that act on localized cells of modes. Their construction is based on Trotterizing Lindbladian dynamics as layers of Gaussian dynamics followed by local non-Gaussian dynamics, and observing if the strengths of incoherent pump exceed a certain value, the Gaussian dynamics gets decoupled. Therefore, the evolved state is effectively close to a product state within each cell. Their proof technique relies heavily on the assumption that non-Gaussian terms are local.

Our results first extend the regime of classical simulation when we consider only Gaussian Hamiltonians. It furthermore shows that this efficient simulability disappears once we remove incoherent pumping. An important open question is if classicality remains when incoherent pumping is applied to non-local non-Gaussian interactions. It is furthermore interesting to understand if this decoupling phenomenon can be maintained when all Lindbladian parameters are retained constant independent of the number of modes.

    \item \textbf{Bounding mixing times.} Many of the hardness constructions in \Cref{sec:Hardness-specific-instances} are based on gadgets with constant number of modes.
    However bounding mixing time for even constant size bosonic systems have inherent difficulties due to infinite dimensional Hilbert spaces. Hamiltonian terms can lead to rapid energy growth and in general even when fixed point is unique there could a seires of infinite eigenvalues of the Lindbladian that approach arbitrarily close to $0$. As a result even when fixed point is unique there is not guarantee that all initial states converge to such a fixed point in finite time. We leave detailed analysis of mixing time for models studied in this work (including \Cref{eq:basic-Kerr-single-mode} and the model in \Cref{sec:1-from-qudartic-loss}) to future work. We expect that the problem of deciding if a bosonic Lindbladian system has a unique fixed point or has finite mixing time to be undecidable in principle, adding an extra layer of complication in bounding mixing times for bosonic Lindbladian dynamics.

    To add to this, our Lindbladian construction in \cref{sec:loss-only-bqp}, has a unique fixed point, but we do not show convergence from arbitrary initial states. Can a $\mathsf{BQP}$-hard construction be made, where arbitrary initial states converge to the unique fixed point?

   \item \textbf{Boosting the complexity using additional side information.}
The relation between cat-state mixtures and noisy boson sampling
suggests studying whether knowledge of the random input can increase
computational power. In the construction of
\Cref{sec:cat-sampling}, the auxiliary photon counts reveal the
even/odd parity of the cats entering the linear optical instance used in the reduction.
These labels are learned together with the output, after the
linear optical operation has already been chosen. This resembles the
$\frac12\mathsf{BQP}$ model studied in \cite{jacobs2024space},
where a random computational-basis input is revealed at the end
of the computation.

If the parity labels were available beforehand, we could select
the odd-cat modes and then apply a suitable linear optical operation.
When the labels are available only at the end, it is natural to
ask whether classical postprocessing can make better use of them
than simply postselecting the all-odd event.
For qubit circuits, \cite{jacobs2024space} shows that the
$\frac12\mathsf{BQP}$ model can solve problems such as factoring,
which are not known to be efficiently solvable in the
one-clean-qubit model~\cite{knill1998power}.
Whether analogous advantages arise for passive linear optics
with cat inputs remains open.

For example, consider a uniformly random parity pattern
$x\in\{0,1\}^m$, where $\ket{C_\alpha^x}$ denotes the product
of even cats at positions with $x_j=0$ and odd cats at positions
with $x_j=1$. From the joint input labels and output photon counts,
we can estimate functions of the form
\[
F(U)=\frac{1}{2^m}
\sum_{x\in\{0,1\}^m}\sum_{y\in\mathbb N_0^m}
f(x,y)\,
\bigl|\bra{y}U\ket{C_\alpha^x}\bigr|^2,
\]
where $f:\{0,1\}^m\times\mathbb N_0^m\to\mathbb C$
is efficiently classically computable and satisfies $|f(x,y)|\le1$.
For a nonuniform parity distribution, including that produced
by our construction, the factor $2^{-m}$ is replaced by the
corresponding probability $p(x)$.
Can solving hard computational problems such as factoring
be reduced to inverse-polynomial additive estimation of such
an $F(U)$?
    
    \item \textbf{The dual role of loss in bosonic systems.} In bosonic Lindbladian systems  dissipation plays two opposite roles. For low degree Hamiltonians such as Gaussian or non-Gaussian Hamiltonians with at most linear active terms, too much loss can push the system into a trivial classical state. But for genuinely non-Gaussian active terms, too little loss may fail to control energy growth leading to extremely slow (or lack of) convergence to desired fixed points. Efficient dissipative preparation needs a balance: enough dissipation to maintain quantum features and control energy. Formalizing this intuition is a nature future direction.

\item \textbf{Does observable estimation from Lindbladian fixed points remain hard under uniform loss?}
For Lindbladian models considered in this work, hardness results are established only for sampling tasks from Gaussian fixed points. 
Sampling hardness does not necessarily imply hardness of
estimating low-degree observables. In particular observable estimation from the aforementioned Gaussian instances are easy.

The $\BQP$-hardness construction for observable estimation in
\Cref{sec:loss-only-bqp} we used linear loss on auxiliary modes,
while leaving the data and clock modes lossless.
Whether this observable hardness persists after adding
inverse-polynomial loss to every data and clock mode
remains open. In Section \Cref{sec:robust-gaussian-lindbladian} we could
use perturbation gadgets to make the non-uniform loss uniform in the case of Gaussian trajectory. These tools do
not however apply directly to the case of higher degree Hamiltonians.

A complementary approach is to design classical algorithms for observable estimation by proving non-negativity for
the Wigner function of fixed points. If a state has a nonnegative
Wigner function that can be sampled efficiently, and the
observable's Weyl symbol can be evaluated efficiently
with polynomially bounded variance, its expectation can
be estimated efficiently
(see \Cref{thm:positive-wigner-observable-estimation}
and \cite{mari2012positive}).
Nonnegativity alone does not guarantee these algorithmic
conditions.

For the single-mode driven Kerr model in
\Cref{eq:basic-Kerr-single-mode}, the exact steady-state
Wigner function is nonnegative
\cite{kheruntsyan1999wigner}, although the transient
evolution can exhibit Wigner negativity
(see \Cref{fig:transient_wigner_min}).
An open question is to identify parameter regimes of
multimode quartic models in which the fixed points have
nonnegative Wigner functions and satisfy the sampling
and variance conditions required for efficient
observable estimation.

\end{enumerate}

\subsection{Acknowledements and disclosures}
 \subsubsection*{Acknowledgments}
 EJK acknowledges financial support from the National Science and Technology Council (NSTC) of Taiwan under Grant No.~NSTC~114-2112-M-A49-036-MY3. SM acknowledges financial support from the national science foundation NSF CCF-2013062. We thank insightful discussion with Bobak Kiani, Varun Upreti, Ulysse Chabaud.

\subsubsection*{AI disclosure}

Throughout this work we used generative LLM's to brainstorm ideas, outline proof ideas, outline proof drafts, edit text, and verify the proofs. All authors take responsibility for the work presented here. 

\subsubsection*{Author contributions}

EJK, AM and SM led the development of technical details of this work (EJK and AM contributed equally). MH and SM supervised this project.

\section{Preliminaries and notation}
\label{sec:preliminaries}

For background on continuous variable quantum information see \cite{weedbrook2012gaussian,braunstein2005quantum}. 

\paragraph{Notation}
The $p$ Schatten norm is denoted by $\|\bullet\|_p$. Matrices are written in bold, e.g.\ $\mathbf{B}$, $\mathbf{M}$, $\boldsymbol{\Omega}$, and $\boldsymbol{\Sigma}$. For a matrix $\mathbf{A}$, $\sigma (\mathbf{A})$ is its spectrum, i.e., the set of eigenvalues of $\mathbf{A}$. For Hermitian operators $A, B$ we say $A \geq B$ if $A - B$ is positive semi-definite. $\mathbb{N}_0 = \mathbb{N} \cup \{0\}$.

\paragraph{Complexity classes.} We assume familiarity with complexity classes $\BPP$, $\BQP$, $\PH$ and $\QMA$. See for instance \cite{aaronson2005complexity} for a review.

\paragraph{Hilbert space.}
We use CV to denote a \emph{continuous-variable} quantum system, namely a system whose underlying Hilbert space is infinite-dimensional.

For a single bosonic mode, the Hilbert space is
$L^2(\mathbb{R};\mathbb{C})$, the space of
square-integrable complex-valued functions.
It is isomorphic to the Fock space
\[
\mathcal{H}_0
=
\ell^2(\mathbb{N}_0;\mathbb{C})
=
\operatorname{span}
\left\{\ket{n}:n\in\mathbb{N}_0\right\},
\]
where $\ket{n}$ denotes the Fock state with
$n$ bosonic excitations.

Throughout, $m$ denotes the number of modes. For an $m$-mode system the Hilbert space is
\[
\mathcal{H} = \mathcal{H}_0^{\otimes m}.
\]

\paragraph{Operators.}
We write bosonic operators with hats. The annihilation and creation operators $\hat a_j$ and $\hat a_j^\dagger$ acting on the $j$-th mode satisfy
\begin{align}
    [\hat a_i,\hat a_j^\dagger]=\delta_{i,j}.
\end{align}
For a multimode Fock basis state $\ket{\mathbf{n}}=\ket{n_1,\dots,n_m}$ (also written $\ket{\mathbf{s}}$ when convenient),
\begin{align}
    \hat a_j\ket{\mathbf{n}}
    &= \sqrt{n_j}\ket{\mathbf{n}-e_j},\\
    \hat a_j^\dagger\ket{\mathbf{n}}
    &= \sqrt{n_j+1}\ket{\mathbf{n}+e_j},
\end{align}
where $e_j$ is the standard basis vector with a $1$ in position $j$.
The number operator of mode $j$ is $\hat n_j= \hat a_j^\dagger \hat a_j$, so that $\hat n_j\ket{\mathbf{n}}=n_j\ket{\mathbf{n}}$, and the total number operator is $\hat N = \sum_{j=1}^m \hat n_j$.
The canonical quadratures are
\begin{align}
\hat q=\frac{\hat a+\hat a^\dagger}{\sqrt2},\qquad
\hat p=\frac{\hat a-\hat a^\dagger}{i\sqrt2},\qquad [\hat q,\hat p]=i.
\end{align}

We collect them into the quadrature vector
\[
\hat{\mathbf{r}} = (\hat q_1,\hat p_1,\ldots,\hat q_m,\hat p_m)^\top,
\]
and write $\hat r_j$ for its components. The (multimode) vacuum is denoted $\ket{\mathrm{vac}}$; a single-mode vacuum is $\ket{0}$.

\paragraph{Lindbladians and rates.} For background on the Lindbladian master equation see \cite{rivas2012open, breuer2002theory}. In our notation,
Lindbladians act with square brackets, $\mathcal{L}[\bullet]$. The dissipator associated with a jump operator $J$ is
\begin{align}
\mathcal{D}(J)[\bullet]
:=
J \bullet J^\dagger - \tfrac12 \{J^\dagger J,\bullet\}.
\end{align}
Using this notation $\mathcal{L}$ in \eqref{eq:basic-Lindbladian} can be written as $\mathcal{L} [\bullet] = - i [H , \bullet] + \sum_j \mathcal{D}(L_j)[\bullet]$.

Photon loss on mode $j$ has rate $\gamma_j$ and jump operator $\sqrt{\gamma_j}\,\hat a_j$; incoherent pumping has rate $\eta_j$ and jump operator $\sqrt{\eta_j}\,\hat a_j^\dagger$. When the rates are mode-independent we drop the index and write $\gamma$ and $\eta$. In that uniform case we also use the combinations
\[
\gamma_{-} := \gamma-\eta, \qquad \gamma_{+} := \gamma+\eta.
\]

A state $\rho_\infty$ is stationary if $\mathcal L[\rho_\infty]=0$.
A spectral gap describes asymptotic decay but, without control of transient
growth and prefactors, does not by itself give a useful uniform mixing time.
For bosonic systems the supremum over all input energies may fail to converge
uniformly even for a stable Gaussian channel. We therefore use
\begin{align}
t_{\mathrm{mix}}(E,\varepsilon)=\inf\left\{t\geq0:
\sup_{\operatorname{Tr}(\rho\hat N)\leq E}
\|e^{t\mathcal L}[\rho]-\rho_\infty\|_1\leq\varepsilon\right\}.
\label{eq:t-mix}
\end{align}
Our perturbation proof bounds the input energy before applying this definition.

\paragraph{Bosonic gates and Gaussian operations.}
A unitary is called \emph{Gaussian} if it is generated by a Hamiltonian that is at most quadratic in the quadrature operators.
  Prominent examples of Gaussian operations include:
\begin{itemize}
    \item \emph{Displacements:} $D(\alpha)=\exp(\alpha \hat a^\dagger - \bar{\alpha} \hat a)$,
    \item \emph{Phase shifts:} $R(\theta) = \exp(i\theta \hat a^\dagger \hat a)$,
     \item \emph{Single-mode squeezing:} $S(\xi) = \exp\!\left(\tfrac{1}{2}(r e^{i\phi} \hat a^{\dagger 2} - r e^{-i\phi} \hat a^2)\right)$ for $\xi = r e^{i\phi}$.

    \item \emph{Beam splitters:} $B(\theta) = \exp(\theta (\hat a^\dagger \hat b - \hat a\hat b^\dagger))$, is a two-mode operation.
   
\end{itemize}

Gaussian operations can be efficiently simulated classically under certain conditions and form the backbone of linear optics. An important example of multi-mode Gaussian operation are linear optical operations. A linear optical network on $m$ bosonic modes is described by a unitary matrix $U\in U(m)$ acting linearly on the annihilation operators. In the Heisenberg picture, the corresponding unitary operator $\hat U$ satisfies
\begin{equation}
    \hat U^\dagger \hat a_j \hat U
    =
    \sum_{k=1}^m U_{jk}\hat a_k,
    \qquad j=1,\ldots,m.
\end{equation}
Such transformations preserve the total photon number $\hat N = \sum_{j=1}^m \hat n_j$. A linear-optical network is generated by a hopping Hamiltonian
\[
\hat H_{\rm hop}=\sum_{j,k}h_{jk}\hat a_j^\dagger\hat a_k,
\]
where $h=h^\dagger$ is a Hermitian matrix. The corresponding
single-particle unitary matrix is $U=e^{-iht}$, while the
unitary acting on Fock space is $\hat U=e^{-i\hat H_{\rm hop}t}$. A linear optical operation can be generated as a sequence of beam splitters and phase shifts.

\paragraph{Symplectic formalism for Gaussian dynamics.}
The quadratures satisfy $[\hat r_j,\hat r_k]=i\boldsymbol\Omega_{jk}$, where
$\boldsymbol\Omega=\bigoplus_{j=1}^m\left(\begin{smallmatrix}0&1\\-1&0\end{smallmatrix}\right)$.
We consistently use the doubled, centered covariance
\[
\mu_j=\langle\hat r_j\rangle,\qquad
\Sigma_{jk}=\left\langle\{\hat r_j-\mu_j,\hat r_k-\mu_k\}\right\rangle.
\]
Thus $\boldsymbol\Sigma_{\mathrm{vac}}=\mathbb I$ and
$\boldsymbol\Sigma+i\boldsymbol\Omega\geq0$ for all Gaussian states.
A Gaussian channel acts as
\[
\boldsymbol\mu\mapsto\mathbf X\boldsymbol\mu+\mathbf b,\qquad
\boldsymbol\Sigma\mapsto\mathbf X\boldsymbol\Sigma\mathbf X^T+\mathbf Y,
\qquad
\mathbf Y+i(\boldsymbol\Omega-\mathbf X\boldsymbol\Omega\mathbf X^T)\geq0.
\]
Here $\mathbf X$ is real and $\mathbf Y$ is real symmetric; the last inequality
is the complete-positivity condition. Gaussian unitaries have $\mathbf Y=0$
and a symplectic matrix $\mathbf X=\mathbf S$, meaning
$\mathbf S\boldsymbol\Omega\mathbf S^T=\boldsymbol\Omega$.
For the displacement operators $\hat D(\mathbf v)=e^{i\mathbf v^T\hat{\mathbf r}}$,
the characteristic function is $\chi_\rho(\mathbf v)=\operatorname{Tr}(\rho\hat D(\mathbf v))$.
For Gaussian states it is
\begin{equation}\label{eq:Gaussian-char}
\chi_\rho(\mathbf v)=\exp\left(-\tfrac14\mathbf v^T\boldsymbol\Sigma\mathbf v
 +i\mathbf v^T\boldsymbol\mu\right).
\end{equation}

\paragraph{Non-Gaussian operations.}
Non-Gaussian unitaries arise from higher-than-quadratic Hamiltonians and are necessary for universal quantum computation in CV systems. A canonical example is the \emph{cross-Kerr interaction}
\[
U_{\mathrm{CK}} = \exp(i U\hat n_1 \hat n_2),
\]
which can be interpreted as a photon-number-dependent phase shift between two modes. More generally a cross Kerr Hamiltonian is
$$
\sum_{j,k} U_{j,k} \hat{n_j} \hat{n}_k.
$$
We define the strength of this cross-Kerr interaction to be
$$
U = \sum_{j,k} |U_{j,k}|
$$
 We say this Hamiltonian is on site if the only nonzero terms are $U_{j,j}$. 
Such interactions generate non-Gaussian states and play a crucial role in complexity-theoretic hardness results.

\paragraph{Active v.s. passive operations.} Bosonic unitaries are generally divided into two classes: (1) \emph{Passive transformations}, generated by Hamiltonians that preserve total photon number, such as linear optical networks, Kerr gates or rotations, and (2) \emph{Active transformations}, which do not preserve photon number and including squeezing operations. These can change the energy of the system and are essential for generating nontrivial CV resources. A bosonic Hamiltonian taking a form of a polynomial in creation and annihilation operators generates a number preserving unitary if it only consists of balanced monomials meaning they have equal number $\hat a^\dagger$ and $\hat a$'s. For instance $(\hat a^\dagger_3)^2\hat a_1 \hat a_2 +\hat a^\dagger _2\hat a^\dagger_1 \hat a_3^2 $ generates a passive operation, whereas $(\hat a^\dagger_3)^3\hat a_1 \hat a_2 +\hat a^\dagger _2\hat a^\dagger_1 \hat a_3^3$ generates an active operation. Active operations tend to blow up energy (expected particle number) rapidly in bosonic dynamics (see \cite{chabaud2025energy}).

\paragraph{Important bosonic states.}
We recall several families of states that appear frequently in both physics and complexity theory:
\begin{itemize}
    \item \emph{Coherent states:} $\ket{\alpha} = \hat D(\alpha)\ket{0}$, which are the $\alpha$ right eigenstates of the annihilation operator. We can explicitly phrase them as
    $$
    \ket{\alpha} = e^{- |\alpha|^2/2} \sum_{n \geq 0} \frac{\alpha^n}{\sqrt{n!}} \ket{n}.
    $$
    \item \emph{Squeezed states:}  for a single bosonic mode, the squeezed vacuum state is
\[
\ket{S_\xi}=\hat S(\xi)\ket{0},
\]
It transforms the mode operator as
\[
\hat S^\dagger(\xi)\hat a\,\hat S(\xi)
=
\hat a\cosh r+e^{i\phi}\hat a^\dagger\sinh r.
\]
    \item \emph{Cat states:} superpositions of coherent states, e.g.\ $\ket{\alpha} + \ket{-\alpha}$, which are highly non-Gaussian and useful for encoding logical qubits.
\end{itemize}

\paragraph{Gaussian computation and boson sampling.}
As indicated, broadly speaking, one can classify bosonic quantum operations (and states) into two classes: Gaussian and non-Gaussian. The former class, which can be viewed as a bosonic analog of Clifford operations for qubits, is efficiently representable on a classical computer, and therefore tasks such as estimating observables from the corresponding output states are efficiently computable classically. By supplementing Gaussian operations with appropriate non-Gaussian resources, such as adaptive photon-number measurements or nonlinear interactions such as cross Kerr, one can instead achieve universal quantum computation~\cite{knill2001scheme}. Thus, the interplay between Gaussian and non-Gaussian resources provides a natural setting in which to study the boundary between classically simulable dynamics and universal quantum computation.

Even within the Gaussian regime, however, one can generate output distributions that are believed to be hard to simulate classically, provided one considers tasks such as sampling in the photon-number basis. Two prominent examples are boson sampling~\cite{aaronson2011computational} and Gaussian boson sampling~\cite{hamilton2017gaussian}. These models provide a weaker, subuniversal notion of quantum computational advantage, but one that can arise from considerably more restricted resources.
One can furthermore make these hardness results robust by showing that, under additional mathematical conjectures, approximate sampling within sufficiently small total variation distance from boson-sampling or Gaussian-boson-sampling distributions remains hard~\cite{aaronson2005quantum, grier2022complexity}.
More specifically, these conjectures include: non-collapse of the polynomial hierarchy~\cite{arora2009computational}, the permanent of Gaussians conjecture (implying that the permanent of i.i.d. complex Gaussian matrices are $\#\mathsf{P}$ hard to approximate on average) and the anticoncentration of random permanents conjecture implying with probability $1-o(1)$ the absolute permanent of $n \times n$ i.i.d. Gaussian matrices is at least $\sqrt{n!}/\poly$ (this conjecture has recently been claimed to have been resolved~\cite{koehler2026anticoncentration}).

Notably, experimental implementations of these platforms have recently reached scales that are believed to be challenging to simulate on classical computers~\cite{madsen2022quantum, zhong2020quantum}. Classical simulation algorithms have also advanced,
including tensor-network methods for lossy Gaussian
boson sampling~\cite{oh2024classical}.

Many of these complexity-theoretic arguments, however, are most naturally formulated for idealized noiseless devices, whereas physical bosonic systems inevitably interact with their surrounding environment. Full fault-tolerant quantum computation can in principle protect against such noise~\cite{matsuura2026continuous}, but doing so requires implementing an explicit and active error-correction architecture together with the associated state-preparation, control, and non-Gaussian resources. It is therefore natural to ask whether one can retain computational hardness, or even embed a robust quantum computation, in noisy bosonic systems without implementing a complete fault-tolerant scheme. More generally, one would like to understand which regimes of noisy bosonic dynamics remain classically simulable and which retain genuinely quantum computational power.

In the context of boson sampling noisy architectures been studied in various works. For instance, \cite{aaronson2016bosonsampling} showed that if $k$ photons in a boson-sampling experiment are lost, then exactly sampling from the output of the device within $1/n^{O(k)}$ total variation distance remains hard. Thus, hardness of approximate sampling can be retained if only a constant $k=O(1)$ number of photons are lost. Similar robustness results to noise were developed in~\cite{go2025sufficient} for Gaussian boson sampling. Generalizing robustness of hardness to noise to other tasks such as observable estimation is an important and interesting research direction.

\paragraph{Vectorization notation.} For a matrix $\mathbf A\in\mathbb{C}^{n\times n}$, we use $\operatorname{vec}(\mathbf A)$ to denote the column-stacking vectorization of $A$. The inverse operation is denoted by $\operatorname{vec}^{-1}$. We will use the standard identity
\[
\operatorname{vec}(\mathbf A\mathbf X\mathbf B)
=
(\mathbf B^{T}\otimes \mathbf A)\operatorname{vec}(\mathbf X),
\]
which allows matrix equations to be rewritten as linear equations on vectorized matrices. 

\paragraph{Phase-space representations.}
The Glauber--Sudarshan $P$ function of a state $\rho$ was defined in \Cref{eq:P-function}. In general, $P_\rho$ need not be a regular probability distribution and may be highly singular. If $P_\rho$ is nonnegative, then $\rho$ is a classical mixture of coherent states. Since multimode coherent states are product states across the modes, positivity of the $P$ function implies that $\rho$ is separable across the mode partition. For a Gaussian state with covariance matrix $\boldsymbol{\Sigma}$, in the convention where the vacuum has covariance matrix $\mathbb{I}$, positivity of the $P$ function is equivalent to
\begin{align}
    \boldsymbol{\Sigma}\geq \mathbb{I}.
\end{align}

The Wigner function $W_\rho(\mathbf{r})$ is a real phase-space quasiprobability distribution on $\mathbb{R}^{2m}$. It may be defined as the Fourier transform of the symmetrically ordered characteristic function,
\begin{align}
    W_\rho(\mathbf{r})
    =
    \frac{1}{(2\pi)^{2m}}
    \int_{\mathbb{R}^{2m}}
    e^{-i\boldsymbol{\xi}^T\mathbf{r}}
    \operatorname{Tr}\!\left[
        \rho\,e^{i\boldsymbol{\xi}^T\hat{\mathbf r}}
    \right]
    d^{2m}\boldsymbol{\xi}.
\end{align}
 It is normalized as
\begin{align}
    \int_{\mathbb{R}^{2m}}
    W_\rho(\mathbf{r})\,d^{2m}\mathbf{r}
    =1.
\end{align}
Unlike an ordinary probability distribution, the Wigner function may take negative values. Wigner negativity is therefore a sufficient witness of nonclassicality. The converse does not hold: nonclassical states, such as squeezed Gaussian states, may have an everywhere nonnegative Wigner function.

\paragraph{Duhamel's formula.}
For two operators $A$ and $B$ for which the corresponding exponentials are well defined, Duhamel's formula gives
\begin{align}
    e^{tA}-e^{tB}
    =
    \int_0^t
    e^{(t-s)A}(A-B)e^{sB}\,ds .
\end{align}

\paragraph{Matrix norms, logarithmic norms, and stability.}
\label{sec:logarithmic-norms}
Three quantities play different roles in this paper. For a finite matrix
$\mathbf A$, its operator norm $\|\mathbf A\|_\infty$ is its largest singular
value, whereas its spectral abscissa is
$\alpha(\mathbf A)=\max_{z\in\sigma(\mathbf A)}\operatorname{Re}z$.
For a non-Hermitian matrix we use $\alpha$, not $\lambda_{\max}$.

The logarithmic norm measures the instantaneous growth allowed by a chosen
vector norm. For the Euclidean vector norm, it is
\begin{equation}\label{eq:logarithmic-norm-definition}
\mu(\mathbf A)=\lim_{h\downarrow0}
 \frac{\|\mathbb I+h\mathbf A\|_\infty-1}{h}
=\lambda_{\max}\!\left(\frac{\mathbf A+\mathbf A^\dagger}{2}\right).
\end{equation}
Despite its name, a logarithmic norm can be negative and is not an ordinary
norm. Higham gives a useful introduction to these definitions.\footnote{Nicholas
J. Higham, \emph{What Is the Logarithmic Norm?}, 18 January 2022,
\url{https://nhigham.com/2022/01/18/what-is-the-logarithmic-norm/}.}

To see its role here, let $x(t)=e^{t\mathbf A}x(0)$. Differentiating gives
\[
\frac{d}{dt}\|x(t)\|_2^2
=x(t)^\dagger(\mathbf A+\mathbf A^\dagger)x(t)
\leq2\mu(\mathbf A)\|x(t)\|_2^2.
\]
Consequently,
\begin{equation}\label{eq:logarithmic-semigroup-bound}
\alpha(\mathbf A)\leq\mu(\mathbf A),\qquad
\|e^{t\mathbf A}\|_\infty\leq e^{t\mu(\mathbf A)}\quad(t\geq0).
\end{equation}
A matrix is \emph{Hurwitz} when $\alpha(\mathbf A)<0$. In finite dimension
this is equivalent to $e^{t\mathbf A}\to0$; for every
$0<\lambda<-\alpha(\mathbf A)$ there is a finite $C_{\mathbf A,\lambda}$ with
\[
\|e^{t\mathbf A}\|_\infty\leq C_{\mathbf A,\lambda}e^{-\lambda t}.
\]
The prefactor need not be one: nonnormal matrices may initially amplify
vectors. Jordan blocks can also prevent a bound at the exact spectral rate.
For example, $\left(\begin{smallmatrix}-1&b\\0&-1\end{smallmatrix}\right)$
is Hurwitz for every real $b$, but its logarithmic norm is $-1+|b|/2$.

Hurwitz stability does imply contraction in a suitable weighted norm.
For any $\mathbf Q>0$, the matrix
\[
\mathbf P=\int_0^\infty e^{t\mathbf A^\dagger}\mathbf Qe^{t\mathbf A}\,dt>0
\]
satisfies $\mathbf A^\dagger\mathbf P+\mathbf P\mathbf A=-\mathbf Q$.
In $\|x\|_{\mathbf P}^2=x^\dagger\mathbf Px$, this gives exponential decay.
Returning to the Euclidean norm introduces a factor bounded by
$\sqrt{\|\mathbf P\|_\infty\|\mathbf P^{-1}\|_\infty}$.

\paragraph{Gronwall's inequality \cite{zeng2013gronwall}.}
We use Gronwall's inequality to convert integral or
differential inequalities into explicit bounds.
If a continuous function $u:[0,T]\to[0,\infty)$ satisfies
\[
u(t)\le A+B\int_0^t u(s)\,ds,
\qquad A,B\ge0,
\]
then $u(t)\le Ae^{Bt}$.
We also use the differential comparison form: if $u$
is locally absolutely continuous and
$u'(t)\le-\lambda u(t)+b(t)$ almost everywhere,
where $\lambda>0$ and $b$ is locally integrable, then
\[
u(t)\le e^{-\lambda t}u(0)
+\int_0^t e^{-\lambda(t-s)}b(s)\,ds.
\]

\paragraph{Conservative quantum dynamical semigroups.}
A quantum dynamical semigroup $(\mathcal T_t)_{t\ge0}$
acting on trace-class operators is called
\emph{conservative} if it preserves trace:
\[
\operatorname{Tr}\mathcal T_t(\rho)
=\operatorname{Tr}\rho
\qquad(\rho\ge0,\ t\ge0).
\]
Equivalently, its dual evolution on bounded observables
satisfies $\mathcal T_t^\dagger(I)=I$.
Thus conservativity means preservation of total probability,
not conservation of energy or photon number.
For unbounded bosonic generators, the formal Lindblad
expression alone does not guarantee this property;
we verify it for the realizations used below, following
\cite{chebotarev1998sufficient}.
Conservative completely positive evolution is
trace-norm contractive on Hermitian trace-class operators:
\[
\|\mathcal T_t(X)\|_1\le\|X\|_1,
\qquad X=X^\dagger.
\]

\paragraph{Well-defined bosonic Lindbladian dynamics.}
For a formal Lindblad expression involving potentially
unbounded Hamiltonian and jump operators to define
well-defined quantum dynamics on the full Hilbert space,
we require the following properties:
\begin{enumerate}
    \item \emph{Closed realization:}
The formal Lindblad expression, initially defined on
a dense subspace of the trace-class operators
$\mathfrak T_1(\mathcal H)$, admits an extension
$\mathcal L:D(\mathcal L)\subseteq\mathfrak T_1(\mathcal H)
\to\mathfrak T_1(\mathcal H)$.
This extension is closed, meaning for every sequence
$\rho_n\in D(\mathcal L)$ such that
$\rho_n\to\rho$ and $\mathcal L[\rho_n]\to X$
in trace norm, we have
$\rho\in D(\mathcal L)$ and $\mathcal L[\rho]=X$.

    \item \emph{Complete positivity:}
    The generated maps $\mathcal T_t$ are completely positive
    for every $t\geq 0$.

    \item \emph{Conservativity:}
    The generated maps $\mathcal T_t$ preserve trace,
    as defined above.

    \item \emph{Strong continuity:}
    For every trace-class operator $\rho$,
    \[
        \lim_{t\to 0^+}
        \|\mathcal T_t(\rho)-\rho\|_1=0.
    \]
\end{enumerate}
Together with the semigroup property
$\mathcal T_0=\mathrm{id}$ and
$\mathcal T_{t+s}=\mathcal T_t\mathcal T_s$,
these conditions ensure that the formal Lindblad
expression generates a strongly continuous CPTP semigroup.
For every $\rho\in\operatorname{Dom}(\mathcal L)$,
the evolution satisfies
\[
\frac{d}{dt}\mathcal T_t(\rho)
=
\mathcal L[\mathcal T_t(\rho)].
\]
The closedness of $\mathcal L$ is, in fact, automatic
once it is identified as the generator of a strongly
continuous semigroup; the nontrivial task is to establish
that this generator realizes the prescribed formal
Lindblad expression.

\section{Encoding circuits into Lindbladian fixed points with linear photon loss}
\label{sec:loss-only-bqp}

We now show that local observable estimation from bosonic Lindbladian fixed points is in principle
$\BQP$-hard when the Hamiltonian is a polynomial of constant degree and
the only dissipative terms are single-photon losses. 
Our proof combines several main ingredients. First, we consider a qubit Lindbladian construction of \cite{verstraete2009quantum} encoding the history state of a $\BQP$ circuit in the fixed point of a Lindbladian and revise it in a way which allows us to gain better understanding of the mixing time needed for the remainder of the section (see \Cref{subsec:loss-only-circuit}). We then embed this revised construction into bosonic jump operators of degree at most six (see \Cref{sec:qubit-to-boson-encoding}).
We then replace each such jump by a Hamiltonian coupling to a lossy
ancillary mode undergoing single-photon (degree-$1$) loss (see \Cref{subsec:loss-only-elimination}). This increases the Hamiltonian degree to at most seven.
To show that the latter step (of replacing high degree terms jump with Hamiltonians) approximately preserves the desired fixed point, we develop a
quantitative \textit{adiabatic-elimination theorem} for this setting. This is a powerful tool that (under mild conditions) allows us to replace arbitrary jump terms with single-photon loss. This approach requires ancillary modes, and also uses loss only on a constant fraction of the modes (in particular, just the ancilla modes). We finally add additional terms to the Hamiltonian to ensure uniqueness of the fixed point and prove that starting from the vacuum state the Lindbladian trajectory flows to the fixed point in polynomial time (see \Cref{sec:unique-fixed-point-full-domain} and \Cref{sec:preparation-from-vacuum}).

\subsection{Encoding circiuts into Lindbladians over finite-dimensional registers}
\label{subsec:loss-only-circuit}

The construction of Verstraete, Wolf, and
Cirac~\cite{verstraete2009quantum} associates a Lindbladian with a
quantum circuit $U_T\cdots U_1$. A clock register records the current
circuit time. Propagation jumps move the clock forward or backward
while applying the corresponding gate or its inverse, and
initialization jumps reset the data when the clock is at zero.
The resulting fixed point is a uniform mixture of the intermediate
computational states. In particular, measuring the clock at the final
time gives the output of the circuit. Appending identity gates amplifies
the probability of observing a completed computation from inverse polynomial to, for instance, a constant.

We use a slight modification of this construction for which the
trace-norm mixing estimate is particularly direct. Namely, we use
separate forward and backward propagation jumps, complete reset
channels at clock time zero, and clock dephasing. Let $n,T\geq1$, and
write
\begin{align}
    V_0:=\mathbb I,\quad
    V_t:=U_t\cdots U_1, \quad\ket{\psi_t}:=V_t\ket{0^n}.
    \label{eq:loss-only-circuit-states}
\end{align}
The Hilbert space that we use, is
$(\mathbb C^2)^{\otimes n}\otimes\mathbb C^{T+1}$.
On this space, consider the following collection of jump operators:
\begin{subequations}\label{eq:loss-only-dephasing}
\begin{align}
J_t
&:=
U_t\otimes\ket{t}\bra{t-1},
\qquad J_t^\dagger,
&&t\in[T],
\label{eq:loss-only-propagation}\\
R_{i,a}
&:=
(\ket{0}\bra{a})_i\otimes\proj{0},
&&i\in[n],\quad a\in\{0,1\},
\label{eq:loss-only-reset}\\
Z_t
&:=
\mathbb I\otimes\proj{t},
&&t\in\{0,\ldots,T\}.
\end{align}
\end{subequations}
We highlight a few differences with the original construction of \cite{verstraete2009quantum}: they considered $J_t+J_t^\dag$ as their jump operators, and did not consider decoherence terms (i.e., jump terms associated with $Z_t$). In our construction, however, every displayed operator, including each $J_t^\dagger$, is used as
a separate jump term. We can enumerate this collection as
$F_1,\ldots,F_m$, with
$m=3T+2n+1$, and $\norm{F_j}=1$.
Then, define
\begin{align}
\mathcal L_{\mathrm{circ}}[\rho]
:=
\sum_{j=1}^m\mathcal D(F_j)[\rho].
\label{eq:loss-only-logical-generator}
\end{align}
We will need quantitative control of convergence in trace norm.
For a mixing finite-dimensional Lindbladian $\mathcal L$, define 
\begin{align}
\beta_{\mathcal L}(t)
:=
\sup_{\substack{
    X=X^\dagger,\ \Tr X=0\\
    \norm{X}_1=1
}}
\norm{e^{t\mathcal L}[X]}_1,
\qquad
\tau(\mathcal L)
:=
\int_0^\infty\beta_{\mathcal L}(t)\,\mathrm dt.
\label{eq:loss-only-integrated-mixing}
\end{align}
To see the importance of $\tau(\mathcal L)$, which we will call \emph{trace-norm relaxation time}, note that we can express the inverse function as the integral of exponentiation i.e., $x^{-1} = \int_{t=0}^\infty e^{-tx}\, \mathrm dt$. 
Now, assume that $\mathcal L$ is invertible on traceless Hermitian operators. Let $\mathcal T=\{X: X=X^\dag, \Tr(X)=0\}$ (note that $\mathcal T$ is invariant under the action of any Lindbladian). This inversion trick allows us to write $\mathcal (\mathcal L|_{\mathcal T})^{-1} = -\int_{0}^{\infty}e^{t\mathcal L|_{\mathcal T}}\, \mathrm dt$, and hence, $\norm{(\mathcal L|_{\mathcal T})^{-1}}_1\le \tau(\mathcal L)$. Unlike a spectral gap alone, it
also accounts for the norm of the Lindbladian. In particular, it is straightforward to show that if $\norm{\mathcal L[\sigma]}_1\le \varepsilon$, then $\norm{\sigma-\rho_f}_1 \le \tau(\mathcal L) \varepsilon$, with $\rho_f$ being the unique fixed point of $\mathcal L$.

The construction of \cite{verstraete2009quantum} proved a bound on the gap of the Lindbladian, but that however is not sufficient for establishing a uniform convergence of all input states to the fixed point. In the next theorem we prove a slightly different variant of \cite{verstraete2009quantum} which allows us to establish a rapid and uniform convergence result, rather than merely a lower bound on the Lindbladian gap. We emphasize, however, that this does not imply that an analogous convergence result cannot be established for the construction of \cite{verstraete2009quantum}.

\begin{theorem}
\label{lem:loss-only-circuit-mixing}
The Lindbladian $\mathcal L_{\mathrm{circ}}$ has the unique fixed point
\begin{align}
    \rho_{\mathrm{hist}}
    :=
    \frac{1}{T+1}
    \sum_{t=0}^T
    \proj{\psi_t}\otimes\proj{t}.
    \label{eq:loss-only-history-state}
\end{align}
More specifically, for any initial state $\rho$, it is the case that
\begin{align}
\left\|
e^{t\mathcal L_{\mathrm{circ}}}[\rho]
-\rho_{\mathrm{hist}}
\right\|_1
\leq
2e^{-t}
+
2n\left(1-\frac{1-1/e}{2e}\right)^{\lfloor t/((T+1)^2+1)\rfloor}
+
\sqrt{T+1}\,e^{-4t/(T+1)^2},
\end{align}
Furthermore, we have
\begin{align}
\tau(\mathcal L_{\mathrm{circ}})
\leq
100(n+T+2)(T+1)^2.
\label{eq:loss-only-circuit-tau}
\end{align}
\end{theorem}

\begin{proof}
Similar to the proof of \cite{verstraete2009quantum}, let
\begin{align}
W=\sum_{t=0}^T V_t\otimes\proj{t}
\end{align}
and conjugate by $W$ to remove all circuit gates from the propagation
jumps:
\begin{align}
W^\dagger J_tW
=
I\otimes\ket{t}\bra{t-1}.
\end{align}
Similarly, $J_t^\dag$ gets conjugated to identity on the data register. The reset and dephasing operators are unchanged, since $V_0=I$. This basically means that the specific gates do not matter for our analysis, and we could just consider the identity circuit ($U_t=\mathbb I$ for all $t$). This is the same observation made in \cite{verstraete2009quantum}.
Consequently, it suffices to study the conjugated Lindbladian,
whose proposed fixed point is
\begin{align}
\rho_0
:=
\proj{0^n}\otimes
\frac{1}{T+1}\sum_{t=0}^T\proj{t}.
\end{align}
First, let $\mathcal E$ denote complete dephasing of the clock. In other words, if we vectorize operators, then $\mathcal E$ would be represented by $\sum_t \mathbb I \otimes \ket{t,t}\bra{t,t} - \mathbb I$. The part of the Lindbladian described by conjugated jump terms $\tilde J_t = \mathbb I \otimes \ket{t}\bra{t-1}$ can be represented in this vectorized space by the operator $\tilde{\mathcal J}_t=\mathbb I \otimes \sum_t \ket{t,t}\bra{t-1,t-1} - \frac12 \left(\sum_t \mathbb I \otimes \ket{t-1}\bra{t-1}\otimes \mathbb I + \mathbb I \otimes \mathbb I \otimes \ket{t-1}\bra{t-1}\right)$. Note that this clearly commutes with $\mathcal E$. It is also straightforward to show that parts of our generator that correspond to reset jump terms ($R_{i,a}$) also commute. We work in the conjugated frame, in which propagation acts only
on the clock. Write
\begin{align}
    \widetilde{\mathcal L}
    =
    \mathcal E+\mathcal P+\mathcal R,
\end{align}
where $\mathcal P$ contains the clock propagation terms (i.e., terms associated with the conjugated jump terms $\tilde J_t = \mathbb I \otimes \ket{t}\bra{t-1})$),
$\mathcal R$ contains the reset terms, and $\mathcal E$ generates
clock dephasing. Now, let $\Delta$ denote the channel that completely dephases the clock. In other words
\begin{align}
\Delta[\bullet]
:=
\sum_{t=0}^{T}
(\mathbb I\otimes\proj{t})\bullet(\mathbb I\otimes\proj{t}).
\end{align}
Note that $\Delta$ removes coherence between distinct clock
values, while retaining the data state associated with each clock
value. We have
$\mathcal E=\Delta-\mathcal I$. Due to $\Delta^2=\Delta$, we have
\begin{align}
e^{s\mathcal E}
=
\Delta+e^{-s}(\mathcal I-\Delta).
\label{eq:loss-only-dephasing-semigroup}
\end{align}
Intuitively, this means that the dynamics generated by $\mathcal E$ converges exponentially fast to the completely dephasing channel $\Delta$. We will use this property later. As $\mathcal E$ commutes with the rest of the dynamics, it follows that
\begin{align}
e^{s\widetilde{\mathcal L}}
=
e^{s(\mathcal P+\mathcal R)}
\circ e^{s\mathcal E}.
\end{align}
We can now assess the effect of the initial clock coherence.
For any initial state $\sigma$,
\begin{subequations}
\begin{align}
    &\left\|
        e^{u\widetilde{\mathcal L}}[\sigma]
        -
        e^{u\widetilde{\mathcal L}}[\Delta[\sigma]]
    \right\|_1
    \\
    &\qquad =
    \left\|
        e^{u(\mathcal P+\mathcal R)}
        \left[
            e^{u\mathcal E}[\sigma-\Delta[\sigma]]
        \right]
    \right\|_1
    \\
    &\qquad =
    e^{-u}
    \left\|
        e^{u(\mathcal P+\mathcal R)}
        [\sigma-\Delta[\sigma]]
    \right\|_1
    \\
    &\qquad \leq
    e^{-u}\|\sigma-\Delta[\sigma]\|_1
    \leq 2e^{-u}.
\end{align}
\end{subequations}
The second equality follows from
$\Delta[\sigma-\Delta[\sigma]]=0$ and
\eqref{eq:loss-only-dephasing-semigroup}.
The first inequality is trace-norm contractivity of the quantum
channel $e^{u(\mathcal P+\mathcal R)}$ on Hermitian operators.
The last inequality holds because $\sigma$ and $\Delta[\sigma]$
are both states.
Thus, the triangle inequality gives
\begin{align}
\left\|
e^{u\widetilde{\mathcal L}}[\sigma]-\rho_0
\right\|_1
\leq
2e^{-u}
+
\left\|
e^{u\widetilde{\mathcal L}}[\Delta[\sigma]]-\rho_0
\right\|_1,
\label{eq:loss-only-dephasing-reduction}
\end{align}
Thus, it remains to analyze initial states of the form
\begin{align}
    \eta(0)=\sum_{k=0}^{T}\sigma_k(0)\otimes\proj{k},
    \qquad
    \sigma_k(0)\geq0,
    \qquad
    \sum_{k=0}^{T}\Tr\sigma_k(0)=1.
\end{align}
Here $\sigma_k(0)$ is an unnormalized data state:
its trace is the probability that the clock initially equals $k$.
The dynamics preserves this form. The reason we have decomposed $\eta(0)$ in terms of $\sigma_k(0)$ is that, at time $t$, we have that $\eta(t)$ can be written in the same form for some other $\sigma(t)$. Please note that here we are using $u$ for the continuous time, and we will use $k$ for denoting the discrete time (corresponding to the stages of computation in the circuit model). In what follows, we establish the equations of evolution for $(\sigma_k(t))_k$:
\begin{subequations}
\begin{align}
\frac{\mathrm d}{\mathrm du}\sigma_0(u)
&=
\sigma_1(u)-\sigma_0(u)
+
\sum_{i=1}^{n}
\bigl(\mathcal R_i[\sigma_0(u)]-\sigma_0(u)\bigr),
\\
\frac{\mathrm d}{\mathrm du}\sigma_k(u)
&=
\sigma_{k-1}(u)+\sigma_{k+1}(u)-2\sigma_k(u),
\qquad 1\leq k\leq T-1,
\\
\frac{\mathrm d}{\mathrm du}\sigma_T(u)
&=
\sigma_{T-1}(u)-\sigma_T(u),
\end{align}
\end{subequations}
where we have used the shorthand notation$
\mathcal R_i[\bullet]
=
\proj{0}_i\otimes\Tr_i [\bullet]$, and without loss of generality, 
we have assumed $T\geq1$ (one can always append an identity gate).
To verify the reset contribution, the two reset jumps at clock
zero give
\begin{align}\label{eq:define-R_i}
\sum_{a=0}^{1}
\mathcal D(\ket{0}\bra{a}_i)[\bullet]
=
\mathcal R_i[\bullet]-\bullet.
\end{align}
Note that the dynamics generated by reset jumps is equivalent to applying a stochastic reset channel it creates the channel $e^{u(\mathcal R_i -\mathcal I)}=e^{-u} \mathcal I + (1-e^{-u})\mathcal R_i$ after applying it for time $u$. Hence, we can indeed interpret this continuous dynamics as applying a reset channel at times following a Poisson process (note that since $\mathcal R_i^2=\mathcal R_i$, we can write $e^{-u} \mathcal I + (1-e^{-u})\mathcal R_i=\sum_i e^{-u}\frac{u^m}{m!}\mathcal R_i^m$, and that $e^{-u}\frac{u^m}{m!}$ is the probability of having $m$ events until time $u$ from a Poisson process).

Next, note that
taking traces in the evolution equations and writing
$p_k(u):=\Tr\sigma_k(u)$ gives
\begin{subequations}\label{eq:evolution-of-ps}
\begin{align}
    \dot p_0&=p_1-p_0,\\
    \dot p_k&=p_{k-1}+p_{k+1}-2p_k,
    \qquad 1\leq k\leq T-1,\\
    \dot p_T&=p_{T-1}-p_T.
\end{align}
\end{subequations}
The reset terms disappear because they preserve trace.
These are precisely the equations for a continuous-time
symmetric random walk on $\{0,\ldots,T\}$, with rate one across
each edge. Transition rates do not depend on the data state. More explicitly, these equations give us a new interpretation of the dynamics: the joint dynamics can be generated by first
sampling this clock trajectory. Then, while the clock is at zero,
each qubit is independently reset at rate one; while it is
elsewhere, the data remain unchanged in the conjugated frame.
This description also allows initial correlations between the
data and the clock: sample the initial clock value $k$ with
probability $p_k(0)$, and use the conditional data state
$\sigma_k(0)/p_k(0)$ whenever $p_k(0)>0$. We first bound the time needed for the clock distribution to
approach uniformity. Let $\mathbf Q$ be the real symmetric matrix
specified by the \eqref{eq:evolution-of-ps}. We have
$\mathbf p(u)=e^{u\mathbf Q}\mathbf p(0)$. Also, it is a standard results that its eigenvalues are
\begin{align}
    -2\left(1-\cos\frac{j\pi}{T+1}\right),
    \qquad j=0,\ldots,T.
\end{align}
For example, these follow by substituting vectors with entries
$\cos(j\pi(k+\tfrac12)/(T+1))$ into the equations for $\mathbf Q$.
The eigenvalue zero corresponds to the uniform distribution
$\pi_k=1/(T+1)$. Every other eigenvalue is at most
$-\lambda_C$, where
\begin{align}
    \lambda_C
    =
    2\left(1-\cos\frac{\pi}{T+1}\right)
    =
    4\sin^2\frac{\pi}{2(T+1)}
    \geq
    \frac{4}{(T+1)^2}.
\end{align}
The last inequality uses
$\sin x\geq2x/\pi$ for $0\leq x\leq\pi/2$. Furthermore, since $\mathbf p(0)-\boldsymbol\pi$ has entries summing to zero, it has no component
along the eigenvector with eigenvalue zero. Expanding it in an
orthonormal eigenbasis of $\mathbf Q$ therefore gives
\begin{align}
    \|\mathbf p(u)-\boldsymbol\pi\|
    \leq
    e^{-\lambda_Cu}\|\mathbf p(0)-\boldsymbol\pi\|.
\end{align}
Moreover,
$\|\mathbf p(0)-\boldsymbol\pi\|^2
=\sum_kp_k(0)^2-1/(T+1)\leq1$.
Using Cauchy--Schwarz to pass from $\ell_2$ to $\ell_1$, we obtain
\begin{subequations}
\begin{align}
\|\mathbf p(u)-\boldsymbol\pi\|_{1}
&\leq
\sqrt{T+1}\,\|\mathbf p(u)-\boldsymbol\pi\|\\
&\leq
\sqrt{T+1}\,e^{-4u/(T+1)^2}.
\label{eq:loss-only-clock-mixing}
\end{align}
\end{subequations}
We next bound the probability that some qubit has not yet been
reset. To do so, let $S$ denote the first hitting time of clock value zero. Let $(C_u)_u$ be a stochastic process defined by the continuous-time Markov chain defined in \eqref{eq:evolution-of-ps}. In other words, $(C_u)_u$ models the status of the clock register. 
Starting from clock value $k$, and let us use the shorthand notation
\begin{align}
s_k:=\mathbb E[S|C_0=k]
\end{align}
to define the expectation values of $S$. Note that $s_0=0$, and $s_k=\frac12 +\frac12(s_{k+1}+s_{k-1})$ for all $k=1,\cdots,T-1$, and $s_{T}=1+s_{T-1}$. Solving this system of equation gives
\begin{align}
s_k=
\frac{k(2T+1-k)}{2}
\leq
\frac{T(T+1)}{2}.
\end{align}
Hence, for any initial distribution of $C_0$ we have $\mathbb E[S]\le \frac{T(T+1)}{2}$. Now, using Markov inequality (for any distribution over $C_0$), we get
\begin{align}\label{eq:bound-S}
\Pr\bigl(S\leq(T+1)^2\bigr)\geq 1 - \frac{\mathbb E[S]}{(T+1)^2}
=1-\frac{T}{2(T+1)}\ge \frac12.
\end{align}
After reaching zero, the clock remains there for one time unit
with probability $e^{-1}$, since its departure rate is one.\footnote{To see this, let $P_s:=\Pr\left[  C_t=0, \forall t\in[t_0,t_0+s]| C_{t_0}=s\right]$, and note that $\partial_s P_s = -P_s$, which follows directly from \ref{eq:evolution-of-ps}. Hence, $P_s = e^{-s}$.}
Conditional on this stay, a specified qubit is reset at least
once with probability $1-e^{-1}$ (due to discussions right after \eqref{eq:define-R_i}).
Thus, within an interval of length $(T+1)^2+1$, a specified qubit is
reset with probability at least
\begin{align}
q_{\mathrm{reset}}=\frac12\times e^{-1}\times (1-e^{-1}),
\end{align}
where the factor of $\frac12$ is due to \eqref{eq:bound-S} (i.e., we reach the clock state $0$), $e^{-1}$ is due to the probability that we stay at clock register $0$ for a unit time, and $1-e^{-1}$ the probability of having a reset channel applied in that unit time interval. Note that this bound holds, no matter what was the initial distribution of $C_0$ (as established in \eqref{eq:bound-S}). Moreover, this estimate remains valid conditional on the history before
the interval, because the clock is Markovian and the reset
processes have independent Poisson increments.
Applying it to successive intervals of length $(T+1)^2+1$ gives
\begin{align}
\Pr\left[\text{qubit }i\text{ has not been reset by time }u\right]
\leq
(1-q_{\mathrm{reset}})^{\lfloor u/((T+1)^2+1)\rfloor}.
\end{align}
By a union bound over all qubits, we obtain
\begin{align}
\Pr\left[\text{some qubit has not been reset by time }u\right]
\leq
n(1-q_{\mathrm{reset}})^{\lfloor u/((T+1)^2+1)\rfloor}.
\label{eq:loss-only-reset-tail}
\end{align}
Once every qubit has been reset, the data state is exactly
$\proj{0^n}$.
To combine the clock and reset estimates without assuming
independence, define
\begin{align}
    \zeta(u)
    :=
    \proj{0^n}\otimes
    \sum_{k=0}^{T}p_k(u)\proj{k}.
\end{align}
Compare the actual process with one having the same clock
trajectory and reset events, but with data initially in
$\proj{0^n}$. After every qubit has been reset, their data states
coincide. On any other trajectory, their trace-norm difference
is at most two. Averaging over trajectories gives
\begin{align}
    \|\eta(u)-\zeta(u)\|_1
    \leq
    2n(1-q_{\rm reset})^{\lfloor u/((T+1)^2 + 1)\rfloor}.
\end{align}
Finally, we use
\begin{align}
\|\zeta(u)-\rho_0\|_1
=
\sum_{k=0}^{T}|p_k(u)-\pi_k|
=
\|\mathbf p(u)-\boldsymbol\pi\|_{1},
\end{align}
and hence, putting things together with
\eqref{eq:loss-only-dephasing-reduction} proves
\begin{align}
\left\|
e^{u\widetilde{\mathcal L}}[\sigma]-\rho_0
\right\|_1
\leq
2e^{-u}
+
2n(1-q_{\mathrm{reset}})^{\lfloor u/((T+1)^2+1)\rfloor}
+
\sqrt{T+1}\,e^{-4u/(T+1)^2}.
\end{align}
Conjugating back to the initial reference, we achieve
\begin{align}
\left\|
e^{u\mathcal L_{\mathrm{circ}}}[\rho]
-\rho_{\mathrm{hist}}
\right\|_1
\leq
2e^{-u}
+
2n(1-q_{\mathrm{reset}})^{\lfloor u/((T+1)^2+1)\rfloor}
+
\sqrt{T+1}\,e^{-4u/(T+1)^2}.
\label{eq:loss-only-full-mixing}
\end{align}
In particular, every initial state converges to
$\rho_{\mathrm{hist}}$, which proves uniqueness of the stationary
state. Next, we show that this fast mixing feature can show finiteness of $\tau_{\mathcal L}$. To this end, recall the definition of $\beta_{\mathcal L}$ from \eqref{eq:loss-only-integrated-mixing}, and consider any operator $X$ that is included in that supremum.
For such an $X$, write $X=X_+-X_-$ where $X_{\pm}$ are both positive and are orthogonal.
Its positive and negative parts each have trace $\|X\|_1/2$,\footnote{This is due $X$ being traceless and that $\norm{X}_1 = \Tr(X_+) + \Tr(X_-)$.}
so that
\begin{align}
    X=\frac{\|X\|_1}{2}(\rho_+-\rho_-)
\end{align}
for states $\rho_+$ and $\rho_-$.
If $f(u)$ denotes the right-hand side of
\eqref{eq:loss-only-full-mixing}, the triangle inequality gives
\begin{align}
\begin{split}
\frac{\|e^{u\mathcal L_{\mathrm{circ}}}[X]\|_1}{\|X\|_1}
    &\leq
    \frac12\left(
        \|e^{u\mathcal L_{\mathrm{circ}}}[\rho_+]
            -\rho_{\mathrm{hist}}\|_1
        +
        \|e^{u\mathcal L_{\mathrm{circ}}}[\rho_-]
            -\rho_{\mathrm{hist}}\|_1
    \right)\\
    &\leq f(u).
\end{split}
\end{align}
Thus $\beta_{\mathcal L_{\mathrm{circ}}}(u)\leq f(u)$.
Using
\begin{align}
\begin{split}
    \int_0^\infty(1-c)^{\lfloor u/B\rfloor}\,du
    =
    B\sum_{m=0}^{\infty}(1-c)^m
    =
    \frac{B}{c},
\end{split}
\end{align}
we conclude that
\begin{align}
\begin{split}
    \tau(\mathcal L_{\mathrm{circ}})
    &:=
    \int_0^\infty\beta_{\mathcal L_{\mathrm{circ}}}(u)\,du\\
    &\leq
    2+\frac{2n((T+1)^2+1)}{q_{\mathrm{reset}}}+\frac{(T+1)^{5/2}}{4}\\
    &\leq
    100(n+T+2)(T+1)^2.
\end{split}
\end{align}
\end{proof}

The modifications we made above to \cite{verstraete2009quantum} made the trace-norm convergence estimate
particularly transparent: explicit clock dephasing controls
initial coherence, and the paired reset jumps implement exact
resets at state-independent Poisson times. 

The  
the trace-norm relaxation time
$\tau(\mathcal L_{\mathrm{circ}})$ will be useful for our subsequent perturbation arguments.
A spectral-gap estimate alone does not generally bound this
quantity with a polynomial prefactor. One must also control
transient behavior and the conversion to trace norm.
The present proof supplies that control directly, without a
factor depending exponentially on the number of data qubits. This is the reason for using the variant here.

\subsection{Encoding circuits into bosonic Lindbladians}
\label{sec:qubit-to-boson-encoding}

In this section, we address how the construction of \cref{lem:loss-only-circuit-mixing} can be encoded into a bosonic Lindbladian with only pure loss for the dissipative dynamics. To do so, we first construct a canonical encoding that uses dissipative terms of degree $6$ in \cref{subsec:loss-only-encoding}. Then, in \cref{subsec:loss-only-elimination}, we prove an adiabatic elimination theorem that completes the proof.

\subsubsection{Polynomial bosonic encoding}
\label{subsec:loss-only-encoding}

We next embed the qubit Lindbladian construction from the previous section into bosonic modes.
The relevant point is that finite-dimensional matrix units have simple polynomial
representatives on a single-photon sector. This gives an exact
encoding of the logical dynamics before any adiabatic
approximation is made.

For each data qubit $i$, introduce two modes
$\hat a_{i,0},\hat a_{i,1}$, and encode
\begin{align}\label{eq:dual-rail}
\ket{\bar{x}}_i
=
\hat a_{i,x}^\dagger\ket{00},
\qquad x\in\{0,1\}.
\end{align}
This is the dual-rail encoding, introduced in \cite{knill2001scheme}. This basically means $\ket{\bar{0}} = \ket{01}$ and $\ket{\bar{1}}=\ket{10}$.\footnote{The extra index $i$ is added in \eqref{eq:dual-rail}, to account for multiple qubits in a circuit. In other words, for each qubit location $i$, we have two bosonic modes $a_i,b$ for $b\in\{0,1\}$.} Hence, each qubit is encoded into two bosonic modes. Furthermore, we can represent Pauli operators as
\begin{subequations}
\begin{align}
\bar{Z}_i&:=a_{i,0}^\dag a_{i,0} - a_{i,1}^{\dag} a_{i,1},\\
\bar{Y}_i &:= -i(a_{i,0}^\dag a_{i,1} - a_{i,1}^\dag a_{i,0}),\\
\bar{X}_i&:=a_{i,0} a_{i,1}^\dag + a_{i,1} a_{i,0}^\dag.
\end{align}
\end{subequations}
For the clock register, we introduce $T+1$ clock modes $\hat c_0,\ldots,\hat c_T$, with
\begin{align}
    \ket{t}_C
    =
    \hat c_t^\dagger\ket{0^{T+1}}.
\end{align}
Define the register number operators
\begin{align}
    \hat N_i
    &:=
    \hat a_{i,0}^\dagger\hat a_{i,0}
    +
    \hat a_{i,1}^\dagger\hat a_{i,1},\\
    \hat N_C
    &:=
    \sum_{t=0}^T\hat c_t^\dagger\hat c_t.
\end{align}
Hence, we use the dual-rail encoding for the computational registers, and a one-shot encoding for the clock register. This means that if $U_t$ (in the circuit we want to simulate) acts on two qubits $i,j$, the jump operator we use becomes
\begin{align}
\overline J_t = \overline U_t \otimes \hat c_{t}^\dag \hat c_{t-1},
\end{align}
where
\begin{align}
\overline U_t
:=
\sum_{x,y,u,v\in\{0,1\}}
(U_t)_{xy,uv}
\bigl(\hat a_{i,x}^\dagger\hat a_{i,u}\bigr)
\bigl(\hat a_{j,y}^\dagger\hat a_{j,v}\bigr).
\label{eq:loss-only-gate-polynomial}
\end{align}
It is also straightforward to write down the reset and dephasing terms
\begin{align}
\begin{split}
\bar R_{i,a}
&:=
\hat a_{i,0}^\dagger\hat a_{i,a}
\hat c_0^\dagger\hat c_0,\\
\bar Z_t
&:=
\hat c_t^\dagger\hat c_t.
\label{eq:loss-only-encoded-dephasing}
\end{split}
\end{align}
Thus, degree-six jump polynomials already reproduce the
dissipative computation exactly inside $\mathcal C$.
The remaining task is to move these polynomials into the
Hamiltonian while retaining only linear loss as physical
dissipation.

\subsubsection{Replacing high-degree jump terms with pure photon loss using adiabatic elimination}
\label{subsec:loss-only-elimination}

We first establish an adiabatic elimination result for an arbitrary
finite-dimensional system. It will later be applied to the finite-dimensional data subspace
$\mathcal C$ (defined in \Cref{thm:loss-only-bqp-hardness}) of the bosonic Hilbert space. Although the system is finite-dimensional here, the
ancillary oscillators remain infinite-dimensional.

Let $ F_1,\ldots, F_m$ act on a \textit{finite-dimensional} space
$\mathcal H_S$, and define
\begin{align}
    \mathcal L[\rho]
    :=
    \sum_{j=1}^m\mathcal D(F_j)[\rho],
    \label{eq:loss-only-effective-data}
\end{align}
along with the norms
\begin{align}
f_1:=\sum_{j=1}^m\norm{F_j},\quad
f_2^2:=\sum_{j=1}^m\norm{F_j}^2.
\end{align}
For auxiliary bosonic modes $\hat b_1,\ldots,\hat b_m$, let
\begin{align}
\begin{split}
\hat N_B
&:=
\sum_{j=1}^m\hat b_j^\dagger\hat b_j,\\
\hat G
&:=
\frac12\sum_{j=1}^m
\left(
    \hat b_j^\dagger F_j+\hat b_jF_j^\dagger
\right),\\
\mathcal K_\gamma [\bullet]
&:=
-i[\hat G,\bullet]
+
\gamma\sum_{j=1}^m\mathcal D(\hat b_j)[\bullet].
\label{eq:loss-only-ancilla-generator}
\end{split}
\end{align}
Note $\mathcal{K}_\gamma$ is an intermediate construction which involves qubit-oscillator interactions. We will later replace this Lindbladian with a construction consisting of oscillator-oscillator interactions. 

In the remainder of this section we first show that, as an operator acting on an infinite-dimensional Hilbert space, the Lindbladian $\mathcal{K}_\gamma$ defines a well-posed, physically valid quantum Markovian evolution. This is established in \Cref{lem:K-well-defined}. We then show that the energy of ancillary modes of the fixed points of this Lindbladian are suitably bounded. This is achieved in \Cref{lem:loss-only-ancilla-moments} where besides justifying the
use of unbounded oscillator operators,  guarantees
existence of a fixed point of the full bosonic dynamics. Adiabatic elimination theorem is developed in \Cref{thm:loss-only-stationary-elimination}. Finally \Cref{thm:loss-only-bqp-hardness} combines these ingredients to encode the history state of a $\BQP$ computation in one of the fixed points of $\mathcal{K}_\gamma$. 

Before going to the details of the proof, we would like to borrow a notation from \cite{PhysRevA.98.013809,laiho2012characteristics} and use a technique of theirs that significantly simplifies calculations involving loss. For integers $n,p\in\mathbb N$, let
\begin{align}
(n)_p:=n(n-1)\cdots(n-p+1) = \frac{n!}{(n-p)!}.
\end{align}
We can similarly define $(\hat N)_p=\hat N(\hat N-1)\cdots (\hat N-p+1)$. This function has a particular property that
\begin{align}\label{eq:particular-np}
n((n)_p - (n-1)_{p})=p(n)_p.
\end{align}
Note how a uniform loss Lindbladian $\mathcal A=\sum_j \mathcal D[a_j]$ satisfies\footnote{This is due to the fact that $f(\hat N) a_j=a_j f(\hat N-\mathbb I)$ and that $\mathcal D(a_j)^\dag[\bullet] = a_j^\dag \bullet a_j - \frac12\{\hat n_j, \bullet\}$.}
\begin{align}\label{eq:differential-of-fN}
\mathcal A^\dag[f(\hat N)]= \hat N\left[ f(\hat N-\mathbb I) - f(\hat N) \right].
\end{align}
Using the identity \eqref{eq:particular-np} for the $(\bullet)_p$ function, we obtain
\begin{align}\label{eq:np-is-nice}
\mathcal A^\dag[(\hat N)_p] = -p(\hat N)_p.
\end{align}
This is a very interesting identity! We can view $\mathcal A^\dag$ on $f(\hat N)$ as a sort of differentiation (defined by \eqref{eq:differential-of-fN}). Then, in this form of differentiation, the functions $f(\bullet) = (\bullet)_p$ act as ``exponential forms,'' since their differentiation, gives the same function back. This allows us to simplify the evolution of moments of the number operator. Lastly, we highlight that for a single mode, the expression $(\hat n_i)_p=(a^\dag_i)^p a_i^p$ admits a simple form.
 
We now specify the domain on which $\mathcal K_\gamma$ admits a conservative realization (meaning it preserves trace-class operators) of $\mathcal K_\gamma$
before using its stationary equation. Let
\begin{align}
\mathcal D_{\mathrm{fin}} = \mathcal H_S \otimes \mathcal F_{\mathrm{fin}},
\end{align}
where
\begin{align}\label{def:f_fin}
\mathcal F_{\mathrm{fin}} = \mathrm{span}\left\{ \ket{n_1, \cdots, n_m}: n_j\in\mathbb N \right\}.
\end{align}
We now prove $\mathcal{K}_\gamma$ is a well-defined Lindbladian in the sense of the language we specified in \Cref{sec:preliminaries}.
\begin{lemma}
\label{lem:K-well-defined}
The Lindbladian $\mathcal K_\gamma$ admits a closed realization on the trace-class operators that generates a completely positive, trace-preserving semigroup \((\mathcal T_t)_{t\ge0}\), strongly continuous in trace norm.
\end{lemma}

\begin{proof}
All initial calculations are on
$\mathcal H_S\otimes\mathcal F_{\mathrm{fin},B}$. We first show that the Hamiltonian $\hat G$ is essentially self-adjoint. This can be achieved via the following bounds (where we have used the ladder-operator bounds and the Cauchy--Schwarz):
\begin{align}\label{eq:cs-simple}
\begin{split}
\|\hat G\psi\|
&\le f_1\|(N_B+I)^{1/2}\psi\|,\\
\bigl|\langle\psi,i[\hat G,N_B]\psi\rangle\bigr|
&\le \sum_j|\langle b_j\psi,F_j\psi\rangle|
\le f_2\|\psi\|\langle\psi,N_B\psi\rangle^{1/2}.
\end{split}
\end{align}
In particular, the second expression is at most
$\frac{f_2}{2}\langle\psi,(N_B+I)\psi\rangle$.
Since the finite-Fock space is a core for $N_B+I$, these inequalities
verify Nelson's commutator criterion
\cite{reed1975ii}.
Thus $\hat G$ is essentially self-adjoint on this space.
We use the same notation for its self-adjoint closure.
The first inequality also extends to $\mathrm{Dom}(N_B)$ by
photon-number cutoff approximation.

To apply the conservative-semigroup construction, we first show that
the no-jump operator
\begin{align}
\widetilde G_\gamma=-i\hat G-\frac{\gamma}{2}N_B,
\qquad
\operatorname{Dom}(\widetilde G_\gamma)
=\operatorname{Dom}(N_B),
\end{align}
generates a strongly continuous contraction semigroup (meaning that there exists the evolution map $(V_t)_t$, such that $\norm{V_t \ket{u}}\le \norm{\ket u}$ for all states $\ket u$).
The first inequality in \eqref{eq:cs-simple} implies that, for every $\delta>0$,
\begin{align}
\|\hat G\psi\|
\le \delta\|N_B\psi\|+c_\delta\|\psi\|,
\qquad
c_\delta=f_1+\frac{f_1^2}{4\delta}.
\end{align}
Taking $\delta=\gamma/4$, elementary applications of the triangle inequality give
\begin{align}
\|\widetilde G_\gamma\psi\|
\le \frac{3\gamma}{4}\|N_B\psi\|
+c_{\gamma/4}\|\psi\|,\quad
\frac{\gamma}{4}\|N_B\psi\|
\le \|\widetilde G_\gamma\psi\|
+c_{\gamma/4}\|\psi\|.
\end{align}
Hence the graph norms
$\|\psi\|+\|\widetilde G_\gamma\psi\|$ and
$\|\psi\|+\|N_B\psi\|$ are equivalent.
It follows that $\widetilde G_\gamma$ is closed and has the
finite-Fock space as a core \cite{kato1966perturbation}. Moreover, for $\ket\psi\in\mathrm{Dom}(N_B)$ we have
\begin{align}
2\operatorname{Re}
\langle\psi,\widetilde G_\gamma\psi\rangle
=-\gamma\langle\psi,N_B\psi\rangle\le0,
\end{align}
which means $\widetilde{G}_\gamma$ is a dissipative operator (in the operator theory language).
For positive real $y$, let
$R_y=(y+\gamma N_B/2)^{-1}$.
The first inequality in \eqref{eq:cs-simple} gives
\begin{align}
\|\hat G R_y\|
\le f_1\sup_{n\ge0}
\frac{\sqrt{n+1}}{y+\gamma n/2}
\le f_1\left(\frac1y+\frac1{\sqrt{2\gamma y}}\right)
\to0.
\end{align}
Hence, for sufficiently large $y$, the operator
$I+i\hat G R_y$ invertible. The factorization
\[
y-\widetilde G_\gamma
=(I+i\hat G R_y)(y+\gamma N_B/2)
\]
shows that $y-\widetilde G_\gamma$ is onto.
The Lumer--Phillips theorem \cite{lumer1961dissipative} now implies that
$\widetilde G_\gamma$ generates a strongly continuous contraction
semigroup $V_t$. Now, set $L_j=\sqrt{\gamma}\,b_j$.
Their domains contain $\operatorname{Dom}(\widetilde G_\gamma)$,
and
$
2\operatorname{Re}
\langle\psi,\widetilde G_\gamma\psi\rangle
+\sum_j\|L_j\psi\|^2=0
$.
By the polarization identity \cite{kato1966perturbation} we get the corresponding identity for pairs of vectors.
These are the hypotheses of the minimal quantum dynamical semigroup
construction in \cite[Section~2]{chebotarev1998sufficient}.

We verify conservativity using \cite[Corollary~4.5]{chebotarev1998sufficient}, with
\[
C=(1+\gamma)(N_B+I),
\qquad
\Phi=\gamma N_B.
\]
The operators $C$ and $\widetilde G_\gamma$ have the same domain.
For $P_R=\mathbf1_{\{N_B\le R\}}$ and every
$\ket\psi\in\operatorname{Dom}(N_B)$, the graph-norm estimates above give
$
P_R\ket\psi\to\ket\psi$, $
CP_R\ket\psi\to C\ket\psi$, and $
\widetilde G_\gamma P_R\ket\psi\to\widetilde G_\gamma\ket\psi
$.
Therefore, the finite-Fock vectors provide the simultaneous approximation
required by the corollary. On these vectors,
\[
-2\operatorname{Re}
\langle\psi,\widetilde G_\gamma\psi\rangle
=\langle\psi,\Phi\psi\rangle
\le\langle\psi,C\psi\rangle.
\]
Also, each $L_j$ preserves finite Fock support and therefore maps
this core into $\operatorname{Dom}(C)$.
Finally, direct calculation and the commutator bound give
\begin{align}
&2\operatorname{Re}
\langle C\psi,\widetilde G_\gamma\psi\rangle
+\sum_j\langle L_j\psi,CL_j\psi\rangle
\nonumber\\
&\quad=(1+\gamma)
\left(
\langle\psi,i[\hat G,N_B]\psi\rangle
-\gamma\langle\psi,N_B\psi\rangle
\right)
\nonumber\\
&\quad\le(1+\gamma)
\left(
-\frac{\gamma}{2}\langle\psi,N_B\psi\rangle
+\frac{f_2^2}{2\gamma}\|\psi\|^2
\right)
\le\frac{f_2^2}{2\gamma}\langle\psi,C\psi\rangle.
\end{align}
Here we used
$f_2ab\le\frac{\gamma}{2}b^2+
\frac{f_2^2}{2\gamma}a^2$.
All conditions of the corollary are therefore satisfied, so the
minimal semigroup is conservative. Its predual
$(\mathcal T_t)_{t\ge0}$ on trace-class operators is consequently
completely positive and trace preserving.

To verify trace-norm continuity, let $\rho=|u\rangle\langle u|$.
The minimal construction contains the no-jump contribution, so $
\mathcal T_t(\rho)\ge |V_tu\rangle\langle V_tu|$.
Since the difference is positive and $\mathcal T_t$ preserves trace $
\|\mathcal T_t(\rho)-\rho\|_1
\le
\bigl\||V_tu\rangle\langle V_tu|-|u\rangle\langle u|\bigr\|_1
+\|u\|^2-\|V_tu\|^2
\to0$.
Linearity, density of finite-rank operators, and trace-norm
contractivity extend this convergence to every trace-class operator.

The defining weak evolution equation of the minimal construction
\cite[Eq.~(2.4)]{chebotarev1998sufficient} identifies its generator with the displayed
Lindblad expression on finite-Fock rank-one operators.
Since the generator of a strongly continuous semigroup is closed,
this gives the claimed closed realization of $\mathcal K_\gamma$.
\end{proof}

\begin{lemma}[Fixed point moments]
\label{lem:loss-only-ancilla-moments}
The generator
$\mathcal K_\gamma$ has a fixed point with finite moments of every
order. Every fixed point $\sigma$ with
\begin{align}
    \Tr\bigl(\sigma(\hat N_B+I)^3\bigr)<\infty
\end{align}
satisfies
\begin{align}
    \Tr\bigl(\sigma(\hat N_B)_p\bigr)
    &\leq \left(\frac{f_2}{\gamma}\right)^{2p},
    \qquad p=1,2,3,
    \label{eq:loss-only-factorial-bound}\\
    \Tr\bigl(\sigma(\hat N_B+I)^3\bigr)
    &\leq
    1+7\left(\frac{f_2}{\gamma}\right)^2+6\left( \frac{f_2}{\gamma}\right)^4+\left( \frac{f_2}{\gamma}\right)^6.
    \label{eq:loss-only-third-moment}
\end{align}
\end{lemma}
Note that as a consequence of the lemma, we have that if $\gamma \geq f_2$, then the third moment in
\eqref{eq:loss-only-third-moment} is at most $15$.

\begin{proof}
Let us use the shorthand notation $q=\frac{f_2^2}{\gamma^2}$ throughout. Write
\begin{align}
    \mathcal B
    :=
    \sum_j\mathcal D(\hat b_j),
    \qquad
    M_p:=\Tr\bigl(\sigma(\hat N_B)_p\bigr),
    \qquad M_0:=1.
\end{align}
Due to \eqref{eq:np-is-nice}, we can write
\begin{align}
\mathcal B^\dagger[(\hat N_B)_p]
=
-p(\hat N_B)_p.
\label{eq:loss-only-factorial-loss}
\end{align}
In what follows, we use the operator $(\hat N_B)_p$ as a witness. This means, we want to set
\begin{align}
\Tr(\sigma \mathcal K^\dag((\hat N_B)_p))=0.
\end{align}
However, we cannot simply write this, as $\Tr(\sigma \mathcal K^\dag((\hat N_B)_p))$ might not coincide with the derivative of $\Tr(e^{\mathcal K t}[\sigma] (\hat N_B)_p)$, due to unboundedness of $(\hat N_B)_p$. But let us for now assume that $\Tr(\sigma \mathcal K^\dag((\hat N_B)_p))$ exists and is zero. We will later establish that it should exist.

For $\hat Q=(\hat N_B)_{p-1}$, the Hamiltonian contribution obeys
\begin{align}\label{eq:loss-only-factorial-drive}
\begin{split}
\left|
\Tr\bigl(\sigma [\hat G,(\hat N_B)_p]\bigr)
\right|
&\leq
p\sum_j
\left|
\Tr(\sigma\hat b_j^\dagger\hat QF_j)
\right|\\
&\leq
p\sum_j\norm{F_j}
\sqrt{
\Tr(\sigma\hat b_j^\dagger\hat Q\hat b_j)
\Tr(\sigma\hat Q)
}\\
&\leq
pf_2\sqrt{M_pM_{p-1}}.
\end{split}
\end{align}
The second inequality is Cauchy--Schwarz in the state $\sigma$.
For the last inequality, we used Cauchy--Schwarz over $j$ and
\begin{align}
    \sum_j
    \hat b_j^\dagger(\hat N_B)_{p-1}\hat b_j
    =
    (\hat N_B)_p.
\end{align}
Stationarity therefore gives
\begin{align}\label{eq:moment-recursion}
    p\gamma^2 M_p
    \leq
    p\gamma\,f_2\sqrt{M_pM_{p-1}},
\end{align}
which implies
\begin{align}
    M_p\leq qM_{p-1}.
\end{align}
Iteration proves~\eqref{eq:loss-only-factorial-bound}.
The identity
\begin{align}
    (\hat N_B+I)^3
    =
    (\hat N_B)_3+6(\hat N_B)_2+7\hat N_B+I
\end{align}
then proves~\eqref{eq:loss-only-third-moment}.

Now we get back to justifying the use of the unbounded operators
$(\hat N_B)_p$, for $p=1,2,3$, as witnesses in the stationary identities:
choose a smooth function $\chi:[0,\infty)\to[0,1]$ which
equals one on $[0,1]$ and zero on $[2,\infty)$, and define
\begin{align}
    f_R(n):=(n)_p\chi(n/R).
\end{align}
The observable $f_R(N)$ is bounded and supported on finitely
many ancillary occupation levels. It is straightforward to show that $\mathcal K^\dag_\gamma[f_R(\hat N_B)]$ is a bounded operator. 
Stationarity therefore gives
\begin{align}
    \langle\mathcal K_\gamma^\dagger[f_R(N)]
    \rangle_\sigma=0.
\end{align}
We show that this identity converges to the desired identity
for $(N)_p$ (or in other words, it is meaningful to take the limit of $R\to\infty$). To do so, for the loss contribution, we have
$
\mathcal B^\dagger[f_R(\hat N_B)]
=
N\bigl[f_R(\hat N_B-\mathbb I)-f_R(\hat N_B)\bigr]$ (where the value of $f_R(-1)$ is immaterial).
The smooth cutoff ensures that, uniformly in $R$,\footnote{Indeed, we one show that $C_p\le p+2\max_{x\in[n/R, (n+1)/R]} |\chi'(x)|$.}
\begin{align}\label{eq:growth-of-g}
    n\abs{f_R(n-1)-f_R(n)}
    \leq C_p(n+1)^p.
\end{align}
For every fixed occupation $n$, this expression agrees with
the uncut expression once $R$ is sufficiently large.
Since $\langle(\hat N_B+ \mathbb I)^p\rangle_\sigma<\infty$ (for $p=1,2,3$), dominated
convergence gives\footnote{To more elaborate, we have $g_R(n)=n[f_R(n-1)-f_R(n)]$ as a sequence of functions $g_R$. We have established in \eqref{eq:growth-of-g} that $|g_R(n)|\le h(n)$ for a summable function $h$ (i.e., $\sum_n p(n) h(n) \le \infty$, where $p(n)$ is the probability of measuring total $n$ photons when doing a number basis measurement on all modes of $\sigma$). Now, we can use the dominated convergence theorem to show that $\lim_{R\to\infty}\sum_n p(n) g_R(n) = \sum_n p(n) (\lim_{R\to\infty}g_R(n))$.}
\begin{align}
    \lim_{R\to\infty}
    \langle\mathcal B^\dagger[f_R(N)]\rangle_\sigma
    =
    -p\langle(N)_p\rangle_\sigma.
\end{align}
We need to show a similar identity for conjugation by the, Hamiltonian. Define
\begin{align}
e_R(n)
:=
f_R(n+1)-f_R(n)-p(n)_{p-1},
\end{align}
as an error.
This error vanishes for $n+1\leq R$ and satisfies
\begin{align}
    \abs{e_R(n)}
    \leq C_p(n+1)^{p-1}
\end{align}
uniformly in $R$.
Because the ancillary coupling is linear in $b_j$ and
$b_j^\dagger$, its commutator error is a finite linear
combination of $b_j^\dagger e_R(N)F_j$ and their adjoints.
Weighted Cauchy--Schwarz gives
\begin{align}
\abs{\langle b_j^\dagger e_R(N)F_j\rangle_\sigma}
\leq
\norm{F_j}
\sqrt{
\langle b_j^\dagger\abs{e_R(N)}b_j\rangle_\sigma
\langle\abs{e_R(N)}\rangle_\sigma
}.
\end{align}
Here
\begin{align}
    b_j^\dagger\abs{e_R(N)}b_j
    =
    \hat n_j\abs{e_R(N-I)}
    \leq C_p(N+I)^p,
\end{align}
and $\abs{e_R(N)}\leq C_p(N+I)^{p-1}$.
Both expectations therefore tend to zero by dominated
convergence (i.e., a similar argument as above). This proves convergence of the Hamiltonian
contribution as well.

We next prove existence of the fixed point. Let
\begin{align}
    \Pi_R:=\mathbb I_S\otimes\Pi_R,
\end{align}
where $\Pi_R=\sum_{\mathbf n\in\mathbb N^m: \norm{n}_1\le R} \ket{\mathbf n}\bra{\mathbf n}$,
and compress all ancillary lowering operators to $\Pi_R$.
The resulting generator acts on a finite-dimensional space and
therefore has a fixed point $\sigma_R$.
Equations~\eqref{eq:loss-only-factorial-loss} and
\eqref{eq:loss-only-factorial-drive} remain valid for the
compressed lowering operators. Consequently,
\begin{align}
    \Tr\bigl(\sigma_R(\hat N_B)_p\bigr)\leq q^p
    \qquad
    \text{for every integer }p\geq1,
    \label{eq:loss-only-uniform-cutoff-moments}
\end{align}
uniformly in $R$.

In particular, via Quantum Markov Inequality (\cite{chabaud2026bosonic}, or aka the gentle measurement lemma), we get
\begin{align}
\norm{\sigma_R-\Pi_K\sigma_R\Pi_K}_1
\leq
2\sqrt{\frac{q}{K+1}}.
\label{eq:loss-only-tightness}
\end{align}
For fixed $K$, the compressed operators lie in a compact
finite-dimensional set. Choosing convergent subsequences on
successively larger $\Pi_K$ and using
\eqref{eq:loss-only-tightness}, we obtain a subsequence of
$\sigma_R$ converging in trace norm to a state $\sigma$.

If an observable is supported on finitely many occupation levels,
the adjoint compressed generator agrees with
$\mathcal K^\dagger$ on that observable for all sufficiently
large $R$. Passing to the limit in its stationary expectation
therefore gives
\begin{align}
    \Tr\bigl(\sigma\mathcal K^\dagger[A]\bigr)=0
\end{align}
for every such observable. The moment bounds imply that the
Hamiltonian, anticommutator, and jump terms acting on $\sigma$
are trace class. Finite-occupation compressions also converge
in their graph norms, so $\sigma$ belongs to the generator
domain and satisfies $\mathcal K[\sigma]=0$.

The first-moment bound (i.e., $p=1$) also gives, on the finite-particle core,
\begin{align}
\mathcal K^\dagger[\hat N_B]
\leq
-\frac{\gamma}{2}\hat N_B+\frac{f_1^2}{2\gamma}\mathbb I.
\end{align}
\end{proof}

We next relate a fixed point of the full dynamics to the fixed
point of the effective generator. This concept is called the \textit{adiabatic elimination} in the literature, and have been used in other works such as \cite{reiter2012effective,kessler2012generalized,azouit2016adiabatic}. The proof works directly at
stationarity. In particular, it does not require exchanging
the limits of long evolution time and large $\gamma$.

\begin{theorem} [Adiabatic elimination]
\label{thm:loss-only-stationary-elimination}
Suppose that $\mathcal L$ has a unique fixed point $\rho_f$ and
$\tau(\mathcal L)<\infty$.
Let $\sigma_\gamma$ be a fixed point of $\mathcal K_\gamma$ with
$
    M_3
    =
    \Tr\bigl(\sigma_\gamma(\hat N_B+I)^3\bigr)
    <\infty
$. Let us also write $\overline\rho_\gamma:=\Tr_B\sigma_\gamma.$ to denote the reduced state of the fixed point on the data-clock modes.
Then
\begin{align}
    \norm{\overline\rho_\gamma-\rho_f}_1
    \leq
    48\,\tau(\mathcal L)f_1^3
    \sqrt{\frac{M_3}{\gamma^2}}.
    \label{eq:loss-only-stationary-error}
\end{align}
\end{theorem}
Before moving on to the proof, we highlight that the bound on $M_3\le 15$, established in \cref{lem:loss-only-ancilla-moments} under the assumption $f_2\le \gamma$, together with the above theorem, gives that 
\begin{align}
\norm{\overline\rho_\gamma-\rho_f}_1
\leq
\frac{200\,\tau(\mathcal L)f_1^3}{\gamma},
\label{eq:loss-only-stationary-error-simple}
\end{align}
where we have upper bounded $48\sqrt{15}$ by $200$ for convenience. We now introduce the proof.

\begin{proof}
Let $\mathcal V^\dag[\bullet]=i[\hat G, \bullet]$ represent the coherent part of $\mathcal K^\dag$ (indeed, $\mathcal K^\dag = \mathcal V^\dag +\gamma \mathcal B^\dag$).
A system observable $A$ is identified with $A\otimes \mathbb I_B$.
For such an observable, introduce the corrections
\begin{align}
\label{eq:loss-only-observable-correctors}
\begin{split}
    A_1
    &:=
    2\mathcal V^\dagger[A],\\
    A_2
    &:=
    2(\mathcal V^\dagger)^2[A]
    -
    \mathcal L^\dagger[A],
\end{split}
\end{align}
(it will become transparent later, why we have subtracted $\mathcal L^\dag$ in the second equation). It is transparent that the operator $A_1$ is linear in the ancillary ladder operators.
Since
\begin{align}
    \mathcal B^\dagger[\hat b_j]
    =
    -\frac12\hat b_j,
    \qquad
    \mathcal B^\dagger[\hat b_j^\dagger]
    =
    -\frac12\hat b_j^\dagger,
\end{align}
we have
\begin{align}
    \mathcal B^\dagger[A_1]
    = -\frac12 A_1 =
    -\mathcal V^\dagger[A].
    \label{eq:loss-only-first-corrector}
\end{align}
To analyze the second correction, note that
\begin{align}
    2(\mathcal V^\dagger)^2[A]
    =
    -2[\hat G,[\hat G,A]].
\end{align}
Let us expand $(\mathcal V^\dag)^2[A]$ (for a system operator $A$). This gives
\begin{align}\label{eq:wick}
2(\mathcal V^\dag)^2=-2[\hat G, [\hat G, A]] = C_0\otimes \mathbb I_B + \sum_{j,k}\left( (C_1)_{j,k} \otimes \hat b_j^\dag \hat b_k + (C_2)_{jk} \otimes \hat b_j \hat b_k + (C_3)_{jk} \hat b_j^\dag \hat b_k^\dag \right),
\end{align}
where $(C_{i})_{jk}$ are some system operators. Note that we have used the fact that any operator can be written in Wick's normal form to achieve \eqref{eq:wick}.\footnote{This means putting every creation operator on the left of any annihilation operator.} Because of this ordering, we can readily compute $C_0$ by taking vacuum expectations of \eqref{eq:wick}:
\begin{align}\label{eq:loss-only-vacuum-contraction}
\begin{split}
&\bra{0}_B
\bigl(-2[\hat G,[\hat G,A]]\bigr)
\ket{0}_B\\
&\qquad=
\sum_j
\left(
F_j^\dagger AF_j
-
\frac12\{F_j^\dagger F_j,A\}
\right)
=
\mathcal L^\dagger[A].
\end{split}
\end{align}
After $C_0$ is subtracted, $A_2$ is obtained (\eqref{eq:loss-only-observable-correctors}). Each such monomial (i.e., $\hat b_j^\dag \hat b_k$, $\hat b_j \hat b_k$, or $\hat{b}_j^\dag \hat b_k^\dag$) is an eigen-operator
of $\mathcal B^\dagger$ with eigenvalue $-1$. Hence
\begin{align}
\mathcal B^\dagger[A_2]=-A_2.
\label{eq:loss-only-second-corrector}
\end{align}
Combining~\eqref{eq:loss-only-first-corrector} and
\eqref{eq:loss-only-second-corrector} gives the exact identity\footnote{The unbounded corrected observable in
\eqref{eq:loss-only-exact-corrector-identity} is justified by
inserting smooth cutoffs in $\hat N_B$ on both sides.
Its oscillator degree is at most two, and the remainder has
degree at most three. The cutoff boundary terms have
expectations bounded by a constant, for fixed $A,F_j,\gamma$,
times
$\Tr\left(
\sigma_\gamma(\hat N_B+I)^3
\mathbf1_{\{\hat N_B\geq R-3\}}
\right)$,
which tends to zero. Thus, the finite third moment suffices to
remove the cutoffs in the stationary identity.}
\begin{align}
\begin{split}
\mathcal K^\dagger
\left[
A+\gamma^{-1}A_1+\gamma^{-2}A_2
\right]
&= (\mathcal V^\dag + \gamma \mathcal B^\dag)[A+\gamma^{-1}A_1+\gamma^{-2}A_2]\\
&=\underbrace{\mathcal V^\dag[A]}_{\frac12 A_1} + \underbrace{\mathcal B^\dag[A_1]}_{-\frac12 A_1} + \underbrace{\gamma^{-1} \mathcal V^\dag[A_1]}_{\gamma^{-1} A_2 + \gamma^{-1}\mathcal L^\dag[A]} + \gamma^{-2} \mathcal V^\dag[A_2] + \underbrace{\gamma^{-1} \mathcal B^\dag[A_2]}_{-\gamma^{-1} A_2}\\
&=\gamma^{-1}\mathcal L^\dagger[A]
+\gamma^{-2}\mathcal V^\dagger[A_2].
\label{eq:loss-only-exact-corrector-identity}
\end{split}
\end{align}
Taking its expectation in $\sigma_\gamma$, makes the left hand side zero, and yields
\begin{align}
    \Tr\bigl(A\mathcal L[\overline\rho_\gamma]\bigr)
    =
    -\gamma^{-1}
    \Tr\bigl(\sigma_\gamma\mathcal V^\dagger[A_2]\bigr).
    \label{eq:loss-only-stationary-residual-identity}
\end{align}
We now bound the right-hand side. Expanding the commutators gives
\begin{align}
    \mathcal V^\dagger[A_2]
    =
    -2i[\hat G,[\hat G,[\hat G,A]]]
    -
    i[\hat G,\mathcal L^\dagger[A]].
    \label{eq:loss-only-cubic-remainder}
\end{align}
In the expansion of $\hat G$, the sum of the norms of the system
coefficients is $f_1$. Note that we can write $-2i[\hat G, [\hat G, \hat [G, A]]] = -2i\left( \hat G^3 A-3\hat G^2 A \hat G + 3\hat G A\hat G^2 -A \hat G^3\right)$. Moreover, we can write each term, say $\hat G^3 A$, as $\sum_{\alpha,\beta,\gamma} \tilde F_\alpha \tilde F_\beta \tilde F_\gamma A \otimes (\hat \ell_\alpha \hat\ell_\beta \hat \ell_\gamma)$, where $\tilde F_\alpha$ iterate over $\frac{1}{2}F_j, \frac{1}{2}F_j^\dag$, and $\ell_\alpha$ are creation/annihilation operators. Note that we can bound the norm of the system part (e.g., for the case of $\hat G^3 A$, it would be $\sum_{\alpha,\beta,\gamma} \norm{\tilde F_\alpha \tilde F_\beta \tilde F_\gamma A}$), by $\left(\sum_j(\norm{F_j} + \norm{F_j^\dag})\right)^3 \norm{A} = f_1^3 \norm{A}$. Therefore, the first term in
\eqref{eq:loss-only-cubic-remainder} consists of words of length
three in the ancillary ladder operators, with the total coefficient
norm at most $16f_1^3\norm{A}$\footnote{The factor $16$ comes due to the sum of the terms in the expansion of $[\hat G, [\hat G, \hat [G, A]]]$ (i.e., $1+3+3+1=8$), together with the prefactor $2$ that the first term in \eqref{eq:loss-only-cubic-remainder} carries.}.
To analyze the second term, note that
\begin{align}\label{eq:bound-norm-LA}
\begin{split}
\|\mathcal L^\dagger[A]\|
&\le \sum_j\left(
\|F_j^\dagger A F_j\|
+\frac12\|F_j^\dagger F_j A\|
+\frac12\|A F_j^\dagger F_j\|
\right)\\
&\le \sum_j\left(1+\frac12+\frac12\right)
\|F_j\|^2\|A\|\\
&=2f_2^2\|A\|,
\end{split}
\end{align}
which allows us to write $
-i[\widehat G,\mathcal L^\dagger[A]]
=\sum_j\left(G^{(0)}_j\otimes b_j^\dagger+G^{(1)}_j\otimes b_j\right)
$, with $G^{(0)}_j=-\frac{i}2[F_j, \mathcal L^\dag[A]], G_j^{(1)}=-\frac{i}2[F^\dag_j, \mathcal L^\dag[A]]$, and we can obtain obtain
\begin{align}\label{eq:bound-term-2}
\begin{split}
\sum_j\bigl(\|G^{(0)}_j\|+\|G^{(1)}_j\|\bigr)
&\le 2\sum_j\|F_j\|\,\|\mathcal L^\dagger[A]\|\\
&\le 4f_1f_2^2\|A\|\\
&\le 4f_1^3\|A\|.
\end{split}
\end{align}
Now, let us introduce an intermediary lemma that becomes useful later as well: 
\begin{lemma}[Lemma A.4 of \cite{chabaud2025energy}]\label{lem:word-w}
Consider bosonic modes $a_1, \cdots, a_m$ with $\hat N \ sum_i \hat n_i$. For any ladder word $w$ (i.e., a monomial in creation and annihilation operators), we have
\begin{align}
\norm{w (\hat N+1)^{-d/2}} \le \sqrt{d!}.
\end{align}
\end{lemma}
\begin{proof}
For any system ladder-operator word $w$ we have \(w=b_d\cdots b_1\), where each \(b_r\) is some \(a_j\) or \(a_j^\dagger\). For a Fock basis vector \(|\mathbf n\rangle\). Now, consider applying $w$ on $\ket{\mathbf n}$. Right before applying the $r$-th creation/annihilation operator, we have at most $|\mathbf n| + r-1$ photons overall. This means the $r$-th term can increase the norm by at most $\sqrt{|\mathbf n|+r}$. As a result, 
\begin{align}
\norm{w\ket{\mathbf n}} \le \prod_{r=1}^d \sqrt{|\mathbf n|+r}.
\end{align}
This also provides
\begin{align}
\left\|w(N+1)^{-d/2}|\mathbf n\rangle\right\|^2
\le
\prod_{r=1}^{d}\frac{|\mathbf n|+r}{|\mathbf n|+1}
\le
\prod_{r=1}^{d}r
=d!,
\end{align}
On the other hand, note that $w$ sends initially orthogonal Fock basis states to orthogonal Fock states. Hence, if $c_{\mathbf n}:= \norm{w\ket{\mathbf n}}$, for any finite Fock-support state $\ket\psi = \sum_{\mathbf n} \psi_{\mathbf n} \ket{\mathbf n}$, we have
\begin{align}
\left\|w(\hat N+1)^{-d/2}\psi\right\|^2
=
\sum_{\mathbf n}|\psi_{\mathbf n}|^2
\frac{c_{\mathbf n}^2}{(|\mathbf n|+1)^d}
\le d!\,\|\psi\|^2.
\end{align}
By density, the operator extends uniquely to a bounded operator satisfying
$\left\|w(\hat N+1)^{-d/2}\right\|\le\sqrt{d!}
$. We highlight that the bound is tight as it is achieved for $w=(a_1^{\dag})^d$ acting on vacuum.
\end{proof}

Using \cref{lem:word-w}, together with the bounds on the system-operator parts of the first and the second term \eqref{eq:bound-term-2}, for $d=1,3$, we obtain
\begin{align}\label{eq:VA2}
\begin{split}
\norm{
\mathcal V^\dagger[A_2](\hat N_B+I)^{-3/2}
}
&\leq
(16\sqrt6+4)f_1^3\norm{A}\\
&\leq
48f_1^3\norm{A}.
\end{split}
\end{align}
Cauchy--Schwarz in the state $\sigma_\gamma$ now gives
\begin{align}
\left|
\Tr\bigl(\sigma_\gamma\mathcal V^\dagger[A_2]\bigr)
\right|
\leq
48f_1^3\norm{A}\sqrt{M_3}.
\end{align}
Taking the supremum over Hermitian observables with
$\norm{A}\leq1$ in
\eqref{eq:loss-only-stationary-residual-identity}, we conclude
\begin{align}
    \norm{\mathcal L[\overline\rho_\gamma]}_1
    \leq
    48f_1^3\sqrt{\frac{M_3}{\gamma^2}}.
    \label{eq:loss-only-residual-bound}
\end{align}
Finally, $\mathcal L[\overline\rho_\gamma]$ is traceless and
Hermitian. Since the effective semigroup converges to $\rho_f$,
\begin{align}
    \overline\rho_\gamma-\rho_f
    =
    -\int_0^\infty
    e^{u\mathcal L}
    \left[\mathcal L[\overline\rho_\gamma]\right]
    \,\mathrm du.
\end{align}
Together with~\eqref{eq:loss-only-integrated-mixing} and
\eqref{eq:loss-only-residual-bound}, this proves
\eqref{eq:loss-only-stationary-error}.
If $\gamma^2\geq f_1^2$, then
\cref{lem:loss-only-ancilla-moments} gives $M_3\leq15$.
The inequality $48\sqrt{15}<200$ proves
\eqref{eq:loss-only-stationary-error-simple}.
\end{proof}

The preceding theorem controls every fixed point with the stated
moment bound. We do not show the uniqueness of the fixed point in this section.

We now combine the encoding and elimination results. We have the subspace
\begin{align}
\mathcal C = \mathrm{span}\left\{ \left( \prod_{i=1}^n \hat a_{i,x_i}^\dag\right) \hat c_t^\dag \ket{0^{2n+T+1}}: x_i\in\{0,1\}, t=0,\cdots, T \right\},
\end{align}
and the encoding isometry
\begin{equation}
 E\bigl(|x_1\cdots x_n\rangle\otimes|t\rangle\bigr)
 =\left(\prod_{i=1}^n a_{i,x_i}^\dagger\right)c_t^\dagger
       |0^{2n+T+1}\rangle.
 \label{eq:uniq-encoding}
\end{equation}
We study the fixed points that are supported
on $\mathcal C$ together with the ancillary Fock space, and
having finite third ancillary photon-number moment.

\begin{proposition}
\label{thm:loss-only-bqp-hardness}
Given an $n$-qubit circuit with $T$ one- and two-qubit gates and
an accuracy parameter $\varepsilon>0$, one can efficiently
construct a bosonic Lindbladian
\begin{align}
    \overline{\mathcal K}_\gamma[\sigma]
    =
    -i[\hat H,\sigma]
    +
    \gamma\sum_{j=1}^m
    \mathcal D(\hat b_j)[\sigma]
    \label{eq:loss-only-final-generator}
\end{align}
with the following properties:
\begin{enumerate}
    \item
    The number of modes and monomials is $O(n+T)$.
    The Hamiltonian $\hat H$ has degree at most seven,
    and every jump operator is linear single-photon loss.

    \item
    The encoded data-clock sector $\mathcal C$ is invariant.
    There is a fixed point supported on
    $\mathcal C\otimes\mathcal F_B$ with finite moments of every
    order.

    \item
    Every fixed point $\sigma_\gamma$ supported on this sector
    and having finite third ancillary moment satisfies
    \begin{align}
        \norm{
            \Tr_B\sigma_\gamma
            -
            E\rho_{\mathrm{hist}}E^\dagger
        }_1
        \leq\varepsilon.
        \label{eq:loss-only-final-accuracy}
    \end{align}

    \item
    All coefficients and loss rates are bounded by
$\poly(n,T,\varepsilon^{-1})$.
The ancillary moments satisfy
\begin{align}
\Tr(\sigma_\gamma\hat N_B)\leq1,
\quad
\Tr\bigl(\sigma_\gamma(\hat N_B+I)^3\bigr)\leq15.
\label{eq:loss-only-final-energy}
    \end{align}
\end{enumerate}
\end{proposition}

\begin{proof}
Let $\overline F_1,\ldots,\overline F_m$ be the polynomials constructed by the dual-rail encoding from
$F_1,\cdots, F_m$, and introduce one
ancillary mode $\hat b_j$ for every jump. Define
\begin{align}
    \hat H
    :=
    \frac{1}{2}
    \sum_{j=1}^m
    \left(
        \hat b_j^\dagger\overline F_j
        +
        \hat b_j\overline F_j^\dagger
    \right).
    \label{eq:loss-only-final-hamiltonian}
\end{align}
Since $\deg(\overline F_j)\leq6$, the Hamiltonian has degree at
most seven. The only dissipative terms in
\eqref{eq:loss-only-final-generator} are
$\gamma\mathcal D(\hat b_j)$.

Every $\overline F_j$ and its adjoint preserve $\mathcal C$.
The loss operators act only on the additional ancillary modes.
Hence $\mathcal C\otimes\mathcal F_B$ is invariant.
On this subspace,
\eqref{eq:loss-only-final-generator} is precisely the generator
of~\eqref{eq:loss-only-ancilla-generator} for the logical
operators $F_j$. In particular,
\begin{align}
f_2^2=m,
\qquad
f_1=m,
\qquad
m=3T+2n+1.
\end{align}
Choose $\gamma$ to satisfy
\begin{align}
    \gamma^2
    \geq
    \max\left\{
        m,\,
        \left(
            \frac{200\tau_0m^3}{\varepsilon}
        \right)^2
    \right\},
    \qquad
    \tau_0=100(n+T+2)(T+1)^2.
    \label{eq:loss-only-kappa-choice}
\end{align}
Existence follows from
\cref{lem:loss-only-ancilla-moments}.
Applying
\cref{lem:loss-only-circuit-mixing,thm:loss-only-stationary-elimination}
gives~\eqref{eq:loss-only-final-accuracy}.
The same moment lemma gives
\eqref{eq:loss-only-final-energy}.

The encoded data and clock contain exactly $n+1$ photons.
Thus their energy restriction is exact, and the total photon
number has polynomially bounded moments through order three.
The value of $\gamma$ in
\eqref{eq:loss-only-kappa-choice} is polynomial in
$n,T,\varepsilon^{-1}$, as are the Hamiltonian coefficients.
One may choose $\gamma$ to be a sufficiently large perfect
square, so that $\gamma/2$ has a simple finite
description. A fixed universal gate set gives efficiently
describable gate coefficients.
\end{proof}

\subsection{A unique fixed point with linear loss}
\label{sec:unique-fixed-point-full-domain}
The construction in \cref{thm:loss-only-bqp-hardness} does not show that the fixed point encoding the history state is unique. In this section, we alter the construction, by adding extra lossy modes, and coupling these new ancillas with data and clock modes. In particular, we add one lossy mode for each data register (which contains two modes) and for the clock (i.e., just one ancilla for all the clock modes).
The additional couplings force every stationary state into the
one-photon encoding. Moreover, we end up adding a diagonal Hamiltonian to ensure that the
modified dynamics is well-defined on the entire Fock space. Concretely, for data modes $a_{i,0},a_{i,1}$ and clock modes $c_0,\ldots,c_T$, set
\begin{equation}
 N_i=a_{i,0}^\dagger a_{i,0}+a_{i,1}^\dagger a_{i,1},\qquad
 N_C=\sum_{t=0}^T c_t^\dagger c_t,\qquad
 \mathcal R=\{1,\ldots,n,C\}.
\end{equation}
Here $\mathcal R$ indexes the $n$ two-mode data registers and the
whole of $(T+1)$-mode clock register.

The original circuit operators $\overline F_1,\ldots,\overline F_m$
satisfy $\overline F_jE=E F_j$, preserve every $N_r$ (for $r\in\mathcal R$), and have degree
at most six. We retain
\begin{equation}
 \overline H=\frac12\sum_{j=1}^m
       (b_j^\dagger\overline F_j+b_j\overline F_j^\dagger).
\end{equation}

For each $r\in\mathcal R$, introduce a mode $s_r$ and put
\begin{equation}
d_{i,0}=a_{i,0},\qquad d_{C,0}=c_0,\qquad
A_r=d_{r,0}^\dagger(1-N_r),
\label{eq:uniq-fill}
\end{equation}
and we define the additional coupling Hamiltonian
\begin{align}
H_{\mathrm{add}}=g\sum_{r\in\mathcal R}
(s_r^\dagger A_r+s_rA_r^\dagger) = g\sum_{i=1}^n\left[s_r^\dag a_{i,0}^\dag(1-N_i) + h.c.  \right] + g\left[ s_C^\dag c_0^\dag(1-N_C)+h.c. \right]
\end{align}
where $g>0$ is fixed. We have studied this form of Hamiltonian with loss in \cref{ex:2-mode-1-engineering}, showing that it preserves the state $\ket{10}$ on the data-ancilla modes. Hence, the intuition behind addition of these terms it to ensure the fixed point lies in $\mathcal C$. 

\begin{figure}
    \centering
    \begin{tikzpicture}[x=1cm,y=1cm, text=figInk]
  \begin{scope}[on background layer]
    \path[panel, fill=auxViolet!2] (-.55,.25) rectangle (2.15,7.35);
    \path[panel, fill=dataTeal!1] (3.75,.25) rectangle (9.45,7.35);
    \path[panel, fill=loadRose!2] (11.65,.25) rectangle (14.05,7.35);
  \end{scope}

  \node[group title] at (.80,8.00) {Adiabatic elimination ancilla modes};
  \node[group count] at (.80,7.64) {$m$ modes};
  \node[group title] at (6.60,8.00) {System modes: data + clock};
  \node[group count] at (6.60,7.64) {$2n+T+1$ modes};
  \node[group title] at (12.85,8.00) {Loading/uniqueness ancilla modes};
  \node[group count] at (12.85,7.64) {$n+1$ modes};

  \node[small label, text=dataTeal!85!black] at (6.60,6.93)
    {DATA \quad $2n$ modes};
  \foreach \row/\idx in {6.10/1,4.92/2,2.95/n}{
    \draw[register] (6.60,\row) ellipse[x radius=1.91cm,y radius=.54cm];
    \node[data mode] (a\idx0) at (5.55,\row) {$a_{\idx,0}$};
    \node[data mode] (a\idx1) at (7.65,\row) {$a_{\idx,1}$};
    \node[loading mode] (s\idx) at (12.85,\row) {$s_{\idx}$};
    \draw[loading link] (8.51,\row) -- (s\idx.west);
  }
  \node at (5.55,3.96) {$\vdots$};
  \node at (7.65,3.96) {$\vdots$};
  \node at (12.85,3.96) {$\vdots$};

  \draw[figLine, line width=.5pt] (4.12,2.21) -- (9.08,2.21);
  \node[small label, text=clockGold!85!black] at (6.60,1.95)
    {CLOCK \quad $T+1$ modes};
  \draw[register] (6.60,1.16) ellipse[x radius=2.40cm,y radius=.54cm];
  \node[clock mode] (c0) at (4.92,1.16) {$c_0$};
  \node[clock mode] (c1) at (6.00,1.16) {$c_1$};
  \node at (7.17,1.16) {$\cdots$};
  \node[clock mode] (cT) at (8.28,1.16) {$c_T$};
  \node[loading mode] (sC) at (12.85,1.16) {$s_C$};
  \draw[loading link] (9.00,1.16) -- (sC.west);
  \node[small label] at (12.85,.49) {loss rate $\Gamma$};

  \node[auxiliary mode] (b1) at (.80,6.10) {$b_1$};
  \node[auxiliary mode] (b2) at (.80,4.92) {$b_2$};
  \node at (.80,3.96) {$\vdots$};
  \node[auxiliary mode] (bm) at (.80,2.95) {$b_m$};
  \node[small label, align=center] at (.80,1.70)
    {one per jump\\[3pt]$\overline F_j$};
  \node[small label] at (.80,.49) {loss rate $\gamma$};

  \draw[computation link] (2.15,4.42) -- (3.75,4.42);
  \node[text=auxViolet!85!black] at (2.95,4.79) {$\overline H$};
  \node[text=loadRose!85!black] at (10.55,6.93) {$H_{\mathrm{add}}$};

  \draw[figLine, line width=.6pt] (-.55,-.13) -- (14.05,-.13);
  \node[anchor=west, text=auxViolet!85!black, font=\small]
    at (-.45,-.80)
    {$\displaystyle \overline H=\frac12\sum_{j=1}^{m}
      \bigl(b_j^\dagger\overline F_j+\mathrm{h.c.}\bigr)$};
  \node[anchor=east, small label, align=right] at (13.95,-.80)
    {$\overline F_j$ acts on data and clock};
  \node[anchor=west, text=loadRose!85!black, font=\small]
    at (-.45,-1.95)
    {$\displaystyle H_{\mathrm{add}}=g\left[
      \sum_{i=1}^{n}s_i^\dagger a_{i,0}^\dagger(I-N_i)
      +s_C^\dagger c_0^\dagger(I-N_C)+\mathrm{h.c.}\right]$};
\end{tikzpicture}
    \caption{The graph of intreraction for the $\mathsf{BQP}$-hard instance in \cref{sec:unique-fixed-point-full-domain}, and \cref{eq:uniq-generator}. We highlight that $m$ is the number of jump terms in the qubit construction introduced in \cref{eq:loss-only-logical-generator}, according to which $m=3T+2n+1$. The overall mode count is $5n+4T+3$. The figure shows different interactions considered in our construction: $\overline{H}$ is the interaction between the data/clock and lossy modes on the left. This comes from the \textit{adiabatic elimination}, which we had introduced in \cref{eq:loss-only-ancilla-generator}. Finally, we added ancillary modes $\hat s_r$ for $r\in\{1,\cdots,n,C\}$ to ensure the uniqueness of the fixed point, and to make sure that the fixed point stays within $\mathcal C$. Our calculations in \cref{sec:preparation-from-vacuum} explicitly show how interactions with these ancillas (or more precisely, the interaction $H_{\mathrm{add}}$) results in one photon in each pair of $a_{i,0}, a_{i,1}$ modes, and also the single photon in the clock registers. We highlight that on top of these interactions, we have the local interactions $\sum_{i,b} (a_{i,b}^\dag a_{i,b})^6 + \sum_{i=0}^m (c_{i}^\dag c_i)^6$ to ensure that our Lindbladian generates a conservative Markov semigroup (which is a technical constraint to have a well-defined evolution associated with our Lindbladian).}
    \label{fig:BQP-hard}
\end{figure}
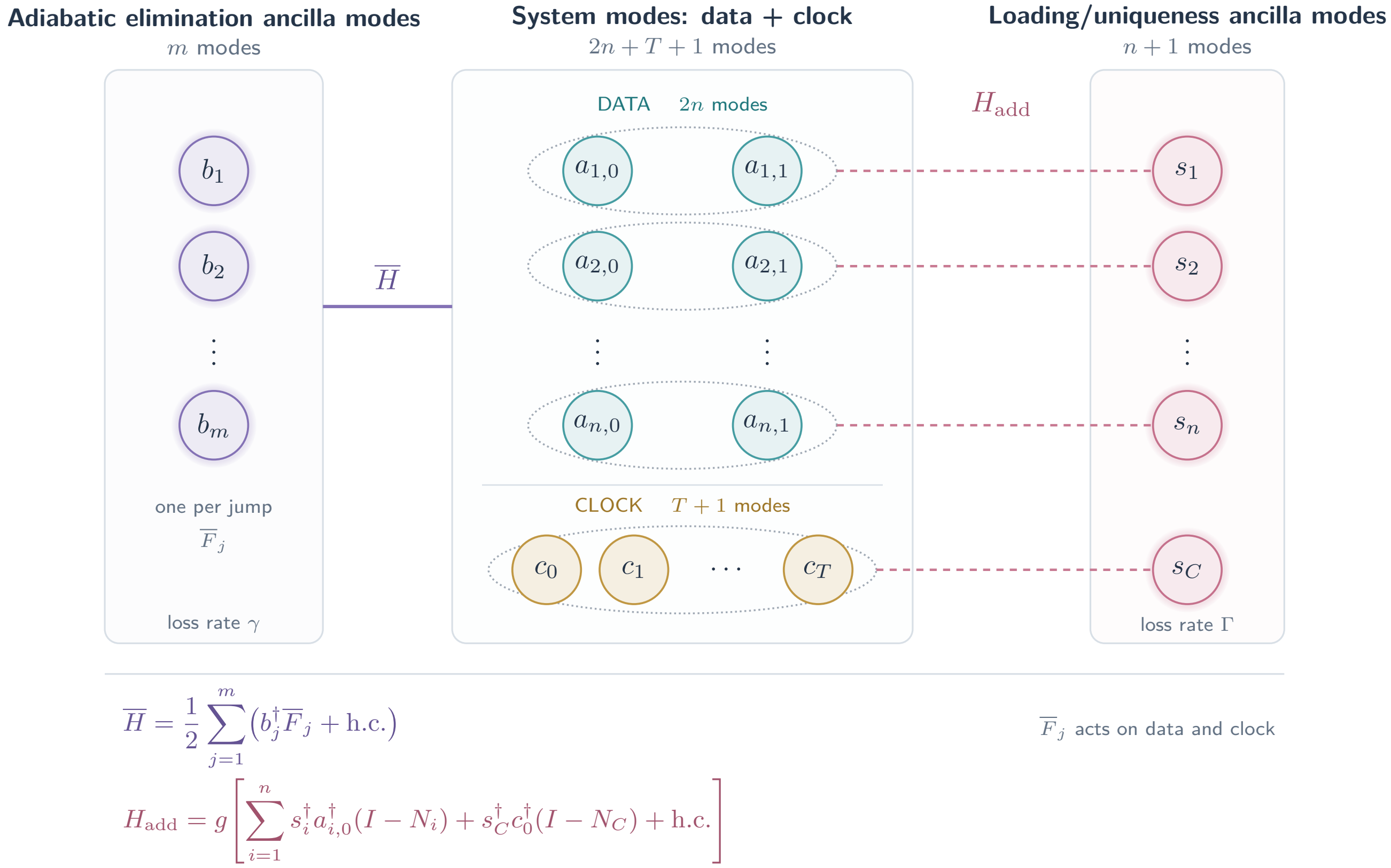

Write $B$ for the original auxiliary modes,
$S$ for the new modes, and
\begin{equation}
 N_B=\sum_j \hat b_j^\dagger \hat b_j,\qquad
 N_{\mathrm{new}}=\sum_r \hat s_r^\dagger \hat s_r,\qquad
 N_{\mathrm{aux}}=N_B+N_{\mathrm{new}}.
\end{equation}
Separately enumerate all $M=2n+T+1$ data and clock modes as
$\hat d_1,\ldots,\hat d_M$, and define
\begin{equation}
 \mathsf W=\mathbb I+\sum_{\ell=1}^{M}(\hat d_\ell^\dagger \hat d_\ell)^6.
 \label{eq:uniq-confining-weight}
\end{equation}
We now propose the generator $\widetilde{\mathcal K}_{\kappa,\eta}$ defined as follows:
\begin{align}
\begin{split}
H_\eta&=\overline H+H_{\mathrm{add}}+\eta\mathsf W,\\
\widetilde{\mathcal K}_{\gamma,\eta}(\rho)
&=-i[H_\eta,\rho]
+\gamma\sum_{j=1}^m\mathcal D[\hat b_j](\rho)
+\Gamma\sum_{r\in\mathcal R}\mathcal D[\hat s_r](\rho),
 \label{eq:uniq-generator}
\end{split}
\end{align}
where $\Gamma>0$ is fixed. We specify $\eta$ below.
Note that the Hamiltonian has degree at most twelve, every loss operator is
linear, and the number of modes and monomials remains $O(n+T)$. We now prove that a sufficiently large, polynomially bounded choice of \(\eta\) makes the proposed dynamics (generated by $\widetilde{\mathcal K}_{\gamma,\eta}$) well-defined and trace-preserving on the full Fock space. We refer to \cref{fig:BQP-hard} for an illustration of the full construction in \cref{eq:uniq-generator}.

\begin{lemma}
\label{lem:uniq-conservative}
For a sufficiently large, explicitly selectable $\eta>0$,
\eqref{eq:uniq-generator} defines a conservative quantum Markov
semigroup on the full Fock space. For fixed $g,\Gamma>0$, this
choice of $\eta$ is polynomially bounded in the circuit parameters
and in $\gamma+\gamma^{-1}$.
\end{lemma}

\begin{proof}
In this proof, let $\mathcal F_{\mathrm{fin}}$ denote the linear span of the joint
Fock basis of all system modes and both sets of auxiliary modes, and set
\begin{equation}
 H_{\mathrm{int}}=\overline H+H_{\mathrm{add}},\qquad
 D=\gamma N_B+\Gamma N_{\mathrm{new}}.
\end{equation}
Every interaction is linear in one auxiliary mode. Write
$H_{\mathrm{int}}=\sum_\alpha
(z_\alpha^\dagger T_\alpha+z_\alpha T_\alpha^\dagger)$,
where $z_\alpha$ runs through the $\hat b_j$ and $\hat s_r$, and
$T_\alpha$ is respectively $\overline F_j/2$ or $gA_r$.
Each $T_\alpha$ is a system polynomial of degree at most six.

There is a computable, polynomially bounded $K\geq1$ such that,
on $\mathcal F_{\mathrm{fin}}$,
\begin{align}
 \|H_{\mathrm{int}}\ket\psi\|
 &\leq K\|\mathsf W^{1/2}(N_{\mathrm{aux}}+I)^{1/2}\ket\psi\|,
 \label{eq:uniq-operator-bound}\\
 |\langle\psi,X\psi\rangle|
 &\leq K\langle\psi,\mathsf W\psi\rangle^{1/2}
          \langle\psi,N_{\mathrm{aux}}\psi\rangle^{1/2},
 \label{eq:uniq-form-bound}
\end{align}
for $X\in\{H_{\mathrm{int}},\overline H,H_{\mathrm{add}},
i[H_{\mathrm{int}},N_{\mathrm{aux}}]\}$.
We justify the two estimates as follows. Write
$N_{\mathrm{sys}}=\sum_\ell \hat d_\ell^\dag \hat d_\ell$.
Applying \cref{lem:word-w}, for anny ladder word operator of degree $d$, we have:
\begin{align}
\|w(N_{\mathrm{sys}}+1)^{-d/2}\|\le\sqrt{d!},
\end{align}
Also, $(N_{\mathrm{sys}}+1)^6\le32M^5\mathsf W$.
Since every word in $T_\alpha$ has length at most six, these
bounds imply
\[
\max\{\|T_\alpha\psi\|,\|T_\alpha^\dagger\psi\|\}
\le c_\alpha\|\mathsf W^{1/2}\psi\|,
\qquad
c_\alpha=\sqrt{32M^5\,6!}\sum_w|t_{\alpha,w}|,
\]
where $T_\alpha=\sum_w t_{\alpha,w}w$. Next, system and auxiliary operators commute, and
$
\|z_\alpha\phi\|,\ \|z_\alpha^\dagger\phi\|
\le\|(N_{\mathrm{aux}}+I)^{1/2}\phi\|
$.
Applying these bounds to each interaction term gives
\begin{align}
\|H_{\mathrm{int}}\psi\|
\le 2\sum_\alpha c_\alpha\,
\|\mathsf W^{1/2}(N_{\mathrm{aux}}+I)^{1/2}\psi\|.
\end{align}
For the quadratic-form estimate, we instead use
\begin{align}
\langle\psi,H_{\mathrm{int}}\psi\rangle
=2\operatorname{Re}\sum_\alpha
\langle z_\alpha\psi,T_\alpha\psi\rangle.
\end{align}
Cauchy--Schwarz therefore gives
\begin{align}
|\langle\psi,H_{\mathrm{int}}\psi\rangle|
\le 2\sum_\alpha c_\alpha\,
\langle\psi,\mathsf W\psi\rangle^{1/2}
\langle\psi,N_{\mathrm{aux}}\psi\rangle^{1/2}.
\end{align}
The same bound holds for either interaction sum separately.
It also holds for $i[H_{\mathrm{int}},N_{\mathrm{aux}}]$,
whose expectation replaces the real part above by the imaginary part.
Thus both claimed inequalities hold with any
$K\ge2\sum_\alpha c_\alpha$. In particular, we may choose
\begin{align}
K=1+400(M+1)^3\sum_{\alpha,w}|t_{\alpha,w}|.
\end{align}
Let $r_-=\min\{\gamma,\Gamma\}, r_+=\max\{\gamma,\Gamma\}$, and $\mu=4(1+r_+)$ and choose 
\begin{equation}
 \eta=16(1+K)^2(1+r_-^{-1}).
 \label{eq:uniq-eta-choice}
\end{equation}
Define the diagonal comparison operator and its perturbation by
\begin{equation}
 R=\eta\mathsf W+\mu(N_{\mathrm{aux}}+I),\qquad
 C=R+H_{\mathrm{int}}=H_\eta+\mu(N_{\mathrm{aux}}+I).
\end{equation}
Equation~\eqref{eq:uniq-operator-bound} gives
$\norm{H_{\mathrm{int}}\ket\psi}
\leq K(2\sqrt{\eta\mu})^{-1}\norm{R\ket\psi}$,
with relative bound less than $1/2$.
The Kato--Rellich theorem (c.f. \cite[Theorem 6.4]{teschl2014mathematical}) therefore makes $C$ self-adjoint on
$\operatorname{Dom}(R)$, with core $\mathcal D_{\mathrm{fin}}$.
Equation~\eqref{eq:uniq-form-bound} gives
\begin{equation}
 C\geq\tfrac12R,\qquad C\geq D,\qquad C\geq I.
 \label{eq:uniq-C-positive}
\end{equation}
Furthermore,
$[H_\eta,C]=\mu[H_{\mathrm{int}},N_{\mathrm{aux}}]$.
Equations~\eqref{eq:uniq-operator-bound}--\eqref{eq:uniq-C-positive}
verify Nelson's commutator criterion with comparison operator $C$.
Thus $H_\eta$ is essentially self-adjoint on
$\mathcal F_{\mathrm{fin}}$ \cite[Theorem~X.36]{reed1975ii}.

To construct the no-jump evolution, put
\begin{equation}
 G_0=-i\eta\mathsf W-\tfrac12D,\qquad
 G=G_0-iH_{\mathrm{int}},\qquad
 \operatorname{Dom}(G)=\operatorname{Dom}(R).
 \label{eq:uniq-G}
\end{equation}
The diagonal operator $G_0$ generates a contraction semigroup.
For real $z\geq r_-/2$, its diagonal resolvent and
\eqref{eq:uniq-operator-bound} give
\begin{equation}
 \|H_{\mathrm{int}}(z-G_0)^{-1}\|
 \leq\frac{K}{\sqrt{\eta r_-}}<1.
\end{equation}
The scalar inequality behind this bound is
$\eta^2w^2+(r_-/2)^2(a+1)^2\geq\eta r_-w(a+1)$
for $w\geq1$ and $a\geq0$.
A Neumann series proves surjectivity of $z-G$.
The relative bound also proves closedness, equivalence of the graph
norms of $G$ and $R$, and the common core
$\mathcal F_{\mathrm{fin}}$.
Since $2\operatorname{Re}\langle\psi,G\psi\rangle
=-\langle\psi,D\psi\rangle$, $G$ is maximal dissipative.

Finally, compute on the common core
\begin{equation}
 \widetilde{\mathcal K}_{\gamma,\eta}^*(C)
 =\mu i[H_{\mathrm{int}},N_{\mathrm{aux}}]
   -\frac\gamma2\overline H-\frac\Gamma2H_{\mathrm{add}}-\mu D
 \leq cC
 \label{eq:uniq-C-drift}
\end{equation}
for some finite $c$, by \eqref{eq:uniq-form-bound}.
The losses leave $\mathsf W$ unchanged and damp their respective
auxiliary-linear interaction terms at half the corresponding rate.
Now $G$ and $C$ have the same domain and a common core, every loss
maps that core into itself, and $C\geq D$.
Together with \eqref{eq:uniq-C-drift}, these are precisely the
conditions of \cite[Corollary~4.5]{chebotarev1998sufficient}.
The resulting minimal quantum dynamical semigroup is conservative.
We use its closed pre-dual generator for
\eqref{eq:uniq-generator} from now on.
\end{proof}

\begin{theorem}
\label{thm:uniq-linear-loss}
Let $\varepsilon_0=\min\{\varepsilon,1/4\}$ and choose
\begin{equation}
\gamma\geq\max\left\{4\sqrt m,
\frac{200\tau_0m^3}{\varepsilon_0}\right\},
\qquad \tau_0=100(n+T+2)(T+1)^2.
 \label{eq:uniq-kappa-choice}
\end{equation}
For fixed $g,\Gamma>0$, choose $\eta$ as in
\eqref{eq:uniq-eta-choice}.
Then \eqref{eq:uniq-generator} has exactly one stationary density
operator $\widetilde\sigma_{\kappa,\eta}$ on the full Fock space.
It is supported on
$\mathcal C\otimes\mathcal F_B\otimes
\operatorname{span}\{|0^{n+1}\rangle_S\}$ and satisfies
\begin{align}
\left\|\Tr_{B,S}\widetilde\sigma_{\kappa,\eta}
-E\rho_{\mathrm{hist}}E^\dagger\right\|_1
\leq\varepsilon,
 \label{eq:uniq-target-bound}
\end{align}
All coefficients can be chosen polynomially bounded in
$n,T,\varepsilon^{-1}$.
\end{theorem}

\begin{proof}
We first locate the support of an arbitrary stationary density
operator $\widetilde\sigma$.
For each register define
\begin{equation}
 Q_r=N_r-s_r^\dagger s_r,\qquad
 P_{r,k}=\mathbf1_{\{Q_r\geq k\}},\qquad k\in\mathbb Z.
\end{equation}
Both $H_\eta$ and $D$ commute with $Q_r$.
Their common-core construction in Lemma~\ref{lem:uniq-conservative}
shows that the no-jump semigroup $V(u)=e^{uG}$ commutes with all
$P_{r,k}$. Every loss preserves $Q_r$, except $s_r$, which increases
it by one. Thus $P_{r,k}\mathcal H$ is invariant under the full
evolution. This helps us get the following fact.

\begin{lemma}\label{lem:bound-Prk}
We have
\begin{align}
P_{r,k} \le \Phi_t^\dag (P_{r,k}) \le \mathbb I.
\end{align}
\end{lemma}
\begin{proof}
To see this, first note that positivity of $\Phi_t$ along with $P_{r,k}\le \mathbb I$ directly imply $\Phi_t^\dag(P_{r,k})\le \Phi^\dag_t(\mathbb I) = \mathbb I$, where the last equality is due to $\Phi_t$ being TP (or equivalently $\Phi^\dag_t$ being unital). Now, to prove the remaining inequality, pick any $\psi$ within the image of $P_{r,k}$. Since $P_{r,k} \mathcal H$ is invariant under the evolution, we get
\begin{align}
\Tr\left[ P \Phi_t(\ket\psi\bra\psi) \right]=1.
\end{align}
Note that we can equivalently write
\begin{align}
1=\Tr\left[ P_{r,k} \Phi_t(\ket\psi\bra\psi) \right] = \bra{\psi} \Phi_t^\dag(P_{r,k}) \ket\psi,
\end{align}
and hence (since we have so far established $0\le \Phi_t^\dag(P_{r,k})\le \mathbb I$) we get $\ket\psi=\Phi_t^\dag(P_{r,k})\ket\psi$. This is equivalent to $\Phi_t^\dag(P_{r,k}) P_{r,k} = P_{r,k}$, and furthermore, taking a dagger we get $P_{r,k}\Phi_t^\dag(P_{r,k})  = P_{r,k}$. Using these equalities together, we can write
\begin{align}
\Phi_t^\dag(P_{r,k}) - P_{r,k} = (\mathbb I - P_{r,k}) \Phi_t^\dag(P_{r,k}) (\mathbb I -P_{r,k}) \ge 0,
\end{align}
which proves $\Phi_t^\dag(P_{r,k}) - P_{r,k}\ge 0$.
\end{proof}

Now, we use another lemma that tells us all ancilla $S$ modes are forced to be in the vacuum in the fixed point:

\begin{lemma}\label{lem:vacuum-S}
Any stationary state $\widetilde\sigma$, must have no photons in the new ancilla modes ($S$):
\begin{equation}
 \widetilde\sigma=\sigma\otimes
 |0^{n+1}\rangle\langle0^{n+1}|_S.
 \label{eq:uniq-new-vacuum}
\end{equation}
\end{lemma}
\begin{proof}
Fix a register $r\in\{1,\ldots,n,C\}$ and a joint Fock basis vector
$|\nu\rangle$ with $\nu_{s_r}\ge 1$ photons in mode $s_r$. We will
show that $\langle\nu|\widetilde{\sigma}|\nu\rangle=0$. Let $q$ be the eigenvalue of $Q_r=N_r-s_r^\dagger s_r$ on
$|\nu\rangle$.
Also, let $|\nu^-\rangle$ be the Fock basis vector obtained from $|\nu\rangle$
by removing one photon from mode $s_r$ and leaving every other
occupation unchanged. Then
\begin{align}
s_r|\nu\rangle=\sqrt{\nu_{s_r}}\,|\nu^-\rangle,
    \quad
    s_r^\dagger|\nu^-\rangle=\sqrt{\nu_{s_r}}\,|\nu\rangle.
\end{align}
Removing this photon leaves $N_r$ unchanged and hence increases $Q_r$
from $q$ to $q+1$. Thus $|\nu\rangle$ lies below the threshold
$P_{r,q+1}=\mathbf{1}_{\{Q_r\ge q+1\}}$, while $|\nu^-\rangle$
lies in its range. Now consider the part of the stationary state below this threshold,
\begin{align}
\rho_{r,\le q}
:= (\mathbb I-P_{r,q+1})\,\widetilde{\sigma}\,
(\mathbb I-P_{r,q+1}).
\end{align}
We first show that no population from $\rho_{r,\le q}$ ever reaches
the threshold. By \cref{lem:bound-Prk},
\begin{align}
0\le \Phi_t^\dagger(P_{r,q+1})-P_{r,q+1}
  \le \mathbb I-P_{r,q+1}.
\end{align}
The difference is therefore supported on the range of
$\mathbb I-P_{r,q+1}$. Since $P_{r,q+1}\rho_{r,\le q}=0$,
we obtain
\begin{align}
\Tr\left[P_{r,q+1}\Phi_t(\rho_{r,\le q})\right]
=
\Tr\left[
\bigl(\Phi_t^\dagger(P_{r,q+1})-P_{r,q+1}\bigr)
\widetilde{\sigma}\right] =
\Tr\left[
P_{r,q+1}\bigl(\Phi_t(\widetilde{\sigma})
-\widetilde{\sigma}\bigr)\right]
=0.
\end{align}
In particular, since $|\nu^-\rangle\langle\nu^-|\le P_{r,q+1}$ we have
\begin{align}
\langle\nu^-|\Phi_t(\rho_{r,\le q})|\nu^-\rangle=0.
\end{align}
We now compute the rate at which an $s_r$-loss feeds population into
$|\nu^-\rangle$. Test the master equation at $t=0$ against
$|\nu^-\rangle\langle\nu^-|$.\footnote{This is a valid weak test
without a photon-number moment assumption. The finite-Fock vectors
form a core for the no-jump generator, so conservativity implies that
the span of the finite-Fock matrix units is a core for the predual
generator. On this core, the formal adjoint generator applied to
$|\nu^-\rangle\langle\nu^-|$ is a bounded finite-rank operator,
since the Hamiltonian and the loss operators are polynomials in the
ladder operators. The identity therefore extends from the core by
graph-norm continuity.}
Because $\rho_{r,\le q}|\nu^-\rangle=0$ and
$\langle\nu^-|\rho_{r,\le q}=0$, the Hamiltonian commutator and all
anticommutator terms have zero $(\nu^-,\nu^-)$ matrix element.
A jump operator $L$ contributes
$\langle L^\dagger\nu^-|\rho_{r,\le q}|L^\dagger\nu^-\rangle$.
For every loss other than $\sqrt{\Gamma}\,s_r$, the adjoint
$L^\dagger$ preserves $Q_r$, so $L^\dagger|\nu^-\rangle$ still has
charge $q+1$ and its contribution is zero. Consequently,
\begin{align}
    0
    =
    \left.\frac{\mathrm d}{\mathrm dt}
      \langle\nu^-|\Phi_t(\rho_{r,\le q})|\nu^-\rangle
    \right|_{t=0^+} =
    \Gamma\,
    \langle\nu^-|s_r\,\rho_{r,\le q}\,s_r^\dagger|\nu^-\rangle =
    \Gamma\,\nu_{s_r}\,
    \langle\nu|\widetilde{\sigma}|\nu\rangle.
\end{align}
The last equality uses $s_r^\dagger|\nu^-\rangle
=\sqrt{\nu_{s_r}}\,|\nu\rangle$ and the fact that $|\nu\rangle$
lies below the threshold. Since $\Gamma>0$ and $\nu_{s_r}\ge 1$,
we conclude that $\langle\nu|\widetilde{\sigma}|\nu\rangle=0$.

This holds for every joint Fock basis vector with a photon in any
mode of $S$. By positivity,
$\langle\nu|\widetilde{\sigma}|\nu\rangle
=\|\widetilde{\sigma}^{1/2}|\nu\rangle\|^2$, so each such vector
lies in the kernel of $\widetilde{\sigma}$. Hence
$\widetilde{\sigma}$ is supported on the joint vacuum of $S$.
That vacuum subspace is one-dimensional on $S$, which gives
\eqref{eq:uniq-new-vacuum}.
\end{proof}

By \cref{lem:vacuum-S},
$\widetilde{\sigma}=\sigma\otimes
|0^{n+1}\rangle\langle 0^{n+1}|_S$.
We now show that $\sigma$ has exactly one photon in each register.
Fix a register $r$, and write
$|1_r\rangle_S=s_r^\dagger|0^{n+1}\rangle_S$ for the state with
one photon in mode $s_r$ and vacuum in every other mode of $S$.
For finite-Fock vectors $u,v$ of the system and $B$, take the
matrix element of the stationary equation between
$|v\rangle\otimes|1_r\rangle_S$ on the left and
$|u\rangle\otimes|0^{n+1}\rangle_S$ on the right. This is an
admissible weak test because the corresponding finite-Fock
rank-one observable has a bounded, finite-rank adjoint-generator
image. The only contribution comes from $g s_r^\dagger A_r$
acting on the left of $\widetilde{\sigma}$ in the Hamiltonian
commutator. Hence
\begin{align}
0 =
\langle v,1_r|
\widetilde{\mathcal K}_{\gamma,\eta}(\widetilde{\sigma})
|u,0^{n+1}\rangle
= -ig\,\langle A_r^\dagger v,\sigma u\rangle.
\end{align}
Since $g>0$, we obtain
\begin{align}
  \langle A_r^\dagger v,\sigma u\rangle=0.
  \label{eq:uniq-weak-kernel}
\end{align}
Now let $|\boldsymbol{\nu}\rangle$ be a Fock basis vector with
$N_r|\boldsymbol{\nu}\rangle=k|\boldsymbol{\nu}\rangle$, where
$k\ne1$. Let $e_{r,0}$ denote the occupation vector for one photon
in the distinguished mode of register $r$. Since
$A_r^\dagger=(1-N_r)d_{r,0}$,
\begin{align}
A_r^\dagger|\boldsymbol{\nu}+e_{r,0}\rangle
=(1-k)\sqrt{\nu_{r,0}+1}\,|\boldsymbol{\nu}\rangle.
\end{align}
The coefficient is nonzero. Taking
$v=|\boldsymbol{\nu}+e_{r,0}\rangle$ and
$u=|\boldsymbol{\nu}\rangle$ in \eqref{eq:uniq-weak-kernel}
therefore gives
$\langle\boldsymbol{\nu}|\sigma|\boldsymbol{\nu}\rangle=0$.
By positivity, $|\boldsymbol{\nu}\rangle\in\ker\sigma$.
Applying this to every register gives
$\operatorname{supp}\sigma\subseteq\mathcal C\otimes\mathcal F_B$. Conversely, on this sector with the new modes in vacuum,
$H_{\mathrm{add}}$ and all new losses vanish. Each system mode has
occupation zero or one and the total system occupation is $n+1$;
therefore
\begin{equation}
 \mathsf W|_{\mathcal C}=(n+2)\, \mathbb I_{\mathcal C}.
 \label{eq:uniq-scalar-weight}
\end{equation}
Thus the restricted semigroup is exactly the original encoded
semigroup. Stationary states of the two models correspond exactly
by \eqref{eq:uniq-new-vacuum}. In particular, existence follows from
\cref{lem:loss-only-ancilla-moments}. To apply \cref{lem:loss-only-ancilla-moments} to an arbitrary stationary state $\sigma$, we
first verify its moment assumption. Working on
$\mathcal C\otimes\mathcal F_B$, define
\begin{align}
P_k=\mathbb I_{\mathcal C}\otimes\mathbf1_{\{N_B=k\}},
\qquad
p_k=\operatorname{Tr}(\sigma P_k),
\qquad
T_k=\mathbb I_{\mathcal C}\otimes\mathbf1_{\{N_B\ge k\}}.
\end{align}
For each $k\ge1$, $T_k$ is an admissible weak stationary test:
$I-T_k$ has finite rank with finite-Fock range, and its
adjoint-generator image is bounded. Thus no moment assumption
is needed to test stationarity against $T_k$. The loss
contribution is $-\gamma kP_k$, while the Hamiltonian commutator
connects only the occupation sectors $k-1$ and $k$. By
Cauchy--Schwarz and $\sum_j\|F_j\|^2\le m$, we obtain
\begin{align}
\gamma k p_k
=\Tr\left(\sigma\,i[\overline H,T_k]\right)
\le \sqrt{m k p_kp_{k-1}},
\qquad\text{hence}\qquad
k p_k\le \frac{m}{\gamma^2}p_{k-1}.
\end{align}
By repeated application of this inequality gives, for every integer $r\ge1$ we get
\begin{align}
\Tr\left[\sigma(N_B)_r\right]
=\sum_{k\ge r}(k)_r p_k
\le \left(\frac{m}{\gamma^2}\right)^r
   \sum_{k\ge r}p_{k-r}
=\left(\frac{m}{\gamma^2}\right)^r.
\end{align}
In particular, every stationary $\sigma$ has finite third
ancillary moment, as required by \cref{lem:loss-only-ancilla-moments}. The same lemma applies to every such stationary $\sigma$, and gives
\begin{equation}
 \Tr[\sigma(N_B)_p]\leq q^p,
 \qquad q=\frac{m}{\gamma^2}\leq\frac1{16}.
 \label{eq:uniq-old-moments}
\end{equation}
We use $m$ as an upper bound for $\sum_j\|F_j\|^2$.
In particular, all required stationary identities in the elimination
theorem are legitimate. Its estimate via \eqref{eq:uniq-kappa-choice} gives
\begin{equation}
 \|\sigma_A-\overline{\rho}_{\mathrm{hist}}\|_1
 \leq\frac{200\tau_0m^3}{\gamma}\leq\varepsilon_0,
 \qquad \sigma_A=\operatorname{Tr}_B\sigma,\quad
 \overline{\rho}_{\mathrm{hist}}=E\rho_{\mathrm{hist}}E^\dagger.
 \label{eq:uniq-elimination}
\end{equation}

Now, we get to showing the uniqueness: Let
$\Pi=\mathbb I_{\mathcal C}\otimes|\mathrm{vac}\rangle
\langle\mathrm{vac}|_B$ and
$p_B=\operatorname{Tr}[\sigma(1-\Pi)]\leq\operatorname{Tr}(\sigma N_B)
\leq q$.
The gentle measurement lemma (aka Quantum Markov Inequality) gives
$\|\sigma-\Pi\sigma \Pi\|_1\leq2\sqrt{p_B}$.
Also, $\sigma_A-\langle\mathrm{vac}|\sigma|\mathrm{vac}\rangle_B$
is positive and has trace $p_B$. Hence
\begin{equation}\label{eq:gentle}
\|\sigma-\sigma_A\otimes
|\mathrm{vac}\rangle\langle\mathrm{vac}|_B\|_1
\leq2\sqrt q+q.
\end{equation}
Every stationary state of the modified model therefore satisfies
\begin{equation}
 \|\widetilde\sigma-\overline \rho_{\rm hist} \otimes \ket{\mathrm{vac}}\bra{\mathrm{vac}}_{B,S}\|_1
 \leq\varepsilon_0+2\sqrt q+q\leq\frac{13}{16}<1.
 \label{eq:uniq-common-ball}
\end{equation}
We now show that any two distinct fixed points, result in two fixed points that are maximally separated (i.e., their $\norm{\cdot}_1$ distance is $2$, contradicting \eqref{eq:uniq-common-ball} by a triangle inequality): To do so, consider that two distinct stationary density operators
$\sigma_1,\sigma_2$ existed, and set $X=\sigma_1-\sigma_2$.
By \cite[Proposition~3.4]{carbone2016irreducible}, applied to channel $\Phi_t$, the operator $|X|$ is also a fixed point.
Hence
\begin{align}
X_\pm=\frac{|X|\pm X}{2}
\end{align}
are stationary positive operators with orthogonal supports.
Since $X\neq0$ and $\operatorname{Tr}X=0$, their traces are equal
and nonzero. Thus
\begin{align}
\rho_\pm=\frac{X_\pm}{\operatorname{Tr}X_\pm}
\end{align}
are orthogonal stationary density operators, giving
$\|\rho_+-\rho_-\|_1=2$.
This contradicts \eqref{eq:uniq-common-ball}, which bounds their
trace-norm distance by $13/8<2$, and proves uniqueness.
Equation~\eqref{eq:uniq-target-bound} is
\eqref{eq:uniq-elimination} for this unique state.\footnote{Ref.~\cite{burgarth2013ergodic} also establishes this fact that ``if a Lindbladian has two fixed points, then it also has two fixed points with maximal trace distance'' in their Corollary 1 as well, but in a finite-dimensional setting.}
\end{proof}

The unique stationary state has the same approximate mixed-history
marginal as in the original reduction. The one-photon encoding and
the vacuum of the new modes hold exactly. The regulator is part of
the displayed Hamiltonian, so no separate well-posedness assumption
is made.

\subsection{Polynomial-time preparation from vacuum}
\label{sec:preparation-from-vacuum}

We now show that the construction of the preceding subsection
can prepare its unique stationary state from vacuum in polynomial
time. We choose the loss rate $\gamma$ somewhat larger than
required for stationary-state accuracy alone.

\begin{theorem}
\label{thm:bqp-mixing-time}
Fix $g,\Gamma>0$. For every circuit on $n$ qubits with $T$ gates
and every $0<\varepsilon\le 1/4$, the parameters in
$\widetilde{\mathcal K}_{\gamma,\eta}$ (defined in \cref{eq:uniq-generator}) can be chosen polynomially bounded in
$n,T,\varepsilon^{-1}$ so that
\begin{equation}
\left\|
e^{t_{\mathrm{prep}}\widetilde{\mathcal K}_{\gamma,\eta}}
[\ket{\mathrm{vac}}\bra{\mathrm{vac}}]
-\widetilde{\sigma}_{\gamma,\eta}
\right\|_1
 \le\varepsilon,
 \label{eq:vacuum-preparation}
\end{equation}
for some $t_{\mathrm{prep}}=\operatorname{poly}
(n,T,\varepsilon^{-1})$.
Here $\ket{\mathrm{vac}}$ is vacuum on all modes, and
$\widetilde{\sigma}_{\kappa,\eta}$ is the unique stationary
state of the full dynamics.
\end{theorem}

\begin{proof}
As before, we set $
 r=n+1,
 m=3T+2n+1,
 \tau_0=100(n+T+2)(T+1)^2$.
We require $\gamma\ge\Gamma$. As a result, the choice of
$\eta$ in \eqref{eq:uniq-eta-choice} becomes
$\eta=16(1+K)^2(1+\Gamma^{-1})$,
which is independent of $\gamma$. We may therefore first fix
$\eta$ and subsequently increase $\gamma$. Going forward, the proof has two stages: First, we prove that the added couplings fill the
data and clock registers properly with photons (i.e., we reach $\mathcal C$). Second, the encoded circuit dynamics
approaches its stationary state.

So, first, let us show that the subspace $\mathcal C$ is reached from vacuum and its moments. Let us define subspaces of data/clock modes as follows: For each $r\in\{1,\cdots, n, C\}$, we define
\begin{align}
\mathcal H_{r,k}^{\text{data-clk}} = \ker(N_r - k\, \mathbb I),
\end{align}
with the projection $\Pi_{r,k}^{\text{data}}$. We also define subspaces $\mathcal V_{r}^{\text{data-clk-s}}$ which is a joint subspace of a data/clock registers and a uniqueness ancilla mode $\hat s_r$:
\begin{align}
\mathcal V_{r}^{\text{data-clk-s}} = \left( \mathcal H_{r,0}^{\text{data-clk}} \otimes \mathrm{span}\{\ket0_{s_r}\} \right) \bigoplus \left( \mathcal H_{r,1}^{\text{data-clk}} \otimes \mathrm{span}\{\ket0_{s_r}, \ket1_{s_r}\} \right).
\end{align}
Let $P_{r}^{\text{data-clk-s}}$ project onto it:
\begin{align}
P_{r}^{\text{data-clk-s}} := \Pi_{r,0}^{\text{data-clk}} \otimes \ket{0}\bra{0}_{s_r} + \Pi_{r,1}^{\text{data-clk}} \otimes (\ket{0}\bra{0}_{s_r} + \ket{1}\bra{1}_{s_r}).
\end{align}
Finally, we define
\begin{align}\label{eq:def-P}
P := \bigotimes_{r\in\mathcal R} P_r^{\text{data-clk-s}} \otimes \mathbb I_B.
\end{align}
We also let $
\mathcal V=\left(\bigotimes_{r\in\mathcal R}\mathcal V_r^{\text{data-clk-s}}\right)
\otimes\mathcal F_B$.In what follows, we show that if a state lies in the support of $P$, our evolution (generated by $\widetilde{\mathcal K}_{\gamma, \eta}$) keeps in that subspace. The intuition here is that the ``added'' Hamiltonian $H_{\mathrm{add}}$ can generate photons (from vacuum) in pairs between a register system $r$ and its corresponding ancilla $\hat s_r$. However, if we have a single photon in a register system $r$, there can be no more photons created due to the factor $(1-N_r)$. Similarly, photon loss can only reduce the photon number, and hence, we have considered $\Pi_{r,1}^{\text{data-clk}} \otimes \ket1\bra1_{s_r}$ in the expression for $P_{r}^{\text{data-clk-s}}$ (and $\mathcal V_{r}^{\text{data-clk-s}}$). The following lemma explicitly studies this.
\begin{lemma}[Invariance of $\mathcal V$ under $\widetilde{\mathcal K}_{\gamma,\eta}$]\label{lem:space-preserving}
For any state $\rho_0$ such that $P\rho_0 P = \rho_0$, with $P$ defined in \cref{eq:def-P}, we have
\begin{align}
\Phi_t[\rho_0] = P\Phi_t[\rho_0] P,
\end{align}
for all $t$, where $\Phi_t$ is the conservative semigroup constructed in \cref{lem:uniq-conservative}.
\end{lemma}
\begin{proof}
We show preservation term by term on the Fock core $\mathcal F_{\mathrm{core}}$:
\begin{itemize}
    \item Starting from $\overline{H}$, we have that
    \begin{align}
    [\hat b_j^\dag \hat F_j + \hat b_j \hat F_j^\dag, P] =0,
    \end{align}
    since the dual-rail encoding, and our clock construction both preserve the number of photons in their own registers (and that these operators act trivially on modes $(s_r)_{r\in\mathcal R}$). This invariance was discussed before in the proof of \cref{thm:loss-only-bqp-hardness} as well.
    
    \item Next, it is transparent that the diagonal term $\mathsf{W}$ commutes with $P$:
    \begin{align}
    [\mathsf{W}, P]=0.
    \end{align}
    \item Moving on to the Hamiltonian $H_{\mathrm{add}}$, we use a notation to help us simplify the calculation: we let $\hat d_{i,\mu}=\hat a_{i,\mu}$ for $i\in[n]$ and $\mu\in[0,1]$ and $\hat d_{i,\mu} = \hat c_\mu$ when $i=C$. This helps us write $H_{\mathrm{add}}$ concisely as $H_{\mathrm{add}}= \sum_{r\in\mathcal R} (\hat s_r^\dag \hat d_{r,0}^\dag(1-N_r) + \mathrm{h.c.})$. Then, for each register define the ``logical'' states
    \begin{align}
    \ket{\bar \mu}_r = \hat d_{r,\mu}^\dag \ket{\mathrm{vac}}_r,
    \end{align}
    where $\ket{\mathrm{vac}}_r$ denotes the vacuum state over the register $r\in\mathcal R$. Note that these states form a basis for $\mathcal H_{r,1}^{\text{data-clk}}$. Note that
    \begin{align}\label{eq:h-add}
    \begin{split}
    (\hat s_r^\dag \hat d_{r,0}^\dag(1-N_r) + \mathrm{h.c.}) (\ket{\mathrm{vac}}_r \otimes \ket{0}_{s_r}) &= \ket{\bar0}_r\otimes \ket{1}_{s_r},\\
    (\hat s_r^\dag \hat d_{r,0}^\dag(1-N_r) + \mathrm{h.c.}) (\ket{\mathrm{\bar\mu}}_r \otimes \ket{0}_{s_r}) &= 0,\\
    (\hat s_r^\dag \hat d_{r,0}^\dag(1-N_r) + \mathrm{h.c.}) (\ket{\mathrm{\bar\mu}}_r \otimes \ket{1}_{s_r}) &= \delta_{\mu,0} \ket{\mathrm{vac}_r}\otimes\ket{0}_{s_r}.
    \end{split}
    \end{align}
    Note that \cref{eq:h-add} shows $(\hat s_r^\dag \hat d_{r,0}^\dag(1-N_r) + \mathrm{h.c.}) \mathcal V_r^{\text{data-clk-s}} \subseteq \mathcal V_r^{\text{data-clk-s}}$. Since other terms in $H_{\mathrm{add}}$ act trivially on $\mathcal V_r^{\text{data-clk-s}}$, we get that $H_{\mathrm{add}} \mathcal V \subseteq \mathcal V$. Since $H_{\mathrm{add}}$ is symmetric on the core, we have that
    \begin{align}
    [H_{\mathrm{add}}, P] = 0.
    \end{align}
    \item Finally, we see what do losses do to a state supported on $\mathcal V$. Note that the auxiliary losses have no effect as they act trivially on $\mathcal V$ (i.e., $b_j P = P b_j, [b^\dag_j b_j, P]=0$). Next, for losses on $(\hat s_r)_{r\in\mathcal R}$, we have
\begin{align}
\begin{split}
\hat s_r\bigl(|\mathrm{vac}\rangle_r\otimes|0\rangle_{s_r}\bigr)&=0,\\
\hat s_r\bigl(|\bar\mu\rangle_r\otimes|0\rangle_{s_r}\bigr)&=0,\\
\hat s_r\bigl(|\bar\mu\rangle_r\otimes|1\rangle_{s_r}\bigr)
&=|\bar\mu\rangle_r\otimes|0\rangle_{s_r}.
\end{split}
\end{align}
This concludes
\begin{align}
(1-P)\hat s_r P=0, \quad [\hat s_r^\dag \hat s_r, P]=0.
\end{align}
Therefore, if $\sigma = P\sigma P$, we can write 
\begin{align}
\begin{split}
\mathcal D(\hat s_r)[\sigma] &= \hat s_r \sigma \hat s_r^\dag - \frac12 \{\hat s_r^\dag \hat s_r, \sigma\}\\
&= (1-P+P) \hat s_r P \sigma P \hat s_r(1-P+P) -\frac12 \{\hat s_r^\dag \hat s_r, P\sigma P\}\\
&= P \hat s_r \sigma \hat s_rP -P \frac12 \{\hat s_r^\dag \hat s_r, \sigma\}P.
\end{split}
\end{align}
\end{itemize}
All in all, we have shown that if $\sigma = P\sigma P$, then $\widetilde{\mathcal K}_{\gamma,\eta}[\sigma] = P \widetilde{\mathcal K}_{\gamma,\eta}[\sigma] P$. Lastly, to pass rigorously to the semigroup, consider the closed no-jump generator
\begin{align}
G=-iH_\eta-\frac12\left(
\gamma\sum_j \hat b_j^\dag \hat b_j+
\Gamma\sum_r \hat s_r^\dag \hat s_r
\right).
\end{align}
The preceding identities give $GP=PG$ on $\mathcal F_{\mathrm{fin}}$, and $P\mathcal F_{\mathrm{fin}}\subseteq\mathcal F_{\mathrm{fin}}$. Because this is a core for $G$, closedness extends the identity to $\mathrm{Dom}(G)$, including
\begin{align}
P\operatorname{Dom}(G)\subseteq\operatorname{Dom}(G).
\end{align}
It follows that $e^{tG}P=Pe^{tG}$. Moreover, every jump
\begin{align}
L_\alpha\in
\{\sqrt\gamma\,\hat b_j\}_j\cup\{\sqrt\Gamma\,\hat s_r\}_r
\end{align}
maps $\mathcal V$ to $\mathcal V$. Every term of the no-jump/jump expansion defining the minimal semigroup therefore preserves support in $\mathcal V$. This property passes to its trace-norm limit, proving the claim.
\end{proof}

Thus the system and ancilla modes $(\hat s_r)_{r\in\mathcal R}$ remain in a finite-dimensional
subspace. On this subspace,
\begin{align}
 \|\overline F_j P\|\le1,
 \qquad
 \mathsf{W} P=(\mathbb I+\sum_{r\in\mathcal R}N_r)P.
\end{align}
Only the original ancillary modes $B$ remain unbounded. However, \cref{lem:loss-only-ancilla-moments} gives a bound on moments of $B$ along
the evolution. This time, defining the \textit{time-dependent} moments as
\begin{align}
 M_p(t)=\operatorname{Tr}[\rho_t(N_B)_p],
 \qquad M_0(t)=1,\qquad q=\frac{m}{\gamma^2},
\end{align}
where $\rho_t$ starts with $B$ in vacuum. Taking steps similar to the ones used in the derivation of \cref{eq:moment-recursion}, we obtain
\begin{align}
 M_p'(t)
 \le-p\gamma M_p(t)
    +p\sqrt m\,\sqrt{M_p(t)M_{p-1}(t)}.
\end{align}
Applying this inequality to $\sqrt{M_p(t)}$ and inducting on
$p$ gives\footnote{These calculations can first be performed with total ancillary
occupation restricted to $N_B\le R$. The bounds are uniform
in $R$. The difference between the compressed and full
generators consists only of the omitted Hamiltonian transition
from occupation $R$ to $R+1$. On the compressed evolving state,
its trace norm is at most
$
 m\sqrt{(R+1)\Pr(N_B=R)}
 \le m\sqrt{\frac{(R+1)q^2}{(R)_2}}
 \longrightarrow0.
$
Duhamel's formula and trace-norm contractivity therefore give
convergence of the compressed evolutions on every bounded
time interval. Passing to the limit proves
\eqref{eq:dynamic-ancillary-moments} for the conservative
full evolution.}
\begin{equation}
 M_p(t)\le q^p
 \qquad(t\ge0,\ p\ge1).
 \label{eq:dynamic-ancillary-moments}
\end{equation}
In particular, if $\gamma\ge m$, then
\begin{equation}
 \operatorname{Tr}[\rho_t(N_B+I)]\le2,\quad
 \operatorname{Tr}[\rho_t(N_B+I)^2]\le5,\quad
 \operatorname{Tr}[\rho_t(N_B+I)^3]\le15.
 \label{eq:dynamic-three-moments}
\end{equation}
As a consequence of $M_1\le q$, with a similar argument as the one used for \eqref{eq:gentle}, we obtain
\begin{equation}
 \left\|\rho_t-
  (\operatorname{Tr}_B\rho_t)
       \otimes|\mathrm{vac}\rangle\langle\mathrm{vac}|_B
 \right\|_1
 \le 2\sqrt q+q
 \le\frac{3m}{\gamma}.
 \label{eq:dynamic-vacuum-ancilla}
\end{equation}

We now move on to the next part of our proof, where we want to show how fast does our state reach the encoded subspace $\mathcal C$ (on the data/clock registers). Define the encoded initial state
\begin{align}
 \rho_{\mathrm{in}}
 =E\bigl(
 |0^n\rangle\langle0^n|\otimes|0\rangle\langle0|
 \bigr)E^\dagger,
 \qquad
 \Omega_{\mathrm{in}}
 =\rho_{\mathrm{in}}\otimes
 |\mathrm{vac}\rangle\langle\mathrm{vac}|_{B,S}.
\end{align}
We first bound how long it takes to get to this initialization. For a moment, let us remove $\overline H$ from the dynamics, but retain
$H_{\mathrm{add}}+\eta W$ and the losses on the auxiliary modes $(\hat s_r)_{r\in\mathcal R}$.
Starting from vacuum, this simplified dynamics fills the
$r$ registers independently. For one register, the relevant states are
\begin{align}
\ket{e} = \ket{\mathrm{vac}}_r\otimes \ket{0}_{s_r}, \quad \ket{\bar0, b}=\hat d_{r,0}^\dag \ket{\mathrm{vac}}_r\otimes \ket{b}_{s_r}, b=0,1.
\end{align}
As shown in \eqref{eq:h-add}, the Hamiltonian couples $|e\rangle$ and $|\bar0,1\rangle$ with
strength $g$ and detuning $\eta$, while loss transfers
$|\bar0, 1\rangle$ to the absorbing state $|\bar 0, 0\rangle$ at rate $\Gamma$. Let us use the notations introduced in \cref{lem:space-preserving}. It is a standard technique (c.f. \cite[Equation (2.5)]{chebotarev1998sufficient}), to write
\begin{align}
\Phi^{(r)}_t[\rho_0] = e^{Gt} \rho_0 e^{tG^\dag} + \Gamma \int_{t'=0}^t \Phi^{(r)}_{t-t'}\left[ \hat s_r e^{t' G} \rho_0 e^{t' G^\dag} \hat s_r^\dag \right],
\end{align}
where $(\Phi^{(r)}_r)_r$ denotes the decoupled dynamics of the $r$-th register and its auxiliary mode $\hat s_r$. Note that applying $\hat s_r\bullet \hat s_r^\dag$ on $e^{t' G} \rho_0 e^{t' G^\dag}$ brings our state to its fixed point $\ket{\bar 0, 0}\bra{\bar0,0}$. This is desired. If this happens, we have a single photon in the $r$-th register, and we have achieved the encoding of $\rho_{\mathrm{in}}$. We now go on to bound the probabily that this event won't happen. Let the random variable denoting the probability of having a loss jump (i.e., the ``good'' event here), to be $T$. We have $\Pr(T\ge t) = \norm{e^{Gt} \rho_0 e^{G^\dag t'}}_1$. We now calculate this norm.
Considering the no-jump evolution, the amplitudes obey
\begin{align}
 \dot u_{e}=-igu_{\bar0,1},\qquad
 \dot u_{\bar0,1}=-igu_e-(\Gamma/2+i\eta)u_{\bar0,1},
 \qquad u_e(0)=1,\quad u_{\bar0,1}(0)=0.
\end{align}
The failure probability is $\Pr(T\ge t)=|u_e(t)|^2+|u_{\bar0,1}(t)|^2$.
Both amplitudes decay: an undamped eigenvector of this
two-dimensional evolution would have $v=0$, and then $u=0$.
We can solve this system of equation (with the initial condition $u_{e}(0)=1, u_{\bar0,1}(0)=0$), to obtain the value of $\int_{t=0}^\infty (|u_{e}(t)|^2 + |u_{\bar0,1}(t)|^2) \, \mathrm dt = \frac2\Gamma+\frac{\Gamma}{4g^2}
+\frac{\eta^2}{\Gamma g^2}$, which yileds
\begin{equation}
 \mathbb E[T] = \int_0^\infty \Pr[T\ge t]\,\mathrm dt
 =\frac2\Gamma+\frac{\Gamma}{4g^2}
       +\frac{\eta^2}{\Gamma g^2}
 =:\tau_{\mathrm{fill}}.
 \label{eq:mean-filling-time}
\end{equation}
Then, a simple Markov inequality gives
$\Pr[T\ge t]\le\tau_{\mathrm{fill}}/t$.
A union bound over the registers therefore bounds the
simplified evolution's state's distance from the fully filled state by
$2r\tau_{\mathrm{fill}}/t$.
Now it is time to bring back $\overline H$. Let
$\rho_t^{AS}=\operatorname{Tr}_B\Phi_t(\Omega_{\mathrm{vac}})$,
and denote the comparison generator on $AS$ by
$\mathcal L_{\mathrm{fill}}$. Its evolution equation is
\[
 \frac{d}{dt}\rho_t^{AS}
 =\mathcal L_{\mathrm{fill}}[\rho_t^{AS}]+R_t,
 \qquad
 R_t=-i\operatorname{Tr}_B[\overline H,\rho_t].
\]
Cauchy--Schwarz and
\eqref{eq:dynamic-ancillary-moments} imply
\[
 \|R_t\|_1
 \le2\sum_j\|\overline F_j\|
       \sqrt{\operatorname{Tr}(\rho_t b_j^\dagger b_j)}
 \le2\sqrt m\sqrt{\operatorname{Tr}(\rho_tN_B)}
 \le\frac{2m}{\gamma}.
\]
Duhamel's formula, contractivity, and
\eqref{eq:dynamic-vacuum-ancilla} consequently give
\begin{equation}
\|\Phi_t(\Omega_{\mathrm{vac}})-\Omega_{\mathrm{in}}\|_1
\le
\frac{2r\tau_{\mathrm{fill}}}{t}
+\frac{2mt}{\gamma}
+\frac{3m}{\gamma}.
\label{eq:initialization-estimate}
\end{equation}

We now move on to show that if we were to start from a state in $\mathcal C$ (with auxiliary $(\hat s_r)_{r\in\mathcal R}$ modes in vacuum), we would then reach to the fixed point in polynomial time.
Note that on $\mathcal C$ with auxiliary modes $(\hat s_r)_{r\in\mathcal R}$ in vacuum, the added couplings vanish
and the regulator is proportional to identity. Hence, the restricted dynamics coincides with $\mathcal K_\gamma$ from \cref{subsec:loss-only-elimination}.
Recall $\mathcal L_{\mathrm{circ}}$ defined in \eqref{eq:loss-only-logical-generator}. We use $\overline{\mathcal L}_{\mathrm{circ}}$ to denote the boson-encoded version (i.e., $\overline{\mathcal L}_{\mathrm{circ}} = \sum_{j=1}^m \mathcal D[\overline{F}_j]$). We then show a stronger version of \cref{thm:loss-only-bqp-hardness}:
\begin{lemma}\label{lem:adiabatic-strong}
For any data-clock state $\rho$ supported on $\mathcal C$, and $\gamma\ge m$, it is the case that
\begin{align}
\left\|
\operatorname{Tr}_B e^{\gamma s\mathcal K_\gamma}
[\rho\otimes|\mathrm{vac}\rangle\langle\mathrm{vac}|_B]
-e^{s\overline{\mathcal L}_{\mathrm{circ}}}[\rho]
\right\|_1
\le\frac{40m+300s m^3}{\gamma}.
 \label{eq:preparation-finite-time-elimination}
\end{align}
\end{lemma}
\begin{proof}
Our proof strategy is as follows: first, let us define
\begin{align}
\begin{split}
\rho_{\mathrm{real}}(u)&=\Tr_B\left( e^{\gamma u \mathcal K_\gamma}\left[ \rho\otimes \ket{\mathrm{vac}}\bra{\mathrm{vac}} \right] \right),\\
\rho_{\mathrm{ideal}}(u) &= e^{u\overline{\mathcal L}_{\mathrm{circ}}} [\rho].
\end{split}
\end{align}
Our goal is to show that $\rho_{\mathrm{ideal}}(s)$ and $\rho_{\mathrm{real}}(s)$ are close. We show the closeness of these states by bounding the difference in their expectation with respect to a bounded data-clock operator, say $A$. To do so, we decine the interpolation function
\begin{align}
F(u)=\Tr( A e^{(s-u)\overline{\mathcal L}_{\mathrm{circ}}}\left[ \rho_{\mathrm{real}}(u) \right]),
\end{align}
which is, to calculate the expectation of the state, when it is first evolved under the real evolution (here $\mathcal K_{\gamma}$), for time $\gamma u$, and then evolved under the idealized Lindbladian $\overline{\mathcal L}_{\mathrm{circ}}$, for the rest of the time $s-u$. Note that we want to bound the difference $\abs{F(0)-F(s)}$. To study this object, let us define
\begin{align}
A(u) = e^{(s-u)\overline{\mathcal L}^\dag_{\mathrm{circ}}}[A].
\end{align}
As a result
\begin{align}
F(u) = \Tr((A(u)\otimes \mathbb I_{B})\sigma_{\mathrm{real}}(u)),
\end{align}
where $\sigma_{\mathrm{real}}=e^{\gamma u \mathcal K_\gamma}\left[ \rho\otimes \ket{\mathrm{vac}}\bra{\mathrm{vac}} \right]$ is the joint state under the evolution.
As it turns out, we need to add a small correction to $F(u)$ to keep its derivative small. We define
\begin{align}
H(u)=\Tr(J(A(u))\sigma_{\mathrm{real}}(u)),
\end{align}
where $
J(\bullet)=\bullet\otimes \mathbb I_B+\gamma^{-1}A_1(\bullet)+\gamma^{-2}A_2(\bullet)$, with $A_1,A_2$ defined as in the proof of \cref{thm:loss-only-stationary-elimination}. We then show that $H$ and $F$ are close:
\begin{lemma}\label{lem:F-H}
For any operator $A$ such that $\norm{A}=1$, and with $\gamma \ge m$, it is the case that
\begin{align}
|H(u)-F(u)|
\le \frac{6m}{\gamma}+\frac{32m^2}{\gamma^2}
\le\frac{40m}{\gamma}.
\end{align}
\end{lemma}
\begin{proof}
By the definition of \(J\),
\begin{align}\label{eq:H-F}
H(u)-F(u)
=
\gamma^{-1}\operatorname{Tr}\bigl(\sigma_{\mathrm{real}}(u)A_1(A(u))\bigr)
+\gamma^{-2}\operatorname{Tr}\bigl(\sigma_{\mathrm{real}}(u)A_2(A(u))\bigr).
\end{align}
Thus, we only need to bound the expectations of the two correctors.
Write $Q=N_B+\mathbb I$. The moment bounds established earlier in \eqref{eq:dynamic-three-moments}, hold uniformly along the actual evolution, and hence also at the rescaled physical time \(\gamma u\):
\begin{align}\label{eq:moments}
\operatorname{Tr}(\sigma_{\mathrm{real}}(u)Q)\le2,
\qquad
\operatorname{Tr}(\sigma_{\mathrm{real}}(u)Q^2)\le5.
\end{align}
We will use the weighted Cauchy–Schwarz inequality
\begin{align}\label{eq:weighted-cs}
\left|\operatorname{Tr}(\sigma_uY)\right|
\le
\|YQ^{-d/2}\|
\sqrt{\operatorname{Tr}(\sigma_uQ^d)}.
\end{align}
For the first corrector, recall that
\begin{align}
A_1(\bullet)=2i[\widehat G,\bullet\otimes I_B].
\end{align}
Writing $\widehat G=\sum_\alpha \tilde F_\alpha\otimes\ell_\alpha$, with
$\sum_\alpha\|C_\alpha\|=m
$, gives
$
A_1(\bullet)=2i\sum_\alpha[\tilde F_\alpha,\bullet]\otimes\ell_\alpha
$.
Its total coefficient norm is therefore at most
$
2\sum_\alpha\|[\tilde F_\alpha,\bullet]\|
\le4m\|\bullet\|$.
Every \(\ell_\alpha\) is a length-one ladder word, so the ladder-word bound of \cref{lem:word-w} gives
\[
\|A_1(X)Q^{-1/2}\|\le4m\|X\|.
\]
Combining this with \eqref{eq:moments} and \eqref{eq:weighted-cs} gives
\begin{align}\label{eq:bound-A1}
\left|\Tr(\sigma_uA_1(\bullet))\right|
\le4\sqrt2\,m\|X\|
\le6m\norm{\bullet},
\end{align}
which handles the norm of the first corrector.
For the second corrector,
\begin{align}
A_2(\bullet)
=-2[\widehat G,[\widehat G,\bullet\otimes I_B]]
-\mathcal L^\dagger[\bullet]\otimes \mathbb I_B,
\end{align}
where $\mathcal L=\overline{\mathcal L}_{\mathrm{circ}}$ on $\mathcal C$.
Expanding the double commutator gives
$
-2[\widehat G,[\widehat G,X]]
=-2\widehat G^2X+4\widehat GX\widehat G-2X\widehat G^2$,
suppressing tensor identities. Each product has length-two ancillary words, and its coefficient norms sum to at most \(m^2\|X\|\) before the numerical prefactors. Thus the total is at most
$
(2+4+2)m^2\|X\|=8m^2\|X\|
$.
By another use of \cref{lem:word-w}, the length-two ladder-word bound contributes a factor of $\sqrt2$. Also,
\begin{align}\label{eq:L-dag-bound-A}
\|\mathcal L^\dagger[\bullet]\|\le2m\|\bullet\|,
\qquad \|Q^{-1}\|\le1,
\end{align}
where the first inequality follows from \eqref{eq:bound-norm-LA}, and the second one is trivial.
As a result, we get
\begin{align}
\|A_2(X)Q^{-1}\|
\le(8\sqrt2+2)m^2\|X\|.
\end{align}
Again by using \eqref{eq:moments} and \eqref{eq:weighted-cs}, we get
\begin{align}\label{eq:bound-A2}
\begin{split}
\left|\operatorname{Tr}(\sigma_{\mathrm{real}}(u)A_2(\bullet))\right|
&\le(8\sqrt2+2)\sqrt5\,m^2\|\bullet\|\\
&\le32m^2\|\bullet\|.
\end{split}
\end{align}
Finally, the ideal adjoint semigroup is unital and completely positive, so
\[
\|A(u)\|\le\|A\|.
\]
Using \eqref{eq:bound-A1} and \eqref{eq:bound-A2} with $\bullet$ replaced by $A(u)$, and plugging into \eqref{eq:H-F} we obtain
\begin{align}
|H(u)-F(u)|
\le
\left(\frac{6m}{\gamma}+\frac{32m^2}{\gamma^2}\right)\|A\|.
\end{align}
Since \(\gamma\ge m\), we can write $
\frac{6m}{\gamma}+\frac{32m^2}{\gamma^2}
\le\frac{38m}{\gamma}
\le\frac{40m}{\gamma}$.
\end{proof}
Next, we show that $H(0)=F(0)$ and that $|H'(u)|$ is small. These ingredients will then show the closeness of $F(s)$ to $F(0)$.

To see why $F(0)=H(0)$, note that
\begin{align}
\langle\mathrm{vac}|_BA_1(X)|\mathrm{vac}\rangle_B=0,
\qquad
\langle\mathrm{vac}|_BA_2(X)|\mathrm{vac}\rangle_B=0,
\end{align}
where the first identity holds because $A_1(X)$ is linear in ladder operators, and the second one holds because the subtraction of \(\mathcal L^\dagger[X]\) in \(A_2(X)\) removes its vacuum matrix element (c.f. \eqref{eq:wick} and \eqref{eq:loss-only-vacuum-contraction}). Since the state is initialized with vacuum in the ancillary modes, we get $F(0)=H(0)$.

To show that $|H'(u)|$ is small, a straightforward and direct calculation reveals
\begin{align}\label{eq:H'}
\begin{split}
H'(u)
={}&\frac1\gamma
\Tr\left[
\sigma_{\mathrm{real}}(u)\left(
\mathcal V^\dagger[A_2(A(u))]
-A_1(\mathcal L^\dagger[A(u)])
\right)
\right]-\frac1{\gamma^2}
\Tr\left[
\sigma_{\mathrm{real}}(u)A_2(\mathcal L^\dagger[A(u)])
\right].
\end{split}
\end{align}
We can now use \eqref{eq:VA2}, \eqref{eq:bound-A1}, \eqref{eq:bound-A2}, and \eqref{eq:L-dag-bound-A}, to obtain bounds on each term of \eqref{eq:H'}. Putting all together, we establish
\begin{align}
|H'(u)|\le\frac{276m^3}{\gamma}.
\end{align}
As a result, we obtain
\begin{align}
|H(u) - H(0)| \le \frac{276m^3 u}{\gamma}.
\end{align}
Putting this together with \cref{lem:F-H} gives
\begin{align}
|F(s)-F(0)|
\le\frac{40m}{\gamma}+\frac{300sm^3}{\gamma},
\end{align}
hence, proving our bound (where for simplicity, we have upper bounded $276<300$).
\end{proof}
Now we just need to stick the pieces together: using \cref{lem:loss-only-circuit-mixing}, we can choose a polynomially bounded time $s
=O\!\left((T+1)^2
\log\frac{n+T+2}{\varepsilon}\right)$ such that
\begin{align}
\left\|
e^{s\overline{\mathcal L}_{\mathrm{circ}}}[\rho_{\mathrm{in}}]
-\overline{\rho}_{\mathrm{hist}}
\right\|_1
\le\frac{\varepsilon}{4},
\end{align}
where $\overline{\rho}_{\mathrm{hist}}=E\rho_{\mathrm{hist}}E^\dag$, is the boson-encoded history state.

Starting from \(\Omega_{\mathrm{in}}\), the new auxiliary modes $(\hat s_r)_{r\in\mathcal R}$ remain in vacuum. Using the finite-time elimination result of \cref{lem:adiabatic-strong}, together with 
\eqref{eq:dynamic-vacuum-ancilla} gives
\begin{align}
&\left\|
\Phi_{\gamma s}(\Omega_{\mathrm{in}})
-
e^{s\overline{\mathcal L}_{\mathrm{circ}}}[\rho_{\mathrm{in}}]
\otimes|\mathrm{vac}\rangle\langle\mathrm{vac}|_{B,S}
\right\|_1\le
\frac{40m+300sm^3}{\gamma}
+\frac{3m}{\gamma}.
\end{align}
Likewise, the inequalities \eqref{eq:uniq-elimination} and \eqref{eq:gentle} in the proof of \cref{thm:uniq-linear-loss} give
\begin{align}
\left\|
\widetilde{\sigma}_{\gamma,\eta}-\overline{\rho}_{\mathrm{hist}} \otimes \ket{\mathrm{vac}}\bra{\mathrm{vac}}_{BS}
\right\|_1
\le
\frac{200\tau_0m^3+3m}{\gamma}.
\end{align}
A simple triangle inequality then shows
\begin{align}\label{eq:no-idea}
\left\|
\Phi_{\gamma s}(\Omega_{\mathrm{in}})
-\widetilde{\sigma}_{\gamma,\eta}
\right\|_1
\le
\frac{\varepsilon}{4}
+\frac{46m+300sm^3+200\tau_0m^3}{\gamma}\le
\frac{\varepsilon}{4}
+\frac{500m^3(1+s+\tau_0)}{\gamma}.
\end{align}
Therefore, setting
\begin{align}
t_I=\frac{8m\tau_{\mathrm{fill}}}{\varepsilon},
\end{align}
and choosing
\begin{align}
\gamma=
\left\lceil
\max\left\{
\Gamma,\ m,\
\frac{16mt_I}{\varepsilon},\
\frac{2000m^3(1+s+\tau_0)}{\varepsilon}
\right\}
\right\rceil,
\end{align}
we satisfy the hyphotheses for \cref{thm:uniq-linear-loss}, and make the right-hand side of \cref{eq:no-idea} at most $\varepsilon/2$.
For the initialization stage, recall that there are \(r=n+1\le m\) registers, and our initialization bound \eqref{eq:initialization-estimate} gives
\begin{align}
\begin{split}
\left\|
\Phi_{t_I}(\Omega_{\mathrm{vac}})-\Omega_{\mathrm{in}}
\right\|_1
&\le
\frac{2r\tau_{\mathrm{fill}}}{t_I}
+\frac{2mt_I}{\gamma}
+\frac{3m}{\gamma}\\
&\le
\frac{\varepsilon}{4}
+\frac{\varepsilon}{8}
+\frac{\varepsilon}{8}
=\frac{\varepsilon}{2}.
\end{split}
\end{align}
Finally, the semigroup property and trace-norm contractivity imply
\begin{align}
\begin{split}
&\left\|
\Phi_{t_I+\gamma s}(\Omega_{\mathrm{vac}})
-\widetilde{\sigma}_{\gamma,\eta}
\right\|_1\\
&\quad\le
\left\|
\Phi_{\gamma s}\!\left(\Phi_{t_I}(\Omega_{\mathrm{vac}})\right)
-\Phi_{\gamma s}(\Omega_{\mathrm{in}})
\right\|_1
+
\left\|
\Phi_{\gamma s}(\Omega_{\mathrm{in}})
-\widetilde{\sigma}_{\gamma,\eta}
\right\|_1\\
&\quad\le
\left\|
\Phi_{t_I}(\Omega_{\mathrm{vac}})-\Omega_{\mathrm{in}}
\right\|_1
+
\left\|
\Phi_{\gamma s}(\Omega_{\mathrm{in}})
-\widetilde{\sigma}_{\gamma,\eta}
\right\|_1
\le\varepsilon.
\end{split}
\end{align}
Since \(\tau_{\mathrm{fill}},\tau_0,m,s,t_I,\gamma\) are polynomially bounded in \(n,T,\varepsilon^{-1}\), the preparation time
$
t_{\mathrm{prep}}=t_I+\gamma s
$
is at most polynomially large.
\end{proof}

\section{Classical algorithms for the fixed point problem}
\label{sec:classical-algorithms}

In this section, we study when bosonic Lindbladian fixed
points admit efficient classical sampling and whether
this remains possible under weak non-Gaussian perturbations.
We first revisit a previous approach by~\cite{nicacio2016determining} which establishes sufficient condition for having positivity in the $P$ function representation for the fixed point of Gaussian dynamics under photon loss and
incoherent pumping, and generalize this result to the entire Gaussian trajectory using the phase-space framework
of~\cite{rahimi2016sufficient}. This implies an efficient sampling algorithm whenever the $P$ positivity condition holds. We observe that the sufficient condition for the Gaussian map being $P$ positive preserving is stronger than $P$ positivity happening at the fixed point, i.e., there are instances where $P$ positivity holds at time $t =0$ and at the fixed point but not at the transient phase. We then study the stability of this classical algorithm under weak non-Gaussian perturbations. We first establish a finite time stability using basic application of the Duhamel formula. This result does not however extend to the fixed point as mixing time is in general sensitive to non-Gaussian perturbations.
Our main contribution in this section is to extend
this analysis to weak Kerr perturbations directly
at the level of stationary states
(\Cref{thm:kerr-perturbative}).
This avoids assuming a mixing-time bound for the
perturbed dynamics and yields approximate classical
sampling whenever the Gaussian reference fixed point
is classically sampleable.

\subsection{Classical algorithm for sampling from the Gaussian dynamics and fixed points}
\subsubsection{Dynamics of covariance matrix for Gaussian interactions under loss and pumping}
\label{sec:covariance-dynamics}

We start by reviewing the formulation for Gaussian Lindbladian dynamics. We consider the quadratic Hamiltonian
\begin{equation}\label{eq:Gaussian-Hamiltonian-def}
H_G=\tfrac12\hat{\mathbf r}^T\mathbf M\hat{\mathbf r}
 +(\boldsymbol\Omega^{-1}\mathbf d)^T\hat{\mathbf r},
\qquad \mathbf M=\mathbf M^T\in\mathbb R^{2m\times2m}.
\end{equation}
With the conventions in \Cref{sec:preliminaries}, its contribution to
$\dot{\boldsymbol\mu}$ is $\boldsymbol\Omega\mathbf M\boldsymbol\mu+\mathbf d$.

Furthermore, assume a loss rate $\gamma_i \geq 0$ and a pumping rate $\eta_i \geq 0$ for mode $i$ (jump operators $\sqrt{\gamma_i}\,\hat a_i$ and $\sqrt{\eta_i}\,\hat a_i^\dagger$). This means that the Lindbladian is
\begin{align}\label{eq:GaussianLindbladian}
\mathcal{L}[\bullet] = -i[H_G,\bullet] +  \sum_{i} \gamma_i\mathcal{D}(\hat a_i)[\bullet] +  \sum_{i} \eta_i\mathcal{D}(\hat a^\dag_i)[\bullet]
\end{align}
Define
\begin{equation}
\boldsymbol\Gamma_{-}
:=
\bigoplus_{j=1}^{m}
\bigl(\gamma_j-\eta_j\bigr)
\mathbb{I}_{2},
\qquad
\boldsymbol\Gamma_{+}
:=
\bigoplus_{j=1}^{m}
\bigl(\gamma_j+\eta_j\bigr)
\mathbb{I}_{2},
\label{eq:Gamma-plus-minus}
\end{equation}
where the direct sums use the quadrature ordering corresponding to:
\begin{equation*}
\hat{\mathbf{r}}=(\hat q_1,\hat p_1,\ldots,\hat q_m,\hat p_m)^{T}.
\end{equation*}
We also introduce the drift matrix
\begin{equation}
\mathbf A:=\boldsymbol{\Omega} \mathbf{M}-\frac{1}{2}\boldsymbol\Gamma_{-}.
\label{eq:mode-dependent-drift}
\end{equation}

\begin{theorem}[Existence and uniqueness of the attracting Gaussian fixed point]
\label{thm:gaussian-fixed-point}
Every initial state converges in trace norm to a common Gaussian fixed point
if and only if the drift in \eqref{eq:mode-dependent-drift} is Hurwitz:
\begin{equation}\label{eq:mode-dependent-stability}
\alpha(\mathbf A)=\max_{z\in\sigma(\mathbf A)}\operatorname{Re}z<0.
\end{equation}
In this case the fixed point is unique, with
\begin{align}
\boldsymbol\mu_f&=-\mathbf A^{-1}\mathbf d,
\label{eq:mode-dependent-fixed-mean}\\
\boldsymbol\Sigma_f&=\int_0^\infty e^{s\mathbf A}\boldsymbol\Gamma_+
 e^{s\mathbf A^T}\,ds
=\operatorname{vec}^{-1}\!\left[\mathbf K^{-1}\operatorname{vec}(\boldsymbol\Gamma_+)\right],
\label{eq:mode-dependent-fixed-covariance-integral}\\
\mathbf K&=-(\mathbb I\otimes\mathbf A+\mathbf A\otimes\mathbb I)
=\tfrac12(\mathbb I\otimes\boldsymbol\Gamma_-+\boldsymbol\Gamma_-\otimes\mathbb I)
 -\mathbb I\otimes\boldsymbol\Omega\mathbf M-\boldsymbol\Omega\mathbf M\otimes\mathbb I.
\label{eq:mode-dependent-K}
\end{align}
Equivalently, the covariance is the unique solution of
\begin{equation}\label{eq:mode-dependent-Lyapunov}
\mathbf A\boldsymbol\Sigma_f+\boldsymbol\Sigma_f\mathbf A^T+\boldsymbol\Gamma_+=0.
\end{equation}
\end{theorem}
\begin{proof}
The moment equations and their solutions are
\begin{align}
\dot{\boldsymbol\mu}&=\mathbf A\boldsymbol\mu+\mathbf d,
\label{eq:mode-dependent-mean-evolution}\\
\dot{\boldsymbol\Sigma}&=\mathbf A\boldsymbol\Sigma+\boldsymbol\Sigma\mathbf A^T+\boldsymbol\Gamma_+,
\label{eq:mode-dependent-covariance-evolution}\\
\boldsymbol\mu(t)&=e^{t\mathbf A}\boldsymbol\mu(0)
 +\int_0^t e^{(t-s)\mathbf A}\mathbf d\,ds,
\label{eq:mode-dependent-mean-solution}\\
\boldsymbol\Sigma(t)&=e^{t\mathbf A}\boldsymbol\Sigma(0)e^{t\mathbf A^T}
 +\int_0^t e^{s\mathbf A}\boldsymbol\Gamma_+e^{s\mathbf A^T}\,ds.
\label{eq:mode-dependent-covariance-solution}
\end{align}
For mode-dependent rates the damping term is the anticommutator
$-(\boldsymbol\Gamma_-\boldsymbol\Sigma+\boldsymbol\Sigma\boldsymbol\Gamma_-)/2$;
commutation of these two matrices is not assumed. The channel matrices are
\begin{equation}\label{eq:mode-dependent-Gaussian-channel}
\mathbf X(t)=e^{t\mathbf A},\qquad
\mathbf Y(t)=\int_0^t e^{s\mathbf A}\boldsymbol\Gamma_+e^{s\mathbf A^T}\,ds.
\end{equation}
If $\mathbf A$ is Hurwitz, the exponential estimate of
\Cref{sec:logarithmic-norms} makes the covariance integral convergent and
$\mathbf X(t)\to0$. For every input state, continuity of its characteristic
function at zero gives convergence to the stated Gaussian characteristic
function. The quantum continuity theorem then gives trace-norm convergence
(see also \cite{girotti2025gaussian}); this conclusion is stronger than
continuity of Gaussian moments alone. Conversely, apply convergence to
Gaussian inputs of the same covariance and arbitrary different means.
Their Gaussian output characteristic functions force $e^{t\mathbf A}v\to0$
for every $v$, hence $\mathbf A$ is Hurwitz. Finally, vectorizing the Lyapunov
equation gives the expression involving $\mathbf K$.
\end{proof}

For the remainder of this section the rates are uniform,
so $\boldsymbol\Gamma_\pm=(\gamma\pm\eta)\mathbb I$.
The exact stability condition
\eqref{eq:mode-dependent-stability} becomes
\[
\gamma-\eta>2\alpha(\boldsymbol\Omega\mathbf M).
\]
The stronger sufficient condition
$\gamma-\eta>\lambda_{\max}([\boldsymbol\Omega,\mathbf M])$
also gives
\[
\|e^{t\mathbf A}\|_\infty
\leq
\exp\!\left[
-\frac{t}{2}
\left(\gamma-\eta
-\lambda_{\max}([\boldsymbol\Omega,\mathbf M])\right)
\right],
\]
by \eqref{eq:logarithmic-semigroup-bound}.
In general $\boldsymbol\Sigma_f$ is a Lyapunov integral, not
$(\gamma+\eta)((\gamma-\eta)\mathbb I-[\boldsymbol\Omega,\mathbf M])^{-1}$.

For uniform rates, the finite-time Gaussian channel has
\begin{equation}\label{eq:Gaussian-channel}
\boldsymbol\Sigma(t)=\mathbf X(t)\boldsymbol\Sigma(0)\mathbf X(t)^T+\mathbf Y(t),
\qquad \mathbf X(t)=e^{t\mathbf A},\qquad
\mathbf Y(t)=\gamma_+\int_0^t e^{s\mathbf A}e^{s\mathbf A^T}\,ds.
\end{equation}
The transpose and the order of the two exponentials are essential: in general
$\mathbf A$ and $\mathbf A^T$ do not commute, so the integrand cannot be
replaced by $e^{2s\mathbf A}$.
\subsubsection{Classical simulation for the transient state and the fixed point problem}
\label{sec:classical-simulation}

We now follow the idea introduced in \cite{rahimi2016sufficient}, in describing conditions on $\mathbf{M}$ and $(\gamma, \eta)$ that allow classical simulability. The idea is that states with positive Gaulber $P$ function \cite{glauber1963photon} are described by a Gaussian distribution over coherent states. In particular, let $\rho_f$ be the fixed point of the Gaussian Lindbladian with loss and pumping from the previous section. We can write
\begin{align}
\rho_f = \int_{\boldsymbol\alpha\in\mathbb{C}^m} P(\boldsymbol\alpha) \, \ket{\boldsymbol\alpha}\bra{\boldsymbol\alpha}\, \mathrm d\boldsymbol\alpha. 
\end{align}
If $\rho_f$ is a Gaussian state with $\boldsymbol\Sigma_f \ge \mathbb I$, we can show that $P(\boldsymbol\alpha)$ is a Gaussian distribution. In that case, it is straightforward to show that $\rho_f$ is classically sampalable in photon number basis since one can start by sampling $\boldsymbol{\alpha}$ from $P(\boldsymbol\alpha)$, and then sample the product state $\ket{\boldsymbol\alpha} = \ket{\alpha_1}\cdots\ket{\alpha_m}$ in number basis (
by drawing $m$ independent Poisson variables with means $|\alpha_j|^2$
). Following this idea, we are able to show

Related Lyapunov conditions for classicality of Gaussian
steady states were studied in
\cite[Eq.~(18)]{nicacio2016determining}.
Here we characterize when Gaussian $P$-classicality is
preserved throughout the evolution and derive the
corresponding consequences for efficient Fock-basis
sampling.
\begin{proposition}[Classical Gaussian trajectories and fixed points]
\label{thm:gaussian-classical-trajectory}\label{prop:main-classical-simulation}
Let $\mathbf A=\boldsymbol\Omega\mathbf M-\gamma_-\mathbb I/2$,
$\mathbf C=[\boldsymbol\Omega,\mathbf M]$ and $\kappa_M = \norm{C}$. The evolution preserves
nonnegative $P$ representations for all classical Gaussian inputs exactly when
\[
\mathbf C+2\eta\mathbb I\geq0
\quad\Longleftrightarrow\quad \kappa_M\leq2\eta.
\]
If $\mathbf A$ is also Hurwitz, its unique fixed point is classical simulable under this
condition. Given its mean and covariance to the required
precision, its Fock-basis distribution can be sampled
by the Gaussian-mixture and Poisson procedure described above. More generally,
for any Hurwitz $\mathbf A$, the exact fixed-point classicality criterion is
\[
\int_0^\infty e^{s\mathbf A}(\mathbf C+2\eta\mathbb I)e^{s\mathbf A^T}\,ds\geq0.
\]
Thus a convenient sufficient pair for convergence and classicality is
\[
\kappa_M\leq2\eta,\qquad \kappa_M<\gamma-\eta,
\]
or equivalently $-2\eta\mathbb I\leq\mathbf C<\gamma_-\mathbb I$.
The strict upper inequality ensures stability; it is not needed merely to
preserve classicality at finite times.
\end{proposition}
\begin{proof}
In our quadrature convention the ordered characteristic functions are
\begin{equation}\label{eq:def-of-s-param-char}
\chi_s(\mathbf v)=\chi_\rho(\mathbf v)e^{s\|\mathbf v\|_2^2/4},\qquad
\chi_1(\mathbf v)=\exp\left(-\tfrac14\mathbf v^T
 (\boldsymbol\Sigma-\mathbb I)\mathbf v+i\mathbf v^T\boldsymbol\mu\right).
\end{equation}
Consequently $P$ is a nonnegative Gaussian measure precisely when
$\mathbf W:=\boldsymbol\Sigma-\mathbb I\geq0$, including singular measures
on the boundary. The covariance equation gives
\[
\dot{\mathbf W}=\mathbf A\mathbf W+\mathbf W\mathbf A^T+\mathbf Q_P,
\qquad \mathbf Q_P=\mathbf C+2\eta\mathbb I,
\]
and therefore
\begin{equation}\label{eq:gaussian-classicality-trajectory}
\mathbf W(t)=e^{t\mathbf A}\mathbf W(0)e^{t\mathbf A^T}
 +\int_0^t e^{s\mathbf A}\mathbf Q_Pe^{s\mathbf A^T}\,ds.
\end{equation}
Both terms are positive semidefinite if $\mathbf W(0),\mathbf Q_P\geq0$.
Conversely, starting from vacuum gives $\dot{\mathbf W}(0)=\mathbf Q_P$;
a negative direction would immediately violate classicality. The paired
spectrum of $\mathbf C$ from \Cref{sec:logarithmic-norms} gives the equivalence
with $\kappa_M\leq2\eta$. If $\mathbf A$ is Hurwitz, taking $t\to\infty$
proves the fixed-point criterion. Positivity of this integral does not require
positivity of its integrand at every time. Sampling follows from the Gaussian
mixture of coherent states described above.
\end{proof}

\paragraph{Interpreting the separability criterion.}
The sufficient pair $\kappa_M\leq2\eta$ and $\kappa_M<\gamma-\eta$ bounds
the squeezing contribution rather than the full Gaussian interaction.
For example, $\mathbf M=\lambda\mathbb I$ gives
$[\boldsymbol\Omega,\mathbf M]=0$ for arbitrarily large $\lambda$.
Classicality is then preserved for all nonnegative rates, and convergence
holds when $\gamma>\eta$. Large number-preserving interactions are therefore
compatible with the criterion.

For the following block representation, reorder the quadratures
as $(q_1,\ldots,q_m,p_1,\ldots,p_m)^T$ and express $\mathbf M$
in the same ordering. Then the symplectic form is
\begin{align}
    \boldsymbol{\Omega}
    =
    \begin{pmatrix}
        0 & \mathbb I_m\\
        -\mathbb I_m & 0
    \end{pmatrix},
\end{align}
the real matrices commuting with \(\boldsymbol{\Omega}\) are precisely those of
the form
\begin{align}
    \mathbf{M}
    =
    \begin{pmatrix}
        A & B\\
        -B & A
    \end{pmatrix}.
\end{align}
When \(\mathbf{M}=\mathbf{M}^T\), this further imposes
\begin{align}
    A=A^T,
    \qquad
    B=-B^T .
\end{align}
Equivalently, such matrices describe complex-linear, or symplectically
compatible, Gaussian transformations on \(\mathbb{C}^m\).  Geometrically, these
interactions preserve the underlying complex structure and do not generate
the active squeezing directions responsible for leaving the positive-\(P\)
Gaussian region.

This gives a parameter regime complementary to the separability-based
simulation mechanism of Ref.~\cite{gonzalez2025dynamical}.  Their method can
tolerate strong onsite nonlinearities by preserving separability, allowing
locally non-Gaussian states.  By contrast, our Gaussian criterion identifies
regimes in which the trajectory or fixed point remains positive-\(P\), and
therefore directly classically samplable in the Fock basis.

\subsubsection{Classical endpoints do not imply a classical trajectory}
\label{sec:non-classical-trajectory-vs-end-points}
The condition
\begin{align}
    \mathbf{Q}:=[\boldsymbol{\Omega},\mathbf{M}]+2\eta \mathbb {I}\geq0
\end{align}
is sufficient for preservation of positive-\(P\) Gaussianity along the entire
trajectory, but it is not necessary for the fixed point alone to be
\(P\)-positive.  It is possible for both the initial state and the Gaussian
fixed point to be classically samplable, while some intermediate-time states
fail to satisfy
\begin{align}
    \boldsymbol{\Sigma}(t)\geq \mathbb {I} .
\end{align}

Indeed, after setting \(\mathbf{X}(t)=\boldsymbol{\Sigma}(t)-\mathbb {I}\), we have
\begin{align}
    \mathbf{X}(t)
    =
    e^{\mathbf{A}t}\mathbf{X}(0)e^{\mathbf{A}^Tt}
    +
    \int_0^t
    e^{\mathbf{A}u}\mathbf{Q}e^{\mathbf{A}^Tu}\,du .
\end{align}
If \(\mathbf{Q}\geq0\) and \(\mathbf{X}(0)\geq0\), then \(\mathbf{X}(t)\geq0\) for all \(t\ge0\).
However, the converse is false.  Even if \(\mathbf{Q}\) is not positive semidefinite,
the infinite-time Lyapunov integral
\begin{align}
    \mathbf{X}_f
    =
    \int_0^\infty
    e^{\mathbf{A}u}\mathbf{Q}e^{\mathbf{A}^Tu}\,du
\end{align}
can still be positive semidefinite, because the propagator \(e^{\mathbf{A}u}\) rotates,
mixes, and damps phase-space directions before they are integrated.

We illustrate this with a single-mode covariance dynamics.  Let
\begin{align}
    \mathbf{A}
    =
    \begin{pmatrix}
        -3 & 5\\
        -5 & -1
    \end{pmatrix},
    \qquad
    \mathbf{Q}
    =
    \begin{pmatrix}
        -1 & 0\\
        0 & 3
    \end{pmatrix}.
\end{align}
The matrix \(\mathbf{A}\) is Hurwitz, since
\begin{align}
    \sigma(\mathbf{A})
    =
    \{-2\pm 2\sqrt{6}\,i\}.
\end{align}
Although \(\mathbf{Q}\) is not positive semidefinite, solving
\begin{align}
    \mathbf{A}\mathbf{X}_f+\mathbf{X}_f\mathbf{A}^T+\mathbf{Q}=0
\end{align}
gives
\begin{align}
    \mathbf{X}_f
    =
    \begin{pmatrix}
        23/112 & 25/112\\
        25/112 & 43/112
    \end{pmatrix}.
\end{align}
This matrix is positive definite, since
\begin{align}
    \det(\mathbf{X}_f)=\frac{13}{448}>0,
    \qquad
    \operatorname{Tr}(\mathbf{X}_f)=\frac{33}{56}>0 .
\end{align}
Hence
\begin{align}
    \boldsymbol{\Sigma}_f=\mathbb {I}+\mathbf{X}_f\geq \mathbb {I} ,
\end{align}
so the fixed point is \(P\)-positive and classically samplable.

On the other hand, take the initial state to be the coherent-state boundary,
\begin{align}
    \boldsymbol{\Sigma}(0)=\mathbb {I},
    \qquad
    \mathbf{X}(0)=0 .
\end{align}
Then
\begin{align}
    \dot{\mathbf{X}}(0)=\mathbf{Q} .
\end{align}
Since \(\mathbf{Q}\) has a negative eigenvalue, for sufficiently small \(t>0\),
\begin{align}
    \mathbf{X}(t)=t\mathbf{Q}+O(t^2)
\end{align}
has a negative eigenvalue.  Therefore
\begin{align}
    \boldsymbol{\Sigma}(t)\not\geq \mathbb {I}
\end{align}
for intermediate times close to \(t=0\).  Thus the trajectory transiently
leaves the positive-\(P\) Gaussian region even though both the initial state
and the fixed point are classically samplable.

This example shows that \(\mathbf{Q}\geq0\) should be interpreted as a
trajectory-level sufficient condition.  It guarantees preservation of
positive-\(P\) Gaussianity for all times, but it is not necessary for the
fixed point alone to be \(P\)-positive.

\begin{figure}
    \centering
    \includegraphics[width=0.5\linewidth]{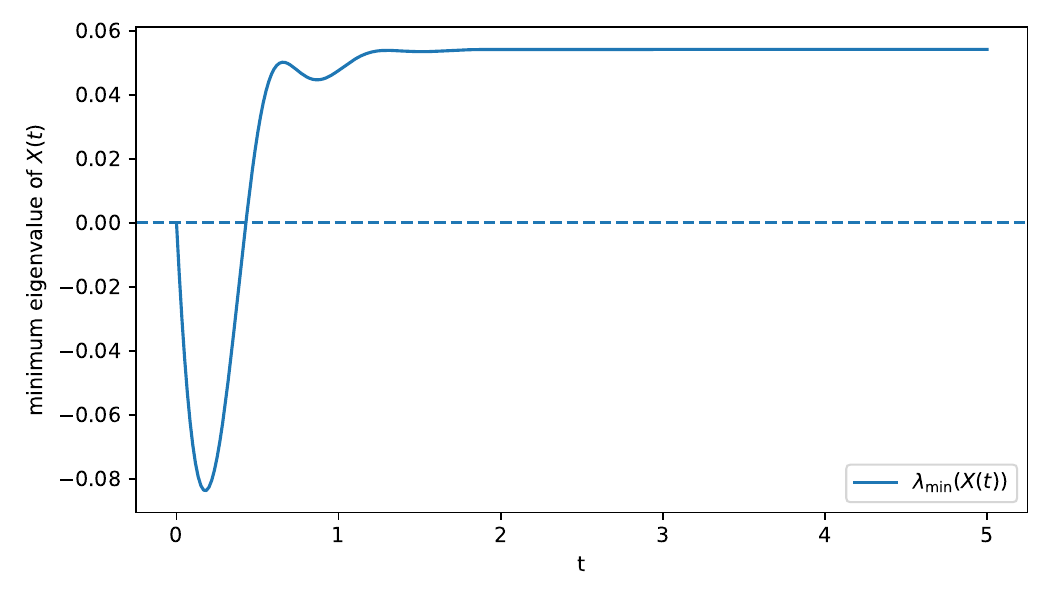}
    \caption{
    A single-mode example in which the initial state and the Gaussian fixed
    point are \(P\)-positive, while the intermediate-time state is not.  The
    plot shows the minimum eigenvalue of \(\mathbf{X}(t)=\boldsymbol{\Sigma}(t)-\mathbb {I}\).  Negative
    values indicate a transient violation of \(\boldsymbol{\Sigma}(t)\geq \mathbb {I}\).
    }
    \label{fig:counter}
\end{figure}

\begin{remark}[First-moment estimate]
For a quadratic Hamiltonian
$H_G=\frac12\hat{\mathbf r}^{T}\mathbf M\hat{\mathbf r}$,
the identity
\[
(\boldsymbol\Omega\mathbf M)^T
+\boldsymbol\Omega\mathbf M
=[\boldsymbol\Omega,\mathbf M]
\]
allows the Hamiltonian contribution to particle-number
growth to be bounded using
$\|[\boldsymbol\Omega,\mathbf M]\|_\infty$.
This retains cancellations that may be lost in
term-by-term absolute-value bounds. In particular,
$[\boldsymbol\Omega,\mathbf M]=0$ implies conservation
of total particle number by $H_G$. This is only first-moment observation; it does not establish
the higher-moment bounds needed to extend the
classical simulation result of \cite{gonzalez2025dynamical}.
\end{remark}

\subsection{Stability under non-Gaussian perturbations}

How well does the classical simulation algorithm outlined in \Cref{sec:classical-simulation} perform when nonzero $U$ terms are turned on?
Due to the infinite-dimensional Hilbert space the answer to this question requires a bit of care, specially for the fixed point problem as additional non-Gaussian terms might alter the long time trajectory significantly.
We first study the stability for transient behavior and
prove a finite-time stability statement for weak Kerr perturbations of a
classically simulable Gaussian trajectory.  Unlike a steady-state perturbation
theorem, the argument does not require a reduced inverse, fast mixing, or a
photon-number cutoff Lindbladian.  The relevant small parameter is the Kerr
strength accumulated over the finite time interval.

\subsubsection{Finite time stability of Gaussian trajectory}
Let
\begin{align}
    \mathcal{L}_U
    =
    \mathcal{L}_0+\mathcal{K},
    \qquad
    \mathcal{K}[\bullet]
    :=
    -i[H_{\rm Kerr},\bullet],
\end{align}
where
\begin{align}
    H_{\rm Kerr}
    =
    \sum_{j,k}U_{jk}\hat n_j\hat n_k.
\end{align}
Recall that $\mathcal L_0$ consists of the Gaussian Hamiltonian $H_G$, together with a uniform loss rate of $\gamma$ and a pumping rate of $\eta$. We now want to talk about the channel $e^{t\mathcal L_U}$, for which we need to prove favorable properties such as being trace-preserving (due to unboundedness of $\mathcal L_U$, this is a non-trivial fact). This is achieved in the following lemma. Similar to what was done in \cref{sec:loss-only-bqp} and following \cref{def:f_fin}, we write $\mathcal F_{\rm fin}$ for the finite span of Fock basis vectors, and
$\mathfrak T_{\rm fin}$ for the finite span of their matrix units. More concretely, we have
\begin{align}
\mathcal F_{\mathrm{fin}} = \left\{ \sum_{\mathbf n\in S} c_{\mathbf n} \ket{\mathbf n}:  S \subset \mathbb N_0^m, |S|< \infty, c_{\mathbf n}\in\mathbb C \right\}, \quad \mathfrak T_{\mathrm{fin}}=\left\{ \sum_{\mathbf k, \mathbf j\in S} A_{\mathbf k, \mathbf j} \ket{\mathbf k}\bra{\mathbf j}: S \subset \mathbb N_0^m, |S|< \infty, A_{\mathbf k, \mathbf j}\in\mathbb C \right\}.
\end{align}
We initially define $\mathcal L_U$ on finite Fock matrices and call it $\mathcal L_{U,\rm fin}$.
For every such matrix $A$, the formal adjoint
$\mathcal L_{U,\rm fin}^\dagger[A]$ also has finite Fock support and is
bounded. We note established conservative conditions for similar infinite-dimensional Lindbladians were also studied in here \cite{becker2026quantum}. For completeness we include a proof for the particular setting in this work. 

\begin{lemma}
$\mathcal L_{U,\rm fin}$ is closable on trace class, and its closure generates a strongly continuous, completely positive, trace-preserving semigroup. In particular, $\mathfrak T_{\mathrm{fin}}$ is a core for its generator.
\end{lemma}
\begin{proof}
We first construct the evolution between jumps, then prove that adding the jumps produces a trace-preserving evolution, and finally identify its generator with the closure of $\mathcal L_{U,\mathrm{fin}}$. The main tool is a family of photon-number moment bounds. These exploit the fact that the Kerr Hamiltonian commutes with total photon number, while the loss, pumping, and quadratic Hamiltonian terms have controlled moment growth. Here the quadraticity of $H_G$ is important for technical reasons. Besides controlling growth, these bounds provide an invariant domain where we can apply the conservativity.

We use $Q = \mathbb I + \hat N$, and on the finite-Fock space ($\mathcal F_{\rm fin}$), we consider the no-jump operator
\begin{align}
G=-i(H_G+H_{\mathrm{Kerr}})
-\frac12\bigl((\gamma+\eta)\hat N+m\eta \mathbb I\bigr).
\end{align}
We say an operator $A$ is $B$-bounded, if $\norm{A\ket\psi}\le c \norm{B\ket\psi}$ for some set of states $\ket\psi$ (e.g., the domain of $B$). We note that for any $\ket\psi\in\mathcal F_{\mathrm{fin}}$, we have that $H_G$ and the dissipative part are $Q$-bounded, while $H_{\rm Kerr}$ is $Q^2$-bounded. This results in the overall bound
\begin{align}\label{eq:master-1}
\norm{G\ket\psi}\le C\|Q^2\ket\psi\|,
\end{align}
for all $\ket\psi\in\mathcal F_{\rm fin}$ (i.e., $G$ is $Q^2$-bounded).
Note that it is straightforward to show that $G$ is a dissipative operator. Next, we need to upper bound $
2\Re\langle Q^p\psi,G\psi\rangle
+\sum_\alpha\|Q^{p/2}L_\alpha\psi\|^2$ with some multiple of $\bra\psi Q^2\ket\psi$ to satisfy the conditions for the Nelson theorem. To do so, note that $[H_{\mathrm{Kerr}},Q]=0$, and hence, the contribution of $H_{\rm Kerr}$ to this bound is zero. We now want to upper bound the quantity $\bra\psi i[H_G, Q^p]\ket\psi$ by a multiple of $\bra\psi Q^2\ket\psi$. To do so, let us use $\Pi_n$ to denote the total number $n$ projection (i.e., projector onto $\ker(\hat N - n\cdot \mathbb I)$). the nonzero number-changing blocks of $H_G$ satisfy
\begin{align}
\|\Pi_{n+q}H_G\Pi_n\|\le C(n+1),
\end{align}
for any $0<|q|\le2$. Also, note that (say, by the midpoint theorem) we have
\begin{align}
|(n+q+1)^p-(n+1)^p|\le C_p(n+1)^{p-1}.
\end{align}
Now, we can write $\ket{\psi_n}=\Pi_n\ket\psi$ to conclude
\begin{align}
\begin{split}
|\langle\psi,i[H_G,Q^p]\psi\rangle|
&\le C_p
\sum_{\substack{n\ge0,\;0<|q|\le2\\n+q\ge0}}
(n+1)^p\|\psi_{n+q}\|\,\|\psi_n\|\\
&\le C'_p\sum_{n\ge0}(n+1)^p\|\psi_n\|^2\\
&=C'_p\langle\psi,Q^p\psi\rangle.
\end{split}
\end{align}
where the second inequality is using $2ab\le a^2+b^2$.
Finally, the contribution of the dissipative part to the quantity $
2\Re\langle Q^p\psi,G\psi\rangle
+\sum_\alpha\|Q^{p/2}L_\alpha\psi\|^2$ is
$
\gamma\hat N\bigl[(Q-I)^p-Q^p\bigr]
+\eta(\hat N+mI)\bigl[(Q+I)^p-Q^p\bigr]
$ (sandwiched between $\bra\psi \cdots \ket\psi$).
The 
contribution of loss is non-positive, and the pumping term is bounded above by a constant times $Q^p$, since $(Q+I)^p-Q^p$ has degree $p-1$ in $Q$. Putting all pieces together, we get
\begin{align}\label{eq:master-2}
2\Re\langle Q^p\psi,G\psi\rangle
+\sum_\alpha\|Q^{p/2}L_\alpha\ket\psi\|^2
\le c_p\|Q^{p/2}\ket\psi\|^2,
\end{align}
for any $\ket\psi\in\mathcal F_{\mathrm{fin}}$ and $p\ge2$, with some $c_p$ (depending only on $p$). We next construct the semigroup generated by $G$. Dropping the nonnegative sum (the second term on the LHS) in \eqref{eq:master-2} and setting $p=2$ gives
\begin{align}
2\Re\langle Q^2\psi,G\psi\rangle
\le c_2\langle\psi,Q^2\psi\rangle.
\end{align}
Together with \eqref{eq:master-1}, and the fact that $\mathcal F_{\mathrm{fin}}$ is a core for $Q^2$ (and the fact that $G$ is dissipative), this verifies the hypotheses of the dissipative Nelson theorem, and hence, the graph closure denoted by $\overline G$ generates a contraction semigroup
$V(t)=e^{t\overline G}$,
and $\mathrm{Dom}(Q^2)\subseteq\operatorname{Dom}(\overline G)$ by \cite[Theorem B.1]{van2025energy}. Furthermore, For every $p\ge2$, \cref{eq:master-1} also gives $Q^p$-boundedness of $G$. Applying \cite[Proposition 3.11 and Lemma 3.9]{van2025energy} to \eqref{eq:master-2} (again with the nonnegative sum on LHS removed) gives
\begin{align}\label{eq:master-3}
\|Q^{p/2}V(t)\psi\|
\le e^{c_pt/2}\|Q^{p/2}\psi\|,
\end{align}
for any $\ket\psi\in\mathrm{Dom}(Q^{p/2})$,
and it also results in strong continuity in the corresponding graph norm. As a direct result the space
\begin{align}
\mathcal S:=\bigcap_{r\ge1}\mathrm{Dom}(Q^r)
\end{align}
is invariant under $V(t)$, and $V(t)$ is strongly continuous on $\mathcal S$ for all the norms $\ket\psi\mapsto\|Q^r\ket\psi\|$. Polynomial ladder operators preserve \(\mathcal S\). In particular, for every integer $r\ge 0$ and any $\ket\psi\in\mathcal S$, we have
\begin{align}\label{eq:master-4}
\begin{split}
\|Q^r\overline G\psi\|&\le C_r\|Q^{r+2}\psi\|,\\
\|Q^rL_\alpha\psi\|&\le C_r\|Q^{r+1}\psi\|.
\end{split}
\end{align}
This implies $\mathcal S$ is a core for $\overline G$, because it contains $\mathcal F_{\mathrm{fin}}$, and it lies in the domain of every power of $\overline G$, because $\overline G\mathcal S\subseteq\mathcal S$ (as shown in \cref{eq:master-4}). Taking Fock cutoffs extends \eqref{eq:master-2} to $\mathcal S$.

We now verify conservativity. First, note that $2\Re(\bra\psi G \ket\psi) = -\sum_\alpha \norm{L_\alpha \ket\psi}^2$. This, together with $\pm\Re(\bra\psi G \ket\psi) \le \norm{\bra\psi G \ket\psi} \le \norm{G\ket\psi} \cdot \norm{\psi}$, together with the elementary inequality $2ab\le a^2+b^2$, implies
\begin{align}
\sum_\alpha\|L_\alpha\psi\|^2
\le \|\psi\|^2+\|G\psi\|^2.
\end{align}
Hence the jump operators extend continuously from $\mathcal F_{\mathrm{fin}}$ to $\mathrm{Dom}(\overline G)$ in the graph norm. Lastly, note that approximation from vectors in $\mathcal F_{\mathrm{fin}}$ then gives
$2\Re\langle\psi,\overline G\psi\rangle
+\sum_\alpha\|L_\alpha\psi\|^2=0$
for any $\ket\psi\in\operatorname{Dom}(\overline G)$.

We now want to use a result of \cite[Theorem 2.1]{chebotarev2001interaction} to prove conservitivity. To satisfy the conditions they require (in particular the condition $\Phi(I)\ge I$ in the notation of that paper), we append the harmless identity jump $L_0=I$ and replace $\overline G$ by $\widetilde G=\overline G-\tfrac12 I$ (
this leaves the Lindblad expression unchanged since $\mathcal D[I]=0$). To explain how this condition works, let us borrow the notation of \cite{chebotarev2001interaction} and write
\begin{align}
\Phi(B):=B+\sum_\alpha L_\alpha^\dagger B L_\alpha
\end{align}
denote the completely positive map associated with $(L_\alpha)_\alpha$ augmented with the identity jump. \
In order to apply \cite[Theorem 2.1]{chebotarev2001interaction}, we need a positive self-adjoint reference operator
$\Lambda$ that dominates $\Phi(I)$ and satisfies an upper bound on
its growth under the full Lindblad expression (we refer the reader to \cite{chebotarev2001interaction} for the introduction of these symbols and an exposition of these conditions. Here, we just show they are satisfied to achieve our statement). We choose
$\Lambda=cQ^2$, with $c>0$ sufficiently large that
\begin{align}\label{eq:master-5}
I
\le \Phi(I)
=I+(\gamma+\eta)\hat N+m\eta I
\le \Lambda.
\end{align}
Such a choice is possible because $\Phi(I)$ grows only linearly
with photon number. As a result, $\Lambda$ controls the total jump rate,
including the added identity jump.
We next verify the required growth bound for \cite[Equation (2.1)]{chebotarev2001interaction}. By
\cref{eq:master-2} with $p=2$, extended to $\mathcal S$, we have
\begin{align}\label{eq:master-6}
\begin{split}
&2\operatorname{Re}\langle\Lambda\psi,\widetilde G\psi\rangle
+\|\Lambda^{1/2}\ket\psi\|^2
+\sum_\alpha\|\Lambda^{1/2}L_\alpha\ket\psi\|^2\\
&\qquad=
c\left(
2\operatorname{Re}\langle Q^2\psi,\overline G\psi\rangle
+\sum_\alpha\|QL_\alpha\ket\psi\|^2
\right)\\
&\qquad\le c_2\|\Lambda^{1/2}\ket\psi\|^2,
\end{split}
\end{align}
for any $\ket\psi\in\mathcal S$.
Here the contribution of the identity jump cancels exactly the
contribution of the shift
$\widetilde G=\overline G-\tfrac12 I$.
Therefore, \cref{eq:master-6} is precisely the
reference-operator inequality required by the theorem.

It remains to check that the domain and continuity assumptions
allow us to use this inequality. We take $\mathcal S$ as the
common invariant core in the reference. Indeed, the bounded
shift from $\overline G$ to $\widetilde G$ preserves its domain
and cores, and
$e^{t\widetilde G}=e^{-t/2}V(t)$.
Hence, $\mathcal S$ is invariant under $e^{t\widetilde G}$.
Moreover, \cref{eq:master-4} shows that
$\widetilde G\mathcal S\subseteq\mathcal S$ and
$L_\alpha\mathcal S\subseteq\mathcal S$. In particular,
$\mathcal S$ lies in the domain of every power of
$\widetilde G$, as well as in the domains of $\Lambda$ and
$\Lambda^{1/2}L_\alpha$.

The required weighted continuity also follows from the preceding
estimates. Strong continuity in the $Q$-graph norm gives continuity
of the first map below, while strong continuity in the
$Q^2$-graph norm, together with \cref{eq:master-4} for $r=1$,
gives continuity of the second:
\begin{align}\label{eq:master-7}
t\mapsto\Lambda^{1/2}e^{t\widetilde G}\ket\psi,
\quad
t\mapsto\Lambda^{1/2}L_\alpha
e^{t\widetilde G}\ket\psi,
\end{align}
for any $\ket\psi\in\mathcal S$.
Both maps are therefore bounded and square-integrable on compact
time intervals. Since $W_{s,t}=e^{(t-s)\widetilde G}$, this also
verifies the corresponding continuity and local integrability
conditions for the two-parameter evolution in the reference.

Finally, the normalized coefficient map is
\begin{align}\label{eq:master-8}
\Lambda^{-1/2}\Phi(B)\Lambda^{-1/2}
=
\Lambda^{-1/2}B\Lambda^{-1/2}
+\sum_\alpha
(L_\alpha\Lambda^{-1/2})^\dagger
B(L_\alpha\Lambda^{-1/2}).
\end{align}
Each $L_\alpha\Lambda^{-1/2}$ is bounded, as follows from
\cref{eq:master-5}. Since there are finitely many jumps,
\cref{eq:master-8} defines a bounded normal completely positive
map. It is independent of time, so the required time-continuity
condition is automatic. The additional continuity requirement
on the reference family is also automatic (since $\Lambda_t=\Lambda$ and
$\Lambda^{-1/2}\Lambda_t\Lambda^{-1/2}=I$).

We have now verified the hypotheses of
\cite[Theorem 2.1]{chebotarev2001interaction}. It follows that the
minimal semigroup constructed from $e^{t\widetilde G}$ and the
augmented jumps is unital on bounded operators. Its predual is
therefore a strongly continuous, completely positive,
trace-preserving semigroup on trace class. Denote its generator
by $\mathcal L_U$.

It remains to identify this generator with the closure of the
operator in the statement. The formal conservativity identity
proved above is preserved by the addition of the identity jump:
\begin{align}\label{eq:master-9}
2\operatorname{Re}\langle\psi,\widetilde G\psi\rangle
+\|\psi\|^2+\sum_\alpha\|L_\alpha\psi\|^2=0,
\qquad
\psi\in\operatorname{Dom}(\widetilde G).
\end{align}
Also, $\mathcal F_{\mathrm{fin}}$ is a core for
$\widetilde G$, because it is a core for $\overline G$.
Conservativity and \cite[Lemma 3.18]{van2025energy}
therefore imply that $\mathfrak T_{\mathrm{fin}}$
is a core for $\mathcal L_U$.
On this core, the generator acts by the original Lindblad
expression: the identity-jump term again cancels the two
$-\tfrac12 I$ contributions from $\widetilde G$. Hence
\[
\mathcal L_U\big|_{\mathfrak T_{\mathrm{fin}}}
=\mathcal L_{U,\mathrm{fin}}.
\]
Since $\mathcal L_U$ is closed and
$\mathfrak T_{\mathrm{fin}}$ is a core, this proves that
$\mathcal L_{U,\mathrm{fin}}$ is closable and
\begin{align}\label{eq:master-10}
\mathcal L_U=\overline{\mathcal L_{U,\mathrm{fin}}}.
\end{align}
\end{proof}

For an initial state \(\rho_{\rm in}\), define
\begin{align}
    \rho_0(t)
    &:=
    e^{t\mathcal{L}_0}[\rho_{\rm in}], \quad 
    \rho_U(t):=
    e^{t\mathcal{L}_U}[\rho_{\rm in}] .
\end{align}

\begin{lemma}[Gaussian photon-moment bound]
Let \(\rho_{\boldsymbol{\mu},\boldsymbol{\Sigma}}\) be an \(m\)-mode Gaussian state with mean vector
\(\boldsymbol{\mu}\) and covariance matrix \(\boldsymbol{\Sigma}\).  Define
\begin{align}
    E_G
    :=
    1+\|\boldsymbol{\Sigma}\|+\|\boldsymbol{\mu}\|^2 .
\end{align}
Then there exists a universal constant \(C_{\rm W}>0\) such that, for all
modes \(j,k\),
\begin{align}
    \sqrt{
    \Tr\left(
        (\hat n_j\hat n_k)^2\rho_{\boldsymbol{\mu},\boldsymbol{\Sigma}}
    \right)}
    \le
    C_{\rm W}E_G^2 .
\end{align}
\end{lemma}

\begin{proof}
The operator \(\hat n_j\hat n_k\) is a polynomial of total degree \(4\) in the creation
and annihilation operators, and \((\hat n_j\hat n_k)^2\) is a polynomial of total degree
\(8\).  Since \(\rho_{\boldsymbol{\mu},\boldsymbol{\Sigma}}\) is Gaussian, Wick's theorem expresses
every degree-\(8\) moment as a finite sum of products of first and second
moments.  Each two-point function is bounded by a constant times
\(1+\|\boldsymbol{\Sigma}\|\), while pairs of first moments are bounded by \(\|\boldsymbol{\mu}\|^2\).
Thus every Wick term is bounded by
\begin{align}
    C\bigl(1+\|\boldsymbol{\Sigma}\|+\|\boldsymbol{\mu}\|^2\bigr)^4 .
\end{align}
There are only finitely many Wick contractions at degree \(8\), so the same
bound holds after increasing the constant.  Taking the square root gives the
second claim.
\end{proof}

\begin{theorem}[Finite-time stability with explicit Gaussian energy scale]
\label{thm:finite-time-stability}
Assume that the unperturbed Gaussian trajectory has moments satisfying
\begin{align}
    \dot{\boldsymbol{\mu}}(s)
    &=
    \mathbf{A}\boldsymbol{\mu}(s)+\mathbf{d}, \qquad
    \dot{\boldsymbol{\Sigma}}(s)
    =
    \mathbf{A}\boldsymbol{\Sigma}(s)+\boldsymbol{\Sigma}(s)\mathbf{A}^T+\mathbf{Q} .
\end{align}
Assume that the Gaussian drift $\mathbf{A}$ is stable:
\begin{align}
    \|e^{s\mathbf{A}}\|
    \le
    C_Ae^{-\lambda_A s},
    \qquad
    s\ge0,
\end{align}
for some constants \(C_A\ge1\) and \(\lambda_A>0\).  Define
\begin{align}
    \mu_G
    &:=
    C_A\|\boldsymbol{\mu}(0)\|
    +
    \frac{C_A}{\lambda_A}\|\mathbf{d}\|,\qquad 
    \Sigma_G
    :=
    C_A^2\|\boldsymbol{\Sigma}(0)\|
    +
    \frac{C_A^2}{2\lambda_A}\|\mathbf{Q}\|,
\end{align}
and
\begin{align}
    E_G
    :=
    1+\Sigma_G+\mu_G^2 .
\end{align}
If \(\sum_{j,k}|U_{jk}|\le U\), then
\begin{align}
    \|\rho_U(t)-\rho_0(t)\|_1
    \le
    C_{\rm ft}\,
    U E_G^2t ,
    \label{eq:finite-time-kerr-stability}
\end{align}
where \(C_{\rm ft}>0\) is a universal constant.  Hence
\begin{align}
    U
    \le
    \frac{\epsilon}{C_{\rm ft}E_G^2t} \implies \|\rho_U(t)-\rho_0(t)\|_1
    \le
    \epsilon .
    \label{eq:finite-time-U-condition}
\end{align}
\end{theorem}

\begin{proof}
By Duhamel's formula,
\begin{align}
    \rho_U(t)-\rho_0(t)
    =
    \int_0^t
    e^{(t-s)\mathcal{L}_U}
    \mathcal{K}\bigl(\rho_0(s)\bigr)\,ds .
\end{align}
Since \(e^{(t-s)\mathcal{L}_U}\) is trace-norm contractive on Hermitian
trace-class operators,
\begin{align}
    \|\rho_U(t)-\rho_0(t)\|_1
    &\le
    \int_0^t
    \|\mathcal{K}[\rho_0(s)]\|_1\,ds \\
    &\le
    \sum_{(j,k)\in E}|U_{jk}|
    \int_0^t
    \|[\hat n_j\hat n_k,\rho_0(s)]\|_1\,ds .
\end{align}
For any positive self-adjoint operator \(\mathbf A\) and density matrix \(\rho\),
\begin{align}
    \|[\mathbf{A},\rho]\|_1
    &\le
    \|\mathbf A\rho\|_1+\|\rho \mathbf A\|_1 .
\end{align}
Moreover,
\begin{align}
    \|\mathbf A\rho\|_1
    =
    \|\mathbf A\rho^{1/2}\rho^{1/2}\|_1 \le
    \|\mathbf A\rho^{1/2}\|_2\|\rho^{1/2}\|_2 =
    \sqrt{\operatorname{Tr}(\mathbf A^2\rho)} .
\end{align}
The same upper bound holds for \(\|\rho \mathbf A\|_1\).  Hence
\begin{align}
    \|[\mathbf A,\rho]\|_1
    \le
    2\sqrt{\operatorname{Tr}(\mathbf A^2\rho)} .
\end{align}
Taking \(A=\hat n_j\hat n_k\) and \(\rho=\rho_0(s)\), we obtain
\begin{align}
    \|[\hat n_j\hat n_k,\rho_0(s)]\|_1
    \le
    2
    \sqrt{
    \Tr\left(
        (\hat n_j\hat n_k)^2\rho_0(s)
    \right)} .
\end{align}

It remains to bound the Gaussian moments uniformly in time.  The stable drift
assumption gives
\begin{align}
    \|\boldsymbol{\mu}(s)\|
    \le
    C_Ae^{-\lambda_A s}\|\boldsymbol{\mu}(0)\|
    +
    C_A\int_0^s e^{-\lambda_A(s-r)}\|\mathbf{d}\|\,dr \le
    C_A\|\boldsymbol{\mu}(0)\|
    +
    \frac{C_A}{\lambda_A}\|\mathbf{d}\|
    =
    \mu_G .
    \label{eq:mu-G}
\end{align}
Similarly,
\begin{align}
    \|\boldsymbol{\Sigma}(s)\|
    \le
    C_A^2e^{-2\lambda_A s}\|\boldsymbol{\Sigma}(0)\|
    +
    C_A^2\int_0^s e^{-2\lambda_A(s-r)}\|\mathbf{Q}\|\,dr \le
    C_A^2\|\boldsymbol{\Sigma}(0)\|
    +
    \frac{C_A^2}{2\lambda_A}\|\mathbf{Q}\|
    =
    \Sigma_G .
    \label{eq:Sigma-G}
\end{align}
Therefore the Gaussian photon-moment lemma implies
\begin{align}
    \sqrt{
    \operatorname{Tr}\!\left(
        (\hat n_j\hat n_k)^2\rho_0(s)
    \right)}
    \le
    C_{\rm W}E_G^2
\end{align}
for all \(s\ge0\).  Substituting this into the Duhamel estimate gives
\begin{align}
    \|\rho_U(t)-\rho_0(t)\|_1
    &\le
    2C_{\rm W}
    \sum_{j,k}|U_{jk}|
    \int_0^t E_G^2\,ds \le2C_{\rm W}UE_G^2t .
\end{align}
Renaming \(2C_{\rm W}\) as \(C_{\rm ft}\) proves
\eqref{eq:finite-time-kerr-stability}.
\end{proof}

As a result of this, under the condition
\eqref{eq:finite-time-U-condition}, the Fock-basis output distribution of
\(\rho_U(t)\) is within total variation distance at most \(\epsilon\) of
that of \(\rho_0(t)\) and hence the same algorithm based on \Cref{prop:main-classical-simulation} achieves a suitable approximate sample assuming the Gaussian trajectory is efficiently sampleable.

\begin{remark}
The finite-time estimate can be converted into a steady-state statement if
the nonlinear dynamics itself has unique fixed point and mixes rapidly. To see this, let \(\rho_U^\ast\) denote the
unique fixed point of \(\mathcal{L}_U\). Starting from initial state \(\rho_{\rm in}\) let the mixing time \(t_\varepsilon\), i.e.,
the minimum time at which     
\begin{align}
    \left\|
    e^{t_\varepsilon \mathcal{L}_U}[\rho_{\rm in}]
    -
    \rho_U^\ast
    \right\|_1
    \le
    \varepsilon /2.
\end{align}
Then
\begin{align}
    \|\rho_U^\ast-\rho_0(t_\epsilon)\|_1
    &\le
    \frac{\epsilon}{2}
    +
    C_{\rm ft}U E_G^2t_\epsilon .
\end{align}
Therefore, if $U
    \le
    \frac{\epsilon}
    {2C_{\rm ft}E_G^2t_\epsilon},$ then
$
    \|\rho_U^\ast-\rho_0(t_\epsilon)\|_1
    \le
    \epsilon .
$
Thus sampling from the Gaussian trajectory at the nonlinear mixing time gives
a sampler for the nonlinear fixed point, provided the accumulated Kerr
response remains small. The main issue is that this bound relies on an unknown mixing time. In the next section we prove a perturbation theory that directly applies to the fixed point.

\end{remark}

\subsubsection{Stability of the fixed point under weak Kerr perturbations}

As discussed in the previous section, a basic application of Duhamel's identity implies finite-time stability of the Lindbladian trajectory under non-Gaussian perturbations. However the fixed point needs to be managed more carefully. In other words, when perturbing the Lindbladian, the fixed point will not necessarily be merely perturbed. A small change in parameters of Lindbaldian might in principle lead to drastic changes in the fixed point, as the fixed point is the result of infinite-time evolution under a Lindbladian. That is the main technical difficulty that we address and overcome in this section. 

Let $\mathbf{M}$ and $\mathbf{d}$ be the parameters of the Gaussian Hamiltonian $H_G$ over $m$ modes as defined by \Cref{eq:Gaussian-Hamiltonian-def}. Consider the Hamiltonian 
$$
H_G + \sum_{j,k} U_{j,k} \hat{n}_j \hat{n}_k,
$$
 undergoing uniform loss with rate $\gamma$ and uniform pumping with rate $\eta$. We show that fixed points of Gaussian Lindbladians only have polynomially large sensitivity to the strength of Kerr non-linearity defined as
$$
U = \sum_{j,k} |U_{j,k}|
$$
More concretely, let $\rho_0$ be the fixed point of the unperturbed Lindbladian and $\rho_U$ be the fixed point of the perturbed Lindbladian. 
In this section, we prove

\begin{theorem}\label{thm:kerr-perturbative}
 Let $D = 2m (\eta + \frac12\norm{[\boldsymbol{\Omega}, \mathbf{M}]}_\infty) + \frac{16 \norm{\mathbf d}^2}{\delta}$, where $\delta = \gamma-\eta - \norm{[\boldsymbol{\Omega},\mathbf{M}]}_\infty>0$. Furthermore, let $s^\ast = \min\{\frac12, \frac{\delta}{8D}\}$, and assume $\delta>0$.
Then, for every $0<\varepsilon\leq1$, the condition
\begin{align}
U=
O\left(
\frac{\varepsilon\delta (s^\ast)^2}
{m} \frac{1}{\log(\frac{m(1+D/\delta)}{\varepsilon})}
\right)
\end{align}
implies
\begin{equation}
    \|\rho_U-\rho_0\|_1\leq\varepsilon.
\end{equation}
\end{theorem}

Let $\mathcal L_U$ denote the Lindbladian corresponding to having the interaction $\sum_{j,k} U_{j,k} \hat n_j \hat n_k$ added to the Hamiltonian, and $\mathcal L_0$ be the initial (unperturbed) Gaussian Lindbladian. The proof technique can be summarized as follows: we first show that $\norm{\mathcal L_0[\rho_U]}$ is small (\cref{corol:l0-on-rhoU}). This would then imply that $e^{\mathcal L_0 t}[\rho_U] \approx \rho_U$ for some polynomial amount of time $t$. We then show fast mixing properties for $\mathcal L_0$ implying that a polynomial evolution time suffices to ensure $e^{\mathcal L_0 t}[\sigma]\approx \rho_0$ for a wide range of input states $\sigma$ (\cref{lem:fast-mixing}). Putting the pieces together, we conclude that $\rho_U\approx \rho_0$. In what follows, we rigorously prove these steps.

A technical tool that we make use of, is the ``exponential norm'' defined as follows.
\begin{definition}[Exponential norm]
For any bosonic operator, say $X$, we define the exponential norm, parameterized by a real number $s\in\mathbb R$ as
\begin{align}
\norm{X}_{\mathrm{Exp}, s}:=\norm{e^{s\hat N/2} X e^{s\hat N/2}}_1,
\end{align}
where $\norm{\cdot}_1$ denotes the standard trace norm.
\label{def:exp-norm}
\end{definition}
As defined in \Cref{sec:preliminaries}, $\hat N = \hat n_1 + \ldots + \hat n_m$ stands for total photon number of the bosonic system. Similar norm was utilized in \cite{gondolf2024energy} for perturbation applications to the fixed point of bosonic Lindbladians. Their approach however does reproduce the final output of this section directly.

Note that for any $s'<s$ we have
\begin{align}
\norm{X}_{\mathrm{Exp}, s'} \le \norm{X}_{\mathrm{Exp}, s}.
\end{align}
In addition, we note that $\norm{\cdot}_{\mathrm{Exp},0}$ coincides with the trace norm $\norm{\cdot}_1$. We also note that for any semi-definite positive operator, say $\sigma \geq 0$, we have
\begin{align}\label{eq:norm-s-for-states}
 \norm{\sigma}_{\mathrm{Exp}, s} = \Tr(\Delta_s \sigma),
\end{align}
where we have used the shorthand notation $\Delta_s:= \exp(s \hat N)$.

We extend the notion of exponential norms to maps between bosonic operators in a natural way:

\begin{definition}
for any map, say $\mathcal O$, we define
\begin{align}
\norm{\mathcal O}_{s\to s'} = \sup_X\frac{\norm{\mathcal O(X)}_{\mathrm{Exp}, s'}}{\norm{X}_{\mathrm{Exp}, s}}.
\end{align}
\end{definition}

 Interestingly, for many polynomial Hamiltonians, the $\norm{\cdot}_{s\to s'}$ is finite in cases such as $s'< s$, although the operator norm is infinite. We use the notation
 \begin{align}
\Delta_s := e^{s\hat N}
\label{eq:Delta-s}
 \end{align}
 
 Below, we demonstrate this for a Kerr non-linearity term.

 \begin{lemma}\label{lem:kerr-strength}
Let $\mathcal{K}[\bullet] := -i[H_{\mathrm{Kerr}}, \bullet]$, with $H_{\mathrm{Kerr}} = \sum_{(j,k) \in E} U_{j,k} \hat n_j \hat n_k$. Then whenever $s'<s$
\begin{align}
\norm{\mathcal{K}}_{s\to s'} \le \frac{32U}{e^2(s-s')^2}.
\end{align}

 \end{lemma}
 \begin{proof}
Let
\begin{align}
\Xi_s(\bullet):= e^{s\hat N/2} \bullet e^{s\hat N/2}
\label{eq:Delta-s-op}
\end{align}
and note that $\Xi_s$ and $\mathcal{K}$ commute as they are both diagonal in number basis. Hence, for $s'< s$, we can write
\begin{align}
\begin{split}
\norm{\mathcal{K}[X]}_{s'}&= \norm{(\Xi_{s'}\circ \mathcal{K})(X)}_1\\
&= \norm{\Xi_{s'-s}\circ \mathcal{K}\circ \Xi_{s}(X)}_1\\
&\overset{(i)}{=} \norm{e^{(s'-s)\hat N/2} [H_{\mathrm{Kerr}}, X_s] e^{(s'-s)\hat N/2}}_1\\
&\le  2\norm{H_{\mathrm{Kerr}} e^{(s'-s) \hat N/2}}_\infty \norm{X_s}_1
\end{split}
\end{align}
where in $(i)$ we have used $X_s:=e^{s\hat N/2} X e^{s\hat N/2}$, and we have used $\norm{e^{(s'-s)\hat N/2}} \le 1$ in the last line. Now, use the fact that $|H_{\mathrm{Kerr}}|\le U \hat N^2$ and use the elementary inequality $\sup_{x\ge 0} x^2 e^{-hx/2} = \frac{16}{e^2h^2}$ to achieve
\begin{align}
\norm{H_{\mathrm{Kerr}} e^{(s'-s) \hat N/2}}_\infty \le  \frac{16 U}{e^2 (s-s')^2}
\end{align}

\end{proof}

To proceed, we present some intermediary results that will become useful. Let
\begin{align}
D:= 2m (\eta + \frac12\norm{[\boldsymbol{\Omega}, \mathbf{M}]}) + \frac{16 \norm{d}^2}{\delta},
\end{align}
with
\begin{align}
\delta = \gamma - \eta - \norm{[\boldsymbol{\Omega}, \mathbf{M}]} >0
\end{align}
being a margin that we need throughout our proof. The assumption \(\delta>0\) provides a quantitative dissipative contraction margin that will be used throughout the proof to establish moment bounds and mixing estimates. This condition is logically distinct from the sufficient condition for preservation of \(P\)-classicality established in \Cref{prop:main-classical-simulation}; neither condition implies the other in general.

In what follows, the notation $\operatorname{Dom}(\mathcal L_0) =\{ X \in \mathcal{T}_1 : \mathcal{L}_0 [X] \in \mathcal{T}_1\}$ denotes the
\emph{domain} of the Gaussian generator. For any $X \in \operatorname{Dom} (\mathcal{L}_0)$, $e^{t\mathcal L_0}[X]$ has
a right derivative at $t=0$ in trace norm. For such operators,
this derivative is $\mathcal L_0[X]$. Because the bosonic
Hamiltonian and jump operators are unbounded, this derivative
need not exist for every density operator.
In our later
perturbation argument, we need $\mathcal L_0[\rho_U]$ to be a
well-defined trace-class operator so that we can bound its norm
and integrate the Gaussian evolution starting from $\rho_U$.
The following theorem establishes this domain condition using
an exponential-moment bound. 

\begin{theorem}
\label{thm:bound-rhou}
$\ker\mathcal L_U$ contains a density operator.
Every density operator $\rho_U\in\ker\mathcal L_U$ satisfies
\begin{equation}
\operatorname{Tr}(e^{s\hat N}\rho_U)
=\|\rho_U\|_{{\rm Exp},s}
\leq \exp\left(\frac{32Ds}{7\delta}\right)
\leq e^{4/7},
\qquad 0\leq s\leq \min\left\{\frac12, \frac{\delta}{8D}\right\}.
\label{eq:repair-exponential-moment}
\end{equation}
Moreover,
$\rho_U\in\operatorname{Dom}(\mathcal L_0)$.
\end{theorem}

Note an important feature of the above thoerem is that the upper bound is independent of $U$. 
\begin{proof}
We shall first work with the weak stationary equations
\begin{equation}
\operatorname{Tr}\bigl(\rho\,\mathcal L_U^\dagger[A]\bigr)=0
\qquad\text{for all }A\in\mathfrak T_{\rm fin}.
\label{eq:repair-weak-stationary}
\end{equation}
Every density operator in $\ker\mathcal L_U$ satisfies these equations
by the definition of the closure. We will construct a density operator
satisfying them, prove the asserted moment bounds for every such density
operator, and then show that it belongs to $\ker\mathcal L_U$ in the
strong, trace-class sense. We first prove an auxiliary lemma:

\begin{lemma}\label{lem:L-dag-bound}
Let $\Delta_s$ be defined as \Cref{eq:Delta-s} with $0\le s\le s^*=\min\{1/2,\delta/(8D)\}$. It is the case that 
\begin{align}
\mathcal L_U^\dagger[\Delta_s]\leq s\Delta_s(D\mathbb \, \mathbb I-\frac{7\delta}{32}\hat N).
\label{eq:repair-lyapunov}
\end{align}
\end{lemma}
\begin{proof}
First, up to an irrelevant scalar, write
\begin{align}
H_G={}&\hat{\boldsymbol a}^{\dagger T}\mathbf h\hat{\boldsymbol a}
 +\frac12\left(
 \hat{\boldsymbol a}^{\dagger T}\mathbf G\hat{\boldsymbol a}^{\dagger}
 +\hat{\boldsymbol a}^{T}\overline{\mathbf G}\hat{\boldsymbol a}
 \right)
 +\boldsymbol\xi^*\cdot\hat{\boldsymbol a}
 +\boldsymbol\xi\cdot\hat{\boldsymbol a}^{\dagger},
\end{align}
where $\mathbf h=\mathbf h^\dagger$, $\mathbf G=\mathbf G^T$, and
\begin{equation}
g:=\|\mathbf G\|_\infty=\frac\kappa2,
\qquad \|\boldsymbol\xi\|_2^2=\frac12\|\mathbf d\|_2^2.
\label{eq:repair-conventions}
\end{equation}
These identities follow by substituting
$\hat a_j=(\hat q_j+i\hat p_j)/\sqrt2$ in $H_G$. Since $[H_{\rm Kerr},\hat N]=0$, direct
conjugation gives
\begin{align}\label{eq:repair-exact-drift}
\begin{split}
\Delta_s^{-1/2}\mathcal L_U^\dagger[\Delta_s]\Delta_s^{-1/2}
={}&-\alpha_s\hat N+\beta_s I
+2i\sinh(s/2)
(\boldsymbol\xi^*\cdot\hat{\boldsymbol a}
-\boldsymbol\xi\cdot\hat{\boldsymbol a}^\dagger)\\
&+i\sinh(s)
(\hat{\boldsymbol a}^{T}\overline{\mathbf G}\hat{\boldsymbol a}
-\hat{\boldsymbol a}^{\dagger T}\mathbf G
   \hat{\boldsymbol a}^\dagger),
\end{split}
\end{align}
where $\alpha_s:=\gamma(1-e^{-s})-\eta(e^s-1)$ and $\beta_s:=m\eta(e^s-1)$.
The squeezing term satisfies the quadratic-form bound
\begin{equation}
i(\hat{\boldsymbol a}^{T}\overline{\mathbf G}\hat{\boldsymbol a}
 -\hat{\boldsymbol a}^{\dagger T}\mathbf G
       \hat{\boldsymbol a}^\dagger)
\leq 2g\hat N+mgI.
\label{eq:repair-squeezing}
\end{equation}
To see this, factor
$\overline{\mathbf G}=V^T\operatorname{diag}(\sigma_j)V$ with $V$
unitary, and set $\hat{\boldsymbol b}=V\hat{\boldsymbol a}$.
For a normalized vector, writing
$n'_j=\langle\hat b_j^\dagger\hat b_j\rangle$, Cauchy--Schwarz gives
\[
\left|\left\langle\sum_j\sigma_j\hat b_j^2\right\rangle\right|
\leq \sum_j\sigma_j\sqrt{n'_j(n'_j+1)}
\leq g\left(\langle\hat N\rangle+\frac m2\right),
\]
which proves \eqref{eq:repair-squeezing}.
Similarly, for every $q>0$,
\begin{equation}
2i\sinh(s/2)
 (\boldsymbol\xi^*\cdot\hat{\boldsymbol a}
   -\boldsymbol\xi\cdot\hat{\boldsymbol a}^\dagger)
\leq \frac q2\hat N
 +\frac{8\sinh^2(s/2)\|\boldsymbol\xi\|_2^2}{q}I.
\label{eq:repair-drive}
\end{equation}
To see this, bound the absolute expectation of the left side by
$4\sinh(s/2)\|\boldsymbol\xi\|_2\sqrt{\langle\hat N\rangle}$
and use $ab\leq(a^2+b^2)/2$.
For $0<s\leq s^\ast$, choose
\begin{align}
q_s&:=\alpha_s-2g\sinh s
\nonumber\\
&=(1-e^{-s})\bigl[\delta-(\eta+g)(e^s-1)\bigr]
\geq\frac{7\delta s}{16}>0.
\label{eq:repair-qs}
\end{align}
Here $e^s-1\leq2s$, $1-e^{-s}\geq s/2$, and
$2s(\eta+g)\leq\delta/8$, using $D\geq2m(\eta+g)$ and $m\geq1$.
If $D=0$, the last term vanishes and the same conclusion holds.
Substituting \eqref{eq:repair-squeezing} and \eqref{eq:repair-drive}
in \eqref{eq:repair-exact-drift}, and using
$\sinh s\leq2s$ and $\sinh(s/2)\leq s$, gives
\begin{align}
\Delta_s^{-1/2}\mathcal L_U^\dagger[\Delta_s]\Delta_s^{-1/2}
&\leq -\frac{q_s}{2}\hat N
 +\left[\beta_s+mg\sinh s
  +\frac{8\sinh^2(s/2)\|\boldsymbol\xi\|_2^2}{q_s}\right]I
\nonumber\\
&\leq -\frac{7\delta s}{32}\hat N
 +s\left[2m(\eta+g)
       +\frac{64\|\mathbf d\|_2^2}{7\delta}\right]I
\nonumber\\
&\leq s(D\mathbb I-\frac{7\delta}{32}\hat N),
\label{eq:repair-conjugated-lyapunov}
\end{align}
which yileds the result.
\end{proof}

In what follows, we use the shorthand notation $c = \frac{7\delta}{32}$. Let $P_L=\mathbf1_{\{\hat N\leq L\}}$ and $\Pi_L=\mathbf1_{\{\hat N=L\}}$.
Compress the Hamiltonian and every jump operator:
\begin{align}
H^{[L]}=P_L(H_G+H_{\rm Kerr})P_L,
\qquad a_j^{[L]}=P_L\hat a_jP_L,
\qquad (a_j^\dagger)^{[L]}=P_L\hat a_j^\dagger P_L.
\end{align}
The resulting finite-dimensional Lindblad generator
$\mathcal L_U^{[L]}$ has a fixed point density operator $\rho_L$, since it is a finite-dimensional Lindbladian.\footnote{$\mathcal L^{[L]}_U$ is defined as $-i[H^{[L]}, \bullet] + \gamma \sum_j \mathcal D(a_j^{[L]}) + \eta \sum_j \mathcal D((a_j^{[L]})^\dag)$.} Write $\Delta_s^{[L]}=P_L\Delta_sP_L$. The Hamiltonian and loss contributions to
its adjoint drift are exactly the projections of the full contributions.
This projection removes only the pumping transitions out of the top sector;
their contribution to the full exponential drift is nonnegative. More
explicitly, we can refine the result of \cref{lem:L-dag-bound} to get
\begin{align}\label{eq:repair-truncated-drift}
\begin{split}
(\mathcal L_U^{[L]})^\dagger[\Delta_s^{[L]}]
={}&P_L\mathcal L_U^\dagger[\Delta_s]P_L
 -\eta(e^s-1)e^{sL}(L+m)\Pi_L\\
\leq{}&s\Delta_s^{[L]}(D\, \mathbb I-c\hat N).
\end{split}
\end{align}
Define $f_L(s)=\Tr(\Delta_s\rho_L)$, finite-dimensional
stationarity therefore gives
\[
0\leq s\bigl(Df_L(s)-cf_L'(s)\bigr),
\qquad 0<s\leq s^\ast.
\]
Since $f_L(0)=1$, integration yields the uniform bound
\begin{equation}
f_L(s)\leq e^{Ds/c},\qquad 0\leq s\leq s^\ast.
\label{eq:repair-finite-moments}
\end{equation}
In particular, we can use $(\mathbb I - P_K) e^{s (K+1)} \le e^{s\hat N}$ for any $s$ (this is very similar to a Chernoff inequality), to obtian:
\begin{align}\label{eq:rho_L}
\operatorname{Tr}((\mathbb I-P_K)\rho_L)
\leq e^{Ds^\ast/c}e^{-s^\ast(K+1)}.
\end{align}
Importantly, there is no dependence on $L$ on the right hand side. This shows that, by taking $K$ sufficiently
large, we can make $\|\rho_L-P_K\rho_LP_K\|_1$ arbitrarily small
for every $L$, using the same $K$. For each fixed $K$, the operators $P_K\rho_LP_K$ are matrices
of a common finite size, are positive, and have trace at most
one. They therefore have a convergent subsequence. Hence, we can choose
indices $L_j$ such that $P_K\rho_{L_j}P_K$ converges for every
fixed $K$. The states $\rho_{L_j}$ themselves then converge in trace norm.
Indeed, for any desired accuracy, first choose $K$ so that the
truncation errors are uniformly small. Convergence of the
compressed matrices then ensures that any two sufficiently late
states $\rho_{L_j}$ are close in trace norm. Since the space of
trace-class operators is complete, this sequence has a
trace-norm limit $\rho$. Positivity and trace one are preserved
under this convergence, so $\rho$ is a density operator. For each $A\in\mathfrak T_{\rm fin}$ and all sufficiently large $L$,
$\mathcal L_U^\dagger[A]$ agrees with
$(\mathcal L_U^{[L]})^\dagger[A]$ on the truncated space. This follows
because the Hamiltonian changes total number by at most two and each
jump changes it by one. Thus for each $A$, there is a large enough $L$ such that
\begin{align}
\operatorname{Tr}(\rho_L\mathcal L_U^\dagger[A])=0.
\end{align}
The observable on the left is bounded. Passing to the limit proves
\eqref{eq:repair-weak-stationary} for $\rho$.

Now, let $\rho$ be any density operator satisfying
\eqref{eq:repair-weak-stationary}, without any moment hypothesis.
Fix $0<s\leq s^\ast$, and for $R>0$ define \textit{bounded} observables
\begin{equation}
R_s=(\mathbb I+\Delta_s/R)^{-1},
\qquad F_{s,R}=\Delta_s(\mathbb I+\Delta_s/R)^{-1}=R(\mathbb I-R_s).
\label{eq:repair-bounded-observable}
\end{equation}
Using the identity $[H, A^{-1}] =- A^{-1} [H,A] A^{-1}$, it is straightforward to show that the Hamiltonian part satisfies the identity
\begin{align}
i[H_G+H_{\rm Kerr},F_{s,R}]
=R_s\,i[H_G+H_{\rm Kerr},\Delta_s]R_s.
\end{align}
For the dissipative part, both $\Delta_s$ and $F_{s,R}$ are functions of
$\hat N$. For any function $b$, we can write the effect of dissipation (loss and pumping) on $b(\hat N)$ as
\begin{equation}
\mathcal L_{\rm diss}^\dagger[b(\hat N)]
=\gamma\hat N\bigl(b(\hat N-1)-b(\hat N)\bigr)
 +\eta(\hat N+mI)\bigl(b(\hat N+1)-b(\hat N)\bigr).
\label{eq:repair-birth-death}
\end{equation}
The scalar function $\phi_R(x)=x/(1+x/R)$ is increasing and concave,
with $\phi_R'(x)=(1+x/R)^{-2}$. Its tangent-line inequality
\[
\phi_R(y)-\phi_R(x)\leq\phi_R'(x)(y-x)
\]
applied to each transition in \eqref{eq:repair-birth-death} gives
\[
\mathcal L_{\rm diss}^\dagger[F_{s,R}]
\leq R_s\mathcal L_{\rm diss}^\dagger[\Delta_s]R_s.
\]
Combining the last two facts with \cref{lem:L-dag-bound}, we obtain
\begin{equation}
\mathcal L_U^\dagger[F_{s,R}]
\leq s\,\frac{\Delta_s}{(I+\Delta_s/R)^2}(D \mathbb I-c\hat N).
\label{eq:repair-bounded-drift}
\end{equation}
Both sides of this inequality are bounded.

We justify stationarity for the bounded observable $F_{s,R}$ directly.
On the sector with total number $n$,
\[
|F_{s,R}(n)-R|=\frac{R^2}{R+e^{sn}}\leq R^2e^{-sn}.
\]
Consequently the observables
$R\,\mathbb I +P_L(F_{s,R}-R\,\mathbb I)P_L$ converge in operator norm to $F_{s,R}$,
and their formal adjoint images converge in operator norm to
$\mathcal L_U^\dagger[F_{s,R}]$.
Indeed, the number-preserving Hamiltonian terms commute with these
observables; the remaining Hamiltonian matrix elements and the
dissipative transition rates grow at most linearly in $n$, whereas
the discarded diagonal coefficients decay exponentially. In particular,
the adjoint-image error is bounded by
$O_{s,R}((L+1)e^{-s(L-2)})$ for sufficiently large $L$.
Using \eqref{eq:repair-weak-stationary} on the finite-rank part and
$\mathcal L_U^\dagger[\mathbb I]=0$ gives, after taking the limit,
\begin{equation}
\operatorname{Tr}(\rho\mathcal L_U^\dagger[F_{s,R}])=0.
\label{eq:repair-bounded-stationarity}
\end{equation}
Let $p_n=\operatorname{Tr}(\mathbf1_{\{\hat N=n\}}\rho)$.
Equations \eqref{eq:repair-bounded-drift} and
\eqref{eq:repair-bounded-stationarity} imply
\begin{equation}
\sum_{cn>D}(cn-D)\frac{e^{sn}}{(1+e^{sn}/R)^2}p_n
\leq
\sum_{cn\leq D}(D-cn)\frac{e^{sn}}{(1+e^{sn}/R)^2}p_n.
\label{eq:repair-tail-control}
\end{equation}
The right-hand sum has finitely many terms and is bounded independently
of $R$. The factors on the left increase to $e^{sn}$ as $R\to\infty$.
Monotone convergence first proves
\[
\sum_n e^{sn}p_n<\infty,
\qquad \sum_n n e^{sn}p_n<\infty,
\]
and then yields
\begin{equation}
c\sum_n n e^{sn}p_n\leq D\sum_n e^{sn}p_n.
\label{eq:repair-legitimate-differential-bound}
\end{equation}
To see the first implication explicitly, outside a finite set
$cn-D\geq1$ and $n\leq 2(cn-D)/c$.

It is now legitimate to set
$f(s)=\operatorname{Tr}(e^{s\hat N}\rho)$ and differentiate:
\[
f'(s)=\sum_n n e^{sn}p_n\leq\frac Dc f(s),
\qquad 0<s<s^\ast.
\]
The endpoint integrability just proved gives continuity at both endpoints.
Since $f(0)=1$, integration proves
$f(s)\leq e^{Ds/c}$ for $0\leq s\leq s^\ast$.
Finally, $Ds^\ast\leq\delta/8$, also when $D=0$, gives
$e^{Ds/c}\leq e^{4/7}$.

We record why the moment estimate suffices for all subsequent uses of
the stationary equation. Note that
every polynomial $P$ in creation and annihilation operators satisfies
$\|Pe^{-r\hat N/2}\|_\infty<\infty$.
Factoring $X=e^{-r\hat N/2}X_re^{-r\hat N/2}$, where
$X_r=e^{r\hat N/2}Xe^{r\hat N/2}$, therefore bounds each term in
$\mathcal L_U[X]$.
This shows that
the formal maps $\mathcal L_U$, $\mathcal L_0$, and
$\mathcal K=-i[H_{\rm Kerr},\cdot]$ are bounded from
this weighted trace class to trace class, for every fixed $r>0$.

For $0<r<s^\ast$ and $\rho^{[L]}=P_L\rho P_L$,
the exponential moment already proved gives
\begin{equation}
\|\rho-\rho^{[L]}\|_{{\rm Exp},r}
\leq O(e^{-(s^\ast-r)(L+1)/2})\to0.
\label{eq:repair-weighted-cutoff}
\end{equation}
This follows by applying the finite-rank compression bound to the
positive operator $e^{r\hat N/2}\rho e^{r\hat N/2}$.
Thus $\rho^{[L]}\to\rho$ and all three formal generator images
converge in trace norm. Closedness shows that $\rho$ belongs to
their respective closed domains. Testing the limit
$\mathcal L_U[\rho]$ against every finite Fock matrix and using
\eqref{eq:repair-weak-stationary} shows that this limit is zero.
Moreover, taking limits in the identity on $\rho^{[L]}$ proves
\begin{equation}
\mathcal L_0[\rho]=-\mathcal K[\rho].
\label{eq:repair-strong-residual-identity}
\end{equation}
Applying this to the state constructed in the paragraph right after \eqref{eq:rho_L} proves existence
in the strong sense, and applying it to every weakly stationary state
proves all remaining assertions.
\end{proof}

For technical reasons later, we establish an energy bound for $\rho_U$ as well
\begin{lemma}
\label{lem:rho-u-energy-bound}
Under the assumptions of Theorem~\ref{thm:bound-rhou}, every stationary
density operator satisfies
\begin{equation}
\operatorname{Tr}(\hat N\rho_U)\leq\frac D\delta.
\label{eq:repair-energy}
\end{equation}
\end{lemma}
\begin{proof}
The same squeezing estimate as above, and
\[
i(\boldsymbol\xi^*\cdot\hat{\boldsymbol a}
 -\boldsymbol\xi\cdot\hat{\boldsymbol a}^\dagger)
\leq\frac\delta2\hat N
 +\frac{2\|\boldsymbol\xi\|_2^2}{\delta}\mathbb I,
\]
give
\begin{align}
\begin{split}
\mathcal L_U^\dagger[\hat N]
&\leq-\frac\delta2\hat N
 +\left[m(\eta+g)+\frac{\|\mathbf d\|_2^2}{\delta}\right]\mathbb I\\
&\leq-\frac\delta2\hat N+\frac D2 \mathbb I.
\end{split}
\end{align}
The exponential integrability has now been established independently.
It justifies taking this expectation in a stationary state, for example
by applying the weak stationary equations to $P_L\hat N P_L$ and
passing to the limit. The boundary terms grow only polynomially in
$L$ and vanish against the exponential tail. Hence
$0\leq-\tfrac\delta2\operatorname{Tr}(\hat N\rho_U)+D/2$.
\end{proof}

We are now ready to state the smallness of $\mathcal{L}_0[\rho_U]$:

\begin{corollary}[Residual usable in the Gaussian comparison]
\label{corol:l0-on-rhoU}
For every fixed point $\rho_U$ of $\mathcal{L}_U$ and every $0<s\leq s^\ast$,
\begin{equation}
\|\mathcal L_0[\rho_U]\|_1
\leq\frac{32U}{e^2s^2}
       \exp\!\left(\frac{32Ds}{7\delta}\right).
\label{eq:repair-residual-sharp}
\end{equation}
\end{corollary}
\begin{proof}
The strong identity \eqref{eq:repair-strong-residual-identity} was proved
above. Hence, we can apply $\norm{\mathcal K}_{s\to 0}$ directly from \cref{lem:kerr-strength} to obtain
\begin{align*}
\|\mathcal L_0[\rho_U]\|_1=\|\mathcal K[\rho_U]\|_1
&
\leq\norm{\mathcal K}_{s\to 0} \norm{\rho_U}_{\mathrm{Exp},s}\\
&\leq\frac{32U}{e^2s^2}
       \exp\!\left(\frac{32Ds}{7\delta}\right),
\end{align*}
where the last line follows from both \cref{lem:kerr-strength} and \cref{thm:bound-rhou}.
\end{proof}

Lastly, we show that any low energy state mixes fast to $\rho_0$, under $\mathcal L_0$.

\begin{lemma}[Energy-constrained fast mixing]\label{lem:fast-mixing}
For $0<\varepsilon<1$, there exists
$T=O(\frac{m}{\delta}
\log \frac{2m(1+D/\delta)}{\varepsilon})$
such that for any $t\ge T$, and for any state $\sigma$ satisfying
$\Tr(\hat N\sigma)\le D/\delta$, we have
\begin{align}
\norm{e^{\mathcal L_0 t}[\sigma]-\rho_0}_1\le\varepsilon.
\end{align}
\end{lemma}

\begin{proof}
From \eqref{eq:mode-dependent-Gaussian-channel} we recall that
the XY parametrization of the Gaussian channel $e^{\mathcal L_0 t}$
is given by
\begin{align}\label{eq:recall-XY}
\mathbf X(t)=e^{\mathbf A t},
\qquad
\mathbf Y(t)=(\gamma+\eta)
\int_0^t e^{s\mathbf A}e^{s\mathbf A^T}\,\mathrm ds,
\end{align}
where
$\mathbf A=\boldsymbol\Omega\mathbf M
-\frac{\gamma-\eta}{2}\mathbb I$
is the drift matrix. Using the fact that
$\mathbf A+\mathbf A^T\le-\delta\mathbb I$, we can write
(cf. \cite{dahlquist1963special} or
\cite[Equation (3.7)]{krovi2023improved})
\begin{align}
\norm{\mathbf X(t)}\le e^{-\delta t/2}.
\end{align}
Next, we show that for long enough times, $\mathbf Y(t)$ is
lower bounded by a constant multiple of the identity matrix.
To do so, we start by noticing that
\begin{align}
\begin{split}
\mathbf A+\mathbf A^T
&=-(\gamma-\eta)\mathbb I+[\boldsymbol\Omega,\mathbf M]\\
&\ge-\bigl(\gamma-\eta
+\norm{[\boldsymbol\Omega,\mathbf M]}\bigr)\mathbb I.
\end{split}
\end{align}
Let us define
$c:=\gamma-\eta+\norm{[\boldsymbol\Omega,\mathbf M]}$.
Let $x(s)=e^{\mathbf A^T s}x_0$ for any vector $x_0$.
Note that
\begin{align}
\begin{split}
\frac{\mathrm d}{\mathrm ds}\norm{x(s)}^2
&=x(s)^T(\mathbf A+\mathbf A^T)x(s)\\
&\ge-c\norm{x(s)}^2.
\end{split}
\end{align}
Therefore,
$\norm{x(s)}^2\ge e^{-cs}\norm{x_0}^2$, which implies
\begin{align}
e^{\mathbf A s}e^{\mathbf A^T s}
\ge e^{-cs}\mathbb I.
\end{align}
Plugging into \eqref{eq:recall-XY} gives
\begin{align}
\mathbf Y(t)
\ge(\eta+\gamma)\int_0^t e^{-cs}\mathbb I\,\mathrm ds=\frac{\eta+\gamma}{c}(1-e^{-ct})\mathbb I.
\end{align}
Note that $c\le2(\gamma+\eta)$, which allows us to write
$(\eta+\gamma)/c\ge1/2$.
Also, since $c\ge\delta$, we have
$1-e^{-ct}\ge1-1/e$ for all $t\ge1/\delta$.
As a result,
\begin{align}\label{eq:lower-bound-Y}
\mathbf Y(t)\ge\frac{e-1}{2e}\mathbb I.
\end{align}

We now define the difference
\begin{align}
\begin{split}
\Delta\rho(t)
&=e^{\mathcal L_0 t}[\sigma]-\rho_0\\
&=e^{\mathcal L_0 t}[\sigma-\rho_0].
\end{split}
\end{align}
Set $E_*:=D/\delta$. By assumption and
\cref{lem:rho-u-energy-bound}, both $\sigma$ and $\rho_0$
have energy at most $E_*$.

We find it easier to study $\Delta\rho$ in the characteristic
function picture. This is because the effect of Gaussian channels
on the characteristic function is quite simple
\cite{serafini2017quantum}. In particular,
\begin{align}\label{eq:chi-of-delta}
|\chi_{\Delta\rho(t)}(\mathbf u)|
=e^{-\frac14\mathbf u^T\mathbf Y(t)\mathbf u}
\left|
\chi_{\sigma}(\mathbf X(t)^T\mathbf u)
-\chi_{\rho_0}(\mathbf X(t)^T\mathbf u)
\right|.
\end{align}
The displacement contributes only a phase, which drops out
in the absolute value. Furthermore, the elementary identity
$e^{ixt}=1+i\int_0^t e^{ixt'}x\,\mathrm dt'$
gives
$e^{i\hat Q_{\mathbf n}u}
=\mathbb I+i\int_0^1
e^{i\hat Q_{\mathbf n}\lambda u}
\hat Q_{\mathbf n}u\,\mathrm d\lambda$,
where $\hat Q_{\mathbf n}=\hat{\mathbf r}^T\mathbf n$
and $\mathbf n\in\mathbb R^{2m}$ is a unit vector.
Here we write $\mathbf u=u\mathbf n$.
Therefore, for any finite-energy state $\tau$, we can write
\begin{align}
\begin{split}
|\chi_{\tau}(\mathbf u)-1|
=\left|
\int_0^1
\Tr\bigl(\tau e^{i\hat Q_{\mathbf n}\lambda u}
\hat Q_{\mathbf n}\bigr)u\,\mathrm d\lambda
\right|=\left|
\int_0^1
\langle e^{-i\hat Q_{\mathbf n}\lambda u},
\hat Q_{\mathbf n}\rangle_{\tau}
u\,\mathrm d\lambda
\right|,
\end{split}
\end{align}
where
$\langle A,B\rangle_{\tau}:=\Tr(\tau A^\dagger B)$
is the $\tau$-weighted Hilbert--Schmidt inner product.
Using the Cauchy--Schwarz inequality for this weighted inner
product, we have
\[
\left|
\langle e^{-i\hat Q_{\mathbf n}\lambda u},
\hat Q_{\mathbf n}\rangle_{\tau}
\right|
\le\sqrt{\Tr(\tau\hat Q_{\mathbf n}^2)},
\]
which yields
\begin{align}\label{eq:chi-1-bound}
\begin{split}
|\chi_{\tau}(\mathbf u)-1|
\le\norm{\mathbf u}
\sqrt{\Tr(\tau\hat Q_{\mathbf n}^2)}\le\norm{\mathbf u}
\sqrt{2\Tr(\tau\hat N)+m},
\end{split}
\end{align}
where the last line follows from
$\hat Q_{\mathbf n}^2\le2\hat N+m\mathbb I$. Next, writing
\begin{align}
\begin{split}
&\left|
\chi_{\sigma}(\mathbf X(t)^T\mathbf u)
-\chi_{\rho_0}(\mathbf X(t)^T\mathbf u)
\right|\le
\left|\chi_{\sigma}(\mathbf X(t)^T\mathbf u)-1\right|
+\left|\chi_{\rho_0}(\mathbf X(t)^T\mathbf u)-1\right|,
\end{split}
\end{align}
and using \eqref{eq:chi-1-bound} together with
\eqref{eq:chi-of-delta} and \eqref{eq:lower-bound-Y},
we get
\begin{align}
|\chi_{\Delta\rho(t)}(\mathbf u)|
\le
2\norm{\mathbf u}\sqrt{2E_*+m}\,
e^{-\delta t/2}
e^{-\frac{e-1}{8e}\norm{\mathbf u}^2}.
\end{align}
Using
$
\norm{\Delta\rho(t)}_2^2
=\frac{1}{(2\pi)^m}
\int_{\mathbb R^{2m}}
|\chi_{\Delta\rho(t)}(\mathbf u)|^2\,\mathrm d\mathbf u$, together with the elementary identity
$
\int_{\mathbb R^{2m}}
\norm{\mathbf u}^2e^{-a\norm{\mathbf u}^2}\,\mathrm d\mathbf u
=m\pi^m a^{-(m+1)}
$, we get
\begin{align}\label{eq:delta-rho-bound-2}
\norm{\Delta\rho(t)}_2^2
\le
\frac{16e\,m(2E_*+m)}{e-1}
\left(\frac{2e}{e-1}\right)^m
e^{-\delta t},
\end{align}
for any $t\ge1/\delta$. Finally, we need to transform the $\ell_2$ norm to an $\ell_1$ norm.
For this, we also need an energy bound along the evolution.
The drift estimate used in the proof of
\cref{lem:rho-u-energy-bound},
\[
\mathcal L_0^\dagger[\hat N]
\le-\frac{\delta}{2}\hat N+\frac D2\mathbb I,
\]
and the Gaussian moment equations give
\begin{align}
\Tr\bigl(\hat N e^{\mathcal L_0t}[\sigma]\bigr)
&\le
e^{-\delta t/2}\Tr(\hat N\sigma)
+\frac D\delta(1-e^{-\delta t/2})\\
&\le E_*.
\end{align}
Thus both states in $\Delta\rho(t)$ have energy at most $E_*$.

Let $\Pi_R$ be the projector onto the subspace of fewer than
$R$ photons, where $R\ge1$ is an integer.
Using the quantum Markov inequality \cite{chabaud2026bosonic},
for any state $\tau$ with $\Tr(\hat N\tau)\le E_*$ we have
\begin{align}
\norm{\tau-\Pi_R\tau\Pi_R}_1
\le2\sqrt{\frac{E_*}{R}}.
\end{align}
As a result,
\begin{align}
\begin{split}
\norm{\Delta\rho(t)}_1
&\le
\norm{\Pi_R\Delta\rho(t)\Pi_R}_1
+4\sqrt{\frac{E_*}{R}}\\
&\le
\sqrt{{m+R\choose m}}\,
\norm{\Pi_R\Delta\rho(t)\Pi_R}_2
+4\sqrt{\frac{E_*}{R}}\\
&\le
\sqrt{{m+R\choose m}}\,
\norm{\Delta\rho(t)}_2
+4\sqrt{\frac{E_*}{R}}.
\end{split}
\end{align}
Setting
\[
R_\theta:=
\max\left\{1,
\left\lceil\frac{64E_*}{\varepsilon^2}\right\rceil
\right\}
\]
makes the second term at most $\varepsilon/2$.
Using
${m+R_\theta\choose m}\le(m+R_\theta)^m$
and \eqref{eq:delta-rho-bound-2}, the first term is at most
$\varepsilon/2$ for all $t\ge T$, with
\[
T=O\left(
\frac m\delta
\log\frac{2m(1+E_*)}{\varepsilon}
\right).
\]
This gives the desired result.
\end{proof}

We are now ready to prove the main result.

\begin{proof}[Proof of \cref{thm:kerr-perturbative}]
Since $\rho_U\in\operatorname{Dom}(\mathcal L_0)$, for every $t\geq0$,
\begin{align}
e^{t\mathcal L_0}[\rho_U]-\rho_U
=
\int_0^t
e^{r\mathcal L_0}
\mathcal L_0[\rho_U]\,dr.
\end{align}
Since $e^{r\mathcal L_0}$ is a quantum channel and channels can only
reduce the trace norm of Hermitian operators,
\cref{corol:l0-on-rhoU} gives
\begin{align}
\norm{
e^{t\mathcal L_0}[\rho_U]-\rho_U
}_1
\leq
\frac{32e^{4/7}U}{e^2(s^\ast)^2}\,t.
\label{eq:Duhamel-fixed-point}
\end{align}
By the energy bound \cref{lem:rho-u-energy-bound},
\cref{lem:fast-mixing} applies to $\sigma=\rho_U$.
Taking $T$ (with $\varepsilon\to\varepsilon/2$ in
\cref{lem:fast-mixing}) and using the triangle inequality,
\begin{align}
\begin{split}
\norm{\rho_U-\rho_0}_1
&\leq
\norm{
\rho_U-e^{T\mathcal L_0}[\rho_U]
}_1
+
\norm{
e^{T\mathcal L_0}[\rho_U]-\rho_0
}_1\\
&\leq
\frac{32e^{4/7}U}{e^2(s^\ast)^2}\,T
+
\frac{\varepsilon}{2}.
\end{split}
\end{align}
The first term is at most $\varepsilon/2$ provided that
\begin{align}
U\leq
\frac{e^2\varepsilon(s^\ast)^2}{64e^{4/7}T},
\end{align}
proving the claim under this condition.
\end{proof}

Finally, we note that \cref{thm:kerr-perturbative} immediately implies that the fixed point is unique, with $\varepsilon<1$ and any parameter $U$ chosen as the condition in \cref{thm:kerr-perturbative}. This is due to the fact that, as established in the proof of \cref{thm:uniq-linear-loss}, if a Lindbladian has two fixed points, then it also has teo fixed points that are maximally separated in trace distance (i.e., there are fixed points $\rho_{f,1}$ and $\rho_{f,2}$ such that $\norm{\rho_{f,1} - \rho_{f,2}}_1 = 2$). However, as established in \cref{thm:kerr-perturbative}, we have that $\norm{\rho_{f,i} - \rho_0} \le \varepsilon$ for some $\varepsilon\in[0,1)$. Therefore, a simple triangle inequality prevents $\rho_{f,1}$ and $\rho_{f,2}$ from being maximally separated, and hence, the fixed point must be unique. We summarize this in the following corollary:

\begin{corollary}
For any parameter $\varepsilon<1$, the choice of $U$ in \cref{thm:kerr-perturbative} ensures that the fixed point ($\rho_U$) is unique. 
\end{corollary}

\section{Robust hardness of sampling from Gaussian dynamics undergoing uniform linear loss}
\label{sec:Gaussian-hardness-sampling}

We establish hardness of sampling results from the fixed point of Gaussian dissipative systems when an incoherent pumping is removed. We have two results in this section. We first show hardness of exact sampling assuming non-collapse of the polynomial hierarchy. We can show that exact sampling remains hard even if the (uniform) loss rates are exponentially greater than coefficients of the Gaussian Hamiltonian terms of our dynamics. This result complements our classical simulation theorem (\cref{thm:gaussian-classical-trajectory}), where we could only establish easiness of sampling in the regime where pumping rates are non-zero. Next, we establish hardness of approximate sampling from the fixed point of Gaussian Lindbladians. Our construction first uses an idea similar to adiabatic elimination to establish robust hardness of sampling from fixed point when loss rates on half of the photons is zero. We then use additional gadgets to obtain uniform loss everywhere. Both constructions heavily rely on existence of squeezing terms in the Hamiltonian.

\subsection{Exact sampling becomes hard when incoherent pumping is removed}

In this section, we show that sampling the fixed point of Lindbladins with quadratic Hamiltonians and pure loss (not considering incoherent pumping) is not classically feasible (unless $\mathsf{PH}$ collapses to the third level). 

It is a standard result that sampling Gaussian pure states in the photon number basis is classically hard assuming non-collapse of the polynomial hierarchy~\cite{hamilton2017gaussian}. However, we cannot directly apply this result in our settings, because it turns out that the family of Lindbladians with pure linear loss which we consider in this work does not accept squeezed states as their fixed point. We first prove this fact.

\begin{proposition}
For a Gaussian Hamiltonian with uniform loss and pumping and $\gamma+\eta>0$,
a pure Gaussian fixed point must be a coherent state, and requires
$\eta=0$. In particular, a nontrivially squeezed pure state cannot be stationary.
\end{proposition}
\begin{proof}
The stationary covariance equation, multiplied by
$\boldsymbol\Sigma_f^{-1}$ and traced, gives
\[
\operatorname{Tr}(\boldsymbol\Sigma_f^{-1})
=\frac{2m\gamma_-}{\gamma_+}\leq2m,
\]
because $\operatorname{Tr}(\boldsymbol\Omega\mathbf M)=0$.
For a pure Gaussian state the Bloch--Messiah decomposition gives, in our
vacuum-$\mathbb I$ convention,
\[
\boldsymbol\Sigma_f=\mathbf O
\bigoplus_{j=1}^m\operatorname{diag}(e^{2r_j},e^{-2r_j})\mathbf O^T,
\qquad \operatorname{Tr}(\boldsymbol\Sigma_f^{-1})
=2\sum_j\cosh(2r_j)\geq2m.
\]
Equality is therefore necessary: all $r_j=0$ and $\gamma_- =\gamma_+$.
Thus the state is a displaced vacuum and $\eta=0$.
\end{proof}

We note that the above statement does not refute the possibility of implementing mixed squeezed states. However, the complexity of Gaussian boson sampling based on mixed squeezed states are not fully understood. Hence, we need to develop a hardness of sampling result for a family of mixed Gaussian states (particularly the ones resulting from Lindbladians with quadratic Hamiltonians and linear loss).

For these formulas, group the quadratures as $(q_1,\ldots,q_m,p_1,\ldots,p_m)$, permuting the covariance accordingly.
For a zero-mean Gaussian state define
\[
\mathbf Q=\tfrac12\mathbf W(\boldsymbol\Sigma+\mathbb I)\mathbf W^\dagger,
\qquad
\mathbf W=\tfrac1{\sqrt2}
\begin{pmatrix}\mathbb I&-i\mathbb I\\\mathbb I&i\mathbb I\end{pmatrix}.
\]
We have ordered the complex coordinates as $(\mathbf a^\dagger,\mathbf a)$;
reversing the two groups only permutes the Bargmann matrix. With
$\boldsymbol{\sigma}_x=\left(\begin{smallmatrix}0&\mathbb I\\\mathbb I&0\end{smallmatrix}\right)$,

\begin{equation}\label{eq:Bargmann-matrix}
\mathbf A=\boldsymbol{\sigma}_x (\mathbb I-\mathbf Q^{-1})
=\boldsymbol{\sigma}_x\left[\mathbb I-2\mathbf W(\boldsymbol\Sigma+\mathbb I)^{-1}\mathbf W^\dagger\right]
\end{equation}
(cf.\ \cite[Eq.~(69)]{motamedi2026stellar}, with this coordinate ordering).
For an even number $m$ of modes, the collision-free probability is
\[
\langle1^m|\rho|1^m\rangle
=\frac{\operatorname{Haf}(\mathbf A)}{\sqrt{\det\mathbf Q}}.
\]
We highlight that $\mathrm{det}(\mathbf Q)$ is computable up to a polynomial number of bits of precision (exponentially small additive error) within polynomial time.

We also need the following lemma
\begin{lemma}
    It is $\# \P$-hard under Turing reductions to approximate the squared hafnian $|\operatorname{Haf} (\mathbf{B})|^2$ of a symmetric matrix $\mathbf{B}$ to within inverse polynomial multiplicative error.
    \label{lem:hafnian-hardness}
\end{lemma}

\begin{proof}
Let $\mathbf{X}$ be an arbitrary real matrix and let $\mathbf{B} = \begin{pmatrix}
    0 & \mathbf{X}\\ \mathbf{X}^T & 0
\end{pmatrix}$. Then
$$
\operatorname{Haf} (\mathbf{B}) = \operatorname{Per} (\mathbf{X})
$$
therefore multiplicative approximation to $|\operatorname{Haf}(\mathbf{B})|^2$ implies a multiplicative approximation to $|\operatorname{Per}(X)|^2$ which we know is $\# \P$ hard under Turing reductions~\cite[Theorem 3.4]{aaronson2011computational}. 
\end{proof}

\begin{theorem}[Exact sampling hardness for pure-loss quadratic Lindbladians]\label{thm:hardness-of-gaussian-lindbladian}
There is no $\mathsf{BPP}$ algorithm to exactly sample from the fixed point of a Gaussian dynamics undergoing photon loss with a uniform constant rate assuming $\mathsf{PH}$ does not collapse.
\end{theorem}

\begin{proof}
Our strategy is to embed the approximation of the squared hafnian of a symmetric matrix as an instance of approximating the output probabilities of the fixed point of a Lindbladian from the family we consider and use \Cref{lem:hafnian-hardness} and standard complexity theoretic arguments developed in \cite{aaronson2011computational}.

Let \(\mathbf B\in \mathbb C^{m\times m}\) be any symmeric matrix with \(m\) even
and \(\|\mathbf B\|\le 1\).
The Takagi decomposition of $\mathbf{B}$ is as follows (c.f. \cite[Section IV]{houde2024matrix})
\begin{align}
\mathbf B=\mathbf U\operatorname{diag}(r_1,\ldots,r_m)\mathbf U^T,
\qquad 0\le r_j\le 1.
\end{align}
Let \(\mathbf R_{\mathbf U} \in \mathbb R^{2m\times 2m}\) be the real passive symplectic matrix implementing the linear
optical transformation \(\mathbf U\) on quadratures $\hat{\mathbf{r}}$. We construct
the following Lindbladian with quadratic Hamiltonian undergoing pure loss 
\begin{align}
\mathcal{L}_\lambda[\rho]
=
-i\left[\frac12 \hat{\mathbf{r}}^T \mathbf{M}_\lambda \hat{\mathbf{r}},\rho\right]
+
\gamma\sum_{j=1}^m \mathcal{D}(\hat a_j)[\rho],
\end{align}
where \(\gamma>0\) is the loss rate and
\begin{align}
\mathbf{M}_\lambda
=
\mathbf R_{\mathbf U}
\left[
\bigoplus_{j=1}^m
\frac{\gamma s_j}{2}
\begin{pmatrix}
0&1\\
1&0
\end{pmatrix}
\right]
\mathbf R_{\mathbf U}^T .
\end{align}
Here the parameters \(s_j\in[0,1)\) are chosen as follows:
\begin{equation}
        s_j
        :=
        \begin{cases}
        \dfrac{\sqrt{1+4\lambda^2r_j^2}-1}{\lambda r_j}, & r_j>0,\\[1.2em]
        0, & r_j=0.
        \end{cases}
\end{equation}
Equivalently, \(s_j\) is the unique solution in \([0,1)\) of
\[
        \frac{2s_j}{4-s_j^2}=\lambda r_j.
\]
Taking $0<\lambda<2/3$ ensures $s_j<1$ for every $j$, so the drift is Hurwitz and \Cref{thm:gaussian-fixed-point} applies. The boundary $\lambda=2/3$ must be excluded if some $r_j=1$.

By plugging \Cref{eq:mode-dependent-fixed-covariance-integral} in \Cref{eq:Bargmann-matrix} we find that the Bargmann matrix corresponding to the fixed point of this Lindbladian has the block form 
\begin{align}
\mathbf A_\lambda
=
\begin{pmatrix}
\mathbf B_\lambda & \mathbf C_\lambda\\
\mathbf C_\lambda^T & \mathbf B_\lambda^*
\end{pmatrix},
\end{align}
with
\begin{align}
\mathbf B_\lambda
=
\mathbf U\operatorname{diag}
\left(
\frac{2s_j}{4-s_j^2}
\right)_{j=1}^m
\mathbf U^T
= \lambda \mathbf B, \quad
\mathbf C_\lambda
=\mathbf U\operatorname{diag}
\left(
\frac{s_j^2}{4-s_j^2}
\right)_{j=1}^m
\mathbf U^\dagger .
\end{align}
Moreover,
$
\frac{s_j^2}{4-s_j^2}
=
\frac{\sqrt{1+4\lambda^2r_j^2}-1}{2}
\le \lambda^2\, r_j^2
\le \lambda^2,
$ implies
\begin{align}
\|\mathbf C_\lambda\|\le \lambda^2.
\end{align}
Since \(\|\mathbf B\|\le 1\), every entry of \(\mathbf B_\lambda=\lambda \mathbf B\)
has absolute value at most \(\lambda\), and every entry of \(\mathbf C_\lambda\) has
absolute value at most \(\lambda^2\). We now isolate the desired hafnian contribution. Let
\begin{align}
H:=|\mathrm{Haf}(\mathbf B)|^2 .
\end{align}
Let us consider a graph with adjacency matrix $\mathbf A_\lambda$. Note that our graph has $2m$ vertices. We call the first $m$ nodes, the ``top'' nodes, and we call the rest the ``bottom'' modes. In the hafnian expansion of \(\mathbf A_\lambda\), call an edge a \(C\)-edge if it
connects a top vertex to a bottom vertex. A perfect matching with no
\(C\)-edges pairs all top vertices among themselves using \(B_\lambda\), and
all bottom vertices among themselves using \(B_\lambda^*\). The total
contribution of such matchings is
\[
\operatorname{Haf}(\mathbf B_\lambda)\operatorname{Haf}(\mathbf B_\lambda^*)
=
|\operatorname{Haf}(\lambda \mathbf B)|^2
=
\lambda^{m} |\operatorname{Haf}(\mathbf B)|^2
=
\lambda^{m} H.
\]
Every other perfect matching uses at least two \(C\)-edges. Indeed, if a
matching uses \(k\) top-bottom edges, then \(m-k\) top vertices remain to be
matched among themselves, so \(m-k\) must be even. Since \(m\) is even, \(k\)
is even. Thus \(k=0\) or \(k\ge 2\).

A matching with \(k\ge 2\) \(C\)-edges has \(m-k\) remaining same-block edges.
Therefore its contribution has magnitude at most
\begin{align}
        \lambda^{m-k}(\lambda^2)^k
        =
        \lambda^{m+k}
        \le
        \lambda^{m+2}.
\end{align}
The total number of perfect matchings on \(2m\) vertices is
$(2m-1)!!\le (2m)^m$, which gives
\begin{align}
\operatorname{Haf}(\mathbf A_\lambda)
=
\lambda^m H+E_\lambda,\quad |E_\lambda|
\le
(2m)^m\lambda^{m+2}.
\end{align}

Here, we make use of the fact that the input to our algorithm is a matrix $\mathbf X$ with rational entries, to put a lower bound on the Hafnian of the instances where $\mathrm{Haf}(\mathbf X) \ne 0$. We choose $\mathbf B = \begin{pmatrix}
0 & \mathbf X\\
\mathbf X^T & 0
\end{pmatrix}$, where $X = (x_{ij})_{ij}$ with $x_{ij} = p_{ij}/q_{ij}$ for integers $p_{ij},q_{ij}$. We choose $\mathbf{X}$ (by normalizing arbitrary input with suitable power of $2$) to ensure $\norm{\mathbf{B}} \leq 1$. Now, assume that each entry $x_{ij}$ is given to us with $b$ bits of precision. Hence, we have $|p_{ij}|,|q_{ij}|\le 2^b$. Let $D = \prod_{ij} q_{ij}$. Then, $D \cdot \mathrm{Per}(\mathbf X)\in\mathbb Z$, and hence
\begin{align}
\text{if }\mathrm{Per}(\mathbf X)\neq 0,\quad \text{then } |\mathrm{Per}(\mathbf X)|\ge \frac{1}{D} \ge 2^{-b m^2}.
\end{align}
Which concludes that we have either $H=0$ or $H\ge 2^{-2bm^2}=:\beta$. We now break the problem into two parts:

\begin{itemize}
    \item We first assume $|\mathrm{Haf}(\mathbf B)| \ge 2^{-2bm^2}=\beta > 0$. Now choose \(\lambda\) so that
\begin{align}
0<\lambda\le \frac23,
\qquad
(2m)^m\lambda^2\le \kappa \beta,
\end{align}
where \(\kappa>0\) is a sufficiently small constant. Hence, we pick $\lambda$ to be a dyadic rational number $\lambda=2^{-q}$, where
\begin{align}\label{eq:lambda-choice}
q = \left\lceil \log \max\left\{ \frac{3}{2}, \sqrt{\frac{(2m)^m}{2^{-2bm^2}\, \kappa}} \right\} \right\rceil.
\end{align}
By assumption, we have
\(H\ge \beta\), and therefore
\begin{align}
|E_\lambda|
\le
\kappa \lambda^m \beta
\le
\kappa \lambda^m H.
\end{align}
Thus
\begin{align}
\left|
\frac{\operatorname{Haf}(\mathbf A_\lambda)}{\lambda^m}
-
H
\right|
\le
\kappa H.
\end{align}

Let \(P_\lambda\) be the Fock-basis distribution of the fixed point, and let
\(x_\star=(1,\ldots,1)\). The Gaussian probability formula gives
\begin{align}
P_\lambda(1^m)
=
\frac{\operatorname{Haf}(\mathbf A_\lambda)}{\sqrt{\det \mathbf Q_\lambda}},
\end{align}
where \(\mathbf{Q}_\lambda\) is the Husimi covariance matrix of the fixed point.
Define
\begin{align}
Z_\lambda:=\sqrt{\det \mathbf Q_\lambda}.
\end{align}
We note that $Z_\lambda$ can be computed up to arbitrary polynomial bits of precision in exactly in polynomial time. 

Then
\begin{align}
\left|
\frac{Z_\lambda P_\lambda(1^m)}{\lambda^m}
-
H
\right|
\le
\kappa H.
\end{align}

Suppose first that there is an efficient exact classical sampler for
\(P_\lambda\). By Stockmeyer's approximate counting theorem \cite{stockmeyer1983complexity}, using this
sampler as an oracle, one can compute in \(\mathsf{BPP}^{\mathsf{NP}}\) an
estimate \(\widetilde p\) satisfying
\begin{align}
|\widetilde p-P_\lambda(1^m)|
\le
\nu P_\lambda(1^m),
\end{align}
for any inverse-polynomial \(\nu>0\). Output
\begin{align}
\widehat H
:=
\frac{Z_\lambda \widetilde p}{\lambda^m}.
\end{align}
Then
\begin{align}\label{eq:H>0}
\begin{split}
|\widehat H-H|
&\le
\frac{Z_\lambda}{\lambda^m}
|\widetilde p-P_\lambda(1^m)|
+
\left|
\frac{Z_\lambda P_\lambda(1^m)}{\lambda^m}
-
H
\right|\\
&\le
\nu
\frac{Z_\lambda P_\lambda(1^m)}{\lambda^m}
+
\kappa H\\
&\le
\nu(1+\kappa)H+\kappa H.
\end{split}
\end{align}
Choosing \(\nu\) and \(\kappa\) sufficiently small gives a constant (and even inverse polynomial)
multiplicative approximation to \(H=|\operatorname{Haf}(\mathbf B)|^2\) (which is $\# \P$ hard due to \Cref{lem:hafnian-hardness}) in
\(\mathsf{BPP}^{\mathsf{NP}}\), causing the collapse of the polynomial hierarchy.
\item Now we study the case where $\mathrm{Haf}(\mathbf B) = 0$. Our choice of $\lambda$ in \eqref{eq:lambda-choice}, still allows us to write $|E_\lambda|\le \kappa \lambda^m \beta$. This gives $0 \leq \mathrm{Haf}(\mathbf A_\lambda) \le \kappa \beta \lambda^m$, which results in
\begin{align}
0\le \frac{Z_\lambda P_\lambda(1^m)}{\lambda^m} \le \kappa \beta.
\end{align}
Hence, if we could efficiently and exactly sample from the output distribution of measuring our fixed point, we could obtain the estimate $\hat H$ satisfying
\begin{align}
\hat H \in[0, (1+\nu) \kappa \beta]
\end{align}
within $\BPP^{\NP}$.
We can choose $\nu$ and $\kappa$ small enough to ensure $(1+\nu)\kappa \le \frac 13$. Which this choice, we get
\begin{align}
|\hat H| \le \frac\beta3.
\end{align}
However, we know that if $H\neq 0$ it is the case that $H\ge \beta$, and we can use \eqref{eq:H>0} to write $\hat H \ge H - (\nu+\kappa+\nu\kappa)H\ge \beta (1-\nu-\kappa - \nu \kappa)$. Thus, choosing $\nu,\kappa$ small enough (say $\nu=\kappa=0.01$) we can ensure that if $H= 0$, then $|\hat H|\le \beta/3$ and if $H\ne 0$, then $|\hat H| > \beta/2$. Therefore, the algorithm is to compute $\hat H$, and use this threshold function (i.e., see if $\hat H \le \beta/2$), to output the exact value of $H=0$ in those cases.
\end{itemize}

Lastly, to account for finite precision computation, we add that all rational input entries are encoded exactly, with \(b\) bounding their bit length after normalization. Since \(\lambda\le 1/2\), the constructed drift has a uniform stability margin. Consequently, the stationary Lyapunov equation and the hafnian expansion imply that replacing \(\mathbf M_\lambda\) by a sufficiently close real symmetric rational matrix \(\overline{\mathbf M}_\lambda\) changes the unnormalized probability \(Z_\lambda P_\lambda(1^m)\) by at most
\begin{align}\label{eq:sth}
C\,m(2m)^m\|\overline{\mathbf M}_\lambda-\mathbf M_\lambda\|,
\end{align}
where \(C\) depends only on the fixed loss rate. We compute this rational approximation with error small enough that \eqref{eq:sth} is at most \(\kappa\beta\lambda^m\) (this requires only polynomially many bits and polynomial time, since \(\log_2(1/(\beta\lambda^m))=\mathsf{poly}(m,b)\)). Applying the assumed exact sampler to this rationally specified Lindbladian therefore introduces at most an additional \(\kappa\beta\) error after normalization by \(\lambda^m\). Finally, we compute the normalizing factor of the rounded instance to relative error \(\nu\) in polynomial time, together with Stockmeyer’s estimate, these errors preserve both the multiplicative approximation and the zero-detection threshold after decreasing the constants \(\kappa,\nu\) if necessary.
\end{proof}

We highlight that our construction in \cref{thm:hardness-of-gaussian-lindbladian} does not give a robust hardness of sampling (i.e., hardness of sampling from a distribution with up to $1/\mathrm{poly}(n)$ total-variation distance from the target distribution; this result will be presented in the next section). 
This result is still valuable, because it shows that exact sampling from a system with constant loss rate and coherent dynamics with exponentially small coefficients is hard.   In particular, it shows that once  incoherent pumping is removed we can obtain hardness of exact sampling for a wide range of parameters, complementing~\Cref{thm:gaussian-classical-trajectory}.

We furthermore note that we could have easily made hard-to-sample states, if we had allowed arbitrary linear jump terms. The difficulty in our work arose from the fact that we restricted our scope to realistic models of error such as loss and pumping.
\begin{remark}
We remark that if we allow arbitrary linear jump terms that take the form $\sum_j (\alpha_j \hat a_j + \beta_j \hat a^\dagger_j)$ then we can almost immediately engineer pure squeezed states as fixed points of such Lindbladian systems. To do this consider a Lindbladian as a sum of pure loss terms $\mathcal{L} [\bullet] = \sum_{j=1}^m \mathcal{D}(\sqrt{\gamma} \hat a_j)$ whose unique fixed point is $\ket{\rm vac}$. We then consider the conjugated Lindbladian 
\begin{align}
\mathcal{L}_U[\bullet] = \hat{U}^\dag  \mathcal{L}[\hat{U} \bullet \hat{U}^\dag] \hat{U},
\end{align}
If we pick $U^\dagger = V \bigotimes_j S(r_j)$ for linear optical $V$ the fixed point of $\mathcal{L}_U[\bullet]$ will be equivalent to $V \bigotimes_j S(r_j) \ket{\rm vac}$ (which is equivalent to the output of Gaussian boson sampling), furthermore $\mathcal{L}_U$ will be composed of jump terms with linear form described. The noise model is however artificial and one ideally would like to obtain hardness based on the model with simple loss which we accomplished in this section. 
\end{remark}

The construction in \Cref{thm:hardness-of-gaussian-lindbladian} relies on a Lindbladian in which the coefficients of the Hamiltonian part is exponentially smaller than that of the loss terms. In other words, the strength of loss terms (which we will define more properly below) is exponentially larger than the coherent parts. In this section we show that we can modify this construction to achieve hardness of exact sampling based on constant strength of loss. Note one could adjust the strength by adding $c \mathbb{I}$, however this addition is trivial and is not going to change the transient dynamics. The solution we give involves a nontrivial adjustment of Hamiltonian terms. 

Let us fix our terminology on what we mean by the strength of loss in a Lindbladian system. Imagine we have a Lindbladian on the form
\begin{align}
\mathcal{L}_{H,\gamma} [\bullet] = -i[H, \bullet] + \sum_i \gamma_i \mathcal{D}(\hat a_i)[\bullet].
\end{align}
Note that the fixed points of $\mathcal{L}_{H,\gamma}$ and $\mathcal{L}_{r\cdot H, r\cdot \gamma}$ are the same.  
Hence, it is the ratio of the strengths that matters. Let $c_H$ be the largest absolute value of the coefficient of the Hamiltonian represented in the $\hat q,\hat p$ operators. We denote by $\min_i \gamma_i/c_H$ the ``strength of loss'' corresponding to incoherent terms. For instance, we say that we have an exponentially strong loss if $\min_i \gamma_i/c_H = \exp(\Omega(n))$, or we say we have loss of inverse polynomial strength if $\min_i \gamma_i/c_H = 1/\mathsf{poly(n)}$.  Intuitively, one expects the computational complexity of sampling from the fixed point of a Lindbladian to increase as we decrease the strength of loss. We formally prove this property for the case of Gaussian Lindbladians in the \cref{lem:hardness-eta} below. Note that in some settings, e.g. the circuit model of \cite{go2025sufficient}, establishing this incremental hardness is more challenging.

\begin{lemma}\label{lem:hardness-eta}
Let $\mathcal L_{H,\boldsymbol\gamma}=-i[H,\cdot]
+\sum_j\gamma_j\mathcal D(\hat a_j)$ be Gaussian, with every $\gamma_j>0$,
and suppose every initial state converges to its unique fixed point $\rho_f$.
There is a quadratic Hamiltonian $K$ for which
$\mathcal L_{H+\lambda K,\boldsymbol\gamma}$ has the same unique attracting
fixed point for every $\lambda\geq0$. Consequently, hardness at loss strength
$\varepsilon=\min_j\gamma_j/c_H$ implies hardness at every prescribed
$0<\varepsilon'\leq\varepsilon$, subject to the same efficient parameter
representation used for the family.
\end{lemma}
\begin{proof}
Let $\boldsymbol\mu,\boldsymbol\Sigma$ be the mean and covariance of $\rho_f$,
and let $\mathbf D$ be the original drift. Put
\begin{equation}\label{eq:stabilizing-Hamiltonian}
K=\tfrac12(\hat{\mathbf r}-\boldsymbol\mu)^T
\boldsymbol\Sigma^{-1}(\hat{\mathbf r}-\boldsymbol\mu).
\end{equation}
The new drift is $\mathbf A_\lambda=\mathbf D+lambda\boldsymbol\Omega\boldsymbol\Sigma^{-1}$,
and the new constant drive is
$\mathbf d_\lambda=\mathbf d-\lambda\boldsymbol\Omega\boldsymbol\Sigma^{-1}\boldsymbol\mu$.
Hence $\mathbf A_\lambda\boldsymbol\mu+\mathbf d_\lambda=0$ and
\[
\mathbf A_\lambda\boldsymbol\Sigma+\boldsymbol\Sigma\mathbf A_\lambda^T
=\mathbf D\boldsymbol\Sigma+\boldsymbol\Sigma\mathbf D^T
+\lambda(\boldsymbol\Omega+\boldsymbol\Omega^T)
=-\boldsymbol\Gamma,
\qquad \boldsymbol\Gamma=\bigoplus_j\gamma_j\mathbb I_2>0.
\]
Thus $\rho_f$ remains stationary. To prove attraction, define the similar matrix
$\mathbf B_\lambda=\boldsymbol\Sigma^{-1/2}\mathbf A_\lambda\boldsymbol\Sigma^{1/2}$.
Its Hermitian part satisfies
\[
\mathbf B_\lambda+\mathbf B_\lambda^\dagger
=-\boldsymbol\Sigma^{-1/2}\boldsymbol\Gamma\boldsymbol\Sigma^{-1/2}<0.
\]
For any eigenpair $\mathbf B_\lambda v=zv$,
\[
2\operatorname{Re}z\,\|v\|_2^2
=v^\dagger(\mathbf B_\lambda+\mathbf B_\lambda^\dagger)v<0.
\]
Therefore $\mathbf A_\lambda$ is Hurwitz and \Cref{thm:gaussian-fixed-point}
gives uniqueness and attraction. The real parts of a matrix's eigenvalues
are bounded by its Hermitian part; they are not generally its eigenvalues.
Finally, the largest absolute Hamiltonian coefficient $c_{H+\lambda K}$ is
continuous in $\lambda$, starts at $c_H$, and is unbounded because $K$ has a
nonzero quadratic coefficient. It therefore attains any target value
$\min_j\gamma_j/\varepsilon'\geq c_H$.
\end{proof}

Therefore, with \cref{lem:hardness-eta} at hand, we can extend the hardness result of \cref{thm:hardness-of-gaussian-lindbladian} and conclude that sampling from Lindbladian fixed points is hard for all loss strengths that are at most exponentially large.

In the following section, we develop a robust hardness of sampling result. In other words, we show that sampling from the fixed point of Gaussian Lindbladians in the number basis, up to $1/\mathsf{poly}(n)$ total variation distance, is classically hard.

\subsection{Robust hardness of sampling for Gaussian Lindbladians with small loss}\label{sec:robust-gaussian-lindbladian}

In this section, we engineer instances of Gaussian
Lindbladians with fixed points that hard to sample even up to $1/{\rm poly} (n)$ total-variation distance if we allow inverse polynomial values of loss that are allowed to be different for different sites.
We start from a simple example based on \cite{ikeda2013deterministic}.

\begin{example}\label{example:two-mode}
Consider $H=g(\hat a\hat b^\dagger+\hat a^\dagger\hat b)
=g(\hat q_1\hat q_2+\hat p_1\hat p_2)$ with real $g\ne0$ and loss
$\gamma>0$ on mode $b$ alone. In the ordering $(q_1,p_1,q_2,p_2)$ the drift is
\begin{equation}\label{eq:D-of-example}
\mathbf D=\begin{pmatrix}
0&0&0&g\\0&0&-g&0\\0&g&-\gamma/2&0\\-g&0&0&-\gamma/2
\end{pmatrix}.
\end{equation}
It decomposes into two $2\times2$ blocks with eigenvalues
\[
\lambda_\pm=-\frac\gamma4\pm\sqrt{\frac{\gamma^2}{16}-g^2}.
\]
Both have negative real part for every $g\ne0$, including $|g|=\gamma/4$.
At that critical value a Jordan block contributes a polynomial prefactor,
but the exponential still decays. Since vacuum is stationary,
\Cref{thm:gaussian-fixed-point} shows that it is the unique attracting fixed
point. At $g=0$ the first mode is undamped and uniqueness fails.
\end{example}

Note the important feature of \cref{example:two-mode}: it has only loss applied to the second mode. Hence, we can conjugate the Lindbladian by a squeezing operator on the first mode (with the corresponding annihilation operator $a$), and prepare the state $(\hat{S}\ket{0})\otimes\ket0$. The reason this happens is that the Lindbladian, despite having loss applied on only one of its modes, has a unique fixed point. This might seem unexpected. One way to understand it is that the Hamiltonian in this construction $\hat{H} = g(\hat a^\dag \hat b + \hat a \hat b^\dag)$ is merely a hopping term, so it moves photons around between the two modes, and the loss on the second mode leaks photons to the environment. Hence, the system has no choice but to eventually reach the vacuum state.

We now use a result that establishes the hardness of approximate Boson sampling under standard assumptions/conjectures:
\begin{theorem}[Theorem 1 of \cite{grier2022complexity}]\label{thm:robust-gbs-hardness}
There is a family of Gaussian pure states, that no classical algorithm can even approximately (within inverse polynomial total-variation distance) sample from, assuming the Permanent-of-Gaussians Conjecture and the Permanent Anti-Concentration Conjecture and that the $\mathsf{PH}$ does not collapse.
\end{theorem}

For formal definition of the Permanent-of-Gaussians and Permanent Anti-Concentration conjectures see~\cite[Conjectures 5 and 6]{aaronson2011computational}. We highlight that the latter conjecture was recently claimed to have been resolved in~\cite{koehler2026anticoncentration}.

Building on this result and \cref{example:two-mode} we can show that a Gaussian Lindbladian dynamics that has loss only on half of its modes, admits fixed points that are hard to sample, even approximately. In particular, we can engineer fixed points of the form
\begin{align}
\ket{\psi}\otimes\ket{0^m},
\end{align}
where $\ket{\psi}$ is an arbitrary $m$-mode Gaussian state. The following proposition summarizes this result.

\begin{proposition}\label{prop:robust-hardness-gbs}
Let $\hat U$ be any Gaussian unitary on $m$ data modes, and define
$\hat a'_i=\hat U\hat a_i\hat U^\dagger$.
For $m$ ancillary modes $b_i$, real $g\ne0$, and $\gamma>0$, the Lindbladian
\[
\mathcal L=-i[H,\cdot]+\gamma\sum_i\mathcal D(\hat b_i),\qquad
H=g\sum_i(\hat a'_i\hat b_i^\dagger+\hat a_i'{}^\dagger\hat b_i)
\]
has the unique attracting fixed point
\[
\rho_f=|\psi_f\rangle\langle\psi_f|,\qquad
|\psi_f\rangle=(\hat U|0^m\rangle)\otimes|0^m\rangle.
\]
\end{proposition}
\begin{proof}
Take the sum of $m$ independent generators from \Cref{example:two-mode}.
Its semigroup is the tensor product of their semigroups, and its unique
attracting state is vacuum. Conjugating this generator by
$\hat U\otimes\mathbb I$ changes the data annihilators to $\hat a'_i$, leaves
all ancillary jumps unchanged, and carries vacuum to the stated target.
\end{proof}

We are now ready to present our robust hardness result when there is a loss on all modes. Our idea is simple: put a small amount of loss on the first half of the modes (those who have annihilation operators $\hat a_i$ associated to them in \cref{thm:robust-gbs-hardness}. This will only slightly alter the covariance matrix of the state, and hence, we get a state close enough to the desired fixed point. Since the fixed point of the Lindbladian introduced in \cref{thm:robust-gbs-hardness} is already hard to sample up to $1/\mathsf{poly}$ total variation distance, being inverse-polynomially close to it preserves the approximate hardness of sampling. As an auxiliary lemma, we need a conversion between closeness of covariance matrices to the closeness of states in trace distance in trace norm. We use the tight bound that \cite{bittel2025optimal} establishes:

\begin{lemma}[Zero-mean case of Theorem 3 of \cite{bittel2025optimal}]
\label{lem:cov-to-tr}
For zero-mean Gaussian states with vacuum-$\mathbb I$ covariances,
\begin{align}
\tfrac12\|\rho_1-\rho_2\|_1
&\leq\frac{1+\sqrt3}{8}\operatorname{Tr}\!\left[
|\boldsymbol\Sigma_1-\boldsymbol\Sigma_2|\,
\boldsymbol\Omega^T\frac{\boldsymbol\Sigma_1+\boldsymbol\Sigma_2}{2}\boldsymbol\Omega\right]
\nonumber\\
&\leq\frac{1+\sqrt3}{8}\|\boldsymbol\Sigma_1-\boldsymbol\Sigma_2\|_1
\left\|\frac{\boldsymbol\Sigma_1+\boldsymbol\Sigma_2}{2}\right\|_\infty.
\end{align}
\end{lemma}
If the covariance operator norms are polynomially bounded, sufficiently small
inverse-polynomial covariance error gives inverse-polynomial trace distance;
one also accounts for the matrix dimension when converting the operator norm
to the trace norm. The following lemma verifies the needed bounded squeezing
for the random BipartiteGBS instances directly.

The construction in \cite{grier2022complexity} is based on zero-displaced pure states with a Bargmann $A$ matrix\footnote{
The ket Bargmann matrix has size $m\times m$, while the density-operator matrix has size $2m\times2m$. In the complex-coordinate ordering used in \eqref{eq:Bargmann-matrix}, a pure state has $A_{|\psi\rangle\langle\psi|}=\begin{pmatrix}
A_{|\psi\rangle}& 0\\
0 & \overline{A_{|\psi\rangle}}
\end{pmatrix}$.} corresponding to a bipartite structure (also called bipartiteGBS) of the form
\begin{align}\label{eq:A-of-bipartiteGBS}
\mathbf A = \begin{pmatrix}
0 & \mathbf C\\
\mathbf C^T & 0
\end{pmatrix}
\end{align}
where $\mathbf C$ is chosen randomly from $\mathcal N(0, 1/(\alpha^2 m))_{\mathbb C}^{m\times m}$. Then, they take
\begin{align}
\alpha = \Theta(m^{1/4}),
\end{align}
and prove that sampling from the output distribution of this family is classically hard, assuming the Permanent-of-Gaussians conjecture, Anticoncenteration of Permanent conjecture, and the non-collapse of $\mathsf{PH}$. 

We furthermore need the following bound on the singular values of $A$. Lemma 15 of \cite{grier2022complexity} directly provides this, however, we found a slight issue with their lemma:\footnote{We note that, this would just result in changes in inessential constant factors in \cite{grier2022complexity}, and their main hardness result holds.} their statement is valid only when $\varepsilon\le \frac{2}{\alpha^2}$. However, this condition is correctly considered in Ref. \cite{haagerup2003random} they use. Therefore, we provide a corrected version with a proof, using the result of \cite{haagerup2003random}:
\begin{lemma}\label{lem:bound-on-A-of-bipartiteGBS}
For the bipartite matrix in \eqref{eq:A-of-bipartiteGBS} and every $u\in[0,\frac12]$,
\begin{equation}\label{eq:spectrum-of-A-bound}
\Pr\!\left[\|\mathbf A\|_\infty\leq\frac{2\sqrt{1+u}}{\alpha}\right]
\geq1-me^{-2mu^2}.
\end{equation}
For fixed $u>0$ and $\alpha=\Theta(m^{1/4})$, this event implies, for all
sufficiently large $m$, a normalizable pure Gaussian state with
\[
\|\boldsymbol\Sigma\|_\infty
=\frac{1+\|\mathbf A\|_\infty}{1-\|\mathbf A\|_\infty}
=1+O(m^{-1/4})\leq3.
\]
\end{lemma}
\begin{proof}
Set $
B=\alpha\sqrt m\,C$, so that its entries are independent standard complex Gaussians, and note that
\begin{align}
\norm{A}^2_\infty = \frac{\lambda_{\max}(B^\dag B)}{\alpha^2 m}.
\end{align}
From \cite[Lemma 8.2]{haagerup2003random}, we get that for all $t\in[0,\frac12]$:
\begin{align}
\mathbb{E}\left[\operatorname{Tr}\left(e^{tB^\dagger B}\right)\right]
\le m\,e^{4mt+2mt^2},
\end{align}
and since all eigenvalues of $e^{tB^\dagger B}$ are positive, we can write
$
e^{t\lambda_{\max}(B^\dagger B)}
\le \operatorname{Tr}\left(e^{tB^\dagger B}\right)$ (and hence, $\mathbb E[e^{t\lambda_{\max}(B^\dag B)}] \le m\, e^{4mt+2mt^2}$). Finally, we use a Markov inequality with $t=u$ to get \cref{eq:spectrum-of-A-bound}.

For a pure Gaussian state, the singular values of its ket Bargmann matrix are
$\tanh r_j$, where $r_j\geq0$ are the squeezing parameters. In our covariance
convention $\|\boldsymbol\Sigma\|_\infty=e^{2r_{\max}}$, giving the displayed
identity. Note that with high probability, we have $\norm{\mathbf A}_\infty\le \frac12$.
\end{proof}

We now show that robust sampling hardness persists when the remaining modes also undergo loss at an inverse-polynomial rate. We first justify restricting BipartiteGBS to instances with \(\|\mathbf A\|\le \tfrac12\). In the reduction of \cite[Corollary~8]{grier2022complexity}, abort the permanent estimator whenever the random embedding of the Gaussian input matrix \(X\) produces a Bargmann matrix \(\mathbf A\) with \(\|\mathbf A\|>\tfrac12\). By \cref{lem:bound-on-A-of-bipartiteGBS}, the abort probability is at most \(e^{-\Omega(m)}\). The acceptance event is invariant under the row and column permutations used to hide \(X\), so the hiding argument applies on the accepted instances. On each such instance, an efficient sampler for the Lindbladian fixed points constructed below would yield a BipartiteGBS sampler with the required total-variation accuracy. Allocating failure probability \(\eta/2\) to the permanent estimator gives an overall failure probability of at most \(\eta/2+e^{-\Omega(m)}\le\eta\) for every inverse-polynomial \(\eta\) and all sufficiently large \(m\).
\begin{theorem}[Robust hardness with loss on every mode]
\label{thm:robust-hardness-all-modes}
There is a family of Gaussian Lindbladians on $2m$ modes, with loss rates that are inverse-polynomially large, such that sampling from the photon-number distribution of the fixed point to within inverse-polynomial total variation distance is classically hard, assuming the Permanent-of-Gaussians Conjecture, the Permanent Anti-Concentration Conjecture, and that $\mathsf{PH}$ does not collapse.
\end{theorem}

\begin{proof}
Let $\beta(m)=1/\poly(m)$ be an inverse-polynomial accuracy parameter for which the approximate-sampling hardness result of~\Cref{thm:robust-gbs-hardness} holds. We may assume without loss of generality that $\beta(m)\leq 1$. Let
\begin{align}
\ket{\psi}=U\ket{0^m}
\end{align}
be an instance from the hard, zero-displacement BipartiteGBS family of~\cite{grier2022complexity}. By the Bloch--Messiah decomposition, we can write
\begin{align}
\ket{\psi}
=
V
\left(
\bigotimes_{j=1}^m S_j(\xi_j)
\right)
\ket{0^m},
\label{eq:bloch-messiah-hard-state}
\end{align}
where $V$ is a passive Gaussian unitary on $m$ modes. Based on \cref{lem:bound-on-A-of-bipartiteGBS}, we can restrict to the high-probability event of
\begin{align}
\norm{\mathbf A_m}\leq \frac12,
\end{align}
where $\mathbf A_m$ is the Bargmann matrix of $\ket{\psi_m}$. Since the singular values of $\mathbf A_m$ are $\tanh(\xi_j)$, this gives
\begin{align}
e^{2\xi_{\max}}
=
\frac{1+\tanh(\xi_{\max})}{1-\tanh(\xi_{\max})}
\leq 3,
\label{eq:rmax-bound-simplified}
\end{align}
where $\xi_{\max}:=\max_j \xi_j$. We now apply the perturbation before the passive linear optical circuit $\hat V_m$. For every $j\in[m]$, consider a data mode $\hat a_j$ and an ancillary mode $\hat b_j$, and define
\begin{align}
    \hat h_j
    :=
    \hat a_j\hat b_j^\dagger+\hat a_j^\dagger\hat b_j.
\end{align}
Also, define the squeezed annihilation operator
\begin{align}
    \hat c_j
    :=
    S_j(\xi_j)^\dagger
    \hat a_j
    S_j(\xi_j).
\end{align}
For $\delta>0$, consider the Lindbladian
\begin{align}
    \overline{\mathcal L}_\delta[\bullet]
    :=
    \sum_{j=1}^m
    \left(
        -i[\hat h_j,\bullet]
        +2\mathcal D(\hat b_j)[\bullet]
        +\delta\mathcal D(\hat c_j)[\bullet]
    \right).
    \label{eq:local-perturbed-lindbladian}
\end{align}

We first show that the fixed point remains close to the vacuum when $\delta$ is small. Consider one of the two-mode systems in~\eqref{eq:local-perturbed-lindbladian}. After a permutation of the quadratures, the drift matrix for $\delta=0$ can be written as
\begin{align}
    \mathbf D_{\mathrm{base}}
    =
    \mathbf G\oplus \mathbf G^T,
    \quad
    \mathbf G=
    \begin{pmatrix}
        -1 & -1\\
        1 & 0
    \end{pmatrix}.
\end{align}
A direct calculation gives
\begin{align}
e^{\mathbf Gt}
=
e^{-t/2}
\left[
\cos\left(\frac{\sqrt{3}t}{2}\right)\mathbb I_2
+
\frac{2}{\sqrt{3}}
\sin\left(\frac{\sqrt{3}t}{2}\right)
\left(\mathbf G+\frac12\mathbb I_2\right)
\right].
\end{align}
Since $\norm{G+\mathbb I_2/2}=3/2$, we obtain
\begin{align}
\norm{e^{\mathbf D_{\mathrm{base}}t}}
\leq
3e^{-t/2}.
\label{eq:base-semigroup-bound}
\end{align}

Let $\Pi_a$ denote the orthogonal projection onto the two quadratures of the data mode. Since $\hat c_j$ is a canonical annihilation operator, adding $\delta\mathcal D(\hat c_j)$ changes the drift matrix by
\begin{align}
\mathbf D_{j,\delta}
=
\mathbf D_{\mathrm{base}}-\frac{\delta}{2}\Pi_a.
\label{eq:local-perturbed-drift}
\end{align}
Duhamel's formula gives
\begin{align}
e^{\mathbf D_{j,\delta}t}
=
e^{\mathbf D_{\mathrm{base}}t}
-
\frac{\delta}{2}
\int_0^t
e^{\mathbf D_{\mathrm{base}}(t-s)}
\Pi_a
e^{\mathbf D_{j,\delta}s}
\dd s.
\end{align}
Using~\eqref{eq:base-semigroup-bound} and $\norm{\Pi_a}=1$, we obtain
\begin{align}
e^{t/2}\norm{e^{\mathbf D_{j,\delta}t}}
\leq
3+
\frac{3\delta}{2}
\int_0^t
e^{s/2}\norm{e^{\mathbf D_{j,\delta}s}}
\dd s.
\end{align}
Gronwall's inequality therefore gives
\begin{align}
\norm{e^{\mathbf D_{j,\delta}t}}
\leq
3\exp\left(
-\left(\frac12-\frac{3\delta}{2}\right)t
\right).
\end{align}
Therefore, setting $\delta\leq 1/6$ gives
\begin{align}
\norm{e^{\mathbf D_{j,\delta}t}}
\leq
3e^{-t/4}.
\label{eq:local-perturbed-semigroup-bound}
\end{align}
Hence, $\mathbf D_{j,\delta}$ is Hurwitz (i.e., its spectrum lies on the left side of the complex plane). It follows from \cref{thm:gaussian-fixed-point} that every two-mode Lindbladian in~\eqref{eq:local-perturbed-lindbladian} has a unique fixed point.

Let $\mathbf T_j$ be the symplectic matrix of $\hat S_j(r_j)\otimes\mathbb I_b$, and define
\begin{align}
\mathbf K_j:=\mathbf T_j^{-1}\Pi_a \mathbf T_j^{-T}.
\end{align}
The diffusion matrix contributed by $\mathcal D(\hat c_j)$ is $\mathbf K_j$. By~\eqref{eq:rmax-bound-simplified},
\begin{align}
\norm{\mathbf K_j}
\leq
\norm{\mathbf T_j^{-1}}^2
\leq
e^{2r_{\max}}
\leq 3.
\label{eq:local-diffusion-bound}
\end{align}

Let $\overline{\boldsymbol\Sigma}_{j,\delta}$ be the covariance matrix of the fixed point of the $j$-th two-mode system. We use the convention in which the vacuum covariance matrix is $\mathbb I_4$, and define
\begin{align}
\boldsymbol\Delta_j
:=
\overline{\boldsymbol\Sigma}_{j,\delta}-\mathbb I_4.
\end{align}
Subtracting the stationary Lyapunov equation at $\delta=0$ from the one at $\delta>0$ gives
\begin{align}
\mathbf D_{j,\delta}\Delta_j
+
\boldsymbol\Delta_j \mathbf D_{j,\delta}^T
+
\delta(\mathbf K_j-\Pi_a)
=
0,
\label{eq:local-difference-lyapunov}
\end{align}
Using~\eqref{eq:local-diffusion-bound}, we obtain
$
\norm{\mathbf K_j-\Pi_a}
\leq 4
$.
Since $\mathbf D_{j,\delta}$ is Hurwitz, Equation~\eqref{eq:local-difference-lyapunov} has the integral solution
\begin{align}
\boldsymbol\Delta_j
=
\delta\int_0^\infty
e^{\mathbf D_{j,\delta}t}
(\mathbf K_j-\Pi_a)
e^{\mathbf D_{j,\delta}^Tt}
\dd t.
\end{align}
Using~\eqref{eq:local-perturbed-semigroup-bound}, we conclude that
\begin{align}
\norm{\boldsymbol\Delta_j}
\leq
4\delta
\int_0^\infty
9e^{-t/2}
\dd t
=
72\delta.
\label{eq:local-covariance-bound}
\end{align}
Let $\overline{\rho}_\delta$ be the unique fixed point of $\overline{\mathcal L}_\delta$, and let $\overline{\boldsymbol\Sigma}_\delta$ be its covariance matrix. Since the systems in~\eqref{eq:local-perturbed-lindbladian} act on disjoint pairs of modes,
\begin{align}
\overline{\rho}_\delta
=
\bigotimes_{j=1}^m\overline{\rho}_{j,\delta},
\qquad
\overline{\boldsymbol\Sigma}_\delta
=
\bigoplus_{j=1}^m\overline{\boldsymbol\Sigma}_{j,\delta}.
\end{align}
It follows from~\eqref{eq:local-covariance-bound} that
\begin{align}
\norm{\overline{\boldsymbol\Sigma}_\delta-\mathbb I_{4m}}
&\leq 72\delta,
\\
\norm{\overline{\boldsymbol\Sigma}_\delta-\mathbb I_{4m}}_1
&\leq 4m(72\delta).
\label{eq:global-covariance-bound}
\end{align}
For $\delta\leq 1/144$, Equation~\eqref{eq:global-covariance-bound} also gives
$
\norm{
\frac{\overline{\boldsymbol\Sigma}_\delta+\mathbb I_{4m}}{2}
}
\leq
1+
\frac12
\norm{\overline{\boldsymbol\Sigma}_\delta-\mathbb I_{4m}}
\leq
\frac54
$.
Both fixed points have zero displacement, since the Lindbladians contain no linear drive and their drift matrices are Hurwitz. Therefore, \cref{lem:cov-to-tr} gives
\begin{subequations}
\begin{align}
\frac12
\norm{
\overline{\rho}_\delta
-
\ket{0^{2m}}\bra{0^{2m}}
}_1
&\leq
\frac{1+\sqrt{3}}{8}
\norm{\overline{\boldsymbol\Sigma}_\delta-\mathbb I_{4m}}_1
\norm{
\frac{\overline{\boldsymbol\Sigma}_\delta+\mathbb I_{4m}}{2}
}
\\
&\leq
\frac{1+\sqrt{3}}{8}
\bigl(4m\bigr)
\bigl(72\delta\bigr)
\frac54
\\
&=
45(1+\sqrt{3})m\delta.
\label{eq:local-product-trace-bound}
\end{align}
\end{subequations}

We conclude the proof by conjugating the Linbdladian $\overline{\mathcal{L}}_\delta$ by $R := U \otimes \mathbb{I}_{\rm anc}$, where $\mathbb{I}_{\rm anc}$ is the identity operator on the noisy ancilla modes, and consider the fixed point of the Lindbladian 
$$
R \overline{\mathcal{L}}_\delta [R^\dagger \bullet R] R^\dagger.
$$
The Hamiltonian part remains quadratic and the jump terms are mapped to $R \hat c_j R^\dagger = V \hat a_j V^\dagger$ and $R \hat b_j R^\dagger = \hat b_j$. Because $V$ is passive $\sum_j \mathcal{D} (V \hat a_j V^\dagger) = \sum_j \mathcal{D} (\hat a_j)$. Thus the physical loss rates are $\delta $ on data modes and $2$ on ancillary modes. To conclude the statement of the theorem we can choose
$$
\delta = \min \{1/144, \frac{\beta(m)}{90 (1+ \sqrt{3}) m}\}.
$$
Using this choice, and unitary invariance of trace distance the physical output state is within \(\beta(m)/2\) in trace distance of \(|\psi\rangle\langle\psi|\otimes |0^m\rangle\langle0^m|\). Measurement contraction therefore implies that the corresponding photon-counting distributions are within \(\beta(m)/2\) in total variation distance. If the physical output distribution could itself be sampled within \(\beta(m)/2\) in total variation distance, then, after discarding the ancillary counts, contraction under marginalization and the triangle inequality yield a \(\beta(m)\)-approximate sampler for the hard GBS family.
\end{proof}

We remark that one could have used the construction of \cite{go2025sufficient} directly as well. Indeed, a direct calculation shows that a loss of strength $O(\log m/m^2)$ recreates their hardness result.

We are now ready to prove the lossy version of the above hardness result, where the same amount of loss acts on every mode. Our idea is simple: for every mode $\hat a_i$, instead of introducing a single ancillary mode, we introduce a register of $K$ ancillary modes. We then choose a passive Hamiltonian that rapidly spreads the $\hat a_i$-mode over all $K+1$ modes in its register. In the frame in which the desired state is vacuum, the physical loss acting on the $\hat a$-modes appears as an unwanted Gaussian noise source supported only on the first layer. The passive Hamiltonian spreads this noise among the $K+1$ layers, and hence, the amount of noise that returns to the first layer is suppressed by a factor of order $1/K$. As a result, the reduced state on the original $\hat a$-modes remains inverse-polynomially close to the desired hard Gaussian state, even though all physical modes experience exactly the same loss.

\begin{theorem}[Robust hardness with uniform loss on every mode]\label{thm:robust-gaussian-uniform-loss}
There is a family of Gaussian Lindbladians on a polynomial number of modes,
with the same inverse-polynomially large loss rate on every mode, such that
sampling from the photon-number distribution of the fixed point to within
inverse-polynomial total-variation distance is classically hard, assuming
the Permanent-of-Gaussians Conjecture, the Permanent Anti-Concentration
Conjecture, and that PH does not collapse.
\end{theorem}

\begin{figure}
    \centering
\resizebox{.7\textwidth}{!}{
\begin{tikzpicture}
  \draw[fill=blue!28] (-0.32,3.28) circle (0.12cm);
  \draw[fill=blue!28] (3.24,3.25) circle (0.12cm);
  \draw (-1.02,2.52) circle (0.12cm);
  \draw (-0.54,2.52) circle (0.12cm);

  \draw (0.48,2.52) circle (0.12cm);
  \draw[loosely dotted] (-0.24,2.52) -- (0.2,2.52);
  \draw[green!55!black] (-1.02,2.64) -- (-0.42,3.2);
  \draw[green!55!black] (-0.54,2.64) -- (-0.37,3.16);
  \draw[green!55!black] (0.48,2.64) -- (-0.22,3.2);
  \node[draw=none] at (-1.27,2.3) {\small{\textcolor{teal}{$\hat{h}_{\mathrm{mix},1}$}}};

  \draw (2.53,2.51) circle (0.12cm);
  \draw (3.01,2.51) circle (0.12cm);
  \draw (4.03,2.51) circle (0.12cm);
  \draw[loosely dotted] (3.31,2.51) -- (3.75,2.51);
  \draw[green!55!black] (2.53,2.63) -- (3.13,3.19);
  \draw[green!55!black] (3.01,2.63) -- (3.18,3.15);
  \draw[green!55!black] (4.03,2.63) -- (3.34,3.16);
  \draw[densely dashed] (-2,2.83) -- (4.5,2.81);
  \node[draw=none] at (3.31,1.75) {\small{$m^{\mathrm{th
}}$ block}};
  \node[draw=none] at (-0.35,1.75) {\small{$1^{\mathrm{st}}$ block}};
  \draw[green!55!black] (-0.9,2.51) -- (-0.66,2.51);
  \draw[green!55!black] (2.65,2.5) -- (2.89,2.5);
  \draw[green!55!black] (-1.02,2.4) .. controls (-0.52,2.07) and (-0.02,2.06) .. (0.48,2.39);
  \draw[green!55!black] (-0.54,2.4) .. controls (-0.2,2.24) and (0.14,2.24) .. (0.48,2.39);
  \draw[green!55!black] (2.55,2.39) .. controls (3.05,2.06) and (3.55,2.05) .. (4.05,2.38);
  \draw[green!55!black] (3.03,2.39) .. controls (3.37,2.23) and (3.71,2.23) .. (4.05,2.38);
    \node at (1.5, 3) {$\cdots$};
  \node[draw=none] at (4.65,2.25) {\small{\textcolor{teal}{$\hat{h}_{\mathrm{mix},m}$}}};

\node[draw=none] at (-2.54,1.75) {\small{ancilla modes}};
\node[draw=none] at (-2.68,3.61) {\small{data modes}};
\end{tikzpicture}
}
    \caption{The construction used in the proof of \cref{thm:robust-gaussian-uniform-loss}. For our robust hardness of sampling result with uniform linear loss, we consider $K$ ancillary modes per data mode. Data and ancilla modes interact via $\hat{h}_{\mathrm{mix}}$. Conjugating this interaction by a Gaussian unitary $U$, we obtain the overall Hamiltonian $\hat H$. $U$ is chosen to be the evolution that maps vacuum to a hard instance of Gaussian boson sampling (e.g. BipartiteGBS).}
    \label{fig:uniform-loss-construction}
\end{figure}
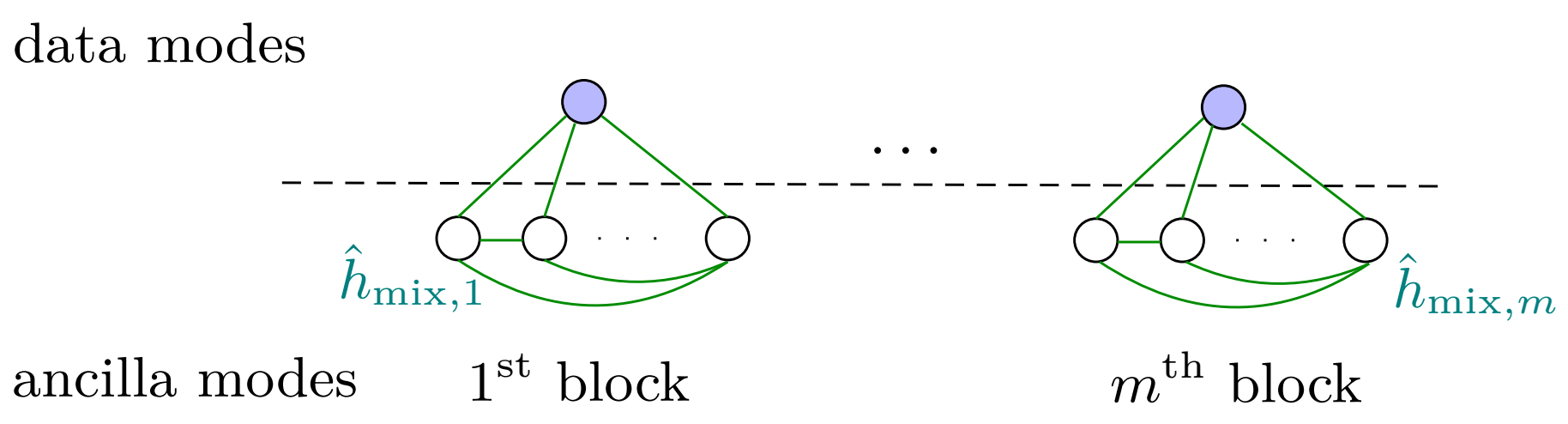

\begin{proof}
Let
\begin{align}\label{eq:hard-gbs-in-pf}
\ket{\psi}
=
\hat U\ket{0}
\end{align}
be an instance from the hard, zero-displacement BipartiteGBS family used in the proof of \cref{thm:robust-gbs-hardness}. We
condition on the high-probability event
$\|\mathbf A\|\leq\frac12$,
where $\mathbf A$ is the Bargmann matrix of $\ket{\psi}$ (which is true with $1-O(1/\exp(m))$ probability according to \cref{lem:bound-on-A-of-bipartiteGBS}). Let $K+1=2^a$ be a power of two whose value will be fixed at the end of the
proof. For every $i\in[m]$, introduce $K$ ancillary modes, and write
\begin{align}
\hat c_{0,i}:=\hat a_i,
\qquad
\hat c_{\ell,i}:=\hat b_{\ell,i},
\qquad
\ell\in[K].
\end{align}
Thus, $\ell=0$ corresponds to the data layer and
$\ell=1,\ldots,K$ correspond to the ancillary layers. Let $\mathbf F:=\mathbf H^{\otimes a}$, with $\mathbf H$ being the $2\times 2$ Hadamard matrix. Therefore,
$\mathbf F$ is a real orthogonal matrix and
\begin{align}
|F_{\ell r}|=\frac{1}{\sqrt{K+1}}
\qquad
\forall \ell,r\in\{0,\ldots,K\}.
\end{align}
For some $\omega>0$, define
\begin{align}
\mathbf G
:=
\mathbf F
\operatorname{diag}
\left(
0,\omega,2\omega,\ldots,K\omega
\right)
\mathbf F^T,
\label{eq:uniform-loss-hn}
\end{align}
and consider the passive Hamiltonian
\begin{align}
    \hat H_{\mathrm{mix}}
    :=
    \sum_{i=1}^m
    \sum_{\ell,r=0}^{K}
    G_{\ell r}
    \hat c_{\ell,i}^\dagger\hat c_{r,i}.
    \label{eq:uniform-loss-mixing-Hamiltonian}
\end{align}
The role of this Hamiltonian is to spread the mode $\hat c_{0,i}$ uniformly
over $K+1$ distinct normal modes. We now conjugate the data layer by the Gaussian unitary $U$ that prepares the
hard instance. In particular, let
\begin{align}
    \hat V
    :=
    \hat U\otimes \mathbb I_{\mathrm{ancilla}},
    \qquad
    \hat H
    :=
    \hat V
    \hat H_{\mathrm{mix}}
    \hat V^\dagger.
    \label{eq:uniform-loss-physical-Hamiltonian}
\end{align}
Note that $\hat U$ acts on $m$ data modes, while $\mathbb I_{\mathrm{ancilla}}$ acts on the $K\times m$ ancilla modes. Since $\hat H_{\mathrm{mix}}$ is quadratic and $\hat V$ is Gaussian,
$\hat H$ is also a quadratic Hamiltonian. Now, we consider the following Lindbladian
\begin{align}
\mathcal L[\bullet]
:=
-i[\hat H,\bullet]
+
\gamma
\sum_{\ell=0}^{K}
\sum_{i=1}^m
\mathcal D(\hat c_{\ell,i})[\bullet].
\label{eq:uniform-loss-Lindbladian}
\end{align}
Importantly, the loss rate in~\eqref{eq:uniform-loss-Lindbladian} is
exactly $\gamma$ on every physical mode. \cref{fig:uniform-loss-construction} schematically represents the construction.

First, we find the quadratic form associated with $H_{\mathrm{mix}}$. That is, up to an additive constant which does not affect the Hamiltonian dynamics, we find the symmetric matrix $\mathbf Q\in\mathbb R^{(2m(K+1))\times (2m(K+1))}$ such that
\begin{align}
\hat H_{\mathrm{mix}} = \frac12 \hat{\mathbf r}^T \mathbf Q \hat{\mathbf r}+\mathrm{const}.
\end{align}
First note that $H_{\mathrm{mix}}$ acts separately on each block of modes. We have
\begin{align}
\hat H_{\mathrm{mix}} = \sum_i \hat h_{\mathrm{mix},i},
\end{align}
where $\hat h_{\mathrm{mix},i}$ are identical copies of a Hamiltonian $\hat h_{\mathrm{mix}}$ acting on different blocks. Each one has the form
\begin{align}
\begin{split}
\hat h_{\mathrm{mix}} &= \sum_{ij} G_{ij} a_i^\dagger a_j\\ 
&= \hat{\mathbf a}^\dagger \mathbf G \hat{\mathbf a}\\
&=\frac12 (\hat{\mathbf q} - i\hat{\mathbf p})^T \mathbf G (\hat{\mathbf q} + i\hat{\mathbf p})\\
&= \frac12 \hat{\mathbf r}^T \begin{pmatrix}
\mathbf G & 0\\
0 & \mathbf G
\end{pmatrix}
\hat{\mathbf r}-\frac12\operatorname{Tr}(\mathbf G).
\end{split}
\end{align}
Here, in the last equality we used that $\mathbf G$ is real and symmetric. The additive constant can be omitted. Note that $\begin{pmatrix}
\mathbf G & 0\\
0 & \mathbf G
\end{pmatrix}$ commutes with the symplectic form $\boldsymbol\Omega$. This is indeed the common feature of all passive Gaussian Hamiltonians. Also, if we write $\mathbf Q$ in the ordering of 
\[\hat{\mathbf r}=(q_{1,0}, \cdots, q_{1,K}, p_{1,0},\cdots, p_{1,K}, \cdots, q_{m,0}, \cdots, q_{m,K}, p_{m,0}, \cdots, p_{m,K})\]
we have that
\begin{align}\label{eq:blocks-of-Q}
\mathbf Q &=\bigoplus_{i=1}^m \begin{pmatrix}
\mathbf H^{\otimes a}\mathrm{diag}(0,\omega,\cdots,K\omega)\mathbf H^{\otimes a} & 0\\
0 & \mathbf H^{\otimes a}\mathrm{diag}(0,\omega,\cdots,K\omega)\mathbf H^{\otimes a}
\end{pmatrix}
\end{align}
which is block-diagonal with respect to the blocks of the modes.

We first show that this Lindbladian has a unique fixed point, and that all initial states converge to that fixed point. Let $\mathbf J=\boldsymbol\Omega \mathbf Q$ denote the drift matrix corresponding to $\hat H_{\mathrm{mix}}$. Since $\hat H_{\mathrm{mix}}$ is passive, we have that $[\boldsymbol\Omega,\mathbf Q]=0$. Since $\mathbf Q$ is symmetric and $\boldsymbol\Omega^T=-\boldsymbol\Omega$, this gives
\begin{align}
\mathbf J^T
=
-\mathbf Q\boldsymbol\Omega
=
-\boldsymbol\Omega\mathbf Q
=
-\mathbf J.
\end{align}
Thus, all eigenvalues of $\mathbf J$ are purely imaginary.

Let $\mathbf S\in\mathrm{Sp}(2m)$ be the symplectic matrix of $\hat U$ and $\mathbf T = \mathbf S\oplus\mathbb I_{2mK}$ be the symplectic matrix of $\hat V$. We note that $\mathbf M = \mathbf T^{-T} \mathbf Q \mathbf T^{-1}$ is the quadratic form of $\hat H$. Thus, the drift matrix of $\mathcal L$ is
\begin{align}\label{eq:D-all-lossy}
\begin{split}
\mathbf D &= \boldsymbol\Omega \mathbf M - \frac\gamma2 \mathbb I\\
&=\boldsymbol\Omega \mathbf T^{-T} \mathbf Q \mathbf T^{-1} - \frac\gamma2\mathbb I\\
&=\mathbf T \boldsymbol\Omega \mathbf Q \mathbf T^{-1}-\frac\gamma2\mathbb I\\
&=\mathbf T\left( \boldsymbol\Omega\mathbf Q-\frac\gamma2\mathbb I \right)\mathbf T^{-1},
\end{split}
\end{align}
where the third equality follows from $\boldsymbol\Omega\mathbf T^{-T}=\mathbf T\boldsymbol\Omega$, which is equivalent to the symplectic property of $\mathbf T$. Therefore, $\mathbf D$ is similar to $\mathbf J-\gamma\mathbb I/2$, and all its eigenvalues have real part $-\gamma/2<0$.
The drift is consequently Hurwitz, and \Cref{thm:gaussian-fixed-point} gives its unique fixed point. More explicitly, since $\mathbf J$ is skew-symmetric,
\begin{align}
\|e^{t\mathbf D}\|_\infty
\leq\|\mathbf T\|_\infty\|\mathbf T^{-1}\|_\infty e^{-\gamma t/2}.
\end{align}
The bounded squeezing imposed above gives
$\|\mathbf T\|_\infty\|\mathbf T^{-1}\|_\infty=e^{2r_{\max}}\leq3$.

Let $\rho_f$ represent this fixed point. We now show that the reduced state of $\rho_f$ on the data layer is close to $\ket{\psi}$ (the hard GBS instance in \eqref{eq:hard-gbs-in-pf}). Let $\boldsymbol\Sigma_f$ be the covariance matrix of $\rho_f$ and define the covariance in the rotated frame by
\begin{align}\label{eq:def-of-sigma-tilde}
\widetilde{\boldsymbol\Sigma}_f
:=
\mathbf T^{-1}
\boldsymbol\Sigma_f
\boldsymbol T^{-T}.
\end{align}
Now, using the Lyapunov equation for our fixed point (\cref{eq:mode-dependent-Lyapunov}) we have
\begin{align}
\mathbf D\boldsymbol\Sigma_f + \boldsymbol\Sigma_f \mathbf D^T + \gamma \mathbb I=0,
\end{align}
and applying \eqref{eq:D-all-lossy}, we obtain
\begin{align}
\mathbf T \left( \boldsymbol\Omega \mathbf Q -\frac\gamma2\mathbb I\right)\mathbf T^{-1}\boldsymbol\Sigma_f + \boldsymbol\Sigma_f \mathbf T^{-T} \left(\boldsymbol\Omega\mathbf Q-\frac\gamma2\mathbb I \right)^T\mathbf T^{T}+\gamma\mathbb I=0.
\end{align}
Multiplying this equation from left by $\mathbf T^{-1}$, and from right by $\mathbf T^{-T}$, and using \eqref{eq:def-of-sigma-tilde} gives
\begin{align}
\left(
\boldsymbol\Omega\mathbf Q-\frac{\gamma}{2}\mathbb I
\right) \widetilde{\boldsymbol\Sigma}_f+\widetilde{\boldsymbol\Sigma}_f\left(
-\boldsymbol\Omega\mathbf Q-\frac{\gamma}{2}\mathbb I
\right)+\gamma \mathbf T^{-1}\mathbf T^{-T}
=
0.
\label{eq:uniform-loss-rotated-Lyapunov}
\end{align}
Using \eqref{eq:mode-dependent-covariance-solution}, we can write our fixed point in the following integral form:
\begin{subequations}
\begin{align}
\widetilde{\boldsymbol{\Sigma}}_f &= \gamma \int_{t=0}^\infty e^{(-\frac\gamma2 \mathbb I +\boldsymbol\Omega\mathbf Q)t} \mathbf T^{-1} \mathbf T^{-T} e^{(-\frac\gamma2\mathbb I -\boldsymbol\Omega\mathbf Q)t}dt\\
&=\gamma \int_{t=0}^\infty e^{-\gamma t} e^{\boldsymbol\Omega \mathbf Qt} \mathbf T^{-1}\mathbf T^{-T} e^{-\boldsymbol\Omega\mathbf Q t}dt.
\end{align}
\end{subequations}
Now, note that $\mathbf T = \mathbf S_{\mathrm{data}} \oplus \mathbb I_{\mathrm{ancilla}}$. Define
\begin{align}
\mathbf W:=\mathbf S^{-1}\mathbf S^{-T}-\mathbb I_{2m}.
\end{align}
Then, writing $\Pi_{\mathrm{data}}$ for the projection onto the data quadratures, we have
\begin{align}
\mathbf T^{-1}\mathbf T^{-T}
=
\mathbb I+\Pi_{\mathrm{data}}^T\mathbf W\Pi_{\mathrm{data}}.
\end{align}
Since $\boldsymbol\Omega\mathbf Q$ is skew-symmetric, $e^{\boldsymbol\Omega\mathbf Qt}e^{-\boldsymbol\Omega\mathbf Qt}=\mathbb I$. We therefore obtain
\begin{align}\label{eq:sigma-f}
\begin{split}
\widetilde{\boldsymbol{\Sigma}}_f
&=
\gamma\int_0^\infty e^{-\gamma t}\mathbb I\,dt
+
\gamma\int_0^\infty e^{-\gamma t}e^{\boldsymbol\Omega\mathbf Qt}
\Pi_{\mathrm{data}}^T\mathbf W\Pi_{\mathrm{data}}
e^{-\boldsymbol\Omega\mathbf Qt}dt\\
&=
\mathbb I
+
\gamma\int_0^\infty e^{-\gamma t}e^{\boldsymbol\Omega\mathbf Qt}
\Pi_{\mathrm{data}}^T\mathbf W\Pi_{\mathrm{data}}
e^{-\boldsymbol\Omega\mathbf Qt}dt,
\end{split}
\end{align}
where
\begin{align}
\Pi_{\mathrm{data}} = \begin{pmatrix}
\mathbb I_{\mathrm{data}} & 0_{\mathrm{ancilla}}
\end{pmatrix}
\end{align}
projects onto the data quadratures. Now, we project $\widetilde{\boldsymbol\Sigma}_f$ onto the data part. Since $(\boldsymbol\Omega\mathbf Q)^T=-\boldsymbol\Omega\mathbf Q$, we have
\begin{align}
\Pi_{\mathrm{data}}e^{-\boldsymbol\Omega\mathbf Qt}\Pi_{\mathrm{data}}^T
=
\left(
\Pi_{\mathrm{data}}e^{\boldsymbol\Omega\mathbf Qt}\Pi_{\mathrm{data}}^T
\right)^T.
\end{align}
It follows from~\eqref{eq:sigma-f} that
\begin{subequations}
\begin{align}
\widetilde{\boldsymbol\Sigma}_{f,\mathrm{data}}
&:=
\Pi_{\mathrm{data}}\widetilde{\boldsymbol\Sigma}_f\Pi_{\mathrm{data}}^T\\
&=
\mathbb I_{2m}
+
\gamma\int_0^\infty e^{-\gamma t}\mathbf P(t)\mathbf W\mathbf P(t)^Tdt,
\label{eq:uniform-loss-projected-covariance}
\end{align}
where
\begin{align}
\mathbf P(t)
:=
\Pi_{\mathrm{data}}e^{\boldsymbol\Omega\mathbf Qt}\Pi_{\mathrm{data}}^T.
\end{align}
\end{subequations}

We now calculate $\mathbf P(t)$. Due to~\eqref{eq:blocks-of-Q}, the restriction of $\boldsymbol\Omega\mathbf Q$ to the block associated with a fixed data mode $i$ is
\begin{align}
\begin{pmatrix}
0&\mathbf G\\
-\mathbf G&0
\end{pmatrix}.
\end{align}
Therefore,
\begin{align}
\exp\left(
t\begin{pmatrix}
0&\mathbf G\\
-\mathbf G&0
\end{pmatrix}
\right)
=
\begin{pmatrix}
\cos(\mathbf Gt)&\sin(\mathbf Gt)\\
-\sin(\mathbf Gt)&\cos(\mathbf Gt)
\end{pmatrix}.
\label{eq:uniform-loss-block-exponential}
\end{align}
Let $\ket{0}$ denote the first standard basis vector in the $(K+1)$-dimensional layer space, and define
\begin{subequations}
\begin{align}
c_K(t)
&:=
\bra0\cos(\mathbf Gt)\ket0
=
\frac1{K+1}\sum_{r=0}^K\cos(r\omega t),\\
s_K(t)
&:=
\bra0\sin(\mathbf Gt)\ket0
=
\frac1{K+1}\sum_{r=0}^K\sin(r\omega t).
\end{align}
\end{subequations}
Here, we used $\mathbf G=\mathbf F\operatorname{diag}(0,\omega,\ldots,K\omega)\mathbf F^T$ and $|F_{0r}|^2=1/(K+1)$. Projecting~\eqref{eq:uniform-loss-block-exponential} onto the data layer gives
\begin{align}
\mathbf P(t)
=
\bigoplus_{i=1}^m
\begin{pmatrix}
c_K(t)&s_K(t)\\
-s_K(t)&c_K(t)
\end{pmatrix}.
\label{eq:uniform-loss-Pt}
\end{align}
Defining
\begin{align}
z_K(t)
:=
c_K(t)-is_K(t)
=
\frac1{K+1}\sum_{r=0}^K e^{-ir\omega t},
\end{align}
each $2\times2$ block in~\eqref{eq:uniform-loss-Pt} satisfies
\begin{align}
\begin{pmatrix}
c_K(t)&s_K(t)\\
-s_K(t)&c_K(t)
\end{pmatrix}
\begin{pmatrix}
c_K(t)&-s_K(t)\\
s_K(t)&c_K(t)
\end{pmatrix}
=
|z_K(t)|^2\mathbb I_2.
\end{align}
Consequently,
\begin{align}
\|\mathbf P(t)\|^2
=
|z_K(t)|^2
=
\frac1{(K+1)^2}
\left|\sum_{r=0}^K e^{-ir\omega t}\right|^2.
\label{eq:uniform-loss-Pt-norm}
\end{align}
Using~\eqref{eq:uniform-loss-projected-covariance}, submultiplicativity of the operator norm, and~\eqref{eq:uniform-loss-Pt-norm}, we obtain
\begin{subequations}
\begin{align}
\|\widetilde{\boldsymbol\Sigma}_{f,\mathrm{data}}-\mathbb I\|
&\leq
\|\mathbf W\|
\gamma\int_0^\infty e^{-\gamma t}\|\mathbf P(t)\|^2dt\\
&=
\|\mathbf S^{-1}\mathbf S^{-T}-\mathbb I\|
\frac{\gamma}{(K+1)^2}
\int_0^\infty e^{-\gamma t}
\left|\sum_{r=0}^K e^{-ir\omega t}\right|^2dt.
\label{eq:uniform-loss-data-bound}
\end{align}
\end{subequations}
We define
\begin{subequations}
\begin{align}
f_K
:=&
\frac{\gamma}{(K+1)^2}
\int_0^\infty e^{-\gamma t}
\left|\sum_{r=0}^K e^{-ir\omega t}\right|^2dt\\
&=
\frac{\gamma}{(K+1)^2}
\sum_{r,s=0}^{K}
\int_0^\infty
e^{-\gamma t}
e^{-i(r-s)\omega t}dt\\
&=
\frac1{K+1}
+
\frac{2}{(K+1)^2}
\sum_{d=1}^{K}
(K+1-d)
\operatorname{Re}\left(
\frac{\gamma}{\gamma+id\omega}
\right)\\
&=
\frac1{K+1}
+
\frac{2}{(K+1)^2}
\sum_{d=1}^{K}
(K+1-d)
\frac{\gamma^2}
{\gamma^2+d^2\omega^2}.
\label{eq:uniform-loss-fn-exact}
\end{align}
\end{subequations}
Using $\frac{K+1-d}{K+1} \le 1$ and $\gamma^2+\omega^2d^2\ge \omega^2 d^2$, we obtain
\begin{subequations}
\begin{align}
f_K
&\leq
\frac1{K+1}
+
\frac{2\gamma^2}{(K+1)\omega^2}
\sum_{d=1}^\infty\frac1{d^2}\\
&=
\frac1{K+1}
\left(
1+
\frac{\pi^2}{3}
\frac{\gamma^2}{\omega^2}
\right).
\label{eq:uniform-loss-fn-bound}
\end{align}
\end{subequations}
Now, we choose
\begin{align}
\gamma=\omega=\frac1{K+1}.
\label{eq:uniform-loss-parameter-choice}
\end{align}
As a result \cref{eq:uniform-loss-fn-bound} gives
\begin{align}
f_K\leq\frac{1+\pi^2/3}{K+1}.
\label{eq:uniform-loss-fn-final}
\end{align}
It remains to bound $\|\mathbf S^{-1}\mathbf S^{-T}-\mathbb I\|$. Let $r_{\max}$ be the largest squeezing
parameter of $\hat U$. Since we had conditioned on $\norm{\mathbf A}\le 1/2$, we have
\begin{align}
    \tanh(r_{\max})\leq\frac12,
\end{align}
and therefore
\begin{align}
    e^{2r_{\max}}
    =
    \frac{1+\tanh(r_{\max})}
         {1-\tanh(r_{\max})}
    \leq 3.
\end{align}
The eigenvalues of $\mathbf S^{-1}\mathbf S^{-T}$ are of the form
$e^{\pm2r_j}$. Hence,
\begin{align}
    \|\mathbf S^{-1} \mathbf S^{-T}-\mathbb I\|\leq2.
    \label{eq:uniform-loss-W-bound}
\end{align}
Combining~\eqref{eq:uniform-loss-data-bound},
\eqref{eq:uniform-loss-fn-final}, and
\eqref{eq:uniform-loss-W-bound}, we conclude that
\begin{align}
    \norm{\widetilde{\boldsymbol\Sigma}_{f,\mathrm{data}}-\mathbb I}
    \leq
    \frac{2(1+\pi^2/3)}{K+1}.
    \label{eq:uniform-loss-covariance-closeness}
\end{align}
Let $\widetilde\rho_{\mathrm{data}}$ be the reduced state of
$\hat V^\dagger\rho_f\hat V$ on the data modes. Since the Lindbladian contains no linear drive and its drift matrix is Hurwitz, its fixed point has zero displacement. Therefore, $\widetilde\rho_{\mathrm{data}}$ is a zero-displacement Gaussian state with covariance matrix $\widetilde{\boldsymbol\Sigma}_{f,\mathrm{data}}$. For
$K\geq2(1+\pi^2/3)-1$, Equation~\eqref{eq:uniform-loss-covariance-closeness}
implies
\begin{align}
\|\widetilde{\boldsymbol\Sigma}_{f,\mathrm{data}}\|\leq2,
\qquad
\|\widetilde{\boldsymbol\Sigma}_{f,\mathrm{data}}+\mathbb I\|\leq3.
\end{align}
Using \Cref{lem:cov-to-tr} and
$\|\widetilde{\boldsymbol\Sigma}_{f,\mathrm{data}}-\mathbb I\|_1
\leq2m\|\widetilde{\boldsymbol\Sigma}_{f,\mathrm{data}}-\mathbb I\|_\infty$,
we obtain
\begin{align}
\tfrac12\|\widetilde\rho_{\mathrm{data}}-|0^m\rangle\langle0^m|\|_1
&\leq\frac{1+\sqrt3}{16}
\|\widetilde{\boldsymbol\Sigma}_{f,\mathrm{data}}-\mathbb I\|_1
\|\widetilde{\boldsymbol\Sigma}_{f,\mathrm{data}}+\mathbb I\|_\infty\nonumber\\
&\leq\frac{3(1+\sqrt3)(1+\pi^2/3)}4\frac m{K+1}.
\label{eq:uniform-loss-trace-distance}
\end{align}
Since $\rho_{\mathrm{data}}=\hat U\widetilde\rho_{\mathrm{data}}\hat U^\dagger$,
unitary invariance gives
\begin{equation}\label{eq:uniform-loss-physical-trace-distance}
\tfrac12\|\rho_{\mathrm{data}}-|\psi\rangle\langle\psi|\|_1
\leq\frac{3(1+\sqrt3)(1+\pi^2/3)}4\frac m{K+1}.
\end{equation}
Hence, we can get inverse polynomially close to the hard-to-sample state of BipartiteGBS.

To complete the reduction, let $0<\beta(m)\leq1$ be an
inverse-polynomial error threshold for the BipartiteGBS hardness
result, and set $C=3(1+\sqrt3)(1+\pi^2/3)/4$. Choose $K+1$ to
be the smallest power of two satisfying
\[
K+1\geq
\max\left\{\frac{4Cm}{\beta(m)},\,2(1+\pi^2/3)\right\}.
\]
Then \eqref{eq:uniform-loss-physical-trace-distance} is at most
$\beta(m)/4$, and both the total number of modes $N=m(K+1)$
and $\gamma^{-1}=K+1$ are polynomial in $m$. Moreover,
$\|\mathbf M\|_\infty\leq
\|\mathbf T^{-1}\|_\infty^2\|\mathbf Q\|_\infty\leq3$,
while the unchanged ancillary diagonal Hamiltonian coefficients
are bounded below by a positive constant. Thus $c_H=\Theta(1)$
and $\gamma/c_H=\Theta((K+1)^{-1})$, as required.
For a finite-bit description, round $\mathbf M$ symmetrically
with operator-norm error $e\leq c\beta(m)\gamma/N$, where $c>0$
is a sufficiently small constant. Indeed, the bound
$\|e^{t\mathbf D}\|_\infty\leq3e^{-\gamma t/2}$ and Duhamel's
formula give
$\|e^{t(\mathbf D+\boldsymbol\Omega\Delta\mathbf M)}\|_\infty
\leq3e^{-\gamma t/4}$ whenever $e\leq\gamma/12$.
The Lyapunov integral therefore bounds the covariance error
by $O(e/\gamma)$, and \cref{lem:cov-to-tr} bounds the resulting
trace distance by $O(Ne/\gamma)\leq\beta(m)/4$.
The coefficients can be computed to this precision in
polynomial time. Consequently, a classical sampler for the
rounded fixed point with total-variation error $\beta(m)/4$,
followed by discarding the ancillary counts, samples the ideal
BipartiteGBS distribution within $3\beta(m)/4$ on the
conditioned instances. The excluded event contributes only
$O(e^{-m})\leq\beta(m)/4$ to the average error for sufficiently
large $m$. This gives a polynomial-time sampler within the
BipartiteGBS hardness threshold, contradicting the stated
complexity assumptions.
\end{proof}

\section{Complexity reductions for specific non-Gaussian instances}
\label{sec:Hardness-specific-instances}

We so far have presented three sets of results: (1) Embedding $\BQP$ in the unique and accessible fixed point of a bosonic Lindbladian system with Hamiltonian degree $12$ and nonuniform simple loss of degree $1$, (2) specified regime of classiclal simulation for the fixed point problem for Gaussian dominated Lindbladian system with cross-Kerr non-Gaussian interactions, (3) hardness of sampling for Gaussian Lindbladian systems. An important missing set of results is understanding the regime of complexity for non-Gaussian Lindbladians with simpler structure and understand how complexity changes by changing the Hamiltonian degree or the degree for incoherent terms. Ideally one desires to prove a complete classification of complexity. We take partial steps towards this goal and relate the complexity of the fixed point to various instances of boson sampling. As a complementary result, we show that for Hamiltonians of loss degree exceeding $5$, the ``existence'' of fixed point problem is in principle undecidable. This section includes a combination of various results. An interesting future direction is to identify a guiding principle in specifying complexity as a function of the properties of a bosonic Lindbladian system.

\subsection{Reduction from noisy boson sampling using single mode quartic gadgets}
\label{sec:Kerr-gadget}

The first example we study regards the situation where Hamiltonian does not have squeezing terms and the only active terms are linear drive. Instead, it includes quartic terms of the form $\hat a_i^{\dagger}\hat a_j^{\dagger} \hat a_k\hat a_l$. The loss model is again linear loss with uniform coefficients. We show that sampling exactly or approximately from the fixed point of this Lindbladian family is at least as hard as approximate sampling from noisy boson sampling  with constant rate of noise; more specifically boson sampling on product of density matrices with binomial distribution which we will see below. We will discuss the issue of mixing time in \Cref{subsec:kerr-moment-bounds}.

Before presenting the result, we need to the definition of the rotated Lindbladian: For a Lindbladian $\mathcal{L}$ with Hamiltonian terms $H$ and jump terms $L_j$, define the rotated Lindbladian with respect to the unitary matrix $U$ to be the same as $\mathcal{L}$ with $H$ replaced with $U H U^\dagger$ and $L_j$ with $U L_j U^\dagger$. Therefore $\rho$ is a fixed point of $\mathcal{L}$ if and only if $U \rho U^\dagger$ is a fixed point of $\mathcal{L}_U$.

\begin{theorem}\label{thm:Quartic}
    Consider Lindbladian dynamics of the form
    $$
    \mathcal{L} = - i [H, \bullet] + \sum_{j} \mathcal{D}(\sqrt{\gamma} \hat a_j)[\bullet]
    $$
    over $m$ modes where
    $$ H=-\Delta\omega\sum_j\hat n_j+\sum_{j,k,s,t}\alpha_{j,k,s,t}\hat a_j^\dagger\hat a_k^\dagger\hat a_s\hat a_t+\sum_j(F_j\hat a_j^\dagger+F_j^*\hat a_j)
    $$
then for any linear optical circuit $C$ (specified with constant bits of precision), any fixed integer $q\ge1$, any $1\le k\le m$, and $0<\delta\le1$, there is a choice of coefficients $\alpha_{j,k,s,t}=O(1)$ and $F_j$, with detuning $\Delta\omega=(q-1)/2$, constructed from local drive amplitude $F=\kappa_q\delta/m$ and uniform loss $\gamma=F^{q+1}$, such that the fixed point of $\mathcal{L}$ is within $\delta$ trace distance of the quantum state $$
C (\sigma_q^{\otimes k} \otimes (\ket{0}\bra{0})^{\otimes m-k}) C^\dagger.
$$
where $\kappa_q>0$ is a sufficiently small constant depending only on $q$, and
$$
\sigma_q = \frac{1}{2^q} \sum_{j = 0}^q {q \choose j} \ketbra{j}{j}.
$$
\end{theorem}

\begin{proof}
 Our approach is to use an exact solution to single-mode driven-dissipative Bose-Hubbard Hamiltonians undergoing linear loss to obtain single-mode fixed points that are mixtures of non-Gaussian states. In particular, for each mode, we consider the Hamiltonian
$$
H^{\mathrm{BH}}_{j} = - \Delta \omega \hat n_j + \frac{U}{2} \hat a_j^{\dagger 2} \hat a_j^2 + F \hat a_j^\dagger + F^* \hat a_j
$$

For \(\gamma>0\), the full quantum master equation has a unique fixed point,
even in parameter regimes where the semiclassical mean-field equations
exhibit bistability \cite{casteels2016powerlaws}.  The exact steady-state
solution was originally obtained using the complex-\(P\) representation by
Drummond and Walls and has been used in later treatments of driven-dissipative
Kerr resonators and Bose-Hubbard lattices
\cite{drummond1980quantum,le2014bose} which indicates that the fixed point for $\mathcal{L}_j$ can be suitably approximated with a simple expression when $\gamma, F$ is small and $U, \Delta \omega$ with certain ratio (the explicit formula was given in the Appendix in \cite{le2014bose}), but for discussion here, we don't need the explicit formulation). 

For $U=1$, $\Delta\omega=(q-1)/2$, and $0<F\le f_q$, $0<\gamma/F^q\le f_q$, \cref{thm:explicit-binomial-bound} (the precise definition of $f_q$ is also given in \cref{thm:explicit-binomial-bound}) gives a fixed point $\rho$ of the single-mode Lindbladian
$$
\mathcal{L}^{\mathrm{BH}} [\bullet] = - i [H^{\mathrm{BH}}, \bullet] + \mathcal{D}(\sqrt{\gamma}\hat a)
$$
such that 
\begin{align}
\mathcal F(\rho,\sigma_q)
\ge
1-O_q\!\left(
|F|^2+\frac{\gamma^2}{|F|^{2q}}
\right).
\end{align}
Here $\mathcal F(\rho,\sigma)=\|\sqrt\sigma\sqrt\rho\|_1$ is the root fidelity between the two quantum states.

We now want to study the fixed point of $\mathcal{L}^{BH}_C = \sum_{j = 1}^m \mathcal{L}_{j,C}^{\mathrm{BH}}$ for a linear optical circuit $C$ which has the quartic form indicated in the statement of the theorem. Choose $F=\kappa_q\delta/m$ and $\gamma=F^{q+1}$. Then the fixed point of $\mathcal{L}^{BH}_C$ is 
$$
C \rho^{\otimes m} C^\dagger
$$
such that 
$$
\frac12\|\rho^{\otimes m}-\sigma_q^{\otimes m}\|_1\le\frac m2\|\rho-\sigma_q\|_1\le2C_qmF\le\delta.
$$
The last inequality holds for $\kappa_q\le\min\{1/4,f_q,(2C_q)^{-1}\}$, where $f_q$ and $C_q$ are the constants in \cref{thm:explicit-binomial-bound}.
Trace distance is invariant under $C$ and upper bounds
the total variation distance after photon-number measurement.Trace distance is invariant under $C$ and upper bounds the total variation distance after photon-number measurement.
We note that we have free access to vacuum states, which are fixed points of the same single-mode model with $F=0$. Therefore, we can approximate states of the form:
$$
C (\sigma_q^{\otimes k} \otimes (\ket{0}\bra{0})^{\otimes m-k}) C^\dagger
$$
\end{proof}

The binomial boson-sampling distributions considered in
this section are closely related to lossy boson sampling
with a constant photon-transmission rate. In particular,
for $q=1$, the binomial input state
\begin{align}
\sigma_1=\frac12
\bigl(|0\rangle\langle 0|+|1\rangle\langle 1|\bigr)
\end{align}
is precisely the state obtained by subjecting a
single-photon input to a loss channel with transmission
probability $1/2$. More generally, $\sigma_q$ is the
state obtained by independently losing each photon
from a $q$-photon Fock input with probability $1/2$.

The computational complexity of lossy boson sampling
depends sensitively on the loss regime and the
notion of approximation. Aaronson and Brod
\cite{aaronson2016bosonsampling} established
conditional hardness when a constant number of
photons is lost. On the other hand, for standard
boson sampling with $N$ single-photon inputs,
Theorem~2 of \cite{garcia2019simulating}
establishes a thermal-state approximation with
total-variation error at most $\varepsilon$ when
the uniform transmission satisfies
$\eta\leq\sqrt{\varepsilon/N}$.
The corresponding thermal output admits efficient
classical sampling. This sufficient condition does
not cover the constant-transmission regime.

More recently, Go and Jeong
\cite{go2026polynomial} established a
polynomial-time approximate classical sampling
algorithm for standard lossy boson sampling with
any fixed photon-survival fraction, with high
probability over Haar-random linear optical operations.
Their result also extends to uniform transmission
loss with any constant transmission rate.
This provides an efficient approximate simulation
algorithm in a regime previously not covered by
the thermal-state approximation. However, it
does not establish efficient simulation for
arbitrary linear optical operations, nor does it directly
address the more general $q>1$ collision-input
families considered here.

These developments highlight the distinction
between exact and approximate sampling hardness
in our construction. Although we can establish
hardness of exact sampling from the limiting
binomial boson-sampling distributions, this
does not imply hardness of approximate sampling
from the corresponding Lindbladian stationary
states. Indeed, the stationary states in our
construction are only guaranteed to approximate
the target binomial boson-sampling distributions
in total-variation distance.

Consequently, establishing approximate sampling
hardness for the relevant binomial boson-sampling
family would also establish hardness for the
corresponding fixed-point sampling problem,
provided the reduction achieves sufficiently
small total-variation error. Conversely, an
efficient approximate classical sampler for
the same family of input states and
linear optical operations would yield an efficient sampler
for the corresponding stationary-state
distributions, with an additional error determined
by the fixed-point approximation.

Thus, our construction relates the complexity
of quartic bosonic Lindbladian fixed-point
sampling to the complexity of lossy boson sampling,
while leaving open the question of approximate
sampling hardness for the full family considered
here.
instance.

\paragraph{Complexity of binomial boson sampling.}
For completeness we now show that exact photon-number sampling from
passive linear-optical circuits with binomial
input states is classically hard unless the
polynomial hierarchy collapses. This result
holds for every fixed integer $q\geq 1$.
Importantly, it does not by itself establish
exact-sampling hardness for the Lindbladian
fixed points constructed above, which only
approximate these binomial input states.

\begin{proposition}
\label{prop:binomial-exact-hardness}
Fix an integer $q\geq 1$. Suppose there is a
polynomial-time classical algorithm that
exactly samples the photon-number distribution
of
\[
C\left(
\sigma_q^{\otimes n}
\otimes
(|0\rangle\langle0|)^{\otimes(m-n)}
\right)C^\dagger
\]
for arbitrary passive linear-optical circuits
$C$ and polynomially bounded $m,n$.
Then the polynomial hierarchy collapses
to its third level.
\label{thm:PostLOk}
\end{proposition}

\begin{proof}
We reduce from postselected ordinary
BosonSampling, which can simulate
$\PostBQP$ computations
\cite{aaronson2011computational}.

Let $\PostLO_1$ denote the class of problems
solvable by polynomial-size passive
linear-optical circuits starting from
single-photon and vacuum inputs, followed
by photon-number measurements and
postselection on specified measurement
outcomes.

More generally, for any fixed integer
$q\geq1$, define $\PostLO_q$ analogously,
except that each initially occupied mode
contains exactly $q$ photons. Thus the
allowed input states have the form
\[
|q\rangle^{\otimes n}
\otimes |0\rangle^{\otimes(m-n)}.
\]
The number of modes and the optical
circuit size are polynomial in the
input length, and the postselection
events are required to have nonzero
probability. As usual, the postselection
measurements may be deferred to the end
of the computation.

Aaronson and Arkhipov showed, using the
linear-optical gadgets of Knill, that
postselected ordinary BosonSampling
captures postselected quantum computation:
\[
\PostLO_1=\PostBQP
\]
\cite{aaronson2011computational,
knill2001scheme}.

We first prove that
\[
\PostLO_1\subseteq\PostLO_q.
\]
Consider an arbitrary computation in
$\PostLO_1$ requiring $N$ single-photon
input modes. For each such mode, introduce
an auxiliary vacuum mode and replace
the single-photon input by the two-mode
state $|q,0\rangle$.

Apply a balanced beam splitter $B$ to
each pair. Its action is
\[
B|q,0\rangle
=
\sum_{j=0}^{q}
\sqrt{\frac{\binom{q}{j}}{2^q}}
\,|j,q-j\rangle.
\]
Postselecting on the outcome $q-1$
photons in the first (auxiliary) mode
leaves exactly one photon in the second
(computational) mode:
\[
{}_a\langle q-1|B|q,0\rangle_{ab}
=
\sqrt{\frac{q}{2^q}}\,|1\rangle_b.
\]
The success probability is therefore
\[
p_{\mathrm{gadget}}=\frac{q}{2^q}>0.
\]

Applying this gadget independently to
all $N$ inputs prepares the required
single-photon state with probability
\[
p_{\mathrm{all}}
=
\left(\frac{q}{2^q}\right)^N.
\]
For fixed $q$, this probability is
inverse-exponential in $N$, as permitted
in postselected computation.

After these gadgets, apply the original
linear-optical circuit to the
computational modes, leaving the
auxiliary modes untouched. Since the
auxiliaries do not interact with the
computational modes again, their
photon-number measurements can be
deferred to the end and included among
the final postselection conditions.
The resulting circuit uses only
$q$-photon and vacuum inputs and has
polynomial size. This proves
\[
\PostLO_1\subseteq\PostLO_q.
\]

Next, we show that exact sampling with
binomial inputs is at least as hard
as sampling from $\PostLO_q$.
Recall that
\[
\sigma_q
=
\frac{1}{2^q}
\sum_{j=0}^{q}
\binom{q}{j}|j\rangle\langle j|.
\]
For $n$ occupied input modes, the
maximum possible total photon number
is $nq$. Since passive linear optics
preserves total photon number,
postselecting on measuring exactly
$nq$ photons at the output selects
the unique input component
\[
|q\rangle^{\otimes n}
\otimes |0\rangle^{\otimes(m-n)}.
\]
This component occurs with probability
\[
p_{\max}=2^{-nq}>0.
\]
Conditioned on this event, the output
distribution is exactly the distribution
obtained from the corresponding pure
$q$-photon input.

Consequently, a polynomial-time exact
classical sampler for the binomial
input family could simulate any
$\PostLO_q$ computation with
postselection. In particular,
\[
\PostLO_1
\subseteq\PostLO_q
\subseteq\PostBPP.
\]
Together with
$\PostLO_1=\PostBQP$, this implies
\[
\PostBQP\subseteq\PostBPP.
\]
Since $\PostBQP=\PP$
\cite{aaronson2005quantum}, the
polynomial hierarchy collapses to
its third level.
\end{proof}

We emphasize that the proposition concerns
exact sampling from the ideal binomial
input family. The stationary states of
our Lindbladian construction approximate
these states in trace distance, but are
not generally identical to them.
Since the reduction relies on
exponentially unlikely postselection
events, inverse-polynomial trace-distance
closeness does not suffice to transfer
the exact-sampling hardness to those
stationary states.

\subsubsection{Effective finite-dimensionality and moment bounds for the driven Kerr model}
\label{subsec:kerr-moment-bounds}

A natural question is to understand if the model studied in the previous section mixes fast. If we were dealing with finite-dimensional systems, since the construction relies on single mode gadgets, uniqueness of the fixed point would imply fast mixing. For infinite-dimensional systems, however, this does not have to be true as we may have gapless Lindbladians with a unique fixed point.

In this subsection we show that the Lindbladian dynamics is effectively finite-dimensional, giving evidence for fast mixing time. We record photon-number moment bounds for the single-mode
driven Kerr oscillator with one-photon loss.  The main point is a
time-dependent estimate: when the system is initialized at the vacuum, the
entire trajectory has controlled photon-number tails.  Taking
\(t\to\infty\) then gives the corresponding steady-state tail bounds.  The
same steady-state bounds can also be derived directly from the exact
complex-\(P\) solution of the driven Kerr oscillator
\cite{drummond1980quantum,le2014bose}.

We consider the single-mode Hamiltonian
\begin{align}
    H
    =
    -\Delta\omega\, \hat{n}
    +
    \frac{U}{2}\hat{a}^{\dagger 2}\hat{a}^2
    +
    F \hat{a}^\dagger+F^*\hat{a}
\end{align}
with one-photon loss
\begin{align}
    \dot\rho
    =
    -i[H,\rho]+\gamma\mathcal D[\hat{a}](\rho),
    \qquad
    \gamma>0.
\end{align}
For \(\gamma>0\), the full quantum master equation has a unique fixed point,
even in regimes where the semiclassical mean-field equations exhibit
bistability \cite{casteels2016powerlaws}.  We use the following
time-dependent moment estimate to show that, from the vacuum initial state,
the trajectory remains effectively finite-dimensional.

\begin{proposition}[Uniform trajectory tail bound from vacuum]
\label{prop:kerr-vacuum-trajectory-tail}
Let $\Phi_t$ be a conservative, strongly continuous quantum
Markov semigroup for the driven Kerr generator, agreeing
with the displayed Lindblad expression on finite Fock
matrix units. Set $\rho(t)=\Phi_t(|0\rangle\langle0|)$.
Define
\begin{align}
    M_j(t)
    :=
    \left\langle
        (\hat n)_j
    \right\rangle_t,
    \qquad
    \mu
    :=
    \frac{4|F|^2}{\gamma^2},
\end{align}
where $(\hat n)_j = \hat n (\hat n-I) \ldots (\hat n -j+I)$ and $\braket{\bullet}_t = \Tr (\bullet \rho(t))$. Then, for every \(j\ge0\),
\begin{align}
    M_j(t)
    \le
    \left[
        \mu\left(1-e^{-\gamma t/2}\right)
    \right]^j .
\end{align}
Consequently, for every \(\theta\ge0\),
\begin{align}
    \left\langle
        e^{\theta \hat{n}}
    \right\rangle_t
    \le
    \exp\!\left(
        \mu\left(1-e^{-\gamma t/2}\right)
        (e^\theta-1)
    \right)
    \le
    \exp\!\left(
        \mu(e^\theta-1)
    \right).
\end{align}
Thus, uniformly for all \(t\ge0\) and all \(M\ge\mu\),
\begin{align}
\Pr_{\rho(t)}(\hat{n}\ge M)
    \le
    \exp\!\left(
        -M\log\frac{M}{\mu}
        +M-\mu
    \right).
\end{align}
\end{proposition}

\begin{proof}
We first perform the calculation with a finite photon cutoff.
Let
\[
P_R=\sum_{n=0}^R|n\rangle\langle n|,\qquad
a_R=P_RaP_R,\qquad H_R=P_RHP_R,
\]
and let $\rho_R(t)$ evolve from vacuum under
$-i[H_R,\cdot]+\gamma\mathcal D[a_R]$.
Define
$M_{j,R}(t)=\operatorname{Tr}((\hat n)_j\rho_R(t))$.
All operators are bounded on this finite-dimensional space.
The following moment identities hold for these compressed
operators, with bounds independent of $R$.
We suppress the cutoff subscript during the calculation.

The detuning and Kerr terms commute with \(\hat{n}\), and hence with every
factorial moment \((\hat{n})_j\).  Therefore only the coherent drive and the
one-photon loss contribute to the growth of \(M_j(t)\).

The loss term gives
\begin{align}
    \frac{d}{dt}M_j(t)\bigg|_{\rm loss}
    =
    -\gamma j M_j(t).
\end{align}
For the drive term,
\begin{align}
    i[
        F \hat{a}^\dagger+F^*\hat{a},
        (\hat{a}^\dagger)^j \hat{a}^j
    ]
    =
    -ijF(\hat{a}^\dagger)^j \hat{a}^{j-1}
    +
    ijF^*(\hat{a}^\dagger)^{j-1}\hat{a}^j .
\end{align}
By Cauchy--Schwarz,
\begin{subequations}
\begin{align}
\left|
\left\langle
(\hat{a}^\dagger)^j \hat{a}^{j-1}
\right\rangle_t
\right|
&\le
\sqrt{M_j(t)M_{j-1}(t)}, \\
\left|
\left\langle
(\hat{a}^\dagger)^{j-1}\hat{a}^j
\right\rangle_t
\right|
&\le
\sqrt{M_j(t)M_{j-1}(t)} .
\end{align}
\end{subequations}
Hence
\begin{align}
    \frac{d}{dt}M_j(t)
    \le
    -\gamma jM_j(t)
    +
    2j|F|\sqrt{M_j(t)M_{j-1}(t)}.
\end{align}
Using Young's inequality,
\begin{align}
    2|F|\sqrt{M_jM_{j-1}}
    \le
    \frac{\gamma}{2}M_j
    +
    \frac{2|F|^2}{\gamma}M_{j-1},
\end{align}
we obtain
\begin{align}
    \dot M_j(t)
    \le
    -\frac{\gamma j}{2}M_j(t)
    +
    \frac{2j|F|^2}{\gamma}M_{j-1}(t).
\end{align}
Equivalently,
\begin{align}
    \dot M_j(t)
    \le
    -\frac{\gamma j}{2}
    \left(
        M_j(t)-\mu M_{j-1}(t)
    \right),
    \qquad
    \mu=\frac{4|F|^2}{\gamma^2}.
\end{align}

We prove the claim by induction on \(j\).  For \(j=0\), \(M_0(t)=1\).  Since
\(\rho(0)=|0\rangle\langle0|\), we have
\begin{align}
    M_j(0)=0
    \qquad
    \text{for every }j\ge1.
\end{align}
Assume that
\begin{align}
    M_{j-1}(t)
    \le
    \left[
        \mu(1-e^{-\gamma t/2})
    \right]^{j-1}.
\end{align}
Let \(y_j(t)\) solve
\begin{align}
    \dot y_j(t)
    =
    -\frac{\gamma j}{2}
    \left(
        y_j(t)
        -
        \mu
        \left[
            \mu(1-e^{-\gamma t/2})
        \right]^{j-1}
    \right),
    \qquad
    y_j(0)=0.
\end{align}
A direct calculation gives
\begin{align}
    y_j(t)
    =
    \left[
        \mu(1-e^{-\gamma t/2})
    \right]^j .
\end{align}
By the comparison principle,
\begin{align}
    M_j(t)\le y_j(t),
\end{align}
which proves the factorial-moment bound.

We now remove the cutoff. For $R\ge2$, the bound for $j=2$
gives
\[
p_R(t):=\langle R|\rho_R(t)|R\rangle
\le \frac{\mu^2}{R(R-1)}.
\]
Embed $\rho_R(t)$ into the full Fock space.
On this state, the full and cutoff generators differ
only through the drive coupling between $R$ and $R+1$.
Duhamel's formula and trace-norm contractivity imply
\[
\begin{aligned}
\|\rho_R(t)-\rho(t)\|_1
&\le 2|F|\sqrt{R+1}\int_0^t\sqrt{p_R(s)}\,ds\\
&\le
2|F|t\mu\sqrt{\frac{R+1}{R(R-1)}}
\longrightarrow0.
\end{aligned}
\]
For each fixed $j$, first test this convergence against
the bounded observable $P_K(\hat n)_jP_K$, and then let
$K\to\infty$. Monotone convergence transfers the
factorial-moment bound to $\rho(t)$.

The exponential moment estimate follows from
\begin{align}
    (1+s)^{\hat{n}}
    =
    \sum_{j=0}^{\infty}
    \frac{(\hat{n})_j}{j!}s^j,
    \qquad
    s\ge0.
\end{align}
Indeed,
\begin{align}
\label{eq:time-dependent-exp-moment}
    \left\langle
        (1+s)^{\hat{n}}
    \right\rangle_t
    \le
    \exp\!\left(
        \mu(1-e^{-\gamma t/2})s
    \right).
\end{align}
Setting \(s=e^\theta-1\) gives
Eq.~\eqref{eq:time-dependent-exp-moment}.  The tail bound follows from
Markov's inequality and the standard Chernoff optimization.
\end{proof}

We now see that the same result holds for the fixed point.
\begin{corollary}[Steady-state tail bound]
\label{cor:kerr-steady-tail-from-time}
Let \(\rho_{\rm ss}\) denote the unique fixed point of the driven Kerr
oscillator under the semigroup assumptions of
Proposition~\ref{prop:kerr-vacuum-trajectory-tail}. Then
\begin{align}
    \left\langle
        (\hat{n})_j
    \right\rangle_{\rm ss}
    \le
    \mu^j,
    \qquad
    \mu=\frac{4|F|^2}{\gamma^2}.
\end{align}
Consequently, for every \(\theta\ge0\),
\begin{align}
    \left\langle
        e^{\theta \hat{n}}
    \right\rangle_{\rm ss}
    \le
    \exp\!\left(
        \mu(e^\theta-1)
    \right),
\end{align}
and for every \(R\ge\mu\),
\begin{align}
    \Pr_{\rho_{\rm ss}}(\hat{n}\ge R)
    \le
    \exp\!\left(
        -R\log\frac{R}{\mu}
        +R-\mu
    \right).
\end{align}
The same steady-state factorial-moment bound can also be derived directly
from the exact complex-\(P\) steady-state solution
\cite{drummond1980quantum,le2014bose}.
\end{corollary}

\begin{proof}
Define the time-averaged state
\[
\bar\rho_T=\frac1T\int_0^T\rho(t)\,dt.
\]
The bound $\operatorname{Tr}(\hat n\bar\rho_T)\le\mu$
makes this family precompact in trace norm:
its weight above photon number $K$ is at most
$\mu/(K+1)$, and finite photon subspaces are finite-dimensional.
Moreover, for every fixed $s\ge0$,
\[
\|\Phi_s(\bar\rho_T)-\bar\rho_T\|_1\le\frac{2s}{T}.
\]
Every accumulation point is therefore stationary.
Uniqueness implies $\bar\rho_T\to\rho_{\rm ss}$.

For fixed $K$, boundedness of $P_K(\hat n)_jP_K$ gives
\[
\operatorname{Tr}\bigl(P_K(\hat n)_jP_K\rho_{\rm ss}\bigr)
=
\lim_{T\to\infty}
\operatorname{Tr}\bigl(P_K(\hat n)_jP_K\bar\rho_T\bigr)
\le\mu^j.
\]
Letting $K\to\infty$ proves the stationary factorial-moment
bound. The exponential-moment and tail bounds follow
by the same nonnegative series expansion and Chernoff
argument as above.
\end{proof}

\paragraph{Effective finite-dimensionality and the remaining mixing problem.}
Corollary~\ref{cor:kerr-steady-tail-from-time} and
Proposition~\ref{prop:kerr-vacuum-trajectory-tail} show that both the steady
state and the vacuum-initiated trajectory have Poisson-type photon-number
tails controlled by
\begin{align}
    \mu=\frac{4|F|^2}{\gamma^2}.
\end{align}
Therefore, for every target tail error \(\varepsilon\), one may choose a
cutoff \(R_\varepsilon\) such that
\begin{align}
    \operatorname{Tr}(\Pi_{>R_\varepsilon}\rho_{\rm ss})
    \le
    \varepsilon,
    \qquad
    \operatorname{Tr}(\Pi_{>R_\varepsilon}\rho(t))
    \le
    \varepsilon
    \quad
    \text{for all }t\ge0.
\end{align}
For instance, the Chernoff bound above implies that
\(R_\varepsilon\) can be chosen on the scale
\begin{align}
    R_\varepsilon
    =
    O\!\left(
        \mu+\log\frac{1}{\varepsilon}
    \right)
\end{align}
up to logarithmic corrections. This is an effective finite-dimensionality statement.  It is weaker than a
mixing-time theorem, but it is not vacuous.  It shows that, starting from
vacuum, the dynamics never develops an uncontrolled high-photon-number tail.
Consequently, the infinite-dimensional mixing problem can be reduced to a
finite-dimensional spectral-gap problem on the relevant cutoff sector.

\paragraph{From tail bounds to a finite-dimensional mixing problem.}
The tail bounds above do not by themselves prove fast mixing, but they reduce
the problem to a finite-dimensional spectral-gap question.  Let
\[
    \Pi_R:=\sum_{n=0}^{R}|n\rangle\langle n|,
    \qquad
    \Pi_{>R}:=I-\Pi_R .
\]
Define the tail probabilities
\[
    \tau_R(t):=\operatorname{Tr}(\Pi_{>R}\rho(t)),
    \qquad
    \tau_R^{\rm ss}:=\operatorname{Tr}(\Pi_{>R}\rho_{\rm ss}) .
\]
For vacuum initial data, Proposition~\ref{prop:kerr-vacuum-trajectory-tail}
and Corollary~\ref{cor:kerr-steady-tail-from-time} imply that
\[
    \tau_R(t)\le \epsilon_{\rm tail}(R),
    \qquad
    \tau_R^{\rm ss}\le \epsilon_{\rm tail}(R),
    \qquad
    \text{for all }t\ge0,
\]
where, for \(R\ge\mu\),
\[
    \epsilon_{\rm tail}(R)
    :=
    \exp\!\left(
        -R\log\frac{R}{\mu}+R-\mu
    \right).
\]
By the gentle measurement lemma,
\[
    \|\rho(t)-\Pi_R\rho(t)\Pi_R\|_1
    \le
    2\sqrt{\tau_R(t)},
\]
and similarly
\[
    \|\rho_{\rm ss}-\Pi_R\rho_{\rm ss}\Pi_R\|_1
    \le
    2\sqrt{\tau_R^{\rm ss}}.
\]

Now let \(\mathcal L_R\) be a suitable truncated Lindbladian on
\(\Pi_R\mathcal H\).  Let
\[
    \rho_R(t):=e^{t\mathcal L_R}(\rho_R(0))
\]
be the truncated evolution, and let \(\rho_{{\rm ss},R}\) be its stationary
state.  Then the full trace distance can be decomposed as
\begin{subequations}
    \begin{align}
    \|\rho(t)-\rho_{\rm ss}\|_1
    &\le
    \|\rho(t)-\Pi_R\rho(t)\Pi_R\|_1                                      \\
    &\quad+
    \|\Pi_R\rho(t)\Pi_R-\rho_R(t)\|_1                                     \\
    &\quad+
    \|\rho_R(t)-\rho_{{\rm ss},R}\|_1                                     \\
    &\quad+
    \|\rho_{{\rm ss},R}-\Pi_R\rho_{\rm ss}\Pi_R\|_1                       \\
    &\quad+
    \|\Pi_R\rho_{\rm ss}\Pi_R-\rho_{\rm ss}\|_1 .
\end{align}
\end{subequations}
The first and last terms are controlled by the tail bounds:
\begin{align}
    \|\rho(t)-\Pi_R\rho(t)\Pi_R\|_1
    +
    \|\Pi_R\rho_{\rm ss}\Pi_R-\rho_{\rm ss}\|_1
    \le
    4\sqrt{\epsilon_{\rm tail}(R)} .
\end{align}
The second and fourth terms are truncation-consistency errors.  Define
\[
    \epsilon_{\rm dyn}(R,t)
    :=
    \|\Pi_R\rho(t)\Pi_R-\rho_R(t)\|_1,
\]
and
\[
    \epsilon_{\rm fp}(R)
    :=
    \|\rho_{{\rm ss},R}-\Pi_R\rho_{\rm ss}\Pi_R\|_1 .
\]
If, in addition, the truncated dynamics has a convergence estimate
\[
    \|\rho_R(t)-\rho_{{\rm ss},R}\|_1
    \le
    C_R e^{-\lambda_R t},
\]
then
\begin{align}
    \|\rho(t)-\rho_{\rm ss}\|_1
    \le
    4\sqrt{\epsilon_{\rm tail}(R)}
    +
    \epsilon_{\rm dyn}(R,t)
    +
    \epsilon_{\rm fp}(R)
    +
    C_Re^{-\lambda_R t}.
    \label{eq:tail-truncation-mixing-decomposition}
\end{align}

Thus the moment bounds prove effective finite-dimensionality: they control
the tail terms in
Eq.~\eqref{eq:tail-truncation-mixing-decomposition}.  They also identify a
natural cutoff scale \(R\) for the vacuum-initiated trajectory and the steady
state.  However, they do not control the finite-dimensional gap
\(\lambda_R\), nor do they by themselves prove the truncation-consistency
bounds \(\epsilon_{\rm dyn}(R,t)\) and \(\epsilon_{\rm fp}(R)\).  Therefore,
to turn the effective finite-dimensionality statement into a genuine
fast-mixing theorem, one still needs two additional ingredients:
\begin{enumerate}
    \item a quantitative approximation theorem comparing the full and
    truncated dynamics, giving small \(\epsilon_{\rm dyn}(R,t)\) and
    \(\epsilon_{\rm fp}(R)\);

    \item a lower bound on the truncated lindbladian gap, for example
    \[
        \lambda_R^{-1}
        \le
        \operatorname{poly}(R,\gamma^{-1},|U|,|F|,\Delta\omega).
    \]
\end{enumerate}
In a fixed finite dimension, uniqueness of the fixed point guarantees
\(\lambda_R>0\), but it does not guarantee a useful lower bound on
\(\lambda_R\) as a function of \(R\) and the physical parameters.  This is the
remaining mixing-time problem.

This distinction is important.  Driven Kerr resonators are known to exhibit
critical slowing down in large-photon-number or semiclassical bistable
regimes, where the lindbladian gap can become very small
\cite{casteels2017critical}.  The results in this subsection should therefore
be interpreted as rigorous tail and truncation estimates, not as a proof of
rapid convergence to the fixed point.

\subsection{Robust hardness using two-mode quartic Hamiltonian gadgets undergoing linear non-uniform loss}
\label{sec:quartic-non-uniform}

In the previous section we embedded approximate noisy boson sampling in the Lindbladian fixed point problem involving quartic terms with uniform loss. A major drawback is that the complexity of approximate sampling from the noisy boson sampling instance discussed is not known. We show in \Cref{app:nk} that using the single mode Lindbladian considered in the previous section engineering fixed points $\rho_{\rm ss}$ with $\bra{1}\rho_{\rm ss}\ket 1 > 1/2$ is not possible. A natural idea is to try two mode Lindbladians with non-Gaussian Hamiltonians.
In this section, we introduce an alternative route which enables us to prove a robust hardness for this problem when non-uniform loss is allowed (some sites can have different loss strength than others). This construction is inspired by a construction studied in \cite{ikeda2013deterministic} for state generation applications for Gaussian Lindbladians (which we used in \cref{sec:robust-gaussian-lindbladian}). We slightly generalize this construction to generate $\ket{1}$ using non-Gaussian quartic Hamiltonians and nonuniform linear loss. The main drawback is that convergence to the fixed point relies in the initial state starting from a specific subspace (e.g. it works for vacuum). 

In particular, the following two-mode example illustrates the photon-loading mechanism used in \cref{sec:loss-only-bqp}. Starting from vacuum, the Hamiltonian coherently couples \(\lvert 0,0\rangle\) to \(\lvert 1,1\rangle\), creating one photon in the data mode and one in the ancilla. Loss of the ancillary photon leaves the system in \(\lvert 1,0\rangle\), where both the Hamiltonian and the loss operator vanish. We therefore analyze the dynamics on the invariant subspace \(\operatorname{span}\{\lvert 0,0\rangle,\lvert 1,1\rangle,\lvert 1,0\rangle\}\), and all claims of uniqueness and convergence in the example refer to this subspace.

\begin{example}
\label{ex:2-mode-1-engineering}
Consider the two-mode Lindbladian defined as
\begin{align}
\mathcal{L}[\bullet] = -i[\underbrace{\hat b^\dag \hat a^\dag(1-\hat n_a)+(1-\hat n_a)\hat a \hat b}_{H}, \bullet] + \mathcal{D}(\hat b)[\bullet].
\end{align}
where $\hat a$ and $\hat b$ are the annihilation operator over the first and second modes, respectively, and $\hat n_a = \hat a^\dagger \hat a$. Denote the corresponding Hilbert spaces with $\mathcal{H}_a$ and $\mathcal{H}_b$. Let us denote $\overline{\sigma}_-:=(1-\hat n_a) \hat a$, which represents an encoded $\ket{0}\bra{1}$ operator into the $\operatorname{span} \{\ket{0}, \ket{1}, \ket{2}\}$ block of the bosonic system (similar to what was done in \cite[Eq. (113)]{chabaud2026bosonic}). Let $\Pi_k^{\rm diff} : \mathcal{H}_a \otimes \mathcal{H}_b \rightarrow \mathcal{H}_a \otimes \mathcal{H}_b$ be defined as
\begin{align}
 \Pi^{\mathrm{diff}}_{k}
:= \sum_{\substack{m\in\mathbb{Z}\\ m\geq \max\{0,-k\}}} \ket{m+k}_a
\bra{m+k}\otimes \ket{m}_b\bra{m},\quad \forall k\in\mathbb Z,
\end{align}
which means that $\Pi_k^{\mathrm{diff}}$ is the projector onto the subspace where mode $a$ has $k$ more photons (or less if $k < 0$) than mode $b$. Notice that
\begin{align}\label{eq:pi-k-diff-id}
\sum_{k\in\mathbb Z} \Pi_{k}^{\mathrm{diff}} = \mathbb I_{\mathcal{H}_a \otimes \mathcal{H}_b}.
\end{align}
It is straightforward to verify that $[H, \Pi_k^{\mathrm{diff}}]=0$. In other words, our Hamiltonian preserves the difference between the photons of the two sites. Moreover, by
\begin{align}
\begin{split}
\hat b^\dag \Pi_{k}^{\mathrm{diff}} \hat b&= \sum_{m} (m+1)\ket{m+k}\bra{m+k}\otimes \ket{m+1}\bra{m+1} = \hat n_b \Pi_{k-1}^{\mathrm{diff}},\\
\{\hat n_b, \Pi_k^{\mathrm{diff}}\} &= 2\hat n_b \Pi_k^{\mathrm{diff}},
\end{split}
\end{align}
we get
\begin{align}\label{eq:pi_k_diff}
\mathcal{D}(\hat b)^\dag[\Pi_k^{\mathrm{diff}}]=\hat n_b \left( \Pi_{k-1}^{\mathrm{diff}} - \Pi_k^{\mathrm{diff}}\right).
\end{align}
Now, let us define the comulative sum of $\Pi_k^{\mathrm{diff}}$ as another operator
\begin{align}
\Pi_{\ge k}^{\mathrm{diff}} := \sum_{\ell\ge k} \Pi_{\ell}^{\mathrm{diff}}.
\end{align}
Plugging into \eqref{eq:pi_k_diff} gives
\begin{align}
\mathcal{D}(b)[\Pi_{\ge k}^{\mathrm{diff}}] = n_b \left(\sum_{\ell\ge k}\Pi_{\ell-1}^{\mathrm{diff}} - \sum_{\ell\ge k}\Pi_{\ell}^{\mathrm{diff}}\right) = \hat n_b \Pi_{k-1}^{\mathrm{diff}}.
\end{align}
Let us now focus on the states supported on $\ket{00},\ket{10},\ket{11}$. The states supported on this space, remain supported on this space. Now, we try to find a fixed point here: for any such fixed point $\rho^\ast$ of $\mathcal{L}$, we can write
\begin{align}\label{eq:rho-ast-pi-k}
\Tr (\mathcal{L}[\rho^*] \Pi_{\ge k}^{\mathrm{diff}}) = 0=\Tr(\rho^\ast\mathcal{L}^\dag[\Pi_{\ge k}^{\mathrm{diff}}])=\Tr(\rho^\ast \hat n_b \Pi_{k-1}^{\mathrm{diff}}).
\end{align}
Combining \eqref{eq:pi-k-diff-id} and \eqref{eq:rho-ast-pi-k}, we get that
\begin{align}
0=\Tr(\rho^\ast \hat n_b \sum_k\Pi_k^{\mathrm{diff}}) = \Tr(\rho^* \hat n_b).
\end{align}
Hence, we have $\rho^\ast = \rho_a\otimes\ket0\bra0$. It is clear that $\mathcal{D}(\hat b)[\rho_a\otimes \ket0\bra0]=0$, hence, we just need to impose $[H, \rho_a\otimes \ket0\bra0]=0$:
\begin{align}
-i[H, \rho_a\otimes\ket0\bra0]=-i (\bar{\sigma}_-^\dag \rho_a) \otimes \ket1\bra0+i (\rho_a \bar{\sigma}_-)\otimes \ket0\bra1,
\end{align}
which implies $\bar\sigma_-^\dag \rho_a= \rho_a \bar\sigma_-=0$. Letting $\rho_a=\sum_{m,n} c_{mn} \ket{m}\bra{n}$, these conditions read as $\sum_{m,n} \sqrt{m+1} (1-m) c_{mn} \ket{m+1}\bra{n} = \sum_{m,n} \sqrt{n+1}(1-n) c_{m,n} \ket{m}\bra{n+1}=0$ which yield $c_{m,n}=0$ whenever $m\ne 1$ and $c_{m,n}=0$ whenever $n\ne1$. In other words, $\rho_0 = \ket{1}\bra{1}$. Hence, the only fixed point of our system is $\ket{1}\bra{1}\otimes \ket0\bra0$.

Lastly, let us bound how fast the state converges to $\ket{10}$, starting from vacuum. In this case, the state remains supported on the span of $\ket{00},\ket{10}$ and $\ket{11}$. So, for simplicity, let us project everything onto this subspace and get the following substitutions in this invariant subspace
\begin{align}
H \mapsto \ket{00}\bra{11} + \ket{11}\bra{00},\quad b\mapsto \ket{10}\bra{11},
\end{align}
and so, writing the Lindbladian equation on this space, starting from $\rho(0)=\ket{00}\bra{00}$, we get the simplified form of evolution as
\begin{align}
\rho(t)=\ket{\psi(t)}\bra{\psi(t)}
       +\bigl(1-\|\psi(t)\|^2\bigr)\ket{10}\bra{10},
\qquad
\ket{\psi(t)}=x(t)\ket{00}-i y(t)\ket{11},
\end{align}
where $(x,y)$ satisfy $\dot x=-y$ and
$\dot y=x-y/2$, with $(x(0),y(0))=(1,0)$. To bound the population
remaining in $\operatorname{span}\{\ket{00},\ket{11}\}$, observe that
$V=x^2+y^2-xy/2$ satisfies $\dot V=-V/2$ and $V(0)=1$. This gives $V(t)=e^{-t/2}$.
Since $V\geq \tfrac34(x^2+y^2)$, we get $x(t)^2 + y(t)^2\le \frac42 e^{-t/2}$. Next, using the fact that $\ket{\psi(t)}$ is orthogonal
to $\ket{10}$, we get
\begin{align}
\bigl\|\rho(t)-\ket{10}\bra{10}\bigr\|_1
=2\|\psi(t)\|^2
\leq \frac{8}{3}e^{-t/2}.
\end{align}
In particular, time $t\geq 2\log\!\bigl(8/(3\varepsilon)\bigr)$
suffices to reach trace-norm error at most $\varepsilon$.
\end{example}

Now let $\mathcal{L} = \sum_{j=1}^m \mathcal{L}_j$ be the Lindblaidan acting on $2m$ sites where each $\mathcal{L}_j$ has the form in the example above. Now let $C$ be a linear optical network acting only on $a$ sites. Then the fixed point of the rotated Linbladian
$\mathcal{L}_C$ is $(C \ket{1^n}) \otimes \ket{0^n}$ which we know is hard to approximately sample assuming the conjectures in \cite{aaronson2011computational}.

We note that the analysis above does not establish reachability of arbitrary initial states to the described fixed point. We however observe that $H$ preserve the subspace spanned by $\{\ket{00},\ket{11},\ket{10}\}$. Therefore the fixed point $\ket{10}$ is accessible starting from $\ket{00}$.

We remark that inspired by adiabatic elimination (see \Cref{thm:loss-only-stationary-elimination}) we expect that at large loss rates the fixed point ($\ket{1}$) of the Lindbladian studied in this section to be approximately close to the fixed point of a structured cubic-loss terms of the form $\hat a_j^\dagger (\hat n_j - \mathbb{I})$. As a matter of fact as we will study in \Cref{sec:cubic-loss} the fixed point for this latter Lindbladian is exactly $\ket{1}$.

\subsection{Exact sampling hardness from vacuum using quadratic dissipation}
\label{sec:cat-sampling}

The linear-loss construction in \Cref{sec:Kerr-gadget} relates
fixed-point sampling to boson sampling with binomial inputs corresponding to noisy boson sampling with constant (around half) fraction of photons lost. A natural direction is to understand the computational complexity of sampling from fixed points of a conjugated family based on single mode state preparation gadgets with a quadratic loss mechanism.

The single-mode gadget starts with a symmetric coherent state mixture
$1/2 (\ket\beta\bra\beta + \ketbra{-\beta}{-\beta})$ and use existing results (e.g., \cite{albert2014symmetries}) to show that the exact fixed point of the Lindbladian family we consider is a mixture of even and odd cat states. We will then conjugate the Linbdladian with a passive operation similar to \Cref{sec:Kerr-gadget}. We first show that at specific values of $\beta$ of the resulting circuit is close to that of noisy boson sampling with around half of the photons lost (very similar to \Cref{sec:Kerr-gadget}). We will then show that allowing postselection we can sample from the output of boson sampling starting from odd cat states (which is known to be hard to sample approximately assuming standard conjectures \cite{chabaud2022quantum,rohde2015evidence}). This implies hardness of $\textsc{SampFP}^{2,2}_0$ assuming that the polynomial hierarchy does not collapse. 

\paragraph{Coherent inputs and the exact stationary state.}
We use the two-photon driven-dissipative model
\cite{mirrahimi2014dynamically,leghtas2015confining}:
\begin{equation}
\frac{d\rho}{dt}
= \mathcal{G}_\alpha [\rho] :=-i[H_\alpha,\rho]
+\kappa\mathcal D(\hat a^2)[\rho], \qquad H_\alpha=\frac{i\kappa}{2}
\left(\alpha^2(\hat a^\dagger)^2-\overline\alpha^{\,2} \hat a^2\right), 
\qquad
\alpha^2=\frac{2\epsilon}{\kappa},\quad \kappa>0.
\label{eq:cat-vacuum-generator}
\end{equation}
The initial amplitude is $\beta$, whereas $\alpha\ne0$
is set by the two-photon drive and determines the cat
amplitude. This Lindbladian yields a well-defined dynamics (see for example
\cite[Theorem~3.3]{azouit2016well}). Define the normalized
even and odd cats by
\[
\ket{C_\alpha^\pm}
=\frac{\ket\alpha\pm\ket{-\alpha}}
       {\sqrt{2(1\pm e^{-2|\alpha|^2})}}.
\]
We can show that starting from the coherent input state $\ket\beta\bra\beta$,
the exact stationary state is
\cite[Eq.~(2) and Appendix~A]{mirrahimi2014dynamically}
\begin{equation}
\begin{split}
\rho_\infty(\beta)
={}&c_{++}\ket{C_\alpha^+}\bra{C_\alpha^+}
+c_{--}\ket{C_\alpha^-}\bra{C_\alpha^-}\\
&+c_{+-}\ket{C_\alpha^+}\bra{C_\alpha^-}
+\overline{c_{+-}}\ket{C_\alpha^-}\bra{C_\alpha^+},
\end{split}
\label{eq:cat-coherent-limit}
\end{equation}
where
\begin{align}
c_{++}&=\frac{1+e^{-2|\beta|^2}}2,
\qquad
c_{--}=\frac{1-e^{-2|\beta|^2}}2,
\nonumber\\
c_{+-}&=\frac{i\alpha\overline\beta\,e^{-|\beta|^2}}
             {\sqrt{2\sinh(2|\alpha|^2)}}
\int_0^\pi e^{-i\varphi}
 I_0\!\left(\left|\alpha^2-\beta^2e^{2i\varphi}\right|\right)
\,d\varphi.
\label{eq:cat-coherent-coefficients}
\end{align}
Here $I_0$ is the modified Bessel function of the first kind.
In particular, the stationary state resulting from a coherent input   
is generally not a classical mixture of even and odd cat states but retains quantum coherence between them.

By choosing instead the symmetric coherent-mixture input
\begin{equation}
\rho_{\mathrm{in}}(\beta)
=\frac12\left(\ket\beta\bra\beta+\ket{-\beta}\bra{-\beta}\right),
\label{eq:cat-symmetric-input}
\end{equation}
we remove this coherence. This initial state is a classical
mixture of coherent states, whose photon-counting distribution
remains efficiently samplable even after passive linear optics.
It is invariant under parity $\Pi=(-1)^{\hat{a}^\dagger  \hat a}$, and
$\mathcal G_\alpha$ preserves the parity sectors. Equivalently,
the formula above gives $c_{+-}(-\beta)=-c_{+-}(\beta)$.
The limiting state is therefore
\begin{equation}
\rho_{\alpha,p}
=(1-p)\ket{C_\alpha^+}\bra{C_\alpha^+}
+p\ket{C_\alpha^-}\bra{C_\alpha^-},
\qquad
p=\frac{1-e^{-2|\beta|^2}}2.
\label{eq:cat-mixture-limit}
\end{equation}

\subsubsection{Relation to noisy boson sampling.}
We now quantify the small-amplitude relation between this
cat mixture and lossy single-photon inputs. For this comparison,
allow any $p\in[0,1]$ in the definition of $\rho_{\alpha,p}$
in \eqref{eq:cat-mixture-limit}, and set
\[
\tau_p=(1-p)\ket0\bra0+p\ket1\bra1.
\]
Here $\tau_p$ is a single-photon input subjected to loss with
transmission probability $p$. Let $P_{\mathrm{cat}}$ and
$P_{\mathrm{loss}}$ be the photon-counting distributions of
$\rho_{\alpha,p}^{\otimes n}$ and $\tau_p^{\otimes n}$,
respectively, with identical vacuum padding and the same
passive linear optical operations applied to them. Write $x=|\alpha|^2$.
The identities
\[
|\langle0|C_\alpha^+\rangle|^2=\frac1{\cosh x},
\qquad
|\langle1|C_\alpha^-\rangle|^2=\frac{x}{\sinh x}
\]
and convexity of $D(\rho,\sigma)=\frac12\|\rho-\sigma\|_1$
give, for sufficiently small $|\alpha|$,
\[
D(\rho_{\alpha,p},\tau_p)
\le(1-p)\sqrt{1-\frac1{\cosh x}}
  +p\sqrt{1-\frac{x}{\sinh x}}
\le K|\alpha|^2,
\]
where $K$ is independent of $p\in[0,1]$.
For $n$ copies, identical vacuum padding and the same passive
linear optical operation, tensor-product telescoping and contractivity
under photon counting imply
\[
d_{\mathrm{TV}}(P_{\mathrm{cat}},P_{\mathrm{loss}})
\le Kn|\alpha|^2.
\]
where $d_{\rm TV}$ is the total variation distance.
Thus a sampler with error $\delta$ for either distribution
has error at most $\delta+Kn|\alpha|^2$ for the other.
Choosing $Kn|\alpha|^2\le\varepsilon/2$ transfers a classical
sampler at error $\varepsilon/2$ in either direction to one
at error $\varepsilon$. Likewise, noisy-boson-sampling
hardness at error $\varepsilon$ would imply cat-mixture
hardness at error $\varepsilon/2$, for the same $p$ and
linear optical family. These algorithmic implications require
efficient parameter representation and polynomial runtime
at the chosen amplitude and accuracy. The coherent-mixture
preparation gives $p<1/2$, so hardness restricted to transmission
approaching one does not directly apply.

\subsubsection{Hardness of exact sampling}

We first note that
for a nontrivial mixture in \eqref{eq:cat-mixture-limit},
conditioning on total photon number $n$ does not select one
odd cat in each of the $n$ input modes. Unlike a single-photon
input subjected to loss, an even cat can contribute two or
more photons. For example, with two modes, total photon
number two receives contributions from $\ket{1,1}$ in the
odd--odd branch and from $\ket{2,0}$ and $\ket{0,2}$ in the
even--even branch. Passive optics preserves this total
number, but generally mixes the individual mode parities.
Thus total-number postselection on the final counts does
not by itself identify the all-odd input branch.
A one-photon herald after a beam splitter also flips both
parities, so applying it to a nontrivial even/odd mixture
would still leave a mixture.

We therefore specialize the initial amplitude to $\beta=0$,
while keeping the drive amplitude $\alpha\ne0$.
Then both the coherent input and the symmetric mixture
reduce to $\ket0\bra0$, and the exact solution gives
$c_{++}=1$ and $c_{--}=c_{+-}=0$: the limit is the pure
even cat $\ket{C_\alpha^+}$.
We will mix this state with a vacuum auxiliary mode and
condition on one auxiliary photon to obtain a pure odd cat.
The beam splitters are absorbed
into the passive unitary instance using the reduction. They leave the global vacuum
unchanged, and all measurements are photon counting at
the end. This avoids the mixed-parity ambiguity above;
the exact-hardness proof does not assume hardness of noisy
boson sampling.
By \cite[Theorem~4.2]{azouit2016well}, the vacuum trajectory
converges in trace norm to a state in the cat dark space.
Since parity is preserved and the vacuum is even, this limit
is $\ket{C_\alpha^+}\bra{C_\alpha^+}$.
Other initial states may select different fixed points.

\paragraph{Bosonic heralding of an odd cat.}
To convert the even-cat fixed point into an odd cat, we use
photon subtraction implemented through a beam splitter and
conditional photon counting \cite{dakna1998quantum}.
Mixing an even cat with vacuum and detecting exactly one
photon in the auxiliary output heralds an odd cat with
reduced amplitude, as shown below.

For a data mode $a$ and an auxiliary mode $b$, define
\begin{equation}
B_{ab}(\theta)
=\exp\!\left[\theta( \hat b^\dagger  \hat a-\hat a^\dagger \hat b)\right],
\qquad
u=\cos\theta,\quad v=\sin\theta,
\quad 0<\theta<\pi/2.
\label{eq:cat-beam-splitter}
\end{equation}
It preserves $\hat a^\dagger \hat  a+\hat  b^\dagger \hat b$ and satisfies this key formula:
\begin{align}
B_{ab}\ket\alpha_a\ket0_b
=\ket{u\alpha}_a\ket{v\alpha}_b
\label{eq:cat-in}
\end{align}\footnote{Using $B_{ab}\ket{0,0}=\ket{0,0}$ and
$B_{ab}\hat a^\dagger B_{ab}^\dagger=u \hat a^\dagger+v \hat b^\dagger$,
the coherent-state expansion gives
$B_{ab}\ket{\alpha,0}
=e^{-|\alpha|^2/2}e^{u\alpha \hat a^\dagger}
e^{v\alpha \hat b^\dagger}\ket{0,0}
=\ket{u\alpha,v\alpha}$,
where we used $[\hat a^\dagger,\hat b^\dagger]=0$ and $u^2+v^2=1$.}
which one can view as splitting the coherent light $\ket{\alpha}$ into two coherent but with less strength.

\begin{lemma}[Single-photon heralding]
\label{lem:odd-postselection}
Let $x=|\alpha|^2>0$. Measuring one photon in mode $b$
after applying $B_{ab}$ to
$\ket{C_\alpha^+}_a\ket0_b$ prepares
$\ket{C_{u\alpha}^-}_a$ exactly, with probability
\begin{equation}
h_\alpha
=\frac{v^2x\sinh(u^2x)}{\cosh x}>0.
\label{eq:cat-herald-probability}
\end{equation}
\end{lemma}
\begin{proof}
Before measurement, the joint state is
\begin{equation}
\ket{\Psi_\alpha}_{ab}
:=B_{ab}\ket{C_\alpha^+}_a\ket0_b
=\frac{\ket{u\alpha}_a\ket{v\alpha}_b
       +\ket{-u\alpha}_a\ket{-v\alpha}_b}
       {\sqrt{2(1+e^{-2x})}}.
\label{eq:cat-pair-fixed-state}
\end{equation}
Using $\langle1|\pm v\alpha\rangle
=\pm v\alpha e^{-v^2x/2}$ gives
\begin{align}
{}_b\langle1|\Psi_\alpha\rangle_{ab}
&=\frac{v\alpha e^{-v^2x/2}}
        {\sqrt{2(1+e^{-2x})}}
  \left(\ket{u\alpha}-\ket{-u\alpha}\right)
 =c_\alpha\ket{C_{u\alpha}^-},
\label{eq:cat-herald-amplitude}\\
c_\alpha
&=v\alpha e^{-v^2x/2}
  \sqrt{\frac{1-e^{-2u^2x}}{1+e^{-2x}}}.
\nonumber
\end{align}
Its squared norm is
$|c_\alpha|^2=v^2x\sinh(u^2x)/\cosh x$.
\end{proof}

We now apply the unitary-conjugation construction used in
\Cref{sec:Kerr-gadget}, including both the heralding beam
splitters and the linear optical operation used in the reduction. To be more precise:
Let $\hat a_1,\ldots,\hat a_m$ be data modes, with $n\le m$ active
modes, and let $\hat b_1,\ldots, \hat b_n$ be auxiliary bosonic modes.
Start every mode in vacuum, and define
\begin{equation}
\mathcal L_0
=\kappa\sum_{j=1}^{n}\mathcal D[\hat a_j^2-\alpha^2]
 +\lambda\sum_{j=n+1}^{m}\mathcal D[\hat a_j^2]
 +\lambda\sum_{j=1}^{n}\mathcal D[\hat b_j^2],
\qquad \lambda>0.
\label{eq:cat-base-generator}
\end{equation}

By eq \ref{eq:cat-coherent-limit}, fixed point of $\mathcal L_0$ is
\begin{equation}
\rho_{*,0}=\ket{\Omega_\alpha}\bra{\Omega_\alpha},
\qquad
\ket{\Omega_\alpha}
=\left(\ket{C_\alpha^+}^{\otimes n}
        \otimes\ket0^{\otimes(m-n)}\right)_a
 \otimes\ket0_b^{\otimes n}.
\label{eq:base-fixed-point}
\end{equation}
For a linear optical operation $U$ acting on all $m$ data modes,
set
\begin{equation}
V=(U\otimes I_b)\prod_{j=1}^{n}B_{a_jb_j}(\theta),
\qquad
\mathcal L_V[\rho]
=V\mathcal L_0[V^\dagger\rho V]V^\dagger.
\label{eq:Ltotprime}
\end{equation}
Both the beam splitters and $U$ preserve the global vacuum,
so $V\ket{\mathrm{vac}}=\ket{\mathrm{vac}}$.
The conservative semigroup for $\mathcal L_V$ is obtained
by unitary conjugation of the product semigroup for
$\mathcal L_0$; in particular,
\begin{equation}
e^{s\mathcal L_V}[\rho_{\mathrm{vac}}]
=V e^{s\mathcal L_0}[\rho_{\mathrm{vac}}]V^\dagger,
\qquad
\rho_{\mathrm{vac}}=\ket{\mathrm{vac}}\bra{\mathrm{vac}}.
\label{eq:cat-conjugated-trajectory}
\end{equation}
The fixed point, before any measurement or
postselection, is the joint pure state
\begin{equation}
\rho_{*,V}=\ket{\Phi_U}\bra{\Phi_U},
\qquad
\ket{\Phi_U}=V\ket{\Omega_\alpha}.
\label{eq:cat-bosonic-fixed-point}
\end{equation}

For clarity, write $\hat d_j=U \hat  a_jU^\dagger$ and
$\widetilde a_j=V \hat a_jV^\dagger=u \hat d_j+v \hat b_j$ for $j\le n$.
Then $V \hat b_jV^\dagger=u \hat b_j-v  \hat d_j$, and
$V\hat a_jV^\dagger= \hat d_j$ for $j>n$. In the Hamiltonian-plus-loss
representation, the generator has
\begin{align}
H_V&=\frac{i\kappa}{2}\sum_{j=1}^{n}
\left(\alpha^2(\widetilde a_j^\dagger)^2
      -\overline\alpha^{\,2}\widetilde a_j^2\right),
\label{eq:cat-rotated-hamiltonian}\\
J_j&=\sqrt\kappa\,\widetilde a_j^2\quad (j\le n),
\qquad
K_j=\sqrt\lambda\,(u \hat b_j-v \hat d_j)^2\quad (j\le n),
\qquad
R_j=\sqrt\lambda\,\hat d_j^2\quad (j>n).
\label{eq:cat-rotated-jumps}
\end{align}
Thus the Hamiltonian has degree two and every loss jump
operator is quadratic; no linear loss is required.
These are collective polynomial jumps, not independent local
$\hat a_j^2$ losses in the physical output modes. We now present our main theorem.

\begin{theorem}[Exact sampling hardness from a vacuum-selected bosonic fixed point]
\label{thm:postselected-hardness}
Fix nonzero $\alpha$, $0<\theta<\pi/2$, and
$\kappa,\lambda>0$ independently of the instance size;
for example, take $\alpha=1$, $\theta=\pi/4$, and
$\kappa=\lambda=1$.
For every passive linear optical $U$ on $m\ge n$ data modes,
the Lindbladian $\mathcal L_V$ above has a Gaussian Hamiltonian
and quadratic loss jump operators, with no linear losses.
Its trajectory from global vacuum converges to
$\rho_{*,V}$ in \eqref{eq:cat-bosonic-fixed-point}.
The initial photon-counting distribution is the deterministic
all-zero distribution.

If a polynomial-time classical algorithm exactly samples
the joint photon-number distribution of $\rho_{*,V}$ for
every linear optical operation in this family, then the polynomial
hierarchy collapses to its third level.

For this exact-sampling result, we use
$\alpha=1$, $\theta=\pi/4$, and $\kappa=\lambda=1$.
The linear optics are specified by polynomial-size
circuits over a fixed finite algebraic optical gate set
supporting the postselected simulation in
\cite{knill2001scheme,aaronson2011linear}.
In this subsection, coefficients are encoded exactly
by rational coordinates in a fixed number field
containing the gate entries and gadget coefficients.
The conjugated Hamiltonian and jump operators therefore
have exact polynomial-length descriptions.
\end{theorem}

\begin{proof}
Stationarity and convergence follow from
\cite{mirrahimi2014dynamically} and
\eqref{eq:cat-conjugated-trajectory}; a mixing time is given below. First, for the initial state, the vacuum is unchanged by $V$, so its photon-counting distribution is trivially samplable.

Now we investigate the sampling complexity of the fixed point. Measure all $m+n$ modes in the photon-number basis and let
$\mathsf H$ be the event that every auxiliary mode $b_j$
contains exactly one photon. Since $U$ acts only on the data
modes $a_j$, \Cref{lem:odd-postselection} implies that
$\Pr(\mathsf H)=h_\alpha^n$ and the conditional state on $a$ mode is
\begin{equation}
U\left(\ket{C_{u\alpha}^-}^{\otimes n}
       \otimes\ket0^{\otimes(m-n)}\right).
\label{eq:cat-heralded-output}
\end{equation}
The auxiliary modes do not interact again after the beam
splitters. Thus their measurements may be deferred to the
end, and this conditioning requires no adaptive optical gates.

Set $y=u^2|\alpha|^2$. Each odd cat has at least one photon.
Consequently, conditioning further on total data photon number
$n$ selects exactly the input component
$\ket1^{\otimes n}\otimes\ket0^{\otimes(m-n)}$.
Passive optics preserves total photon number, so this
condition can also be imposed on the final counts.
Its conditional probability is
\[
q_n=\left(\frac{y}{\sinh y}\right)^n,
\]
and the two conditions together have probability (Let $N_{\mathrm{data}}=\sum_{j=1}^{m}\hat a_j^\dagger \hat a_j$.)
\begin{equation}
\Pr(\mathsf H,\,N_{\mathrm{data}}=n)
=h_\alpha^n q_n
=\left(\frac{u^2v^2|\alpha|^4}
              {\cosh(|\alpha|^2)}\right)^n.
\label{eq:cat-combined-postselection}
\end{equation}
For the fixed parameters in the theorem this is positive
and inverse-exponential in $n$. Under these conditions the
data distribution is exactly ordinary single-photon boson
sampling for $U$.

Now choose $U$ from the postselected linear-optical circuits
used to establish $\PostLO_1=\PostBQP$ in
\Cref{thm:PostLOk}. An efficient exact classical sampler for
the full joint distribution could be postselected on
$\mathsf H$, on $N_{\mathrm{data}}=n$, and on the designated output events
of that computation. The combined event retains probability
at least $2^{-\operatorname{poly}(n,m)}$. This would place
$\PostBQP$ inside $\PostBPP$ and collapse the polynomial
hierarchy to its third level, as in
\cite{aaronson2011computational}.
\end{proof}

\begin{corollary}[Exact-sampling hardness with even-cat inputs]
For any fixed nonzero amplitude $\alpha$, exact photon-number
sampling from passive linear-optical networks supplied with
$n$ copies of $\ket{C_\alpha^+}$ and polynomially many vacuum
modes is classically hard unless the polynomial hierarchy
collapses to its third level.
\end{corollary}

We note that this corollary extends the exact-sampling hardness result for odd-cat
inputs in \cite[Theorem~5]{chabaud2022quantum} to even-cat
inputs with additional vacuum modes. This however involves a nontrivial use of postselection which prepares the odd cat state conditioned on an exponentially rare postselection event
and does not imply hardness of approximate sampling for this instance. In particular, this is a
bosonic instance of $\textsc{SampFP}^{2,2}_0$ with an
vaccum initial state.

We end this section by showing the fast mixing of this model.
\paragraph{Convergence from vacuum.}
We bound the preparation time for the specified vacuum input.
In contrast to the energy-constrained supremum introduced
earlier, no supremum over arbitrary inputs is needed here.
Define
\begin{equation}
\tau_{\mathrm{vac}}^{\mathcal L_V}(\varepsilon)
:=\inf\left\{s\ge0:
\left\|e^{s\mathcal L_V}[\rho_{\mathrm{vac}}]
             -\rho_{*,V}\right\|_1\le\varepsilon\right\}.
\label{eq:cat-vacuum-mixing-definition}
\end{equation}

\begin{lemma}[Unitary invariance of convergence]
Let $\mathcal L'=\operatorname{Ad}_W\circ\mathcal L
\circ\operatorname{Ad}_{W^\dagger}$ for a unitary $W$.
For an initial state $\rho_0$ and a stationary target $\rho_*$,
set $\rho'_0=W\rho_0W^\dagger$ and
$\rho'_*=W\rho_*W^\dagger$. Then, for every $s\ge0$,
\[
\left\|e^{s\mathcal L'}[\rho'_0]-\rho'_*\right\|_1
=\left\|e^{s\mathcal L}[\rho_0]-\rho_*\right\|_1.
\]
Thus the convergence times agree for these corresponding
inputs and targets, and also for corresponding input
families when both families and targets are conjugated.
\end{lemma}
\begin{proof}
Unitary conjugation gives
\[
e^{s\mathcal L'}[W\rho_0W^\dagger]
=W e^{s\mathcal L}[\rho_0]W^\dagger.
\]
Subtracting $W\rho_*W^\dagger$ and using unitary invariance
of the trace norm proves the identity. Taking a supremum
over corresponding input families preserves the equality.
In our construction $W=V$ fixes the global vacuum, so the
initial state itself is unchanged.
\end{proof}
Now we show the key Lemma.

\begin{lemma}[Single-mode Lyapunov estimate]
\label{lem:well}
Let $L_2=a^2-\alpha^2$ and let $P$ project onto $\ker L_2$.
For an initial density operator in the generator domain
of the weighted trace-class space with weight
$S=(\mathbb{I}+L_2^\dagger L_2)^{1/2}$, the conservative evolution
$\rho(s)=e^{s\mathcal G_\alpha}[\rho(0)]$ satisfies
\[
\operatorname{Tr}(L_2\rho(s)L_2^\dagger)
\le V_0 e^{-2\kappa s},
\qquad
V_0=\operatorname{Tr}(L_2\rho(0)L_2^\dagger).
\]
Moreover, for a constant $c_{\mathrm{gap}}>0$ depending only on
$\alpha$,
\[
\operatorname{Tr}((\mathbb{I}-P)\rho(s))
\le c_{\mathrm{gap}} V_0 e^{-2\kappa s}.
\]
\end{lemma}
\begin{proof}
The first estimate is
\cite[Lemma~4.1]{azouit2016well}, with time rescaled by
$\kappa$. The spectrum of $L_2^\dagger L_2$ has a positive
gap above its kernel, as established in the same reference.
Taking $c_{\mathrm{gap}}$ to be the reciprocal of this gap gives
$\mathbb{I}-P\le c_{\mathrm{gap}} L_2^\dagger L_2$ and proves the second
estimate.
\end{proof}

\begin{theorem}[Trace-norm convergence for admissible inputs]
Under the domain assumption of \Cref{lem:well}, let
$\rho_\infty$ be the fixed point given by
\cite[Theorem~4.2]{azouit2016well}. Then
\[
\|\rho(s)-\rho_\infty\|_1
\le4\sqrt{c_{\mathrm{gap}}V_0}\,e^{-\kappa s}.
\]
On any such family with a common bound $V_0\le V_*$,
this estimate has a uniform prefactor. The limiting state
may depend on the input; this is not a bound to a globally
unique fixed point.
\end{theorem}
\begin{proof}
Set $B_s=P\rho(s)P$. Since $L_2P=0$, $B_s$ is stationary,
although it need not have trace one. Trace-norm contractivity
therefore gives, for every $h\ge0$,
\[
\|\rho(s+h)-B_s\|_1\le\|\rho(s)-B_s\|_1.
\]
Taking $h\to\infty$, applying the triangle inequality and
then the gentle-projection bound yields
\begin{align*}
\|\rho(s)-\rho_\infty\|_1
&\le2\|\rho(s)-P\rho(s)P\|_1\\
&\le4\sqrt{\operatorname{Tr}((I-P)\rho(s))}\\
&\le4\sqrt{c_{\mathrm{gap}}V_0}\,e^{-\kappa s}.
\end{align*}
\end{proof}

For the vacuum input, the trajectory remains in the even
sector and converges to the pure state
$\ket{C_\alpha^+}\bra{C_\alpha^+}$, giving the sharper
estimate below.

\begin{theorem}[Convergence of the bosonic construction from vacuum]
\label{thm:fast-mixing-nmodes}
For the construction above, set
$A_\alpha=\max\{1,2\sqrt{c_{\mathrm{gap}}}|\alpha|^2\}$.
Then
\begin{equation}
\left\|e^{s\mathcal L_V}[\rho_{\mathrm{vac}}]
             -\rho_{*,V}\right\|_1
\le n A_\alpha e^{-\kappa s},
\qquad s\ge0,
\label{eq:n-mode-exp}
\end{equation}
and, for $0<\varepsilon<1$,
\begin{equation}
\tau_{\mathrm{vac}}^{\mathcal L_V}(\varepsilon)
\le\frac1\kappa\log\frac{nA_\alpha}{\varepsilon}.
\label{eq:mixing-time-n}
\end{equation}
For fixed $\alpha$ and $\kappa$, this is
$O(\log n+\log(1/\varepsilon))$.
\end{theorem}
\begin{proof}
For a single active mode initialized in vacuum,
$V_0=\|L_2\ket0\|^2=|\alpha|^4$.
The trajectory remains in the even sector, so
\[
\operatorname{Tr}(P\rho(s))
=\langle C_\alpha^+|\rho(s)|C_\alpha^+\rangle.
\]
The trace-distance bound for a pure target and
\Cref{lem:well} therefore give
\begin{align}
\left\|\rho(s)-\ket{C_\alpha^+}\bra{C_\alpha^+}\right\|_1
&\le2\sqrt{1-\langle C_\alpha^+|\rho(s)|C_\alpha^+\rangle}
\nonumber\\
&\le2\sqrt{c_{\mathrm{gap}}}|\alpha|^2 e^{-\kappa s}.
\label{eq:single-mode-exp}
\end{align}
Under $\mathcal L_0$, the $n$ active-mode trajectories
factorize, while all undriven modes remain exactly
in vacuum. Write $\rho_j(s)$ for the active-mode trajectories
and $\sigma_j=\ket{C_\alpha^+}\bra{C_\alpha^+}$ for their
limits. The tensor-product identity
\[
\bigotimes_{j=1}^n\rho_j(s)-\bigotimes_{j=1}^n\sigma_j
=\sum_{r=1}^n
 \left(\bigotimes_{j<r}\sigma_j\right)
 \otimes\bigl(\rho_r(s)-\sigma_r\bigr)
 \otimes\left(\bigotimes_{j>r}\rho_j(s)\right)
\]
and $\|A\otimes B\|_1=\|A\|_1\|B\|_1$ imply
\[
\left\|e^{s\mathcal L_0}[\rho_{\mathrm{vac}}]-\rho_{*,0}\right\|_1
\le\sum_{r=1}^n\|\rho_r(s)-\sigma_r\|_1
\le n A_\alpha e^{-\kappa s}.
\]
Tensoring with the unchanged auxiliary and padding vacua
does not change the trace norm. Finally,
\eqref{eq:cat-conjugated-trajectory} gives
\[
\left\|e^{s\mathcal L_V}[\rho_{\mathrm{vac}}]-\rho_{*,V}\right\|_1
=\left\|e^{s\mathcal L_0}[\rho_{\mathrm{vac}}]-\rho_{*,0}\right\|_1.
\]
This proves \eqref{eq:n-mode-exp} and
\eqref{eq:mixing-time-n}.
\end{proof}

This estimate controls the full state before conditioning.
It does not provide an unrestricted worst-case mixing bound.
\par

\subsection{Quadratic loss: Finite-cutoff construction of a unique \(\ket{1}\) fixed point}
\label{sec:1-from-qudartic-loss}

Quadratic loss alone cannot select $\ket1$ as the unique
fixed point: the dark space of $\hat a^2$, consisting of
states annihilated by this operator, is
\[
\operatorname{span}\{\ket0,\ket1\}.
\]
Can we add additional Hamiltonian terms to specify $\ket{1}$ the unique invariant subspace? We can combine the quadratic loss with the degree-five
Hamiltonian
\[
H=g\left((\hat n-\mathbb{I})\hat a^3
+(\hat a^\dagger)^3(\hat n-\mathbb{I})\right),
\qquad g\in\mathbb R\setminus\{0\}.
\]
In the resulting construction, both $H$ and $\hat a^2$ annihilate $\ket1$, so
$\ket1\bra1$ is stationary. Meanwhile, $H$ couples
$\ket0$ to $\ket3$, from which quadratic loss transfers
population to $\ket1$.
Below, we prove that this mechanism yields a unique
attracting fixed point when we consider finite photon cutoff
$R\ge3$, including for initial states with higher
photon occupations.

While this fixed point is accessible from every initial state assuming any finite cut-off without assuming a cut-off it is not even clear if the fixed point $\ket{1}$ is accessible from a basic state like vacuum.

Let
\begin{align}
    \mathcal H_R
    :=
    \operatorname{span}\{\ket{0},\ket{1},\ldots,\ket{R}\},
    \qquad
    R\ge3,
\end{align}
and let
\begin{align}
    P_R
    :=
    \sum_{n=0}^{R}\ket{n}\bra{n}
\end{align}
be the projection onto the cutoff space.  On \(\mathcal H_R\), consider the
finite-dimensional Lindbladian
\begin{align}
    \mathcal L_R(\rho)
    =
    -i[H_R,\rho]
    +
    \kappa\mathcal D[L_R](\rho),
    \qquad
    \kappa>0,
\end{align}
where
\begin{align}
    \mathcal D[L_R](\rho)
    =
    L_R\rho L_R^\dagger
    -
    \frac12\{L_R^\dagger L_R,\rho\}.
\end{align}
We choose the jump operator
\begin{align}
    L_R
    :=
    P_R\hat a^2P_R
    =
    \sum_{n=2}^{R}
    \sqrt{n(n-1)}\,\ket{n-2}\bra{n},
\end{align}
and the Hamiltonian
\begin{align}
    H_R
    :=
    P_R g\left((\hat n-\mathbb{I})\hat a^3+(\hat a^\dagger)^3(\hat n-\mathbb{I})\right)
    P_R,
    \qquad
    g\neq0.
\end{align}Equivalently,
\begin{align}
    H_R
    =
    \sum_{n=0}^{R-3}
    h_n
    \left(
        \ket{n+3}\bra{n}
        +
        \ket{n}\bra{n+3}
    \right),
\end{align}
with
\begin{align}
    h_n
    =
    g(n-1)\sqrt{(n+1)(n+2)(n+3)}.
\end{align}
The two important matrix elements are
\begin{align}
    h_0=-g\sqrt6\neq0,
    \qquad
    h_1=0.
\end{align}
Thus the Hamiltonian couples \(\ket{0}\) to \(\ket{3}\), while it does not
couple \(\ket{1}\) to \(\ket{4}\).  Since \(a^2\) sends \(\ket{3}\) to
\(\ket{1}\), this creates the absorbing route
\begin{align}
    \ket{0}
    \longleftrightarrow
    \ket{3}
    \longrightarrow
    \ket{1}.
\end{align}
The Hamiltonian changes photon number by \(3\), while the loss changes photon
number by \(2\).  Since \(\gcd(2,3)=1\), the parity obstruction present for
\(\hat a^2\)-loss alone is removed.
We now make this result precise.
\begin{proposition}[Finite-cutoff uniqueness of the single-photon fixed point]
\label{prop:finite-cutoff-one-photon}
For every \(R\ge3\), \(g\neq0\), and \(\kappa>0\), the Lindbladian
\(\mathcal L_R\) has the unique fixed point
\begin{align}
    \rho_\ast=\ket{1}\bra{1}.
\end{align}
Moreover, every initial state supported on \(\mathcal H_R\) converges to
\(\ket{1}\bra{1}\).
\end{proposition}

\begin{proof}
We first show that $\ket{1}\bra{1}$ is a fixed point.  Since
$L_R\ket{1}=0$,
we have
$\mathcal D[L_R](\ket{1}\bra{1})=0$.
Moreover,
\begin{align}
    P_R,
    (\hat n-\mathbb{I})\hat a^3\ket{1}=0,
\quad(\hat a^\dagger)^3(\hat n-I)\ket{1}=0.
\end{align}
Therefore
$H_R\ket{1}=0$,
and hence
$-i[H_R,\ket{1}\bra{1}]=0$.
Thus $\mathcal L_R(\ket{1}\bra{1})=0$.

We now prove uniqueness.  Let \(\rho\) be any fixed point of
\(\mathcal L_R\), and write
\begin{align}
    p_n:=\bra{n}\rho\ket{n}.
\end{align}
Since \(H_R\ket{1}=0\), the Hamiltonian part does not contribute to
\(\dot p_1\).  The dissipative part gives
\begin{align}
    0
    =
    \dot p_1
    =
    \kappa
    \bra{1}L_R\rho L_R^\dagger\ket{1}.
\end{align}
Using
\begin{align}
    L_R^\dagger\ket{1}
    =
    \hat a^{\dagger 2}\ket{1}
    =
    \sqrt6\,\ket{3},
\end{align}
we obtain
$0=6\kappa p_3$.
Hence $ p_3=0$.
Since \(\rho\ge0\), the vanishing of a diagonal entry implies that the
corresponding row and column vanish:
\begin{align}
    \rho_{3m}=\rho_{m3}=0
    \qquad
    \text{for all }m.
\end{align}

Next consider stationarity of \(p_3\).  Because the third row and column of
\(\rho\) vanish, the Hamiltonian contribution to \(\dot p_3\) is zero.  The
only dissipative inflow into \(\ket{3}\) comes from \(\ket{5}\).  Therefore
\begin{align}
    0=\dot p_3=20\kappa p_5,
\end{align}
so $p_5=0.$
Again positivity implies that the fifth row and column vanish.  Repeating the
same argument gives
\begin{align}
    p_5=p_7=p_9=\cdots=0
\end{align}
up to the cutoff \(R\).  Thus all odd populations above \(p_1\) vanish. We now eliminate the even populations.  Since \(H_R\) couples
\(\ket{0}\) only to \(\ket{3}\), we have
\begin{align}
    -i\bra{0}[H_R,\rho]\ket{0}
    =
    -i\left(
        (H_R)_{03}\rho_{30}
        -
        \rho_{03}(H_R)_{30}
    \right).
\end{align}
But the previous step gave \(p_3=0\), and since \(\rho\ge0\), this implies
\begin{align}
    \rho_{3m}=\rho_{m3}=0
    \qquad
    \forall m.
\end{align}
Therefore the Hamiltonian contribution to \(\dot p_0\) vanishes.  The
dissipator \(L_R=a^2\) gives
$   \dot p_0
    =
    2\kappa p_2$.
Since \(\rho\) is stationary,
\begin{align}
    0=\dot p_0=2\kappa p_2,
\end{align}
and hence \(p_2=0\). Similar reason as before shows that all $p_{k}=0$ when $k$ are even except $k=0$.
Hence the only possibly nonzero diagonal entries of
\(\rho\) are \(p_0\) and \(p_1\).

It remains to rule out \(p_0\).  From the previous steps, all diagonal entries
except possibly \(p_0\) and \(p_1\) vanish.  By positivity, this implies that
all rows and columns except possibly those indexed by \(0\) and \(1\) vanish.
In particular,
\begin{align}
    \rho_{30}=0,
    \qquad
    \rho_{3m}=\rho_{m3}=0,
    \qquad
    \rho_{2m}=\rho_{m2}=0,
    \qquad
    \rho_{5m}=\rho_{m5}=0 .
\end{align}

We now impose stationarity of the off-diagonal matrix element
\(\rho_{30}\).  The dissipative contribution vanishes.  Indeed, for
\(L_R=a^2\),
\begin{align}
    \bra{3}L_R\rho L_R^\dagger\ket{0}
\end{align}
can only receive a contribution from \(\rho_{52}\), while the
anticommutator term is proportional to \(\rho_{30}\).  Since the rows and
columns \(2,3,5\) vanish, both contributions are zero.

Therefore
\begin{align}
    0=\dot\rho_{30}
    =
    -i\bra{3}[H_R,\rho]\ket{0}.
\end{align}
Using the matrix elements of \(H_R\), we have
\begin{align}
    H_R\ket{0}=h_0\ket{3},
    \qquad
    h_0=-g\sqrt6\neq0,
\end{align}
and $H_R\ket{1}=0$
because \(h_1=0\).  Hence
\begin{align}
    \bra{3}H_R\rho\ket{0}
    =
    h_0\rho_{00}
    =
    h_0p_0.
\end{align}
On the other hand,
$\bra{3}\rho H_R\ket{0}=0$,
because the third row of \(\rho\) vanishes.  Thus
\begin{align}
    0=\dot\rho_{30}=-ih_0p_0.
\end{align}
Since \(h_0\neq0\), we conclude that $p_0=0.$
Since $\operatorname{Tr}\rho=1$, it follows that $p_1=1$.
Finally, positivity implies that all off-diagonal entries involving indices
different from \(1\) vanish.  Therefore
\begin{align}
    \rho=\ket{1}\bra{1}.
\end{align}
This proves uniqueness of the fixed point.
To prove convergence, let
$\rho(t)=e^{t\mathcal L_R}[\rho(0)]$ and define
\[
\bar\rho_T=\frac1T\int_0^T\rho(t)\,dt.
\]
For every fixed $s\ge0$,
\[
\|e^{s\mathcal L_R}[\bar\rho_T]-\bar\rho_T\|_1
\le\frac{2s}{T}.
\]
Finite-dimensional compactness therefore implies that
every accumulation point of $\bar\rho_T$ is stationary.
By uniqueness, $\bar\rho_T\to\ket1\bra1$.

Moreover, $p_1(t)=\bra1\rho(t)\ket1$ satisfies
$\dot p_1(t)=6\kappa p_3(t)\ge0$.
Since its time average tends to one, monotonicity implies
$p_1(t)\to1$. Consequently,
\[
\|\rho(t)-\ket1\bra1\|_1
\le2\sqrt{1-p_1(t)}\longrightarrow0.
\]
\end{proof}

For each fixed \(R\ge3\), Proposition~\ref{prop:finite-cutoff-one-photon}
proves that \(\mathcal L_R\) has the unique fixed point \(\ket1\bra1\), and
that every initial state in \(\mathcal H_R\) converges to it.  We do not prove
an explicit \(R\)-dependent mixing-time bound. 

For the untruncated Lindbladian, by a fixed point we mean a density operator
\(\rho\) for which \(\mathcal L(\rho)\) is well-defined and
\(\mathcal L(\rho)=0\).  In this sense, the same Fock-basis elimination shows
that \(\ket1\bra1\) is the unique fixed point.  This does not imply, and we do
not claim, convergence or mixing for arbitrary initial states.

\subsubsection{Complexity and mixing time of Lindbladian dynamics with structured cubic jump terms}
\label{sec:cubic-loss}
In this section we consider Lindbladians with structured cubic jump terms. Compared with the quadratic-loss construction, this case is even simpler:
the fixed point can be made exactly equal to the single-photon Fock state, so no parity readout or postselection is required.
As a result, the resulting sampling problem reduces directly to ordinary boson sampling.
Throughout this subsection, the single-mode cubic-jump dynamics is restricted to the invariant subspace $\mathcal H_{01}=\operatorname{span}\{|0\rangle,|1\rangle\}$.
All fixed-point and sampling statements below refer to
the specified invariant subspaces.
\paragraph{Single-mode fixed point.}
Consider the Lindbladian with no Hamiltonian term and jump operator
\begin{equation}
L = \hat a^\dagger (\hat n-\mathbb{I})
\end{equation}

\begin{theorem}
\label{thm:cubic-single-mode-fixed-point}
For the Lindbladian
\begin{equation}
\mathcal{L}[\rho]=\mathcal{D}(L)[\rho],
\qquad
L=\hat a^\dagger(\hat n-\mathbb{I}),
\end{equation}
the unique fixed point on $\mathcal H_{01}$ is
$\rho_\ast=\ketbra{1}{1}.$
\end{theorem}

\begin{proof}
We compute
\begin{equation}
L\ket{m}=(m-1)\sqrt{m+1}\ket{m+1},
\end{equation}
and
\begin{equation}
L^\dagger L=(\hat n-1)^2(\hat n+1).
\end{equation}
Define
\begin{equation}
\lambda_m:=(m-1)^2(m+1).
\end{equation}
Then
\begin{equation}
\mathcal{L}[\ketbra{m}{n}]
=
(m-1)(n-1)\sqrt{m+1}\sqrt{n+1}\,\ketbra{m+1}{n+1}
-\frac{\lambda_m+\lambda_n}{2}\ketbra{m}{n}.
\end{equation}
Since $\lambda_m=0$ if and only if $m=1$, the only basis element with zero diagonal decay rate is $\ketbra{1}{1}$.
A straightforward induction on the matrix elements shows that every other coefficient must vanish at stationarity.
Hence the unique fixed point is $\ketbra{1}{1}$.
\end{proof}

\paragraph{Multimode fixed point and boson sampling.}
We restrict the multimode dynamics to
\[
\mathcal S=
\mathcal H_{01}^{\otimes n}
\otimes\operatorname{span}\{|0\rangle^{\otimes(m-n)}\}.
\]
After conjugation by $U$, the invariant subspace is
$U\mathcal S$.
Now consider $m$ bosonic modes, with the first $n$ modes governed by the above structured cubic-jump Lindbladian and the remaining $m-n$ modes governed by single-photon loss with jumps $\hat a_j$.
For state preparation, all $m$ modes are initialized in the vacuum.
The unique fixed point is
\begin{equation}
\rho_\ast^{(m)}
=
\ketbra{1^{\otimes n},0^{\otimes(m-n)}}{1^{\otimes n},0^{\otimes(m-n)}}.
\end{equation}

Let $U$ be any passive linear-optical unitary on the $m$ modes, and define the conjugated Lindbladian
\begin{equation}
\mathcal{L}_U[\rho]
=
U\,\mathcal{L}[U^\dagger \rho U]\,U^\dagger.
\end{equation}
By unitary covariance of Lindbladians, the unique fixed point of $\mathcal{L}_U$ is
\begin{equation}
U\,\rho_\ast^{(m)}\,U^\dagger
=
U\ketbra{1^{\otimes n},0^{\otimes(m-n)}}{1^{\otimes n},0^{\otimes(m-n)}}U^\dagger.
\end{equation}

This is exactly the ideal boson-sampling output state obtained by feeding $n$ single photons and $m-n$ vacua into the linear-optical network $U$. Notice that similar construction in the experiments can be found in \cite{lingenfelter2021unconditional}.

\begin{theorem}
\label{thm:cubic-boson-hardness}
The exact sampling problem $\textsc{SampFP}^{0,3}_0$, with vacuum input and dynamics restricted to the specified invariant subspace $U\mathcal S$, is at least as hard as
exact boson sampling.
In particular, if there were an efficient classical exact sampler for the output distribution of the above structured cubic-jump Lindbladian, then the polynomial hierarchy would collapse.
\end{theorem}

\begin{proof}
By Theorem~\ref{thm:cubic-single-mode-fixed-point}, the fixed point of the uncoupled multimode dynamics is
\begin{align}
\ketbra{1^{\otimes n},0^{\otimes(m-n)}}{1^{\otimes n},0^{\otimes(m-n)}}.
\end{align}
After conjugation by a passive linear-optical unitary $U$, the fixed point becomes
\begin{align}
U\ketbra{1^{\otimes n},0^{\otimes(m-n)}}{1^{\otimes n},0^{\otimes(m-n)}}U^\dagger.
\end{align}
Since $U$ preserves the vacuum, the conjugated dynamics also starts from the vacuum and converges to this state in time $O(\log n+\log(1/\varepsilon))$ to accuracy $\varepsilon$,
as shown in Corollary~\ref{cor:cubic-multimode-mixing}. Sampling in the photon-number basis from this state is exactly the boson-sampling problem of Aaronson and Arkhipov.
Therefore an efficient classical exact sampler for Lindladians with cubic jump terms would imply an efficient classical exact sampler for boson sampling, which in turn implies collapse of the polynomial hierarchy.
\end{proof}

\cref{prob:FP} (see \ref{def:fixed-conventions}) requires a well-defined, trace-preserving
evolution on the full Fock space. In this example, even though our
initial state starts from vacuum.
The single-jump model above does not satisfy this
requirement. For example, if we starts from $\ket2$, it generates only
upward transitions, with rates
$\lambda_n=(n-1)^2(n+1)$.
The mean waiting time at level $n$ is $1/\lambda_n$,
and
\begin{align}
\sum_{n=2}^{\infty}\frac1{\lambda_n}\leq 3 ( \sum_{n=1}^{\infty}\frac{1}{n^3})<\infty.
\end{align}
Thus infinitely many upward jumps occur in a finite
total time almost surely. To prevent this, we add the cubic jump
\begin{align}
L_-=\sqrt2\,\hat a(\hat n-1),
\end{align}
This introduces downward transitions with rates
$\mu_n=2n(n-1)^2$. The net photon-number drift is
therefore
\begin{align}
\lambda_n-\mu_n=-(n-1)^3\le1.
\end{align}
To show original explosion is excluded, start from $\ket r$
and let $N_t$ denote the photon number at time $t$.
Pick an integer threshold $R>r$, and let $\tau_R$
be the first time the photon number reaches $R$.
Stopping the process when it reaches this threshold,
the drift bound gives
\[
\mathbb E_r[N_{\min(t,\tau_R)}]\le r+t.
\]
If the threshold is reached before time $t$, the stopped
photon number is exactly $R$. Hence
\[
\Pr_r(\tau_R\le t)\le\frac{r+t}{R}.
\]
For each finite $t$, this probability tends to zero
as $R\to\infty$, excluding escape to arbitrarily high
photon numbers before time $t$. Infinitely many jumps
within a bounded range are also impossible, since the
transition rates there are bounded. Thus the process
is nonexplosive. Obviously, $L_-$ annihilates $\{\ket{0}, \ket{1} \}$. Hence, the augmented dynamics also has $\ket1\bra1$ as its
unique stationary state on the full Fock space.

\paragraph{Mixing time.}
We now show rapid convergence on the invariant subspace
$\mathcal H_{01}:=\operatorname{span}\{|0\rangle,|1\rangle\}$,
which contains the vacuum input. All single-mode spectral and mixing statements below refer
to the restriction of $\mathcal L$ to
$\mathcal B(\mathcal H_{01})$.

\begin{corollary}[Spectral gap and single-mode mixing on $\mathcal H_{01}$]
\label{cor:cubic-gap}
For the single-mode Lindbladian generated by
\begin{align}
L=\hat a^\dagger(\hat n-1),\qquad
L_-=\sqrt2\,\hat a(\hat n-1),\qquad H=0.
\end{align}
the unique fixed point is $\rho_\ast=\ketbra{1}{1}$ and the spectral gap of the restricted Lindbladian is
\begin{align}
\gamma=\frac12.
\end{align}
Moreover, on the invariant subspace $\mathcal H_{01}$, the induced semigroup
\begin{align}
\mathcal T_t^{(01)}:=e^{t\mathcal{L}^{(01)}}
\end{align}
mixes exponentially fast:
for every density matrix $\rho$ supported on $\mathcal H_{01}$,
\begin{equation}
\label{eq:cubic-single-mixing}
\bigl\|\mathcal T_t^{(01)}[\rho]-\ketbra{1}{1}\bigr\|_1
\le
2 e^{-t/2}.
\end{equation}
In particular,
\begin{align}
t_{\mathrm{mix}}^{(01)}(\varepsilon)\le 2\log(2/\varepsilon)=O(\log(1/\varepsilon)).
\end{align}
\end{corollary}

\begin{proof}
Since $L_-$ annihilates $\mathcal H_{01}$,
$\mathcal D(L_-)$ vanishes on operators supported
on this subspace, so the restricted generator
is unchanged.
Since $L|0\rangle=-|1\rangle$, $L|1\rangle=0$, and
$L^\dagger L$ preserves $\mathcal H_{01}$, this subspace
is invariant under the dynamics.
In the ordered basis $(|0\rangle,|1\rangle)$, write
\[
\rho=\begin{pmatrix}p&c\\ \bar c&1-p\end{pmatrix}.
\]
The restricted evolution is
\[
\mathcal T_t^{(01)}(\rho)=
\begin{pmatrix}
pe^{-t}&ce^{-t/2}\\
\bar c e^{-t/2}&1-pe^{-t}
\end{pmatrix}.
\]
The restricted generator has eigenvalues
$0,-1,-1/2,-1/2$, hence spectral gap $1/2$.
Using $|c|^2\le p(1-p)$ gives
\[
\|\mathcal T_t^{(01)}(\rho)-|1\rangle\langle1|\|_1
=2\sqrt{p^2e^{-2t}+|c|^2e^{-t}}
\le 2e^{-t/2}.
\]
This proves the stated mixing bound.
\end{proof}

\begin{corollary}[Multimode mixing and preparation of boson sampling states]
\label{cor:cubic-multimode-mixing}
Let
\begin{align}
\mathcal T_t^{(01,n)}
:=\bigl(\mathcal T_t^{(01)}\bigr)^{\otimes n}
\end{align}
denote the product semigroup on $n$ modes, and let
\begin{align}
\rho_{\mathrm{in}}^{(n)}=\rho_1\otimes\cdots\otimes\rho_n
\end{align}
be any product input supported on $\mathcal H_{01}^{\otimes n}$.
Then
\begin{equation}
\label{eq:cubic-multimode-mixing}
\bigl\|
\mathcal T_t^{(01,n)}[\rho_{\mathrm{in}}^{(n)}]
-
\ketbra{1^{\otimes n}}{1^{\otimes n}}
\bigr\|_1
\le
n 2 e^{-t/2}.
\end{equation}
Consequently,
\[
t_{\mathrm{mix}}^{(01,n)}(\varepsilon)
\le
2\log\!\Big(\frac{2n}{\varepsilon}\Big)
=
O(\log n+\log(1/\varepsilon)).
\]
Let $V:=\ketbra{0}{0}^{\otimes(m-n)}$.
Furthermore, for any passive linear-optical unitary $U$,
\begin{equation}
\label{eq:cubic-unitary-mixing}
\bigl\|
U\bigl(\mathcal T_t^{(01,n)}[\rho_{\mathrm{in}}^{(n)}]
\otimes V\bigr)U^\dagger
-
U\bigl(\ketbra{1^{\otimes n}}{1^{\otimes n}}\otimes V\bigr)U^\dagger
\bigr\|_1
\le
n 2 e^{-t/2}.
\end{equation}
Hence the boson-sampling output distribution can be prepared within total variation distance at most $\varepsilon$ in time
\begin{align}
O(\log n+\log(1/\varepsilon)).
\end{align}
\end{corollary}

\begin{proof}
For product inputs, Eq.~\eqref{eq:cubic-single-mixing} together with a telescoping argument gives
\begin{align}
\bigl\|
\mathcal T_t^{(01,n)}[\rho_{\mathrm{in}}^{(n)}]
-
\ketbra{1^{\otimes n}}{1^{\otimes n}}
\bigr\|_1
&\le
\sum_{j=1}^n
\bigl\|
\mathcal T_t^{(01)}[\rho_j]-\ketbra{1}{1}
\bigr\|_1
\le
n 2 e^{-t/2},
\end{align}
which proves \eqref{eq:cubic-multimode-mixing}.

Now let
\begin{align}
\rho_{\mathrm{BS}}
:=
U\ketbra{1^{\otimes n},0^{\otimes(m-n)}}{1^{\otimes n},0^{\otimes(m-n)}}U^\dagger
\end{align}
be the ideal boson-sampling output state.
Since trace norm is invariant under unitary conjugation,
\begin{align}
\|U\rho U^\dagger-U\sigma U^\dagger\|_1=\|\rho-\sigma\|_1,
\end{align}
we immediately obtain \eqref{eq:cubic-unitary-mixing}. Finally, photon-number measurement is a CPTP map, so the induced output distributions satisfy
\begin{align}
\|p_t-p_{\mathrm{BS}}\|_{\mathrm{TV}}
\le
\frac12
\bigl\|
U\bigl(\mathcal T_t^{(01,n)}[\rho_{\mathrm{in}}^{(n)}]
\otimes V\bigr)U^\dagger
-
\rho_{\mathrm{BS}}
\bigr\|_1
\le
n e^{-t/2}.
\end{align}
Choosing
\begin{align}
t\ge 2\log\!\Big(\frac{2n}{2\varepsilon}\Big)
\end{align}
ensures
\begin{align}
\|p_t-p_{\mathrm{BS}}\|_{\mathrm{TV}}\le \varepsilon.
\end{align}
This proves the claim.
\end{proof}

{\color{red}
}

\subsection{Uncomputability for bosonic fixed points}
\label{subsec:uncomputability}

We finally show that deciding existence of fixed point for bosonic Lindbladian systems is in principle undecidable for dissipation degree larger $\geq 5$. A major tool in showing the undecidability of bosonic problems has been the problem of finding a root for a Diophantine equation. Formally, the problem is that we are given a polynomial with integer coefficients, and we should decide if there exists an integer root for the polynomial. This problem is well-studied in the literature, and is known to be undecidable \cite{matijasevivc1971diophantine,davis1973hilbert,davis1976hilbert}, and in fact $\mathsf{RE}$-complete. In the past, \cite{chabaud2026bosonic,chabaud2025energy} have reduced several bosonic problems to the Diophantine equations, showing undecidabilities that naturally arise in infinite-dimensions. Examples of such problems include deciding whether a given CV Hamiltonian of degree-$10$ has a spectrum that is bounded from below, and whether a CV unitary dynamics produces states of infinite energy. Below, we show that the undecidability phenomenon presist even in open-system evolutions.

\begin{theorem}\label{thm:finite-energy-fixed-point-re-hard}
$\exists$\textsc{FP}$^{0,5}$ is $\mathsf{RE}$-hard. $\mathsf{RE}$-hardness holds even if each jump term in the Lindbladian acts on at most $4$ modes.
\end{theorem}

\begin{proof}
We reduce from deciding whether a polynomial with integer
coefficients has a root over $\mathbb N_0$, which is
$\mathsf{RE}$-hard by \cite{matijasevivc1971diophantine,davis1973hilbert,davis1976hilbert}. Using this theorem, \cite[Theorem 2.4]{hillar2013most} establishes undecidability of the following problem: given a set of quadratic polynomials $f_i(\mathbf x)$ over integers, decide whether there is a common solution $\mathbf x^\ast$ such that $f_i(\mathbf x^\ast)=0$ for all $i$. We note that by the addition of auxiliary variables, we can make these polynomials $f_i$ to each act on at most $3$ variables. Moreover, a many-to-one reduction from the Halting problem can be made for the case where the integers $\mathbf x$ are all non-negative \cite{larchey2022hilbert}. Here, for completeness, but also to keep the presentation simple, we present a short Turing reduction only: we can pass $(s_1x_1, \cdots, s_m x_m)$ to all quadratic functions, for some fixed $\mathbf s\in\{\pm 1\}^m$ and then enumerate over all possible $\mathbf s$. Hence, an algorithm for finding common roots over non-negative integers, yields an algorithm for finding a common root over $\mathbb Z$.

Now, for reducing a bosonic Lindbladian problem to this common root finding question, consider the problem of finding a common non-negative root for quadratic functions $f_1,\cdots, f_r$. On $m+1$ modes, with zero Hamiltonian, consider
\begin{align}
\mathcal L
&=
\sum_{j=1}^r
\mathcal D\!\left[
\hat a_{m+1}^\dagger f_j(\hat n_1,\ldots,\hat n_m)
\right]
+\mathcal D[\hat a_{m+1}].
\end{align}
Since each $f_j$ has degree at most two and involves at most
three variables, each gain jump has degree at most five
and acts on at most four modes.

We first check that these jumps define a trace-preserving
evolution. Use their minimal quantum dynamical semigroup.
The first $m$ occupations are preserved. For each fixed
$\mathbf n\in\mathbb N_0^m$, define
\[
R(\mathbf n):=\sum_{j=1}^r f_j(\mathbf n)^2.
\]
The diagonal probabilities in this sector follow a
birth--death process on the last mode, with rates
\[
k\longrightarrow k+1:\quad R(\mathbf n)(k+1),
\qquad
k\longrightarrow k-1:\quad k.
\]
For fixed $\mathbf n$, these rates grow at most linearly
in $k$, so infinitely many transitions cannot occur in
finite time. Thus the minimal evolution preserves trace
on every Fock basis state. Its trace-decreasing property
and positivity then imply trace preservation on all states. Now, we go to analyze the fixed points of $\mathcal L$ for cases where the quadratic system does or does not have a solution:
\begin{enumerate}
\item
If the constraint system has a solution $\mathbf n^\ast$,
every jump annihilates $\ket{\mathbf n^\ast,0}$.
As a result, the state
\[
\rho_\ast
=
\ket{\mathbf n^\ast,0}\bra{\mathbf n^\ast,0}
\]
is a fixed point, with finite energy
$\sum_{i=1}^m n_i^\ast$.

\item
Suppose the constraint system has no solution.
Since $f_j$ have integer coefficients,
\begin{align}
R(\mathbf n)\geq1
\qquad\text{for every }\mathbf n\in\mathbb N_0^m.
\end{align}
Assume that a fixed-point density operator $\rho$ exists,
and write
\[
p_{\mathbf n,k}
:=
\bra{\mathbf n,k}\rho\ket{\mathbf n,k}.
\]
Fix $\mathbf n$ and $K\geq0$. Stationarity of the
probability in the finite set
$\{\ket{\mathbf n,0},\ldots,\ket{\mathbf n,K}\}$
requires the incoming and outgoing boundary flows to agree:
\[
0
=
(K+1)p_{\mathbf n,K+1}
-
R(\mathbf n)(K+1)p_{\mathbf n,K}.
\]
Indeed, the only incoming transition crosses from $K+1$
to $K$, and the only outgoing transition crosses from $K$
to $K+1$. This identity holds even when $\rho$ has
off-diagonal entries: the population equations depend
only on the diagonal entries.

It follows that
\[
p_{\mathbf n,k}
=
R(\mathbf n)^k p_{\mathbf n,0}.
\]
Since $R(\mathbf n)\geq1$ and
$\sum_{k=0}^\infty p_{\mathbf n,k}\leq1$, we must have
$p_{\mathbf n,0}=0$, and hence $p_{\mathbf n,k}=0$
for every $k$. This holds for every $\mathbf n$,
contradicting
\[
1=\operatorname{Tr}(\rho)
=\sum_{\mathbf n,k}p_{\mathbf n,k}.
\]
Thus no fixed-point density operator exists.
\end{enumerate}
\end{proof}

\bibliographystyle{alpha}
\bibliography{Ref}

@article{casteels2017critical,
  title={Critical dynamical properties of a first-order dissipative phase transition},
  author={Casteels, Wim and Fazio, Rosario and Ciuti, Christiano},
  journal={Physical Review A},
  volume={95},
  number={1},
  pages={012128},
  year={2017},
  publisher={APS}
}

@article{becker2026quantum,
  title={Quantum Gibbs Sampling in Infinite Dimensions: Generation, Mixing Times and Circuit Implementation},
  author={Becker, Simon and Rouz{\'e}, Cambyse and Salzmann, Robert},
  journal={arXiv preprint arXiv:2604.01192},
  year={2026}
}

@article{ikeda2013deterministic,
  title={Deterministic generation of Gaussian pure states in a quasilocal dissipative system},
  author={Ikeda, Yusuke and Yamamoto, Naoki},
  journal={Physical Review A—Atomic, Molecular, and Optical Physics},
  volume={87},
  number={3},
  pages={033802},
  year={2013},
  publisher={APS}
}

@article{nicacio2016determining,
  title={Determining stationary-state quantum properties directly from system-environment interactions},
  author={Nicacio, F and Paternostro, M and Ferraro, A},
  journal={Physical Review A},
  volume={94},
  number={5},
  pages={052129},
  year={2016},
  publisher={APS}
}

@article{kessler2012generalized,
  title={Generalized Schrieffer-Wolff formalism for dissipative systems},
  author={Kessler, Eric M},
  journal={Physical Review A—Atomic, Molecular, and Optical Physics},
  volume={86},
  number={1},
  pages={012126},
  year={2012},
  publisher={APS}
}

@article{dakna1998quantum,
  author  = {Dakna, M. and Kn{\"o}ll, L. and Welsch, D.-G.},
  title   = {Quantum state engineering using conditional
             measurement on a beam splitter},
  journal = {The European Physical Journal D},
  volume  = {3},
  pages   = {295--308},
  year    = {1998},
  doi     = {10.1007/s100530050177}
}

@article{reiter2012effective,
  title={Effective operator formalism for open quantum systems},
  author={Reiter, Florentin and S{\o}rensen, Anders S},
  journal={Physical Review A—Atomic, Molecular, and Optical Physics},
  volume={85},
  number={3},
  pages={032111},
  year={2012},
  publisher={APS}
}

@article{oh2024classical,
  title={Classical algorithm for simulating experimental Gaussian boson sampling},
  author={Oh, Changhun and Liu, Minzhao and Alexeev, Yuri and Fefferman, Bill and Jiang, Liang},
  journal={Nature Physics},
  volume={20},
  number={9},
  pages={1461--1468},
  year={2024},
  publisher={Nature Publishing Group UK London}
}

@article{prosen2010quantization,
  title={Quantization over boson operator spaces},
  author={Prosen, Toma{\v{z}} and Seligman, Thomas H},
  journal={Journal of Physics A: Mathematical and Theoretical},
  volume={43},
  number={39},
  pages={392004},
  year={2010}
}

@article{barthel2022solving,
  title={Solving quasi-free and quadratic Lindblad master equations for open fermionic and bosonic systems},
  author={Barthel, Thomas and Zhang, Yikang},
  journal={Journal of Statistical Mechanics: Theory and Experiment},
  volume={2022},
  number={11},
  pages={113101},
  year={2022},
  publisher={IOP Publishing and SISSA}
}

@article{PhysRevX.10.021022,
  title = {Driven-Dissipative Quantum Kerr Resonators: New Exact Solutions, Photon Blockade and Quantum Bistability},
  author = {Roberts, David and Clerk, Aashish A.},
  journal = {Phys. Rev. X},
  volume = {10},
  issue = {2},
  pages = {021022},
  numpages = {27},
  year = {2020},
  month = {Apr},
  publisher = {American Physical Society},
  doi = {10.1103/PhysRevX.10.021022},
  url = {https://link.aps.org/doi/10.1103/PhysRevX.10.021022}
}

@incollection{zeng2013gronwall,
  title={Gronwall inequalities},
  author={Zeng, Fanhai and Cao, Jianxiong and Li, Changpin},
  booktitle={Recent Advances in Applied Nonlinear Dynamics With Numerical Analysis: Fractional Dynamics, Network Dynamics, Classical Dynamics and Fractal Dynamics With Their Numerical Simulations},
  pages={1--22},
  year={2013},
  publisher={World Scientific}
}

@misc{keyl2015schwartz,
  title         = {Schwartz operators},
author = {Keyl, M. and Kiukas, J. and Werner, R. F.},
title = {Schwartz operators},
journal = {Reviews in Mathematical Physics},
volume = {28},
number = {03},
pages = {1630001},
year = {2016},
doi = {10.1142/S0129055X16300016},
URL = { 
        https://doi.org/10.1142/S0129055X16300016
},
eprint = { 
        https://doi.org/10.1142/S0129055X16300016
}
}

@article{casteels2016powerlaws,
  title = {Power laws in the dynamic hysteresis of quantum nonlinear photonic resonators},
  author = {Casteels, W. and Storme, F. and Le Boit{\'e}, A. and Ciuti, C.},
  journal = {Physical Review A},
  volume = {93},
  pages = {033824},
  year = {2016},
  doi = {10.1103/PhysRevA.93.033824}
}

@article{diehl2008quantum,
  title={Quantum states and phases in driven open quantum systems with cold atoms},
  author={Diehl, Sebastian and Micheli, Andrea and Kantian, Adrian and Kraus, B and B{\"u}chler, Hans Peter and Zoller, Peter},
  journal={Nature Physics},
  volume={4},
  number={11},
  pages={878--883},
  year={2008},
  publisher={Nature Publishing Group UK London}
}

@book{serafini2017quantum,
  title={Quantum continuous variables: a primer of theoretical methods},
  author={Serafini, Alessio},
  year={2017},
  publisher={CRC press}
}

@article{weedbrook2012gaussian,
  title={Gaussian quantum information},
  author={Weedbrook, Christian and Pirandola, Stefano and Garc{\'\i}a-Patr{\'o}n, Ra{\'u}l and Cerf, Nicolas J and Ralph, Timothy C and Shapiro, Jeffrey H and Lloyd, Seth},
  journal={Reviews of modern physics},
  volume={84},
  number={2},
  pages={621--669},
  year={2012},
  publisher={APS}
}

@article{bartlett2002efficient,
  title={Efficient classical simulation of continuous variable quantum information processes},
  author={Bartlett, Stephen D and Sanders, Barry C and Braunstein, Samuel L and Nemoto, Kae},
  journal={Physical Review Letters},
  volume={88},
  number={9},
  pages={097904},
  year={2002},
  publisher={APS}
}

@article{chabaud2025energy,
  title={Energy, bosons and computational complexity},
  author={Chabaud, Ulysse and Gharibian, Sevag and Mehraban, Saeed and Motamedi, Arsalan and Naeij, Hamid Reza and Rudolph, Dorian and Sambrani, Dhruva},
  journal={arXiv preprint arXiv:2510.08545},
  year={2025}
}

@article{dahlquist1963special,
  title={A special stability problem for linear multistep methods},
  author={Dahlquist, Germund G},
  journal={BIT Numerical Mathematics},
  volume={3},
  number={1},
  pages={27--43},
  year={1963},
  publisher={Springer}
}

@article{krovi2023improved,
  title={Improved quantum algorithms for linear and nonlinear differential equations},
  author={Krovi, Hari},
  journal={Quantum},
  volume={7},
  pages={913},
  year={2023},
  publisher={Verein zur F{\"o}rderung des Open Access Publizierens in den Quantenwissenschaften}
}

@article{chen2023efficient,
  title={An efficient and exact noncommutative quantum Gibbs sampler},
  author={Chen, Chi-Fang and Kastoryano, Michael J and Gily{\'e}n, Andr{\'a}s},
  journal={arXiv preprint arXiv:2311.09207},
  year={2023}
}

@article{chabaud2026bosonic,
  title={Bosonic quantum computational complexity},
  author={Chabaud, Ulysse and Joseph, Michael and Mehraban, Saeed and Motamedi, Arsalan},
  journal={Quantum},
  volume={10},
  pages={2110},
  year={2026},
  publisher={Verein zur F{\"o}rderung des Open Access Publizierens in den Quantenwissenschaften}
}

@article{motamedi2026stellar,
  title={The stellar decomposition of Gaussian quantum states},
  author={Motamedi, Arsalan and Yao, Yuan and Nielsen, Kasper and Chabaud, Ulysse and Bourassa, J Eli and Alexander, Rafael N and Miatto, Filippo M},
  journal={Quantum},
  volume={10},
  pages={1971},
  year={2026},
  publisher={Verein zur F{\"o}rderung des Open Access Publizierens in den Quantenwissenschaften}
}

@article{kheruntsyan1999wigner,
  title={Wigner function for a driven anharmonic oscillator},
  author={Kheruntsyan, KV},
  journal={Journal of Optics B: Quantum and Semiclassical Optics},
  volume={1},
  number={2},
  pages={225--233},
  year={1999}
}

@misc{aaronson2005complexity,
  title={The complexity zoo},
  author={Aaronson, Scott and Kuperberg, Greg and Granade, Christopher},
  year={2005}
}

@article{lingenfelter2021unconditional,
  title={Unconditional Fock state generation using arbitrarily weak photonic nonlinearities},
  author={Lingenfelter, Andrew and Roberts, David and Clerk, AA},
  journal={Science advances},
  volume={7},
  number={48},
  pages={eabj1916},
  year={2021},
  publisher={American Association for the Advancement of Science}
}

@article{johansson2012qutip,
  title={QuTiP: An open-source Python framework for the dynamics of open quantum systems},
  author={Johansson, J Robert and Nation, Paul D and Nori, Franco},
  journal={Computer physics communications},
  volume={183},
  number={8},
  pages={1760--1772},
  year={2012},
  publisher={Elsevier}
}

@book{breuer2002theory,
  title={The theory of open quantum systems},
  author={Breuer, Heinz-Peter and Petruccione, Francesco},
  year={2002},
  publisher={OUP Oxford}
}

@article{aghaee2025scaling,
  title={Scaling and networking a modular photonic quantum computer},
  author={Aghaee Rad, H and Ainsworth, Thomas and Alexander, Rafael N and Altieri, Brandon and Askarani, Mohsen F and Baby, R and Banchi, Leonardo and Baragiola, Ben Q and Bourassa, J Eli and Chadwick, RS and others},
  journal={Nature},
  volume={638},
  number={8052},
  pages={912--919},
  year={2025},
  publisher={Nature Publishing Group UK, London}
}

@article{larchey2022hilbert,
  title={Hilbert's Tenth Problem in Coq (Extended Version)},
  author={Larchey-Wendling, Dominique and Forster, Yannick},
  journal={Logical Methods in Computer Science},
  volume={18},
  year={2022},
  publisher={Episciences. org}
}

@book{rivas2012open,
  title={Open quantum systems},
  author={Rivas, Angel and Huelga, Susana F},
  volume={10},
  year={2012},
  publisher={Springer}
}

@article{van2025energy,
  title={Energy-limited quantum dynamics},
  author={van Luijk, Lauritz},
  journal={Communications in Mathematical Physics},
  volume={406},
  number={5},
  pages={120},
  year={2025},
  publisher={Springer}
}

@inproceedings{chebotarev2001interaction,
  title={Interaction representation method for Markov master equations in quantum optics},
  author={Chebotarev, Alexander M and Garcia, Julio C and Quezada, Roberto B},
  booktitle={Stochastic Analysis and Mathematical Physics II: 4th International ANESTOC Workshop in Santiago, Chile},
  pages={9--28},
  year={2001},
  organization={Springer}
}

@article{haagerup2003random,
  title={Random matrices with complex Gaussian entries},
  author={Haagerup, Uffe and Thorbj{\o}rnsen, Steen},
  journal={Expositiones Mathematicae},
  volume={21},
  number={4},
  pages={293--337},
  year={2003},
  publisher={Elsevier}
}

@book{kato1966perturbation,
  title={Perturbation theory for linear operators},
  author={Kato, Tosio and Kat{\aa}o, Tosio},
  volume={132},
  year={1966},
  publisher={Springer}
}

@article{lumer1961dissipative,
  title={Dissipative operators in a Banach space.},
  author={Lumer, G{\"u}nter and Phillips, Ralph S},
  year={1961}
}

@inproceedings{aaronson2011computational,
  title={The computational complexity of linear optics},
  author={Aaronson, Scott and Arkhipov, Alex},
  booktitle={Proceedings of the forty-third annual ACM symposium on Theory of computing},
  pages={333--342},
  year={2011}
}

@article{matijasevivc1971diophantine,
  title={Diophantine representation of enumerable predicates},
  author={Matijasevi{\v{c}}, Ju V},
  journal={Mathematics of the USSR-Izvestiya},
  volume={5},
  number={1},
  pages={1--28},
  year={1971}
}

@book{teschl2014mathematical,
  title={Mathematical methods in quantum mechanics},
  author={Teschl, Gerald},
  volume={157},
  year={2014},
  publisher={American Mathematical Soc.}
}

@article{burgarth2013ergodic,
  title={Ergodic and mixing quantum channels in finite dimensions},
  author={Burgarth, Daniel and Chiribella, Giulio and Giovannetti, Vittorio and Perinotti, Paolo and Yuasa, Kazuya},
  journal={New Journal of Physics},
  volume={15},
  number={7},
  pages={073045},
  year={2013},
  publisher={IOP Publishing}
}

@article{borras2026complexity,
  title={On the Complexity of Finding Decoherence Free Subspaces},
  author={Borras, Evan},
  journal={arXiv preprint arXiv:2609.26769},
  year={2026}
}

@article{carbone2016irreducible,
  title={Irreducible decompositions and stationary states of quantum channels},
  author={Carbone, Raffaella and Pautrat, Yan},
  journal={Reports on Mathematical Physics},
  volume={77},
  number={3},
  pages={293--313},
  year={2016},
  publisher={Elsevier}
}

@article{chebotarev1998sufficient,
  title={Sufficient conditions for conservativity of minimal quantum dynamical semigroups},
  author={Chebotarev, Alexander M and Fagnola, Franco},
  journal={journal of functional analysis},
  volume={153},
  number={2},
  pages={382--404},
  year={1998},
  publisher={Elsevier}
}

@book{reed1975ii,
  title={II: Fourier analysis, self-adjointness},
  author={Reed, Michael and Simon, Barry},
  volume={2},
  year={1975},
  publisher={Elsevier}
}

@article{hillar2013most,
  title={Most tensor problems are NP-hard},
  author={Hillar, Christopher J and Lim, Lek-Heng},
  journal={Journal of the ACM (JACM)},
  volume={60},
  number={6},
  pages={1--39},
  year={2013},
  publisher={ACM New York, NY, USA}
}

@article{davis1973hilbert,
  title={Hilbert's tenth problem is unsolvable},
  author={Davis, Martin},
  journal={The American Mathematical Monthly},
  volume={80},
  number={3},
  pages={233--269},
  year={1973},
  publisher={Taylor \& Francis}
}

@article{brenner2025factoring,
  title={Factoring an integer with three oscillators and a qubit},
  author={Brenner, Lukas and Caha, Libor and Coiteux-Roy, Xavier and Koenig, Robert},
  journal={Nature Communications},
  year={2025},
  publisher={Nature Publishing Group UK London}
}

@inproceedings{davis1976hilbert,
  title={Hilbert’s tenth problem: Diophantine equations: positive aspects of a negative solution},
  author={Davis, Martin and Matijasevi{\v{c}}, Yuri and Robinson, Julia},
  booktitle={Proceedings of symposia in pure mathematics},
  pages={323--378},
  year={1976},
  organization={American Mathematical Society}
}

@article{aaronson2011linear,
  title={A linear-optical proof that the permanent is\# P-hard},
  author={Aaronson, Scott},
  journal={Proceedings of the Royal Society A: Mathematical, Physical and Engineering Sciences},
  volume={467},
  number={2136},
  pages={3393--3405},
  year={2011},
  publisher={The Royal Society}
}

@article{bressanini2024gaussian,
  title={Gaussian boson sampling at finite temperature},
  author={Bressanini, Gabriele and Kwon, Hyukjoon and Kim, MS},
  journal={Physical Review A},
  volume={109},
  number={1},
  pages={013707},
  year={2024},
  publisher={APS}
}

@article{gonzalez2025dynamical,
  title={Dynamical complexity of non-Gaussian many-body systems with dissipation},
  author={Gonz{\'a}lez-Garc{\'\i}a, Guillermo and Gorshkov, Alexey V and Cirac, J Ignacio and Trivedi, Rahul},
  journal={Physical Review Letters},
  volume={135},
  number={19},
  pages={190401},
  year={2025},
  publisher={APS}
}

@article{mari2012positive,
  title={Positive Wigner functions render classical simulation of quantum computation efficient},
  author={Mari, Andrea and Eisert, Jens},
  journal={Physical review letters},
  volume={109},
  number={23},
  pages={230503},
  year={2012},
  publisher={APS}
}

@article{drummond1980quantum,
  title={Quantum theory of optical bistability. I. Nonlinear polarisability model},
  author={Drummond, PD and Walls, DF},
  journal={Journal of Physics A: Mathematical and General},
  volume={13},
  number={2},
  pages={725},
  year={1980},
  publisher={IOP Publishing}
}

@article{rohde2015evidence,
  title={Evidence for the conjecture that sampling generalized cat states with linear optics is hard},
  author={Rohde, Peter P and Motes, Keith R and Knott, Paul A and Fitzsimons, Joseph and Munro, William J and Dowling, Jonathan P},
  journal={Physical Review A},
  volume={91},
  number={1},
  pages={012342},
  year={2015},
  publisher={APS}
}

@article{chabaud2022quantum,
  title={Quantum-inspired permanent identities},
  author={Chabaud, Ulysse and Deshpande, Abhinav and Mehraban, Saeed},
  journal={Quantum},
  volume={6},
  pages={877},
  year={2022},
  publisher={Verein zur F{\"o}rderung des Open Access Publizierens in den Quantenwissenschaften}
}

@article{le2014bose,
  title={Bose-Hubbard model: Relation between driven-dissipative steady states and equilibrium quantum phases},
  author={Le Boit{\'e}, Alexandre and Orso, Giuliano and Ciuti, Cristiano},
  journal={Physical Review A},
  volume={90},
  number={6},
  pages={063821},
  year={2014},
  publisher={APS}
}

@article{garcia2019simulating,
  title={Simulating boson sampling in lossy architectures},
  author={Garc{\'\i}a-Patr{\'o}n, Ra{\'u}l and Renema, Jelmer J and Shchesnovich, Valery},
  journal={Quantum},
  volume={3},
  pages={169},
  year={2019},
  publisher={Verein zur F{\"o}rderung des Open Access Publizierens in den Quantenwissenschaften}
}

@article{aaronson2016bosonsampling,
  title={BosonSampling with lost photons},
  author={Aaronson, Scott and Brod, Daniel J},
  journal={Physical Review A},
  volume={93},
  number={1},
  pages={012335},
  year={2016},
  publisher={APS}
}

@article{albert2014symmetries,
  title={Symmetries and conserved quantities in Lindblad master equations},
  author={Albert, Victor V and Jiang, Liang},
  journal={Physical Review A},
  volume={89},
  number={2},
  pages={022118},
  year={2014},
  publisher={APS}
}

@article{leghtas2015confining,
  title={Confining the state of light to a quantum manifold by engineered two-photon loss},
  author={Leghtas, Zaki and Touzard, Steven and Pop, Ioan M and Kou, Angela and Vlastakis, Brian and Petrenko, Andrei and Sliwa, Katrina M and Narla, Anirudh and Shankar, Shyam and Hatridge, Michael J and others},
  journal={Science},
  volume={347},
  number={6224},
  pages={853--857},
  year={2015},
  publisher={American Association for the Advancement of Science}
}

@article{azouit2016well,
  title={Well-posedness and convergence of the Lindblad master equation for a quantum harmonic oscillator with multi-photon drive and damping},
  author={Azouit, R{\'e}mi and Sarlette, Alain and Rouchon, Pierre},
  journal={ESAIM: Control, Optimisation and Calculus of Variations},
  volume={22},
  number={4},
  pages={1353--1369},
  year={2016},
  publisher={EDP Sciences}
}

@article{girotti2025gaussian,
  title={Gaussian quantum Markov semigroups on finitely many modes admitting a normal invariant state},
  author={Girotti, Federico and Poletti, Damiano},
  journal={Journal of Mathematical Analysis and Applications},
  pages={130150},
  year={2025},
  publisher={Elsevier}
}

@article{grier2022complexity,
  title={The complexity of bipartite Gaussian boson sampling},
  author={Grier, Daniel and Brod, Daniel J and Arrazola, Juan Miguel and de Andrade Alonso, Marcos Benicio and Quesada, Nicol{\'a}s},
  journal={Quantum},
  volume={6},
  pages={863},
  year={2022},
  publisher={Verein zur F{\"o}rderung des Open Access Publizierens in den Quantenwissenschaften}
}

@inproceedings{stockmeyer1983complexity,
  title={The complexity of approximate counting},
  author={Stockmeyer, Larry},
  booktitle={Proceedings of the fifteenth annual ACM symposium on Theory of computing},
  pages={118--126},
  year={1983}
}

@article{mirrahimi2014dynamically,
  title={Dynamically protected cat-qubits: a new paradigm for universal quantum computation},
  author={Mirrahimi, Mazyar and Leghtas, Zaki and Albert, Victor V and Touzard, Steven and Schoelkopf, Robert J and Jiang, Liang and Devoret, Michel H},
  journal={New Journal of Physics},
  volume={16},
  number={4},
  pages={045014},
  year={2014},
  publisher={IOP Publishing}
}

@article{houde2024matrix,
  title={Matrix decompositions in quantum optics: Takagi/autonne, bloch--messiah/euler, iwasawa, and williamson},
  author={Houde, Martin and McCutcheon, Will and Quesada, Nicol{\'a}s},
  journal={Canadian Journal of Physics},
  volume={102},
  number={10},
  pages={497--507},
  year={2024},
  publisher={Canadian Science Publishing 123 Slater Street, Suite 610, Ottawa, Ontario~…}
}

@article{rahimi2016sufficient,
  title={Sufficient conditions for efficient classical simulation of quantum optics},
  author={Rahimi-Keshari, Saleh and Ralph, Timothy C and Caves, Carlton M},
  journal={Physical Review X},
  volume={6},
  number={2},
  pages={021039},
  year={2016},
  publisher={APS}
}

@article{glauber1963photon,
  title={Photon correlations},
  author={Glauber, Roy J},
  journal={Physical Review Letters},
  volume={10},
  number={3},
  pages={84},
  year={1963},
  publisher={APS}
}

@article{PhysRevA.98.013809,
  title = {Statistics of photon-subtracted and photon-added states},
  author = {Barnett, Stephen M. and Ferenczi, Gergely and Gilson, Claire R. and Speirits, Fiona C.},
  journal = {Phys. Rev. A},
  volume = {98},
  issue = {1},
  pages = {013809},
  numpages = {13},
  year = {2018},
  month = {Jul},
  publisher = {American Physical Society},
  doi = {10.1103/PhysRevA.98.013809},
  url = {https://link.aps.org/doi/10.1103/PhysRevA.98.013809}
}

@article{knill2001scheme,
  title={A scheme for efficient quantum computation with linear optics},
  author={Knill, Emanuel and Laflamme, Raymond and Milburn, Gerald J},
  journal={nature},
  volume={409},
  number={6816},
  pages={46--52},
  year={2001},
  publisher={Nature Publishing Group UK London}
}

@inproceedings{azouit2016adiabatic,
  title={Adiabatic elimination for open quantum systems with effective Lindblad master equations},
  author={Azouit, R{\'e}mi and Sarlette, Alain and Rouchon, Pierre},
  booktitle={2016 IEEE 55th Conference on Decision and Control (CDC)},
  pages={4559--4565},
  year={2016},
  organization={IEEE}
}

@article{laiho2012characteristics,
  title={Characteristics of displaced single photons attained via higher order factorial moments},
  author={Laiho, Kaisa and Avenhaus, Malte and Silberhorn, Christine},
  journal={New Journal of Physics},
  volume={14},
  number={10},
  pages={105011},
  year={2012},
  publisher={IOP Publishing}
}

@article{aaronson2005quantum,
  title={Quantum computing, postselection, and probabilistic polynomial-time},
  author={Aaronson, Scott},
  journal={Proceedings: Mathematics, Physical and Engineering},
  pages={3473--3482},
  year={2005},
  publisher={JSTOR}
}

@book{arora2009computational,
  title={Computational complexity: a modern approach},
  author={Arora, Sanjeev and Barak, Boaz},
  year={2009},
  publisher={Cambridge University Press}
}

@article{bittel2025optimal,
  title={Optimal estimates of trace distance between bosonic Gaussian states and applications to learning},
  author={Bittel, Lennart and Mele, Francesco Anna and Mele, Antonio Anna and Tirone, Salvatore and Lami, Ludovico},
  journal={Quantum},
  volume={9},
  pages={1769},
  year={2025},
  publisher={Verein zur F{\"o}rderung des Open Access Publizierens in den Quantenwissenschaften}
}

@article{go2025sufficient,
  title={Sufficient conditions for hardness of lossy Gaussian boson sampling},
  author={Go, Byeongseon and Oh, Changhun and Jeong, Hyunseok},
  journal={arXiv preprint arXiv:2511.07853},
  year={2025}
}

@article{verstraete2009quantum,
  title={Quantum computation and quantum-state engineering driven by dissipation},
  author={Verstraete, Frank and Wolf, Michael M and Ignacio Cirac, J},
  journal={Nature physics},
  volume={5},
  number={9},
  pages={633--636},
  year={2009},
  publisher={Nature Publishing Group UK London}
}

@article{kempe2006complexity,
  title={The complexity of the local Hamiltonian problem},
  author={Kempe, Julia and Kitaev, Alexei and Regev, Oded},
  journal={Siam journal on computing},
  volume={35},
  number={5},
  pages={1070--1097},
  year={2006},
  publisher={SIAM}
}

@article{hamilton2017gaussian,
  title={Gaussian boson sampling},
  author={Hamilton, Craig S and Kruse, Regina and Sansoni, Linda and Barkhofen, Sonja and Silberhorn, Christine and Jex, Igor},
  journal={Physical review letters},
  volume={119},
  number={17},
  pages={170501},
  year={2017},
  publisher={APS}
}

@article{knill1998power,
  title={Power of one bit of quantum information},
  author={Knill, Emanuel and Laflamme, Raymond},
  journal={Physical Review Letters},
  volume={81},
  number={25},
  pages={5672},
  year={1998},
  publisher={APS}
}

@article{yuan2025universal,
  title={Universal control in bosonic systems with weak Kerr nonlinearities},
  author={Yuan, Ming and Seif, Alireza and Lingenfelter, Andrew and Schuster, David I and Clerk, Aashish A and Jiang, Liang},
  journal={Physical Review A},
  volume={111},
  number={3},
  pages={032606},
  year={2025},
  publisher={APS}
}

@article{koehler2026anticoncentration,
  title={Anticoncentration of the Permanent in Ginibre Ensembles},
  author={Koehler, Frederic and Leung, Pui Kuen},
  journal={arXiv preprint arXiv:2607.20329},
  year={2026}
}

@article{madsen2022quantum,
  title={Quantum computational advantage with a programmable photonic processor},
  author={Madsen, Lars S and Laudenbach, Fabian and Askarani, Mohsen Falamarzi and Rortais, Fabien and Vincent, Trevor and Bulmer, Jacob FF and Miatto, Filippo M and Neuhaus, Leonhard and Helt, Lukas G and Collins, Matthew J and others},
  journal={Nature},
  volume={606},
  number={7912},
  pages={75--81},
  year={2022},
  publisher={Nature Publishing Group UK London}
}

@article{zhong2020quantum,
  title={Quantum computational advantage using photons},
  author={Zhong, Han-Sen and Wang, Hui and Deng, Yu-Hao and Chen, Ming-Cheng and Peng, Li-Chao and Luo, Yi-Han and Qin, Jian and Wu, Dian and Ding, Xing and Hu, Yi and others},
  journal={Science},
  volume={370},
  number={6523},
  pages={1460--1463},
  year={2020},
  publisher={American Association for the Advancement of Science}
}

@article{gottesman2000encoding,
  title = {Encoding a qubit in an oscillator},
  author = {Gottesman, Daniel and Kitaev, Alexei and Preskill, John},
  journal = {Phys. Rev. A},
  volume = {64},
  issue = {1},
  pages = {012310},
  numpages = {21},
  year = {2001},
  month = {Jun},
  publisher = {American Physical Society},
  doi = {10.1103/PhysRevA.64.012310},
  url = {https://link.aps.org/doi/10.1103/PhysRevA.64.012310}
}

@article{matsuura2026continuous,
  title={Continuous-variable fault-tolerant quantum computation under general noise},
  author={Matsuura, Takaya and Menicucci, Nicolas C and Yamasaki, Hayata},
  journal={Nature Communications},
  volume={17},
  number={1},
  pages={1709},
  year={2026},
  publisher={Nature Publishing Group UK London}
}

@inproceedings{bergamaschi2024quantum,
  title={Quantum computational advantage with constant-temperature Gibbs sampling},
  author={Bergamaschi, Thiago and Chen, Chi-Fang and Liu, Yunchao},
  booktitle={2024 IEEE 65th Annual Symposium on Foundations of Computer Science (FOCS)},
  pages={1063--1085},
  year={2024},
  organization={IEEE}
}

@article{rajakumar2026gibbs,
  title={Gibbs sampling gives quantum advantage at constant temperatures with O (1)-local Hamiltonians},
  author={Rajakumar, Joel and Watson, James D},
  journal={Quantum},
  volume={10},
  pages={1981},
  year={2026},
  publisher={Verein zur F{\"o}rderung des Open Access Publizierens in den Quantenwissenschaften}
}

@article{prosen2008third,
  title={Third quantization: a general method to solve master equations for quadratic open Fermi systems},
  author={Prosen, Toma{\v{z}}},
  journal={New Journal of Physics},
  volume={10},
  number={4},
  pages={043026},
  year={2008}
}

@article{karevski2013exact,
  title={Exact matrix product solution for the boundary-driven Lindblad XXZ chain},
  author={Karevski, Dragi and Popkov, V and Sch{\"u}tz, Gunter M},
  journal={Physical review letters},
  volume={110},
  number={4},
  pages={047201},
  year={2013},
  publisher={APS}
}

@article{kliesch2011dissipative,
  title={Dissipative quantum church-turing theorem},
  author={Kliesch, Martin and Barthel, Thomas and Gogolin, Christian and Kastoryano, Michael and Eisert, Jens},
  journal={Physical review letters},
  volume={107},
  number={12},
  pages={120501},
  year={2011},
  publisher={APS}
}

@article{firanko2024area,
  title={Area law for steady states of detailed-balance local Lindbladians},
  author={Firanko, Raz and Goldstein, Moshe and Arad, Itai},
  journal={Journal of Mathematical Physics},
  volume={65},
  number={5},
  year={2024},
  publisher={AIP Publishing}
}

@article{jacobs2024space,
  title={The space just above one clean qubit},
  author={Jacobs, Dale and Mehraban, Saeed},
  journal={arXiv preprint arXiv:2410.08051},
  year={2024}
}

@article{braunstein2005quantum,
  title={Quantum information with continuous variables},
  author={Braunstein, Samuel L and Van Loock, Peter},
  journal={Reviews of modern physics},
  volume={77},
  number={2},
  pages={513--577},
  year={2005},
  publisher={APS}
}

@article{gondolf2024energy,
  title={Energy preserving evolutions over Bosonic systems},
  author={Gondolf, Paul and M{\"o}bus, Tim and Rouz{\'e}, Cambyse},
  journal={Quantum},
  volume={8},
  pages={1551},
  year={2024},
  publisher={Verein zur F{\"o}rderung des Open Access Publizierens in den Quantenwissenschaften}
}

@article{go2026polynomial,
  title={Polynomial-Time Classical Simulation of BosonSampling under Any Constant-Fraction Photon Loss},
  author={Go, Byeongseon and Jeong, Hyunseok},
  journal={arXiv preprint arXiv:2610.27065},
  year={2026}
}

\appendix

\section{Discussion on sampling v.s. observable estimation at the fixed point}
\label{app:observable-estimation-proofs}
\label{sec:sampling-hardness-vs-observables}

In \Cref{sec:loss-only-bqp} we established $\BQP$-hardness for the Lindbladian fixed point problem. In other words, observable estimation from the fixed point is in principle hard for classical computers (assuming $\BQP\neq \BPP$). The loss model was however non-uniform (i.e. a constant fraction of modes were lossless) and loss rate was polynomially large compared with other coupling constants. These structures were essential in our $\BQP$-hardness construction. Does observable estimation remain hard for uniform loss at a constant rate? 

In \Cref{sec:Gaussian-hardness-sampling} we could establish hardness of sampling for uniform loss when Hamiltonian dynamics was Gaussian. This hardness of sampling however does not imply hardness of observable estimation. As an example, consider product state $\otimes_j \rho_j$ evolving according to a passive linear-optical unitary $C$. Low-degree observable estimation
can be classically efficient. Specifically, let
$\rho=C(\bigotimes_{j=1}^m\rho_j)C^\dagger$, where $C$
is an efficiently specified passive linear-optical circuit.
For a constant-degree polynomial observable $O$ with
polynomially many terms and polynomial-bit coefficients,
$C^\dagger O C$ remains an efficiently representable
constant-degree polynomial and is computable in polynomial time.

One way to study hardness of observable estimation  from fixed points of Lindbladian systems with uniform linear loss is to study Wigner negativity  for the fixed point. 
One can show that assuming the non-negative Wigner density function can be sampled in polynomial time (plus additional natural conditions) this implies that observable estimation problem is easy (see \Cref{app:observable-estimation-proofs}). Interestingly, all the exactly solvable fixed point solutions described in \Cref{eq:basic-Kerr-single-mode} have Wigner non-negativity~\cite{kheruntsyan1999wigner}. In particular, 
Kheruntsyan derived an exact steady-state Wigner function of the form
\begin{align}
    W_{\rm ss}(\alpha)
    =
    \mathcal N
    e^{-2|\alpha|^2}
    \left|
        \frac{
            J_{\lambda-1}\!\left(
                \sqrt{-8\varepsilon\alpha^*}
            \right)
        }{
            (\alpha^*)^{(\lambda-1)/2}
        }
    \right|^2,
    \label{eq:kheruntsyan-steady-wigner}
\end{align}
where \(\mathcal N>0\), \(J_\nu\) is the Bessel function of the first kind,
and \(\varepsilon,\lambda\) are the dimensionless drive and damping
parameters used in Ref.~\cite{kheruntsyan1999wigner}.  Equation
\eqref{eq:kheruntsyan-steady-wigner} is manifestly non-negative:
\begin{align}
    W_{\rm ss}(\alpha)\ge0
    \qquad
    \text{for all }\alpha\in\mathbb C.
\end{align}

We can furthermore show that the balanced binomial limiting state
\begin{align}
    \sigma_k
    :=
    \frac{1}{2^k}
    \sum_{j=0}^{k}
    \binom{k}{j}
    |j\rangle\langle j|.
    \label{eq:balanced-binomial-state}
\end{align}
used in the
hardness construction in \Cref{sec:Kerr-gadget} also has a non-negative Wigner function and allow efficient sampler. We furthermore demonstrate
numerically that the dissipative trajectory may nevertheless develop
transient Wigner negativity before converging to the positive fixed point. In particular,
\begin{align}
    W_{\sigma_k}(\alpha)
    =
    \frac{2^{k+1}}{\pi k!}
    |\alpha|^{2k}e^{-2|\alpha|^2} \geq 0.
    \label{eq:binomial-wigner}
\end{align}

We derive this expression in~\Cref{app:binomial-wigner-proof}.  The
balanced diagonal mixture should not be confused with a pure binomial
superposition.  The positivity in
Eq.~\eqref{eq:binomial-wigner} relies on both the diagonal form and the
balanced binomial weights.

A natural open question is if this Wigner non-negativity in the fixed point for natural non-Gaussian Lindbladian systems with uniform linear loss (such as those with Kerr interaction) is a general principle or if we can find nontrivial instances where Wigner negativity is observed (suggesting hardness of observable estimation). 
\begin{remark}[Exact phase-space sampler]
\label{rem:binomial-wigner-sampler}
The probability density
\(W_{\sigma_k}(\alpha)\,d^2\alpha\) can be sampled exactly.  Draw
\begin{align}
    X
    &\sim
    \operatorname{Gamma}(k+1,1), \quad \Theta\sim
    \operatorname{Unif}[0,2\pi),
\end{align}
independently, and set
    $\alpha
    =
    \sqrt{\frac{X}{2}}\,e^{i\Theta}$.
Then \(\alpha\) is distributed according to \(W_{\sigma_k}\).
\end{remark}

\paragraph{Transient Wigner negativity}

\begin{figure}[ht]
    \centering
    \includegraphics[width=\linewidth]{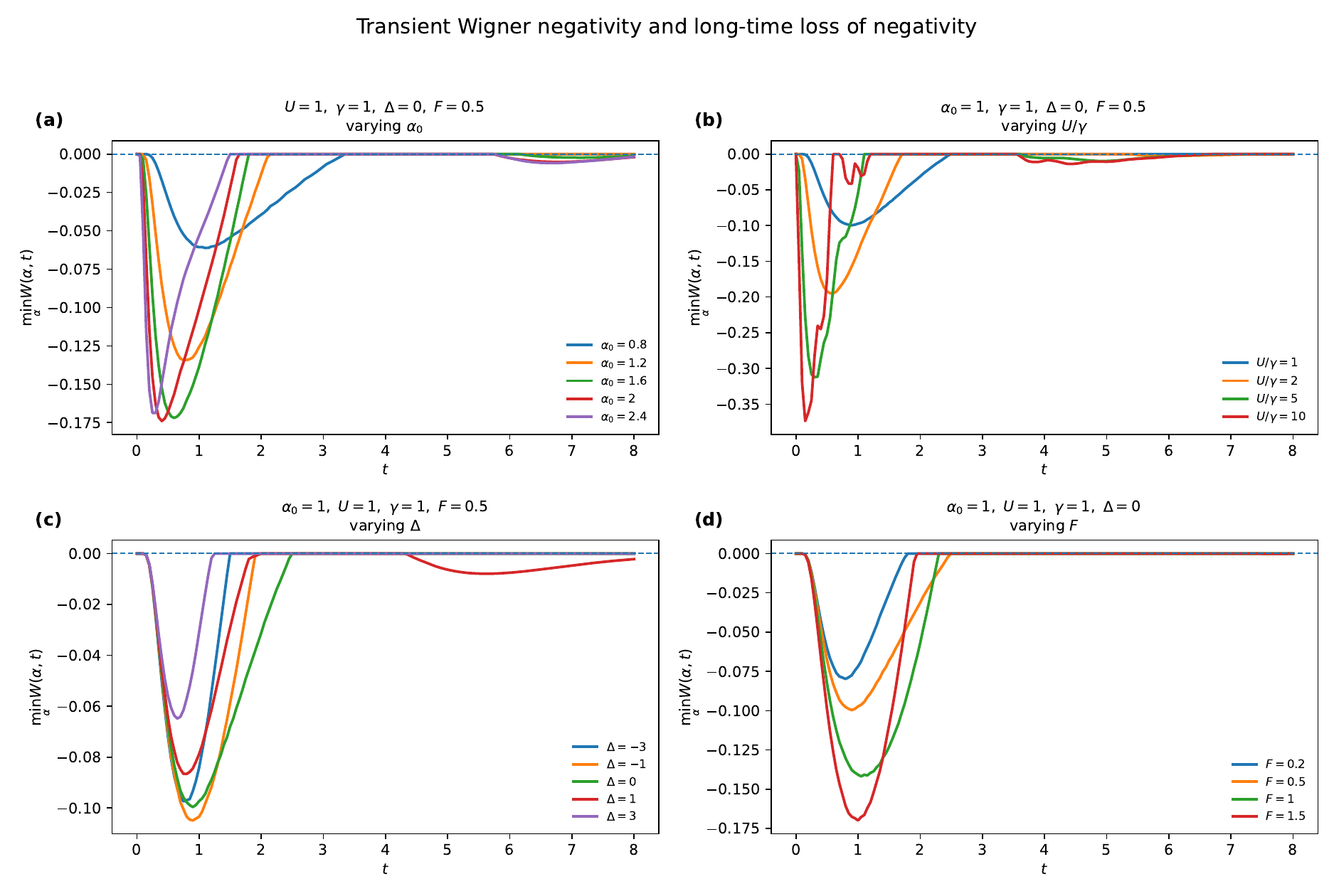}
    \caption{
    Time evolution of the minimum Wigner value over the sampled
    phase-space grid, \(\min_{\alpha\in\mathcal G}W(\alpha,t)\),
    for the single-mode Kerr oscillator with one-photon loss.
    Panel (a) varies the initial coherent amplitude \(\alpha_0\);
    panel (b) varies \(U/\gamma\); panel (c) varies the detuning
    \(\Delta\); and panel (d) varies the drive \(F\).
    Numerical evolution uses QuTiP's \cite{johansson2012qutip} \texttt{mesolve} with the
    BDF integrator, absolute tolerance \(10^{-8}\), relative
    tolerance \(10^{-6}\), and the truncated Fock basis
    \(\{\ket0,\ldots,\ket{99}\}\).
    The Wigner function is evaluated on a uniform \(221\times221\)
    grid \(\mathcal G\) covering
    \(\operatorname{Re}\alpha,\operatorname{Im}\alpha\in[-6,6]\),
    with \(g=2\).
    The sampled times are \(t_j=0.05j\), \(j=0,\ldots,160\),
    with \(\gamma=1\).
    The trajectories exhibit transient Wigner negativity;
    the exact stationary Wigner function is nonnegative
    \cite{kheruntsyan1999wigner}.
    }
    \label{fig:transient_wigner_min}
\end{figure}

As discussed, the exact driven-Kerr fixed point is Wigner
\cite{kheruntsyan1999wigner} and its limiting binomial behavior have non-negative Wigner functions. Neither fact implies that the
entire dissipative trajectory remains Wigner non-negative.  Kerr evolution
can generate phase-space interference at intermediate times before
dissipation drives the system to its positive fixed point. 
  Figure~\ref{fig:transient_wigner_min} shows that the Wigner
function can become negative at intermediate times even though the final
fixed point is non-negative.

As shown in Fig.~\ref{fig:transient_wigner_min}, the transient negativity
typically becomes more pronounced as the initial coherent amplitude
increases, reflecting the occupation of higher Fock sectors.  A similar trend
is observed as \(U/\gamma\) increases, since stronger Kerr shearing produces
more pronounced phase-space interference before dissipation suppresses it.
Increasing the drive amplitude \(F\) also tends to deepen the transient
negativity.  The dependence on detuning \(\Delta\) is less transparent and is
not monotonic over the parameter range explored.
\subsection{Balanced binomial Wigner identity}
\label{app:binomial-wigner-proof}

The Wigner function of the Fock state \(|k\rangle\) is
\begin{align}
    W_k(\alpha)
    =
    \frac{2}{\pi}
    (-1)^k
    L_k(4|\alpha|^2)e^{-2|\alpha|^2},
\end{align}
where \(L_k\) is the Laguerre polynomial.  Set
$x:=4|\alpha|^2$
and define
\begin{align}
    S_n(x)
    :=
    \sum_{k=0}^{n}
    \binom{n}{k}
    (-1)^kL_k(x).
\end{align}
We show that
\begin{align}
    S_n(x)=\frac{x^n}{n!}.
    \label{eq:laguerre-binomial-identity}
\end{align}

The Laguerre generating function is
\begin{align}
    A(t)
    :=
    \sum_{k=0}^{\infty}L_k(x)t^k
    =
    \frac{1}{1-t}
    \exp\!\left(
        -\frac{xt}{1-t}
    \right).
\end{align}
For the inverse binomial transform
\begin{align}
    b_n
    =
    \sum_{k=0}^{n}
    \binom{n}{k}(-1)^k \hat a_k,
\end{align}
the generating function is
\begin{align}
    B(z)
    =
    \frac{1}{1-z}
    A\!\left(
        -\frac{z}{1-z}
    \right).
\end{align}
Substituting the Laguerre generating function gives
\begin{align}
    A\!\left(
        -\frac{z}{1-z}
    \right)
    =
    (1-z)e^{xz},
\end{align}
and therefore
\begin{align}
    B(z)=e^{xz}.
\end{align}
Comparing coefficients proves
Eq.~\eqref{eq:laguerre-binomial-identity}.

By linearity of the Wigner transform,
\begin{align}
    W_{\rho_n}(\alpha)
    &=
    \frac{1}{2^n}
    \sum_{k=0}^{n}
    \binom{n}{k}W_k(\alpha) \\
    &=
    \frac{2}{\pi\,2^n}
    e^{-2|\alpha|^2}
    S_n(4|\alpha|^2) \\
    &=
    \frac{2^{n+1}}{\pi n!}
    |\alpha|^{2n}e^{-2|\alpha|^2}.
\end{align}
The expression is manifestly non-negative.

For completeness, let
\begin{align}
    X:=2|\alpha|^2.
\end{align}
After integrating over the uniform angular variable, the radial density is
\begin{align}
    \frac{x^n e^{-x}}{n!}\,dx,
\end{align}
which is the \(\operatorname{Gamma}(n+1,1)\) distribution.  This proves the
sampler stated in
Remark~\ref{rem:binomial-wigner-sampler}.

\subsection{Wigner--Weyl resolution}
\label{app:wigner-weyl-resolution}

Let
\begin{align}
    \boldsymbol{\alpha}
    =
    (\alpha_1,\ldots,\alpha_m)\in\mathbb C^m,
\end{align}
and define
\begin{align}
    \Omega(\boldsymbol{\alpha})
    :=
    2^m
    D(\boldsymbol{\alpha})
    \Pi
    D(\boldsymbol{\alpha})^\dagger,
\end{align}
where
\begin{align}
    D(\boldsymbol{\alpha})
    &:=
    \prod_{j=1}^{m}
    \exp(
        \alpha_j \hat a_j^\dagger
        -
        \alpha_j^*\hat a_j
    ), \\
    \Pi
    &:=
    (-1)^{\sum_{j=1}^{m}\hat a_j^\dagger \hat a_j}.
\end{align}
For a trace-class operator \(A\), define
\begin{align}
    W_A(\boldsymbol{\alpha})
    :=
    \frac{1}{\pi^m}
    \operatorname{Tr}\!\left[
        A\Omega(\boldsymbol{\alpha})
    \right].
\end{align}

\begin{lemma}[Resolution by displaced parity kernels]
\label{lem:wigner-resolution}
For every trace-class operator \(A\),
\begin{align}
    A
    =
    \int_{\mathbb C^m}
    W_A(\boldsymbol{\alpha})
    \Omega(\boldsymbol{\alpha})
    \,d^{2m}\boldsymbol{\alpha}.
    \label{eq:wigner-inversion}
\end{align}
The inversion formula is understood in the
Hilbert--Schmidt sense; see
\cite[Sec.~2.3]{keyl2015schwartz}.
The integral notation does not assert convergence
in trace norm.

For a polynomial observable $O$, define its Weyl symbol
$O_W$ by writing $O$ in fully symmetric ordering and
replacing $\hat a_j,\hat a_j^\dagger$ by
$\alpha_j,\alpha_j^*$, respectively. In particular,
\[
I_W=1,\qquad
(\hat a_j^\dagger\hat a_j)_W
=|\alpha_j|^2-\frac12.
\]
For a state $\rho$ with a Schwartz integral kernel
\footnote{This means that the position-space kernel
$\langle x|\rho|y\rangle$ is smooth and, together with
all its derivatives, decays faster than any inverse
polynomial in $(x,y)$; see
\cite[Sec.~3.3]{keyl2015schwartz}.},
the Wigner--Weyl pairing gives
\[
\operatorname{Tr}(\rho O)
=
\int_{\mathbb C^m}
W_\rho(\boldsymbol\alpha)
O_W(\boldsymbol\alpha)
\,d^{2m}\boldsymbol\alpha.
\]
In particular, this applies to the passive linear-optical
images of products of balanced-binomial states considered
below: they have finite Fock support, hence Schwartz
kernels and finite moments of every order.
\end{lemma}

\begin{proof}
For one mode,
\begin{align}
    \Omega(\alpha)
    =
    2D(2\alpha)\Pi.
\end{align}
Using the Weyl relations, parity covariance, and the distributional identity
\begin{align}
    \operatorname{Tr}D(\xi)
    =
    \pi\delta^{(2)}(\xi),
\end{align}
one obtains
\begin{align}
    \operatorname{Tr}\!\left[
        \Omega(\alpha)\Omega(\beta)
    \right]
    =
    \pi\delta^{(2)}(\alpha-\beta).
\end{align}
The multimode kernel factorizes, and hence
\begin{align}
    \operatorname{Tr}\!\left[
        \Omega(\boldsymbol{\alpha})
        \Omega(\boldsymbol{\beta})
    \right]
    =
    \pi^m
    \delta^{(2m)}
    (
        \boldsymbol{\alpha}
        -
        \boldsymbol{\beta}
    ).
    \label{eq:wigner-kernel-orthogonality}
\end{align}

Let the right-hand side of
Eq.~\eqref{eq:wigner-inversion} be \(B\).  By
Eq.~\eqref{eq:wigner-kernel-orthogonality},
\begin{align}
    \operatorname{Tr}\!\left[
        B\Omega(\boldsymbol{\beta})
    \right]
    =
    \pi^mW_A(\boldsymbol{\beta})
    =
    \operatorname{Tr}\!\left[
        A\Omega(\boldsymbol{\beta})
    \right].
\end{align}
The displaced parity kernels form a complete operator basis, so \(A=B\).
Taking the trace against \(O\) proves the Wigner--Weyl identity.
\end{proof}

\subsection{Positive-Wigner observable estimation}
\label{app:positive-wigner-estimation-proof}

We use the Wigner--Weyl trace identity
\begin{align}
    \operatorname{Tr}(\rho O)
    =
    \int_{\mathbb C^m}
    W_\rho(\boldsymbol{\alpha})
    O_W(\boldsymbol{\alpha})
    \,d^{2m}\boldsymbol{\alpha},
    \label{eq:wigner-weyl-trace}
\end{align}
where \(O_W\) is the Weyl symbol of \(O\).  Our normalization conventions and
a derivation of Eq.~\eqref{eq:wigner-weyl-trace} are given in
Appendix~\ref{app:wigner-weyl-resolution}.

\begin{theorem}[Non-negative Wigner observable estimation]
\label{thm:positive-wigner-observable-estimation}
Let \(\rho\) be an \(m\)-mode state whose Wigner function is non-negative and
can be sampled classically.  Let \(O\) be a Hermitian observable whose Weyl
symbol \(O_W\) is computable in polynomial time and satisfies
\begin{align}
    \operatorname{Var}_{\boldsymbol{\alpha}\sim W_\rho}
    \left[
        O_W(\boldsymbol{\alpha})
    \right]
    \le
    B^2.
\end{align}
Then \(\operatorname{Tr}(\rho O)\) can be estimated to additive error
\(\epsilon\), with failure probability at most \(\delta\), using
\begin{align}
    O\!\left(
        \frac{B^2}{\epsilon^2}
        \log\frac{1}{\delta}
    \right)
\end{align}
phase-space samples.
\end{theorem}

\begin{proof}
Because \(W_\rho\ge0\) and
\begin{align}
    \int_{\mathbb C^m}
    W_\rho(\boldsymbol{\alpha})
    \,d^{2m}\boldsymbol{\alpha}
    =
    1,
\end{align}
the Wigner function is a probability density.  By
Eq.~\eqref{eq:wigner-weyl-trace},
\begin{align}
    \operatorname{Tr}(\rho O)
    =
    \mathbb E_{\boldsymbol{\alpha}\sim W_\rho}
    \left[
        O_W(\boldsymbol{\alpha})
    \right].
\end{align}
Draw \(M\) independent samples and form
\begin{align}
    \widehat O
    :=
    \frac{1}{M}
    \sum_{\ell=1}^{M}
    O_W(\boldsymbol{\alpha}^{(\ell)}).
\end{align}
Then \(\widehat O\) is unbiased and
\begin{align}
    \operatorname{Var}(\widehat O)
    \le
    \frac{B^2}{M}.
\end{align}
A median-of-means estimator with
\begin{align}
    M
    =
    O\!\left(
        \frac{B^2}{\epsilon^2}
        \log\frac{1}{\delta}
    \right)
\end{align}
achieves additive error \(\epsilon\) with failure probability at most
\(\delta\).
\end{proof}

Now we have the following explicit bound on the variance $B$.
\begin{lemma}[Explicit variance bound for polynomial Weyl symbols]
\label{lem:explicit-weyl-variance-bound}
Let
\[
\rho_{\rm lim}
=C\left(\bigotimes_{j=1}^m\sigma_{n_j}\right)C^\dagger,
\qquad
N_{\rm tot}:=\sum_{j=1}^m n_j,
\]
where $C$ is passive linear optical and each
$n_j\in\mathbb Z_{\geq0}$ is a binomial parameter,
not a number operator.
Let $O$ be an observable whose Weyl symbol is a polynomial
\[
    O_W(\boldsymbol{\alpha},\boldsymbol{\alpha}^*)
    =
    \sum_{\ell=1}^{L}
    c_\ell
    M_\ell(\boldsymbol{\alpha},\boldsymbol{\alpha}^*),
\]
where each monomial \(M_\ell\) has total degree at most \(D\).  Define
\[
    c_{\max}:=\max_{1\le \ell\le L}|c_\ell|.
\]
Then
\[
    \operatorname{Var}_{\boldsymbol{\alpha}\sim W_{\rho_{\rm lim}}}
    \left[
        O_W(\boldsymbol{\alpha})
    \right]
    \le
    B_{\rm poly}^2,
\]
where
\[
    B_{\rm poly}^2
    :=
    L^2c_{\max}^2
    \sum_{r=0}^{D}
    {D\choose r}
    2^{-r}
    (N_{\rm tot}+m)_r .
\]
Here
\[
    (x)_r:=x(x+1)\cdots(x+r-1)
\]
is the rising factorial.  In particular, if \(D=O(1)\), \(L=\operatorname{poly}(m)\),
and \(c_{\max}=\operatorname{poly}(m,N_{\rm tot})\), then
\[
    B_{\rm poly}
    =
    \operatorname{poly}(m,N_{\rm tot}).
\]

To obtain polynomial runtime in the input description
length, we additionally require $N_{\rm tot}$ and the
coefficient magnitudes to be polynomially bounded,
together with efficient phase-space sampling and
symbol evaluation.
\end{lemma}

\begin{proof}
By the explicit Wigner sampler for the product state, before passive mixing
we have
\[
    2|\beta_j|^2
    \sim
    \operatorname{Gamma}(n_j+1,1),
\]
independently over \(j\).  Therefore
\begin{align}
    2\|\boldsymbol{\beta}\|_2^2
    =
    2\sum_{j=1}^{m}|\beta_j|^2
    \sim
    \operatorname{Gamma}(N_{\rm tot}+m,1).
\end{align}
Since the passive linear-optical transformation is unitary,
\begin{align}
    \boldsymbol{\alpha}=V\boldsymbol{\beta}
\end{align}
satisfies
\begin{align} 
    \|\boldsymbol{\alpha}\|_2=\|\boldsymbol{\beta}\|_2.
\end{align}
Hence, for every \(r\ge0\),
\begin{align} 
    \mathbb E
    \left[
        \|\boldsymbol{\alpha}\|_2^{2r}
    \right]
    =
    2^{-r}(N_{\rm tot}+m)_r .
\end{align}
Now each monomial \(M_\ell\) has total degree at most \(D\), so
\begin{align} |M_\ell(\boldsymbol{\alpha},\boldsymbol{\alpha}^*)|^2
    \le
    (1+\|\boldsymbol{\alpha}\|_2^2)^D .
\end{align}
Therefore
\begin{align}
    |O_W(\boldsymbol{\alpha})|^2
    \le
    \left(
        \sum_{\ell=1}^{L}
        |c_\ell|
        |M_\ell(\boldsymbol{\alpha},\boldsymbol{\alpha}^*)|
    \right)^2 \le
    L^2c_{\max}^2
    (1+\|\boldsymbol{\alpha}\|_2^2)^D .
\end{align}
Taking expectation gives
\begin{align}
    \mathbb E|O_W(\boldsymbol{\alpha})|^2
    &\le
    L^2c_{\max}^2
    \mathbb E
    \left[
        (1+\|\boldsymbol{\alpha}\|_2^2)^D
    \right] =
    L^2c_{\max}^2
    \sum_{r=0}^{D}
    {D\choose r}
    \mathbb E
    \left[
        \|\boldsymbol{\alpha}\|_2^{2r}
    \right] =
    L^2c_{\max}^2
    \sum_{r=0}^{D}
    {D\choose r}
    2^{-r}
    (N_{\rm tot}+m)_r .
\end{align}
Since
\[
    \operatorname{Var}(O_W)
    \le
    \mathbb E|O_W|^2,
\]
the claim follows.
\end{proof}

\section{Details of the nonlinear Kerr model}\label{app:nk}

We now consider the nonlinear Kerr driven-dissipative open system for which the fixed point can be solved exactly \cite{le2014bose, drummond1980quantum}. We consider the following coherent driving hamiltonian $H=-\Delta \omega \hat{n}+\frac{U}{2}\hat{n}(\hat{n}-1)+F\hat a^\dagger+F^*\hat a$ with single photon loss decay rate $\gamma.$ The evolution of the density matrix satisfies the following Lindblad master equation $i\partial_{t}\rho=[H,\rho]+i\sum_i \gamma_i(L_i\rho L_i^{\dagger}-1/2\{L_i^{\dagger}L_i,\rho \})$.
There is an exact solution \cite{le2014bose,drummond1980quantum}:
\begin{align}
\rho=\sum_{mn}\rho_{nm}\ket{n}\bra{m}  
\end{align}
For the driven-dissipative Bose Hubbard model for cavity photons \cite{le2014bose}, one can solve the density matrix of the fixed point $\rho_{nm}$ exactly.
\begin{eqnarray}
\rho_{n,m} &=& \frac{1}{\sqrt{n!m!}}\left(\frac{-2F}{U}\right)^n\left(\frac{-2F^{*}}{U}\right)^m 
\frac{\Gamma(c)\Gamma(c^{*})}{\Gamma(c+n)\Gamma(c^{*}+m)}\frac{\mathcal{F}(c+n,c^{*}+m,4|F/U|^2)}{\mathcal{F}(c,c^{*},8|F/U|^2)} \label{rhoan},
\end{eqnarray}
with
\begin{equation}\label{defc}
c = \frac{2(-\Delta \omega -i\gamma/2)}{U}. 
\end{equation}
In Eq.(\ref{rhoan}) $\Gamma(z)$ is the gamma special function which has poles at negative and zero integer values,
whereas $\mathcal{F}$ is an hypergeometric series given by:
\begin{equation}\label{hyp}
\mathcal{F}(c,d,z) = \sum_{k=0}^{\infty} \frac{\Gamma (c) \Gamma (d)}{\Gamma(c+k) \Gamma(d+k)}\frac{z^{k}}{k!}.
\end{equation}

To reconcile the complexity proof in the main text, we first review the proof of 
\cite{le2014bose} that under a suitable scaling limit the fixed point 
reduces to a binomial distribution.

\begin{theorem}[Binomial limit]
Consider the driven Kerr cavity with Hamiltonian
\[
H=-\Delta \omega \hat{n}+\frac{U}{2}\hat{n}(\hat{n}-1)+F\hat a^\dagger+F^*\hat a
\]
and single photon loss rate $\gamma$. 
Fix an integer $q\ge1$ and $U>0$, and set
\[
\Delta\omega=\frac{q-1}{2}U.
\]
Take the scaling limit
\[
\frac{|F|}{U}\longrightarrow0,
\qquad \frac{\gamma/U}{(|F|/U)^q}\longrightarrow0,\qquad \gamma>0.\]
while regarding $U,\Delta\omega = O(1)$ constants. 
Then the fixed point approaches the binomial state
\begin{align}
    \sigma_q
    :=
    \sum_{k=0}^{q}
    \frac{1}{2^q}\binom{q}{k}
    |k\rangle\langle k|.
\label{eq:binom-ap}
\end{align}
\end{theorem}

\begin{proof}

We start from the exact solution (\ref{rhoan}). 
Define
\[
z=\frac{4|F|^2}{U^2}.
\]

Under the resonance condition,
\[
c=\frac{2(-\Delta\omega-i\gamma/2)}{U}
=-(q-1)+\delta,
\]
where
\[
\delta=-i\gamma/U.
\]
In the limit $\gamma\to0$, we have $|\delta|\ll1$. Using the series representation (\ref{hyp}),
\[
\mathcal F(c,c^*,2z)=\sum_{r=0}^{\infty}T_r,
\qquad T_r:=
\frac{\Gamma(c)\Gamma(c^*)}
{\Gamma(c+r)\Gamma(c^*+r)}
\frac{(2z)^r}{r!}.
\]

Because $c\approx -(q-1)$, the Gamma function has poles near the first
$q$ terms of the series. 

For $r<q$, both $\Gamma(c)$ and $\Gamma(c+r)$ are near poles,
and the singularities cancel. 
Consequently
\[
\frac{\Gamma(c)\Gamma(c^*)}{\Gamma(c+r)\Gamma(c^*+r)}
= O_q(1),
\]
so that
\[
T_r = O_q(z^r).
\]

For $r=q$, the denominator $\Gamma(c+q)=\Gamma(1+\delta)$
is regular, while $\Gamma(c)$ still contains a pole.
Thus
\[
\frac{\Gamma(c)\Gamma(c^*)}{\Gamma(c+q)\Gamma(c^*+q)}
\sim \frac{1}{((q-1)!)^2 |\delta|^2}.
\] 

Hence the $r=q$ term scales as
\[
T_q
\sim
\frac{(2z)^q}{q!((q-1)!)^2 |\delta|^2}.
\]

For $r>q$, the Gamma functions are regular and the terms scale as
\[
T_r = T_q \, O_q(z^{\,r-q}).
\]

The terms with $r<q$ satisfy
\[
\sum_{r=0}^{q-1}T_r=O_q(1),
\qquad
T_q\asymp_q\frac{z^q}{|\delta|^2}.
\]
Hence
\[
\frac{\sum_{r=0}^{q-1}T_r}{T_q}
=O_q\!\left(\frac{|\delta|^2}{z^q}\right).
\]
For $s\ge1$, since $\delta$ is purely imaginary,
\[
\frac{T_{q+s}}{T_q}
=
\frac{q!}{(q+s)!}
\frac{(2z)^s}{\prod_{j=1}^{s}|j+\delta|^2}
\le \frac{(2z)^s}{(s!)^2}.
\]
Therefore,
\[
\frac{\sum_{r>q}T_r}{T_q}=O_q(z).
\]
Under the scaling $z\to0$ and $|\delta|^2/z^q\to0$,
the $r=q$ term dominates, and therefore
\[
\mathcal F(c,c^*,2z)
\simeq
\frac{\Gamma(c)\Gamma(c^*)}{\Gamma(c+q)\Gamma(c^*+q)}
\frac{(2z)^q}{q!}.
\]

\medskip

\noindent
\textbf{Step 1: Numerator asymptotics.}

For $0\le k\le q$, applying the same argument to
\[
\mathcal F(c+k,c^*+k,z),
\]
the dominant term occurs at $r=q-k$, giving
\[
\mathcal F(c+k,c^*+k,z)
\simeq
\frac{\Gamma(c+k)\Gamma(c^*+k)}
{\Gamma(c+q)\Gamma(c^*+q)}
\frac{z^{q-k}}{(q-k)!}.
\]

\medskip

\noindent
\textbf{Step 2: Diagonal elements.}

Substituting these asymptotic expressions into
Eq.~\eqref{rhoan} with $n=m=k$, the Gamma factors cancel.
For each $0\le k\le q$, we obtain
\[
\rho_{kk}
=
\frac{z^k}{k!}
\frac{z^{q-k}}{(q-k)!}
\frac{q!}{(2z)^q}
\bigl(1+o(1)\bigr)
=
\frac{1}{2^q}\binom{q}{k}\bigl(1+o(1)\bigr).
\]
Since $\rho$ is a density operator and the limiting
binomial probabilities sum to one,
\[
\sum_{k>q}\rho_{kk}
=
1-\sum_{k=0}^{q}\rho_{kk}
\longrightarrow0.
\]
Thus the photon-number distribution converges
to the binomial distribution.

\medskip

\noindent
\textbf{Step 3: Off-diagonal elements.}

The quantitative estimates established below in the proof
of Theorem~\ref{thm:explicit-binomial-bound}, derived
directly from the exact solution, imply
\[
\|\rho-\sigma_q\|_1
\le C_q\left(
\sqrt z+\frac{|\delta|}{z^{q/2}}
\right)
\]
for sufficiently small $z$ and $|\delta|/z^{q/2}$.
Here $z=4|F|^2/U^2$ and $|\delta|=\gamma/U$;
the general-$U$ statement follows by expressing the
parameters in units of $U$.
Under the stated scaling, both terms tend to zero.
Hence $\rho$ converges to $\sigma_q$ in trace norm.
\end{proof}

\begin{figure}[htbp]
\centering
\begin{tikzpicture}[>=stealth]
\draw[dashed,->] (0,0) --
node[above] {$\frac{2z}{q|\delta|^2}$} (3,0);

\draw[dashed,->] (3,0) --
node[above] {$\frac{2z}{(q+1)|1+\delta|^2}$} (6,0);
\end{tikzpicture}

\caption{Scaling relation between $T_{q-1}$, $T_q$, and $T_{q+1}$.}
\label{fig:scaling_relation}

\end{figure}
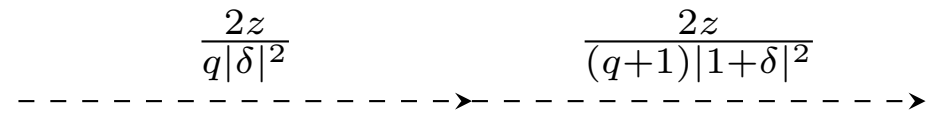

Figure~\ref{fig:scaling_relation} illustrates the neighboring
ratios
\[
\frac{T_q}{T_{q-1}}=\frac{2z}{q|\delta|^2},
\qquad
\frac{T_{q+1}}{T_q}
=\frac{2z}{(q+1)|1+\delta|^2}.
\]
However, comparison with neighboring terms alone is
insufficient: since $T_0=1$, we also need
\[
\frac{T_0}{T_q}
=\Theta_q\!\left(\frac{|\delta|^2}{z^q}\right)\to0.
\]
Together with $z\to0$, this ensures that $T_q$ dominates
the full series.

 Since our complexity-theoretic proof requires a
quantitative bound on the fidelity in terms of the
physical parameters, we must determine the subleading
correction to the leading-order approximation.
The resulting error scales as anticipated from
Fig.~\ref{fig:scaling_relation}. We prove the following result.
\begin{theorem}
\label{thm:explicit-binomial-bound}
Fix an integer $q\ge1$, and set
\[
U=1,\qquad \Delta\omega=\frac{q-1}{2}.
\]
There exist constants $f_q\in(0,1/4]$ and $C_q>0$,
depending only on $q$, such that whenever
\[
0<|F|\le f_q,\qquad
0<\frac{\gamma}{|F|^q}\le f_q,
\]
the exact fixed point $\rho$ satisfies
\[
\mathcal F(\rho,\sigma_q)
\ge 1-C_q\left(
|F|^2+\frac{\gamma^2}{|F|^{2q}}
\right),
\]
\[
\sum_{k\ge0}
\left|\rho_{kk}-\langle k|\sigma_q|k\rangle\right|
\le C_q\left(
|F|^2+\frac{\gamma^2}{|F|^{2q}}
\right),
\]
and
\[
D(\rho,\sigma_q):=\frac12\|\rho-\sigma_q\|_1
\le C_q\left(
|F|+\frac{\gamma}{|F|^q}
\right).
\]
Here $\sigma_q$ is defined in \eqref{eq:binom-ap}, and
$\mathcal F(\rho,\sigma)=\|\sqrt{\sigma}\sqrt{\rho}\|_1$
denotes the root fidelity.
\end{theorem}
\begin{proof}
Under the assumptions of the theorem, write
\[
z:=4|F|^2,\qquad
\delta:=-i\gamma,\qquad
c=-(q-1)+\delta.
\]
We first estimate the diagonal entries.
We set $n=m=k$ in (\ref{rhoan}), the diagonal entries of the fixed point are
\begin{align}
\rho_{kk}
=
\frac{z^k}{k!}
\frac{\Gamma(c)\Gamma(c^*)}{\Gamma(c+k)\Gamma(c^*+k)}
\frac{\mathcal F(c+k,c^*+k,z)}{\mathcal F(c,c^*,2z)},
\qquad 0\le k.
\label{eq:rhokk-start}
\end{align}

We now analyze the denominator and numerator separately. Using the series definition
\[
\mathcal F(c,c^*,2z)
=
\sum_{r=0}^{\infty}
T_r,
\qquad
T_r:=
\frac{\Gamma(c)\Gamma(c^*)}{\Gamma(c+r)\Gamma(c^*+r)}
\frac{(2z)^r}{r!},
\]
we compare the terms around $r=q$. For $r<q$, since
\[
c+r=-(q-1-r)+\delta
\]
is still close to a pole of the Gamma function, the poles in the numerator and denominator cancel. Therefore,
\[
T_r=O_q(z^r).
\]
In particular,
\[
\frac{T_{q-1}}{T_q}
=
\frac{q|\delta|^2}{2z}.
\]
For $r>q$, the Gamma functions are regular and
\[
\frac{T_{q+1}}{T_q}
=
\frac{2z}{q+1}\frac{1}{|1+\delta|^2}
=
O_q(z),
\]
and more generally
\[
T_r=T_q\,O_q(z^{r-q}),\qquad r>q.
\]

We must also control all terms below $r=q$.
Using $T_r=O_q(z^r)$ for $r<q$ and
$T_q\asymp_q z^q/|\delta|^2$, we obtain
\[
\frac{\sum_{r=0}^{q-1}T_r}{T_q}
=O_q\!\left(\frac{|\delta|^2}{z^q}\right).
\]
For the upper tail, the Gamma recursion gives
\[
\frac{T_{q+s}}{T_q}
=
\frac{q!}{(q+s)!}
\frac{(2z)^s}{\prod_{j=1}^{s}|j+\delta|^2}
\le \frac{(2z)^s}{(s!)^2},
\qquad s\ge1.
\]
Consequently,
\[
\frac{\sum_{r>q}T_r}{T_q}
\le e^{2z}-1=O(z).
\]
Combining these estimates yields
\begin{align}
\mathcal F(c,c^*,2z)
&=T_q(1+r_D),\\
0\le r_D
&\le C_q\left(z+\frac{|\delta|^2}{z^q}\right).
\label{eq:denom-asymp}
\end{align}

For fixed $k\in\{0,1,\dots,q\}$, write
\[
\mathcal F(c+k,c^*+k,z)
=
\sum_{r=0}^{\infty}
S_r^{(k)},
\]
where
\[
S_r^{(k)}
:=
\frac{\Gamma(c+k)\Gamma(c^*+k)}{\Gamma(c+k+r)\Gamma(c^*+k+r)}
\frac{z^r}{r!}.
\]
The dominant term now occurs at
\[
r=q-k.
\]
For $0\le k<q$, the terms below $r=q-k$ satisfy
\[
\sum_{r=0}^{q-k-1}S_r^{(k)}=O_q(1),
\qquad
S_{q-k}^{(k)}
\asymp_q\frac{z^{q-k}}{|\delta|^2}.
\]
Hence
\[
\frac{\sum_{r=0}^{q-k-1}S_r^{(k)}}{S_{q-k}^{(k)}}
=O_q\!\left(\frac{|\delta|^2}{z^{q-k}}\right).
\]
For every $0\le k\le q$ and $s\ge1$,
\[
\frac{S_{q-k+s}^{(k)}}{S_{q-k}^{(k)}}
=
\frac{(q-k)!}{(q-k+s)!}
\frac{z^s}{\prod_{j=1}^{s}|j+\delta|^2}
\le \frac{z^s}{(s!)^2}.
\]
Thus the upper tail has relative size at most
$e^z-1=O(z)$.
When $k=q$, there are no lower-order terms and
$S_0^{(q)}=1$.
Since $z\le1$, these estimates give, uniformly for
$0\le k\le q$,
\begin{align}
\mathcal F(c+k,c^*+k,z)
&=S_{q-k}^{(k)}(1+r_{N,k}),\\
0\le r_{N,k}
&\le C_q\left(z+\frac{|\delta|^2}{z^q}\right).
\label{eq:numer-asymp}
\end{align}
with
\begin{align}
S_{q-k}^{(k)}
=
\frac{\Gamma(c+k)\Gamma(c^*+k)}{\Gamma(c+q)\Gamma(c^*+q)}
\frac{z^{q-k}}{(q-k)!}.
\label{eq:Sk-leading}
\end{align}

\medskip
\noindent

Substituting (\ref{eq:denom-asymp}), (\ref{eq:numer-asymp}), and (\ref{eq:Sk-leading}) into (\ref{eq:rhokk-start}), we obtain
\begin{align}
\rho_{kk}
&=
\frac{z^k}{k!}
\frac{\Gamma(c)\Gamma(c^*)}{\Gamma(c+k)\Gamma(c^*+k)}
\frac{S_{q-k}^{(k)}(1+r_{N,k})}{T_q(1+r_D)}.
\end{align}
Using the explicit formulas for $S_{q-k}^{(k)}$ and $T_q$, all Gamma factors cancel, and we get
\begin{align}
\rho_{kk}
&=
\frac{z^k}{k!}
\frac{z^{q-k}}{(q-k)!}
\frac{q!}{(2z)^q}
\frac{1+r_{N,k}}{1+r_D}
\nonumber\\
&=
\frac{1}{2^q}\binom{q}{k}\frac{1+r_{N,k}}{1+r_D}.
\end{align}
Hence
\begin{align}
\rho_{kk}
=
\frac{1}{2^q}\binom{q}{k}\left(1+\varepsilon_k\right),
\label{eq:rhokk-eps}
\end{align}
where
\[
\varepsilon_k:=\frac{1+r_{N,k}}{1+r_D}-1.
\]
Since
$|r_D|,|r_{N,k}|\le C_q(z+|\delta|^2/z^q)$,
and $z+|\delta|^2/z^q\to0$ under the stated scaling,
we may assume $|r_D|\le1/2$ for sufficiently small
parameters. Thus
\[
|\varepsilon_k|
\le C_q\left(z+\frac{|\delta|^2}{z^q}\right).
\]
Therefore,
\begin{align}
\rho_{kk}
=
\frac{1}{2^q}\binom{q}{k}
\left(
1+O_q\!\left(z+\frac{|\delta|^2}{z^q}\right)
\right).
\label{eq:rhokk-final-zdelta}
\end{align}

Finally, recalling that under our normalization
\[
z=4|F|^2,\qquad |\delta|=\gamma,
\]
we conclude that
\begin{align}
\rho_{kk}
=
\frac{1}{2^q}\binom{q}{k}
\left(
1+O_q\!\left(|F|^2+\frac{\gamma^2}{|F|^{2q}}\right)
\right).
\label{eq:rhokk-final-F-gamma}
\end{align}

Equivalently, if we define the ideal binomial diagonal state
\[
\sigma_q:=\sum_{k=0}^q \frac{1}{2^q}\binom{q}{k}\ket{k}\bra{k},
\]
then the diagonal part of $\rho$ satisfies
\[
\rho_{\mathrm{diag}}
=
\sigma_q
+
O_q\!\left(|F|^2+\frac{\gamma^2}{|F|^{2q}}\right)
\]
entrywise on the support $\{0,1,\dots,q\}$.
This completes the analysis of the diagonal contribution.

We now control the off-diagonal entries.
Write $\alpha=-2F$, $Z=\mathcal F(c,c^*,2z)$, and define
\[
(c)_j:=\prod_{h=0}^{j-1}(c+h),\qquad (c)_0:=1.
\]
Expanding the exact solution gives
\[
\rho_{nm}
=
\frac{\alpha^n(\alpha^*)^m}{\sqrt{n!m!}\,Z}
\sum_{r\ge0}
\frac{z^r}{r!\,(c)_{n+r}(c^*)_{m+r}}.
\]
Since $c=-(q-1)+\delta$ and $\delta$ is purely imaginary,
\[
\frac{1}{|(c)_j|}
\le
\begin{cases}
C_q, & 0\le j<q,\\[2pt]
\dfrac{C_q}{|\delta|(j-q)!}, & j\ge q.
\end{cases}
\]
For $0\le m<n\le q$, put $d=n-m$ and split the sum into
\[
r<q-n,\qquad q-n\le r<q-m,\qquad r\ge q-m.
\]
The first range is empty when $n=q$.
Using the bounds above and
$Z\ge T_q\asymp_q z^q/|\delta|^2$, the three ranges give
\begin{align}
|\rho_{nm}|
\le C_q\left(
|\delta|^2z^{(n+m)/2-q}
+|\delta|z^{-d/2}
+z^{d/2}
\right).
\label{eq:offdiag-window}
\end{align}
The last range is summable uniformly for $z\le1$,
using the factorial denominators.

For sufficiently small parameters,
$z\le1$ and $|\delta|^2/z^q\le1$.
Since $n+m\ge1$ and $1\le d\le q$, it follows that
\begin{align}
|\rho_{nm}|
\le C_q\left(
\sqrt z+\frac{|\delta|}{z^{q/2}}
\right),
\qquad 0\le m\ne n\le q,
\label{eq:offdiag-window-bound}
\end{align}
where the case $n<m$ follows by Hermiticity.
Summing over this finite set, and using
$z=4|F|^2$ and $|\delta|=\gamma$, yields
\begin{align}
\sum_{\substack{0\le m,n\le q\\m\ne n}}|\rho_{nm}|
\le C_q\left(
|F|+\frac{\gamma}{|F|^q}
\right).
\label{eq:offdiag-window-sum}
\end{align}

It remains to control the terms with $\max\{m,n\}>q$. 
We first estimate the diagonal tail. For $s\ge 1$, setting $k=q+s$ in (\ref{rhoan}) gives
\begin{align}
\rho_{q+s,q+s}
=
\frac{z^{q+s}}{(q+s)!}
\frac{\Gamma(c)\Gamma(c^*)}{\Gamma(c+q+s)\Gamma(c^*+q+s)}
\frac{\mathcal F(c+q+s,c^*+q+s,z)}{\mathcal F(c,c^*,2z)},
\qquad z=4|F|^2.
\end{align}
Since $c+q=1+\delta$ and $\delta$ is purely imaginary,
\[
|(c+q)_s|
=\prod_{j=1}^{s}|j+\delta|
\ge s!.
\]
Moreover, the series definition gives, uniformly for $s\ge1$,
\[
\mathcal F(c+q+s,c^*+q+s,z)
=
\sum_{r\ge0}\frac{z^r}{r!\,|(1+s+\delta)_r|^2}
\le e^z.
\]
Using $Z=\mathcal F(c,c^*,2z)\ge T_q$, we obtain
\begin{align}
\rho_{q+s,q+s}
&\le
\frac{q!}{2^q}\,
\frac{z^s}{(q+s)!\,|(c+q)_s|^2}
\mathcal F(c+q+s,c^*+q+s,z)
\nonumber\\
&\le
\frac{q!e^z}{2^q}\,
\frac{z^s}{(q+s)!(s!)^2}
\le C_q z^s,
\qquad s\ge1,\quad 0<z\le\tfrac12.
\label{eq:tail-diag}
\end{align}
In particular,
\begin{align}
\sum_{k>q}\rho_{kk}=O_q(|F|^2).
\label{eq:tail-mass}
\end{align}

Next we use positivity of $\rho$. Since $\rho\geq 0$, for all $m,n\ge 0$ one has
\begin{align}
|\rho_{mn}|^2\le \rho_{mm}\rho_{nn}.
\label{eq:psd-entry-bound}
\end{align}

We first consider the cross terms with $0\le m\le q$
and $n=q+s>q$, where $s\ge1$.
Using (\ref{eq:psd-entry-bound}), the bound $\rho_{mm}=O_q(1)$ for $0\le m\le q$, and (\ref{eq:tail-diag}), we obtain
\[
|\rho_{m,q+s}|
\le
\sqrt{\rho_{mm}\rho_{q+s,q+s}}
\le
C_q z^{s/2}.
\]
Summing over the finite set $m\in\{0,\dots,q\}$ and over $s\ge 1$, we get
\begin{align}
\sum_{m=0}^q\sum_{s\ge 1} |\rho_{m,q+s}|
+
\sum_{m=0}^q\sum_{s\ge 1} |\rho_{q+s,m}|
\le
C_q \sum_{s\ge 1} z^{s/2}
=
O_q(|F|).
\label{eq:cross-tail}
\end{align}

Finally, for the pure tail off-diagonal sector, (\ref{eq:psd-entry-bound}) and (\ref{eq:tail-diag}) imply
\[
|\rho_{q+s,q+t}|
\le
\sqrt{\rho_{q+s,q+s}\rho_{q+t,q+t}}
\le
C_q z^{(s+t)/2},
\qquad s,t\ge 1.
\]
Therefore
\begin{align}
\sum_{\substack{s,t\ge 1\\ s\neq t}} |\rho_{q+s,q+t}|
\le
C_q \sum_{s,t\ge 1} z^{(s+t)/2}
=
O_q(|F|^2).
\label{eq:tail-tail}
\end{align}

Combining (\ref{eq:offdiag-window-sum}), (\ref{eq:cross-tail}), and (\ref{eq:tail-tail}), we conclude that the total off-diagonal mass satisfies
\begin{align}
\sum_{m\ne n}|\rho_{mn}|
\le C_q\left(
|F|+\frac{\gamma}{|F|^q}
\right).
\label{eq:offdiag-total}
\end{align}
Let
\begin{align}
\rho_{\mathrm{diag}}:=\sum_{k\ge0}\rho_{kk}\ket{k}\bra{k}
\end{align}
be the diagonal part of $\rho$.
From (\ref{eq:rhokk-final-F-gamma}), for each $0\le k\le q$ we have
\[
\rho_{kk}
=
\frac{1}{2^q}\binom{q}{k}
\left(
1+O_q\!\left(|F|^2+\frac{\gamma^2}{|F|^{2q}}\right)
\right).
\]
Since \(\sum_{k=0}^q 2^{-q}\binom{q}{k}=1\), summing over
\(k=0,\dots,q\) yields
\begin{align}
\sum_{k=0}^q
\left|
    \rho_{kk}
    -
    \frac{1}{2^q}\binom{q}{k}
\right|
\le
C_q\left(|F|^2+\frac{\gamma^2}{|F|^{2q}}\right).
\label{eq:diag-l1}
\end{align}
Combining this with the diagonal tail estimate
\[
    \sum_{k>q}\rho_{kk}=O_q(|F|^2),
\]
we obtain
\begin{align}
\|\rho_{\mathrm{diag}}-\sigma_q\|_1
\le
C_q\left(|F|^2+\frac{\gamma^2}{|F|^{2q}}\right).
\label{eq:diag-trace}
\end{align}

On the other hand,
\[
\rho-\rho_{\mathrm{diag}}
=
\sum_{m\neq n}\rho_{nm}\ket{n}\bra{m},
\]
and therefore by (\ref{eq:offdiag-total}),
\begin{align}
\|\rho-\rho_{\mathrm{diag}}\|_1
\le \sum_{m\ne n}|\rho_{nm}|
\le C_q\left(
|F|+\frac{\gamma}{|F|^q}
\right).
\label{eq:offdiag-trace}
\end{align}

By the triangle inequality,
\begin{align}
\|\rho-\sigma_q\|_1
&\le
\|\rho-\rho_{\mathrm{diag}}\|_1
+\|\rho_{\mathrm{diag}}-\sigma_q\|_1
\nonumber\\
&\le C_q\left(
|F|+\frac{\gamma}{|F|^q}
+|F|^2+\frac{\gamma^2}{|F|^{2q}}
\right).
\label{eq:rho-sigma-trace}
\end{align}
For sufficiently small parameters,
$|F|\le1$ and $\gamma/|F|^q\le1$.
Hence
\begin{align}
\|\rho-\sigma_q\|_1
\le C_q\left(
|F|+\frac{\gamma}{|F|^q}
\right).
\label{eq:rho-sigma-trace-final}
\end{align}
To obtain the sharper root-fidelity bound, define
\[
P:=\sum_{k=0}^{q}|k\rangle\langle k|,
\qquad
t:=\operatorname{Tr}[(I-P)\rho]=O_q(|F|^2),
\qquad
\tau:=\frac{P\rho P}{1-t}.
\]
For sufficiently small parameters, $t<1$.
Write
\[
E:=|F|^2+\frac{\gamma^2}{|F|^{2q}}.
\]
The diagonal estimates and
Eq.~\eqref{eq:offdiag-window-bound} imply
\[
\|\tau-\sigma_q\|_2^2=O_q(E),
\]
where $\|\cdot\|_2$ denotes the Hilbert--Schmidt norm.
Indeed, the diagonal errors are $O_q(E)$, the
off-diagonal entries are $O_q(\sqrt E)$, and
normalization changes the entries by $O_q(t)$.
There are only $(q+1)^2$ entries in this block.

On $P\mathcal H$, the smallest eigenvalue of $\sigma_q$
is $\lambda=2^{-q}$. Set
\[
X:=\sqrt{\tau}-\sqrt{\sigma_q}.
\]
The identity
\[
\sqrt{\tau}\,X+X\sqrt{\sigma_q}=\tau-\sigma_q
\]
gives
\[
\sqrt{\lambda}\,\|X\|_2^2
\le
\operatorname{Tr}\!\left[X(\tau-\sigma_q)\right]
\le
\|X\|_2\,\|\tau-\sigma_q\|_2.
\]
Consequently,
\[
\|X\|_2^2
\le \lambda^{-1}\|\tau-\sigma_q\|_2^2.
\]
Since $\tau$ and $\sigma_q$ are normalized,
\[
\mathcal F(\tau,\sigma_q)
\ge \operatorname{Tr}(\sqrt{\tau}\sqrt{\sigma_q})
=1-\frac12\|X\|_2^2
\ge 1-O_q(E).
\]
Finally, because $\sigma_q=P\sigma_q P$,
\[
\mathcal F(\rho,\sigma_q)
=\sqrt{1-t}\,\mathcal F(\tau,\sigma_q)
\ge
1-O_q\!\left(
|F|^2+\frac{\gamma^2}{|F|^{2q}}
\right).
\]
Choosing $f_q$ sufficiently small and $C_q$
sufficiently large makes all three stated estimates
hold throughout the parameter range of the theorem.
This proves the theorem.
\end{proof}

Finally, can the single-photon population
$\rho_{11}=\braket{1|\rho|1}$ approach one?
The following bound also follows from the representation
of the Kerr stationary state as a pure state passed through
a $50\%$ loss channel~\cite{PhysRevX.10.021022}.
We give a direct proof below.
\begin{theorem}
For the driven Kerr fixed point with $U>0$, $\gamma>0$, and arbitrary $F\in\mathbb C$, one has
\[
\rho_{11}\le\tfrac12
\]
\label{thm:rho-11-Kerr}
\end{theorem}

\begin{proof}
Set $x=4|F|^2/U^2$ and
$c=2(-\Delta\omega-i\gamma/2)/U$.
\begin{align}
    \rho_{11}=\frac{x}{|c|^2}\frac{\mathcal{F}(c+1, c^*+1, x)}{\mathcal{F}(c, c^*, 2x)},
    \end{align}
and
\begin{align}
\mathcal{F}(c, c^*, 2x) &= |\Gamma(c)|^2 \sum_{k=0}^\infty \frac{(2x)^k}{k! |\Gamma(c + k)|^2}, \\
\mathcal{F}(c+1, c^*+1, x) &= |c|^2 |\Gamma(c)|^2 \sum_{k=0}^\infty \frac{x^k}{k! |c + k|^2 |\Gamma(c + k)|^2}.
\end{align}
So we have:
\begin{align}
\braket{1|\rho|1}=\rho_{11}=\frac{x\sum\limits_{k=0}^\infty \frac{x^k}{k! |c + k|^2 |\Gamma(c + k)|^2}}{\sum\limits_{k=0}^\infty \frac{(2x)^k}{k! |\Gamma(c + k)|^2}}
\end{align}
Using the Gamma function recursion \(\Gamma(z+1) = z\Gamma(z)\), for any integer \(k \ge 0\),
\begin{align}
|c+k|^2\,|\Gamma(c+k)|^2 = |\Gamma(c+k+1)|^2.
\end{align}
Hence the numerator can be rewritten as (letting \(j = k + 1\))
\begin{align}
x\sum_{k=0}^\infty \frac{x^k}{k!\,|c+k|^2\,|\Gamma(c+k)|^2}
= \sum_{j=1}^\infty \frac{x^j}{(j-1)!\,|\Gamma(c+j)|^2}.
\end{align}
Define
\begin{align}
d_j := \frac{(2x)^j}{j!\,|\Gamma(c+j)|^2}, \quad (j \ge 0),
\end{align}
so that \(d_0 = \frac{1}{|\Gamma(c)|^2}\), and for \(j \ge 1\),
\begin{align}
\frac{x^j}{(j-1)!\,|\Gamma(c+j)|^2} = \frac{j}{2^j}\,d_j.
\end{align}
Therefore,
\begin{align}
\rho_{11} = \frac{\displaystyle\sum_{j=1}^\infty \frac{j}{2^j}\,d_j}
{\displaystyle d_0 + \sum_{j=1}^\infty d_j }.
\end{align}
Since for all integers \(j \ge 1\), we have \(\frac{j}{2^j} \le \frac{1}{2}\), it follows that
$\sum_{j=1}^\infty \frac{j}{2^j}\,d_j
\le \frac{1}{2} \sum_{j=1}^\infty d_j.$
Thus,
\begin{align}
\rho_{11} \le
\frac{\frac{1}{2}\sum_{j=1}^\infty d_j}{d_0 + \sum_{j=1}^\infty d_j}
\le \frac{1}{2},
\end{align}
because \(d_0 \ge 0\).
\end{proof}

\end{document}